\documentclass[a4paper,12pt]{report}

\usepackage{cite}
\usepackage{cutwin}
\usepackage[utf8]{inputenc}
\usepackage{amsmath,amsfonts,amscd,amssymb,mathrsfs}
\DeclareMathOperator{\tr}{Tr}

\usepackage[T1]{fontenc}
\usepackage[english]{babel}
\usepackage{epsfig}
\usepackage{changebar}
\usepackage{lmodern} 
\usepackage{slashed}
\usepackage[all]{xy}
\usepackage{mathtools}
\usepackage{MnSymbol}
\usepackage{fancybox}
\usepackage{wrapfig}
\usepackage{cite}
\usepackage{hyperref}
\usepackage{fancyhdr}
\usepackage{vmargin} 
\usepackage[nottoc, notlof, notlot]{tocbibind}
\usepackage{graphicx}
\usepackage{caption}
\usepackage{subcaption}
\usepackage{titlesec}
\usepackage{color}
\usepackage{stmaryrd}
\usepackage[most]{tcolorbox}
\usepackage{wrapfig,lipsum,booktabs}
\usepackage{listings}
\usepackage{MnSymbol}
\usepackage[toc,page]{appendix}
\usepackage{tensor}
\definecolor{dkgreen}{rgb}{0,0.6,0}
\definecolor{gray}{rgb}{0.5,0.5,0.5}
\definecolor{mauve}{rgb}{0.58,0,0.82}

\titleformat{\chapter}[display]   
{\normalfont\huge\bfseries}{\chaptertitlename\ \thechapter}{20pt}{\Huge}   
\titlespacing*{\chapter}{0pt}{-50pt}{40pt}

\titlespacing*{\section}{0pt}{0.2\baselineskip}{\baselineskip}
 
\DeclareMathOperator*{\Res}{Res}
\setmarginsrb{2cm}{2cm}{2cm}{2cm}{14mm}{6mm}{0mm}{10mm}
\newcommand{\ii}{{\rm i}}
\newcommand{\n}{\nonumber\\}
\def \k{\kappa}
\DeclareUnicodeCharacter{2212}{-}
\DeclareMathOperator*{\Beta}{\boldsymbol{\beta}}
\DeclareMathOperator*{\sgn}{sign}

\def\be{\begin{eqnarray}  }
   
\def\ee{\end{eqnarray}}
\DeclareMathOperator*{\K}{\kappa}
\newcommand{\p}{\partial}

\newcommand{\Q}{\mathcal{Q}}
\DeclareMathOperator*{\A}{\mathcal{A}}
\DeclareMathOperator*{\B}{\mathcal{B}}
\DeclareMathOperator*{\C}{\mathcal{C}}
\DeclareMathOperator*{\mt}{\mathcal{T}}

\begin{document}
\thispagestyle{empty}

\begin{flushright}
\end{flushright}

\vspace{1cm}
\setcounter{footnote}{0}

\begin{center}

 {\Large\bf  A diagrammatic approach towards the thermodynamics of integrable systems}

\vspace{20mm} 

Dinh-Long Vu \\[7mm]
 
{\it  \small{Universit\'e Paris-Saclay, CNRS, CEA, Institut de physique th\'eorique, 91191, Gif-sur-Yvette, France.}}
 \\[5mm]

\end{center}

\vskip9mm

\vskip18mm

{ \noindent{ We  propose an exact summation method to compute thermodynamic observables in integrable quantum field theories. The key idea is to use the matrix-tree theorem to write the Gaudin determinants that appear in the cluster expansion as a sum over graphs. For theories with a diagonal S-matrix, this method is more powerful than the standard Thermodynamic Bethe Ansatz (TBA) technique as it is exact to all orders of powers in inverse volume. We have obtained using this method the TBA equation, the excited state energies in finite volume, the Leclair-Mussardo formula for one point functions, the finite-temperature boundary entropy and  cumulants of conserved charges in Generalized Gibbs Ensembles. Moreover, the graph expansion  can also be regarded as an alternative to algebraic manipulations involving Gaudin determinants. We have applied this idea to derive the equations of state and other transport properties in Generalized Hydrodynamics. For theories with a non-diagonal S-matrix, the description of a complete set of states is more involved and it is not known how a cluster expansion can be implemented. It is nevertheless possible to apply the direct summation method in reverse and interpret known TBA equations with complex strings in terms of diagrams.} }

\tableofcontents
\newpage
\chapter*{Introduction}
Quantum integrable systems are important for at least two reasons
\begin{itemize}
\item They can be realized in experiments. Here are some examples. The Heisenberg spin 1/2 chain \cite{TAKAHASHI1971325,PhysRevLett.26.1301} is expected to describe magnetic properties of crystals in which  magnetic ions are arranged in one dimensional rays and interaction between rays is screened by large non-magnetic ions \cite{doi:10.1080/00018737600101372}. The dynamical structure factors  (Fourier transform of the spin-spin correlation function) of $\text{KCuF}_3$ measured by inelastic neutron scattering experiments \cite{PhysRevB.52.13368} have been shown to be in astounding agreement with theoretical computation \cite{Maillet:2007pda} using Bethe ansatz (see below). The  Lieb-Liniger gas in repulsive regime \cite{Lieb:1963rt,Yang:1968rm} was replicated in laboratory \cite{PhysRevLett.95.190406} by  ${}^{87}\text{Rb}$ atoms trapped in an optical lattice. The probability of observing two particles at the same position measured by photoassociation conforms with the prediction  \cite{PhysRevLett.90.010401} that the overlap of bosonic wave functions decreases as interaction strength increases, a phenomenon known as the fermionization of bosons. Moreover, the same authors prepared in their famous  \textit{quantum Newton cradle} experiment \cite{article} an out-of-equilibrium initial state using the same atoms. They observed that even after thousands of collisions, the gas does not equilibrate, thus experimentally verified of the lack of thermalization in integrable systems.
\item Exact theoretical computations are available. In 1931, Hans Bethe \cite{Bethe:406121} solved the  Heisenberg spin 1/2 chain by writing its eigenvectors as superpositions of plane waves that properly take into account the exchange of particles. The proposed form of the eigenvectors and the equations that govern the momenta of these plane waves are known respectively as \textit{Bethe's ansatz} and \textit{Bethe's equations}. Although Bethe's idea was later extended to other models \cite{PhysRev.112.309,PhysRev.116.1089,PhysRev.150.321}, it was not known at that time why such a simple ansatz works for these precise models? The answer to this question was found in a coherent framework that can explain the origin of quantum integrability: an algebraic formulation of Bethe's ansatz discovered in the late 70's by Faddeev, Sklyanin and Taktadjan \cite{Faddeev:1979gh,Takhtajan:1979iv}. The central object in their construction is a transfer matrix that  encodes within itself an infinite number of commuting observables the Hamiltonian is one of which. Bethe's ansatz and Bethe's equations can both be written in terms of the elements of this matrix. It should be noted however that the notion of transfer matrix appeared in fact earlier in the work of Baxter \cite{Baxter:1982zz} on two dimensional statistical models. This algebraic formalism works for a wide class of models with higher-rank symmetry groups\cite{KULISH1981246,Kulish_1983}, in that case Bethe's ansatz  is particularly known as \textit{nested Bethe ansatz}. 
\end{itemize}
When applied in the thermodynamic setting of an integrable gas, Bethe's ansatz provides information about the distribution of particle momenta at thermal equilibrium.  This was first illustrated for the Lieb-Liniger model in the seminal work of Yang and Yang \cite{Yang:1968rm}. Forgetting technical details, their idea consists of solving  Bethe equations for the thermodynamic  state that minimizes the free energy of the system, hence the name \textit{thermodynamic Bethe ansatz} (TBA). Accordingly, the functional equation that governs the momenta distribution is called the TBA equation. The validity of this equation has recently been confirmed in an experiment of ultracold atoms\cite{PhysRevLett.100.090402}. Soon after the work of Yang and Yang, the TBA has been extended to the XXZ spin chain \cite{10.1143/PTP.46.401} and critical exponents of its susceptibility and specific heat were subsequently obtained \cite{doi:10.1143/JPSJ.55.2024,PhysRevLett.54.2131}. Compared to the Lieb-Linger model, the TBA of XXZ spin chain is considerably more complicated as it involves an infinite number of bound states with complex momenta that arrange themselves in string patterns, a feature that became known as \textit{ the string hypothesis}. The TBA of other lattice models such as the Hubbard model was also obtained quickly  \cite{10.1143/PTP.47.69}. On the other hand, it took almost twenty years for TBA to find its applications in integrable quantum field theories. With his revolutionary idea of a double Wick rotation \cite{Zamolodchikov:1987zf}, Zamolodchikov pointed out that the ground state energy of the \textit{physical theory} at  volume $R$ is nothing but the TBA free energy density of the \textit{mirror theory} at temperature $1/R$. The twenty-year gap between the work of Yang and Yang and that of Zamolodchikov is partly due to the fact that the first result on factorized quantum S-matrix appeared as late as  in 1978 \cite{Zamolodchikov:1978xm}.  In the context of the AdS/CFT correspondence, Zamolodchikov's idea allows the spectrum of classical strings on $\text{AdS}_5\times \text{S}^5$ and the scaling dimensions in planar $\mathcal{N}=4 \text{ SYM}$ to both be computed by TBA \cite{Ambjorn:2005wa,Arutyunov:2007tc,Arutyunov:2009zu,Arutyunov:2009ur,Bombardelli:2009ns,Gromov:2009bc,Gromov:2009tv,Cavaglia:2010nm,Cavaglia:2011kd,Balog:2011nm,Gromov:2013pga}. On the other hand, the relevance of TBA in higher-point correlation functions in planar $\mathcal{N}=4 \text{ SYM}$ is much less known. Embracing Zamolodchikov's idea, Basso, Komatsu and Viera proposed a two-step program \cite{Basso:2015zoa} to compute three point functions of this theory. The first step consists of cutting them into two hexagon form factors and the second step is to glue them back by inserting a complete basis of mirror states on each one of the three mirror sides. Although there is not much trouble in bootstrapping these form factors, their expression is sufficiently involved. Combined with the not so simple bound state structure of the theory, this makes the summation over mirror particles extremely challenging. The strategy often adopted in the literature is to either compute the leading-order terms or to consider particular limits in which the contribution from mirror particles can be neglected. Given the tremendous success of TBA in the spectral problem of planar $\mathcal{N}=4 \text{ SYM}$, one can nevertheless expect some TBA-like structures in the wrapping correction to the structure constants or to other observables of the theory.
\vspace{0.5cm}

\noindent
Originally motivated by this problem, the purpose of this thesis is to investigate, to what extend a direct thermal summation could reproduce TBA-like quantities. Before working with planar $\mathcal{N}=4 \text{ SYM}$, we must implement this approach in the simpler case of 1+1 dimensional integrable quantum field theories. We first have to make sure that it can recover the well-known TBA equation. Then, we probe its limit by testing it on observables that the original TBA method has shown considerable difficulty to derive. Here are some candidates
\begin{itemize}
\item For the one point function of a local observable, a formula involving the TBA density function and the connected evaluation of diagonal form factors (connected form factors for short) was conjectured by Leclair and Mussardo in \cite{Leclair:1999ys}. Saleur provided some supporting arguments\cite{ Saleur:1999hq} for this conjecture and although he could not prove it, he proposed a relation between connected form factors and diagonal matrix elements in finite volume that could potentially justify the formula. Even with this hint, a proof was not ready until Pozsgay \cite{Pozsgay:2010xd} understood how to apply Saleur proposal in the thermodynamic setting. On a side note, the relation proposed by Saleur has only been recently proved by Bajnok and Wu \cite{Bajnok:2017bfg}. The similarity between the problem of one point functions and the wrapping correction to three point functions in the hexagon approach is that both cases involve a summation over a complete set of states along with some form factors. There is an important difference however, here there is only one theory, the \textit{mirror theory} in the sense that it is of large volume. For one point functions in finite volume, see \cite{Bajnok:2019mpp}.

\item Excited state energies. The double Wick rotation singles out the ground state and the ground state only. To get access to higher energy levels, Dorey and Tateo \cite{Dorey:1996re} proposed an analytic continuation in some parameter of the theory (temperature, mass scale). If one starts with a real value of this parameter and one moves in the complex plane in such a way that the singularity in the TBA equation crosses the real axis then upon coming back one ends up in a different sheet of energy. Here, both the physical and the mirror theory are at play.

\item In situations where $O(1)$ corrections to extensive quantities are needed, the original derivation of TBA shows weaknesses. This is due to its inherent approximations: the length of the interval over which the momentum density is defined can vary between inverse volume and unity,  the Stirling formula used to compute the entropy is also insensitive to $O(1)$ correction. An example is the boundary correction to the free energy, also known as the finite-temperature g-function. It was partially obtained  by Leclair \textit{et al.} in \cite{LeClair:1995uf} using TBA saddle point approximation. Fluctuations around the saddle point coming from correction to the Stirling formula was then added by Woynarovich \cite{Woynarovich:2004gc}. A paradox arose from the work of Woynarovich however as it predicts a similar contribution for periodic systems. Meanwhile, the authors of \cite{Dorey:2004xk} took a different route towards this problem. By means of a cluster expansion, a considerable amount of guess work, and possibly some hints from the result of Woynarovich they obtained a formula for the g-function with similar appearance to a Leclair-Mussardo series. Various UV limit tests suggest that their result is correct, and yet the mismatch with the result of Woynarovich stayed unexplained. This discrepancy was finally understood by Pozsgay as a nontrivial measure that must be taken into account when the discrete quantum number description is replaced by  a continuous distribution of momenta. For a periodic system, this measure precisely cancels the fluctuations around the saddle point, the paradox is resolved. The lesson we can draw from this rather long history is that there could be serious flaw in our understanding of TBA.
\end{itemize}
As we will show throughout this thesis, these quantities can all be obtained by the direct summation method in a uniform fashion. We also obtain with this method an observable that is inaccessible by the original approach: the cumulants of the conserved charges in a generalized Gibbs ensemble (see below).

\vspace{0.5cm}

\noindent
Let us explain how the method works. The first step is to choose a complete basis of  states each one of which is labeled by a set of Bethe quantum numbers. The thermal observable that we seek to compute is then written as a double sum: over the number of particles and over the quantum numbers. The second step is to remove the constraint (if any) between mode numbers. This constraint is model dependent, for instance if Bethe's wave functions are of \textit{fermionic} type, then mode numbers are pairwise different and this constraint can be eliminated by insertion of $1-\delta$ symbol. The next step is to approximate the sums over mode numbers by integrals over the corresponding rapidities. This approximation is exact up to exponential correction in volume.  The Jacobians of this change of variables are Gaudin determinants, which are known in the literature for describing the norm of Bethe's wave functions. Such determinant also appeared in Pozsgay derivation of g-function. We note however that Pozsgay performed a change of variables for the thermodynamic state, thus leading to a thermodynamic version of the Gaudin determinat that we consider here. Before describing the next steps, let us take a moment to make some comments regarding the nature of this approach
\begin{itemize}
\item Compared to the traditional derivation, the direct summation method is certainly more involved. There is only one state in the original approach: the thermodynamic state. Moreover, the determination of this state can be relied on a powerful physics principle in addition to Bethe's ansatz, namely the minimization of the free energy. There is nothing similar to that in our approach as we just write down the thermal trace and perform the infinite sum. The partial absence of physical laws means more freedom however and can sometimes be in favor of our method, attributing to its versatility. For instance, it can derive the three above mentioned observables without the need for fundamental modifications. 
\item The steps described above are "textbook" manipulations of low-temperature expansion and are very common in the literature. Usually the series is truncated and explicit analytic expressions of the first few terms are obtained. With enough supporting evidences one can even guess the entire series based on these  terms. In fact this is exactly how the Leclair-Mussardo series was conjectured and also how Dorey \textit{et al.} obtained their formula of the finite-temperature g-function. The peculiarity in our approach is that we treat the entire series, no truncation is needed.
\end{itemize}
So, how can we perform an exact summation of such infinite series? The trick is not to compute the Gaudin determinants analytically but to consider their diagrammatic expansion. As a direct consequence of Bethe's equations, the Gaudin matrix is the sum of a diagonal matrix and a Laplacian matrix i.e. a matrix in which the elements in each row  sum up to zero. The matrix-tree theorem and its variants from graph theory allow its determinant to be written as a sum over combinatorial objects. The  equation satisfied by the generating function of these objects can be read off from their combinatorial structure prescribed by the  theorem. In the simplest case of the free energy of a periodic system the graphs are trees and this equation is nothing but the well-known TBA equation. For more involved observables, there are additional types of graphs and/or extra structures imposed on them. 

\vspace{0.5cm}

\noindent
Now that we have described the method of direct summation and some of its applications, let us discuss the limitations.
\begin{itemize}
\item In the first step we made two assumptions: first, there is no constraint between the numbers of different types of Bethe roots and second, all rapidities are real. These assumptions hold only for theories with a non-degenerate mass spectrum and the S-matrix is thus purely elastic. In other models at least one of them is violated. 
\item In the second step, unphysical states appear when we remove the constraint between mode numbers. For instance, if we expand the product of  Kronecker symbols then some rapidities are forced to take coinciding values. To perform the sum over these states, we must know how the observable in question acts on them. For the observables that we have considered, such action is a straightforward generalization of the action on physical states. However, there could be cases where this generalization is no longer trivial.
\item Finally, what is the extend of the matrix tree theorem? Can it describe the Gaudin determinant corresponding to more exotic Bethe equations? Up to this point the various forms and corollaries of this theorem provided in the work of Chaiken \cite{Chaiken82acombinatorial} suffice our needs. On the other hand, it would be interesting to find a physically meaningful Bethe equation that leads to a variant of the matrix-tree theorem that has never appeared in the mathematical literature before.
\end{itemize}
The issue with the first step is serious and at this point we do not have a satisfying solution. However, we can have a diagrammatic interpretation of the known TBA equations of models with non-diagonal S-matrix. Let us illustrate this idea for the $SU(2)$ chiral Gross-Neveu model. The Bethe equations of this theory involve physical roots and auxiliary roots, the number of latter should not exceed half the number of the former, as dictated by their algebraic construction. To derive the TBA of this model by the direct summation method, we must perform the sum over the number of particles while respecting this constraint, which is an impossible task. The traditional approach however, succeeded in obtaining the correct TBA using the string hypothesis. The resulting TBA system involves the physical Bethe root and an infinite number of strings of auxiliary roots. Now if we apply the steps described above in the reverse order, we could say that had we started with not two, but an infinite number of Bethe roots without any algebraic constraint between their numbers, then we would end up with the correct TBA equations. Needless to say, both assumptions are incorrect: strings of auxiliary roots only exist in the thermodynamic limit and the inequality between their number and the number of physical root must always be respected. The point is however, that if we incorporate these assumptions into the summation method, then the final result is a correct system of TBA equations. In other words, the original honest sum over physical roots and auxiliary roots with constraint can be mimicked by a sum over physical roots and strings of auxiliary roots without constraint. 
\vspace{-0.2cm}

\noindent
This replacement of the honest summation could either be seen as a mundane manipulation of the known TBA equations or it could be taken more seriously, namely it could work for other observables as well. Both points of view are justified and in this thesis we decide to experiment with the second: suppose the equivalence between the two summations described above, what is the implication on other observables of the theory, such as its g-function? Applying this idea for the $SU(2)$ chiral Gross-Neveu model and more generally for the current-perturbed $SU(2)_k$ WZNW model, we find that the resulting g-function is divergent, at least in the IR and UV limit. This means that we cannot simply replace the honest summation by an unrestricted sum involving magnon strings in the case of theories with boundary. However we find that the two divergences are of the same order and if we normalize the g-function by its IR value then the result is finite and can even match a CFT g-function for even values of $k$. This normalization amounts to removing graphs made entirely of auxiliary Bethe roots.
\vspace{0.5cm}

\noindent
Our conclusion regarding the efficiency  of the method of direct summation in $1+1$ dimensional integrable quantum field theories is the following. For theories with diagonal S-matrix, it can derive any known observable that can be obtained within the original TBA approach. It is in fact more robust than the latter, as the derivation of these observables does not require any significant modification of the method. On the other hand, the original approach appears to be more suitable for theories with non-diagonal S-matrix, with the string hypothesis as an advantage. Nonetheless, our partial result shows that the method of direct summation is not completely impotent in this case and further study is needed for a definitive settlement.
\vspace{1.5cm}

Before I could  investigate the applications of this method in the wrapping problem of $\mathcal{N}=4$ SYM, I stumbled upon some unexpected connections with \textit{Generalized Hydrodynamics} (GHD), a recently proposed framework to describe  transport properties of integrable systems out-of-equilibrium. Quantum transport has attracted much attention lately in view of the groundbreaking  advances in experiments that can now probe the dynamics of one-dimensional many-body systems in a controlled manner \cite{article,Bloch2012Quantum,Gring1318}.  From a theoretical point of view, transports in one dimension is somewhat special in that most of them are expected to be anomalous (non-diffusive) \cite{doi:10.1080/00018730802538522, PhysRevLett.108.180601,RevModPhys.87.593}. An exception being integrable systems where  not only ballistic transport \cite{PhysRevLett.82.1764,PhysRevB.55.11029,PhysRevLett.106.217206}, but  also other type of transports such as diffusive and super-diffusive transports can in fact occur \cite{PhysRevLett.103.216602,10.1038/ncomms16117,PhysRevLett.119.080602,PhysRevLett.121.230602}.  In order to provide a coherent understanding in the transport phenomena in integrable systems, a hydrodynamic approach was proposed \cite{Castro-Alvaredo:2016cdj, PhysRevLett.117.207201} and coined Generalized Hydrodynamics. It was originally capable of describing only the dynamics at the Euler scale (leading contribution of the derivative expansion with respect to the space coordinates), but was later extended to capture the sub-leading i.e. diffusive effect \cite{PhysRevLett.121.160603}. Moreover, it can take into account the presence of an external potential   \cite{SciPostPhys.2.2.014}, which is necessary to simulate ultracold atom gases. To this end, there have been experimental supports \cite{Schemmer2019GeneralizedHO}  for the validity of GHD.

Being a hydrodynamic theory, the main assumption of GHD is that, inside an appropriate time window (the hydrodynamic time scale), the average of local operators can be evaluated in a local state with maximum local entropy. In integrable systems, maximal entropy states are described by \textit{generalized Gibbs ensembles} (GGE): an extension of Gibbs ensembles with an infinite number of inverse temperatures $\beta^i$ coupled to the  conserved charges.  The assumption of local maximal entropy hence reads explicitly
\begin{align}
\langle \mathcal{O}(x,t)\rangle\approx\langle \mathcal{O}(0,0)\rangle_{\vec{\beta}(x,t)}.\label{source}
\end{align}
This equation is the main source of criticism towards GHD: what is its order of exactitude, what is its precise domain of validity, how can it be rigorously proven?... On the other hand, if we ignore these questions then equation \eqref{source} allows GHD to exploit the  power of exact solubility. If the inverse temperatures (GGE profile) are known at every point in space-time then the average of local operators can be computed using TBA\footnote{The TBA of GGE was first derived in \cite{Mossel:2012vp}, there is not much difference compared to the TBA of a Gibbs ensemble, one simply includes in the TBA source term an infinite number of chemical potentials.}. The main problem in GHD is therefore to determine the space-time dependence of the GGE profile. At Euler scale this problem can be solved, provided that we know the average currents carried by the local state. The form of the average currents is a cornerstone of GHD, as all transport properties of the theory are derived from it.

The currents are related to the corresponding conserved charge densities through a continuity equation. Although this equation is simple and the average of the conserved charge densities is well-known in TBA, the task of finding the average currents is not so trivial. For integrable quantum field theories, the first derivation was presented in \cite{Castro-Alvaredo:2016cdj,PhysRevLett.117.207201} using a double Wick rotation. The main idea is to regard the currents as the conserved charge densities of the mirror theory (as a result of the continuity equation) and to use the mirror TBA to recover their average. There are however two issues with this derivation. First, the assumed analyticity of the TBA source term does not necessarily hold for all GGEs. Second, the trick of using a double Wick rotation works solely for quantum field theories and this derivation cannot be extended to spin chains or integrable gases, which constitute a large part of GHD applications. We find a new derivation of the average currents using the Leclair-Mussardo formula and a diagrammatic representation of the relation between connected and symmetric form factors. Our derivation has recently been extended to integrable spin chains by Pozsgay \textit{et al.} \cite{PhysRevX.10.011054}.
\begin{figure}[ht]
\centering
\includegraphics[width=10cm]{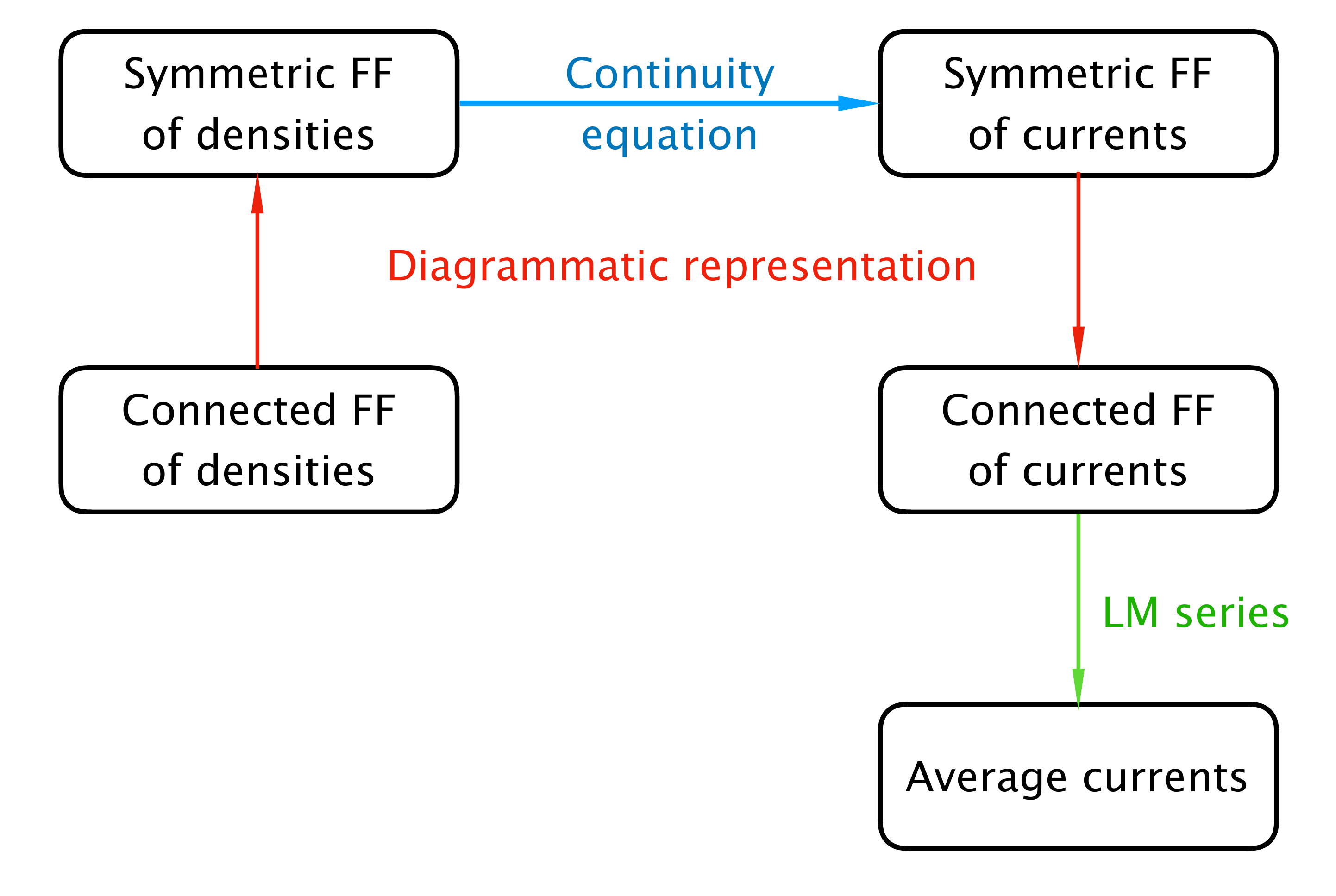}
\caption{The schema of our derivation of the average currents in GHD.}
\end{figure}

Going beyond the average currents, one can study transport fluctuation, manifested by rare events with large deflection from their expected values. The probabilities of these events are encoded in  the cumulants of the time-integrated currents carried by the local state. Although these cumulants were predicted by linear fluctuating hydrodynamics \cite{10.21468/SciPostPhys.8.1.007,Doyon:2019osx}, their  complicated expressions obscure their physical virtues. We conjecture that they are in fact very similar to the cumulants of the corresponding conserved charges. The latter can be obtained by the direct summation method and are represented by a sum over trees with TBA quantities on their vertices and propagators. With only minor modifications, the same diagrams can be used to express the cumulants of the time-integrated currents.  We confirm this proposal by a non-trivial matching with the analytic results of \cite{10.21468/SciPostPhys.8.1.007,Doyon:2019osx} up to the fourth cumulant. Our conjecture highlights a remarkably simple duality between time-integrated currents and conserved charges. 

It should be stressed  that neither the average currents nor higher cumulants were obtained by the direct summation method. They are simply related to quantities which can be derived by this method. Given their diagrammatic representations however, one can apply the method in reverse to find the corresponding matrix element. This has been done in \cite{PhysRevX.10.011054} for the current.
\vspace{1.5cm}

The thesis is organized as follows. In chapter 1 we summarize basic features of integrable quantum field theories. Chapter 2-4 present the applications of the direct summation method: the TBA equation, the Leclair-Mussardo series, the excited state energies and the cumulants of conserved charges in chapter 2, the g-function for theories with diagonal S-matrix in chapter 3 and the g-function for integrable perturbed $SU(2)_k$ WZNW models in chapter 4. Finally, connections with GHD are presented in chapter 5. Except for the derivation of the TBA equation, which has  previously appeared in the literature \cite{doi:10.1063/1.1396836,doi:10.1063/1.1501444,doi:10.1143/JPSJ.73.1171} in an almost identical form, all results from chapter 2 to 5 are original. 

The thesis is based on the following papers 
\begin{itemize}
\item  I. Kostov, D. Serban, D-L. Vu, \textit{"TBA and tree expansion"}, Springer Proc.Math.Stat. 255 (2017) 77-98. (chapter 2)
\item  I. Kostov, D. Serban, D-L. Vu, \textit{"Boundary TBA, trees and loops"},  Nucl.Phys.B 949 (2019) 114817.  (chapter 3)
\item I. Kostov, D. Serban, D-L. Vu, \textit{"Boundary entropy of integrable perturbed $SU(2)_k$ WZNW"},  JHEP 08 (2019) 154. (chapter 4)
\item  D-L. Vu, T. Yoshimura, \textit{"Equations of state in generalized hydrodynamics"}, SciPost Phys. 6 (2019) 2, 023. (chapter 5)
\item D-L. Vu, \textit{"Cumulants of conserved charges in GGE and cumulants of total transport in GHD: exact summation of matrix elements?"}, J.Stat.Mech. 2002 (2020) 2, 023103. (chapter 2 and 5)
\end{itemize}
\newpage
\chapter{Preliminary}
\section{Basics of integrable quantum field theories}
\subsection{Some examples}
\label{iqft-list}
We present in this subsection a list of integrable quantum field theories that will later appear at some point in this thesis. One of the most famous iQFT's is the sine-Gordon model, the Lagrangian density of which is given by
\begin{align}
L_\text{SG}=\frac{1}{2}(\partial_\mu \phi)^2+\frac{m_0^2}{\beta^2}\cos(\beta\phi).\label{sine-Gordon-Lagrangian}
\end{align} 
The existence of classically conserved currents in this theory was established by the inverse scattering method  \cite{Faddeev:1973aj,PhysRevLett.31.125,doi:10.1063/1.522391}. These conservation laws were then shown to survive after quantization by the perturbation theory approach \cite{Kulish:1976ef}. The particle content of the theory includes a soliton, an anti-soliton and their bound states (breathers) with masses  
$m_k=2m\sin(k\gamma/16)$ for $ k=1,2,...<8\pi/\gamma$, where $m$ is the mass of the soliton and $\gamma=\beta^2/[1-\beta^2/(8\pi)\big]$ \cite{PhysRevD.11.3424,Korepin:1975zu} . It was argued in the seminal work \cite{PhysRevD.11.2088} that this theory is actually equivalent to the massive Thirring model with Lagrangian density
\begin{align}
L_\text{Thirring}=i\bar{\psi}\gamma_\mu\partial^\mu\psi-m\bar{\psi}\psi-\frac{g}{2}(\bar{\psi}\gamma_\mu\psi)^2,\label{massive-Thirring-Lagrangian}
\end{align}
where $\psi,\bar{\psi}$ is the Dirac field and the four-fermion coupling constant $g$ is related to $\beta$ in \eqref{sine-Gordon-Lagrangian} as $g/\pi=4\pi/\beta^2-1$. When the Dirac field belongs to the fundamental representation of the $SU(N)$ group, the massive Thirring model is also known as the chiral Gross-Neveu model
\begin{align}
L_{\text{GN}}=\bar{\psi}_a(i\slashed{\partial}-m)\psi_a+g[(\bar{\psi}_a\psi_a)^2-(\bar{\psi}_a\gamma^5\psi_a)^2],\quad a=\overline{1,N}.\label{SU(N)-GN-Lagrangian}
\end{align}
The theory was shown to be  integrable at classical level in \cite{neveu1978} and at quantum level in \cite{Zamolodchikov:1978xm}. Each particle of the theory is in one-to-one correspondence with fundamental representations of the $SU(N)$ group \cite{ANDREI1980106}. The particle corresponding to the Young tableau of one column and $a$ rows is a bound state of $a$ vector particles and has mass $ m_a=m\sin (\pi a/N)/\sin (a/N),\quad a=\overline{1,N-1}$.
In particular the $SU(2)$ chiral Gross-Neveu model is equivalent to the sine-Gordon theory at $\beta^2=8\pi$.\\
A cousin of the sine-Gordon theory is the sinh-Gordon theory
\begin{align}
L_\text{SnhG}=\frac{1}{2}(\partial_\mu \phi)^2-\frac{m_0^2}{\beta^2}\cosh(\beta\phi).
\end{align}
Compared to the sine-Gordon theory, the particle content of the sinh-Gordon theory is much simpler as there is only one particle. The sinh-Gordon theory is an example of affine Toda field theories \cite{Zanon1993}, which form a large family of integrable models. The Lagrangian density of the Toda field theory corresponding to a Lie algebra $\mathfrak{g}$ of rank $r$ is defined as
\begin{align}
L_\text{Toda}=\frac{1}{2}\partial_\mu\phi^a\partial^\mu\phi^a-\frac{m_0^2}{\beta^2}\sum_{i=0}^re^{\beta \vec{\alpha_i}.\vec{\boldsymbol{\phi}}},
\end{align}
where $\phi^a,a=1,...,r$ are $r$ real scalar fields and $\alpha_i,i=1,...,r$ are  positive simple roots of $\mathfrak{g}$ while $\alpha_0$ is the negative of its maximal root. Without the $\alpha_0$ term, the theory is conformal \cite{GERVAIS1983329,MANSFIELD1983419,BRAATEN1983301}, and is referred to as conformal Toda field theories. With the $\alpha_0$ term, conformal invariance is broken but integrability is preserved, one talks about an \textit{integrable perturbation of CFT}. The classical and quantum integrability  of affine Toda field theories was shown in \cite{OLIVE1983470,wilson_1981,mikhailov1981} and \cite{DELIUS1992307} respectively.

The systematic study of integrable perturbations of CFTs was initiated by Zamolodchikov in \cite{Zamolodchikov:1987zf}. The author explicitly constructed the integrals of motion of the three-state Potts model in terms of the operator algebra of the conformal field theory describing its critical point. This CFT is the minimal model $\mathcal{M}_{5,6}$ and the perturbing operator is $\Phi_{1,2}$ of dimension $(2/5,2/5)$. It is the intersection of two families of integrable perturbed CFTs: the first is $\mathcal{M}_{p,p+1}$ perturbed by $\Phi_{1,2}$  operator \cite{Zamolodchikov:1987jf} and the second is $\mathbb{Z}_n$ parafermion perturbed by operator of dimension $(2/(n+2),2/(n+2))$ \cite{Fateev:1990bf,Fateev:1991bv}. The case of perturbed $\mathbb{Z}_4$ parafermion is equivalent to the sine Gordon model at the coupling $\beta^2=6\pi$.  Another example of integrable perturbed CFTs is the scaling Lee Yang model. It is obtained by deforming the minimal model $\mathcal{M}_{2,5}$ in the direction of its only relevant operator namely the $\Phi_{1,3}$ operator. Moreover it was shown that the  $\mathcal{M}_{2,2n+3}$ minimal models perturbed by their $\Phi_{1,3}$ operator \cite{FREUND1989243,SMIRNOV1990156} are also integrable. The general procedure to construct integrals of motion in perturbed CFTs is described in \cite{Zamolodchikov:1989zs}.

The $SU(N)$ chiral Gross-Neveu can itself be regarded as a perturbed CFT. Indeed, using the technique of nonabelian bosonization \cite{Witten:1983ar}, the Lagrangian \eqref{SU(N)-GN-Lagrangian} can be written as the current-perturbed $U(N)_1$ WZNW theory \cite{DiVecchia:1984df},\cite{Knizhnik:1984nr}. The $U(1)$ center is identified with a massless boson that decouples from the rest of the spectrum.



Another source of iQFT comes from sigma models. A sigma model is a field theory where the field takes values in a manifold $\mathcal{M}$ with a metric $g_{ij}$
\begin{align*}
L_\Sigma=g_{ij}(X)\partial_\mu X^i\partial^\mu X^j.
\end{align*}
The theory is classically integrable for a large family of manifolds called a symmetric spaces \cite{EICHENHERR1979381}. A symmetric space is a manifold $G/H$ where $G$ and $H$ are Lie groups, and $H$ is a maximal subgroup of $G$ i.e. no normal sub group other than $G$ itself contains $H$. On the other hand, quantum integrability is more restricted. Two well known examples are \textit{principal chiral model}, where $G=H\times H$ and $H$ is a simple Lie group \cite{Faddeev:1985qu} and the  $O(n)$ model \cite{Brezin:1976qa,Bardeen:1976zh}, where $G = O(n)$ and $H = O(n − 1)$.


\subsection{Consequences of higher conserved charges on the S matrix}
The existence of infinitely many conserved charges in a theory impose  the following constraints on its scattering processes
\begin{itemize}
\item There is no particle production,
\item the sets of initial and final momenta are identical,
\item \textit{factorization}: the $n$-to-$n$ S-matrix is a product of $n(n-1)/2$ two-to-two S-matrices
\end{itemize}
The validity of these properties was first backed up by an heuristic argument of Zamolodchikov and Zamolodchikov \cite{Zamolodchikov:1978xm}. More arguments were given  in \cite{POLYAKOV1977224,PhysRevD.17.2134} and a relatively rigorous proof was presented in \cite{PARKE1980166}. We sketch here the essential ideas of this proof.

In the far past (future) the \textit{in} (\textit{out}) state can be regarded as a collection of non-interacting particles, if we assume that all interactions are short-ranged. In an \textit{in} state, the fastest moving particle is found on the furthest left while the slowest one on the furthest right. In an \textit{out} state, the inversed order applies. The existence of a higher conserved charge $\mathcal{Q}$ implies that
\begin{align}
\langle \text{out}|\mathcal{S}|\text{in}\rangle=\langle \text{out}|e^{i\alpha\mathcal{Q}}\mathcal{S}e^{-i\alpha\mathcal{Q}}|\text{in}\rangle\label{action-of-Q}
\end{align}
As the asymptotic states are tensor products of one-particle wavefunction, the  charge $\mathcal{Q}$ acts on these states component-wise. Due to its higher Lorenz spin, the effect of $\mathcal{Q}$ on an one-particle wavefunction is to shift it in space-time by an amount that depends on the particle momentum. The precise amount of this shift can be obtained by saddle point approximation on a Gaussian wave function peaked around the particle momentum, we refer to the original paper for more details.
\begin{wrapfigure}{r}{0.48\textwidth}
  \begin{center}
    \includegraphics[width=0.46\textwidth]{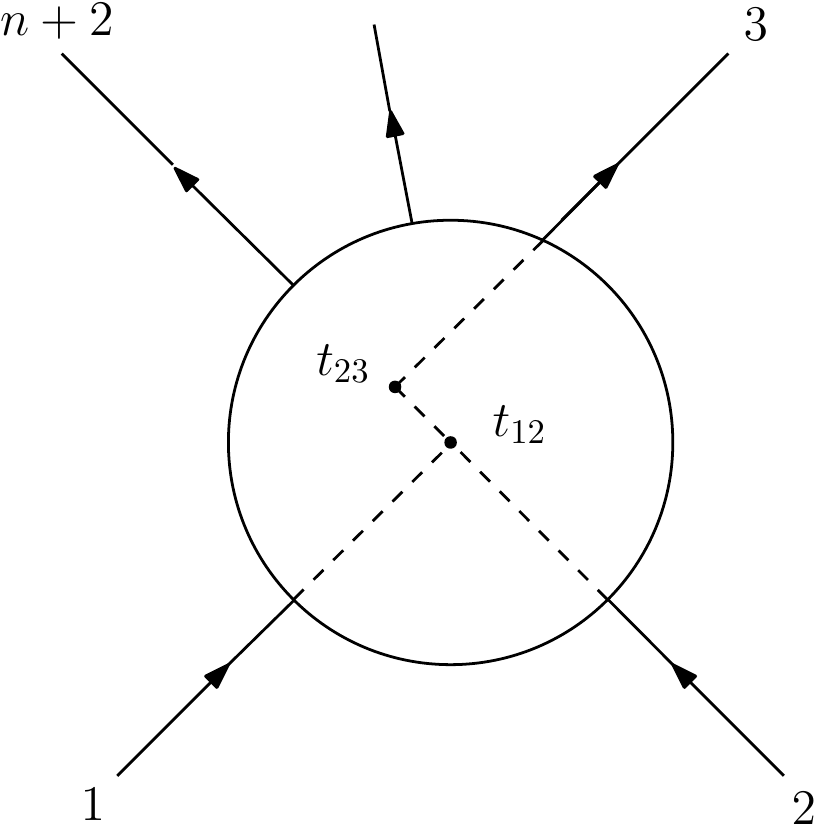}
  \end{center}
  \caption{Space-time diagram of a two-to-n scattering process}
  \label{2-n}
\end{wrapfigure}
With this in mind, consider now  a $2\to n$ scattering process, with the particles labelled as in figure \ref{2-n}.
Let us denote by $t_{12}$ and $t_{23}$ the collison time between the particle 1 and 2 and particle 2 and 3 respectively. According to the principle of macrocausality, the following inequality must be respected
\begin{align}
t_{12}\leq t_{23}
\end{align}
Assume for a moment that the momentum of the particle 3 is not the same as that of the particle 1. According to the above statement, one can choose the parameter $\alpha$ in \eqref{action-of-Q} in such a way that the charge $Q$ shifts the collison time $t_{12}$ arbitrarily high and $t_{23}$ arbitrarily low. This violation of macrocausality leads to the conclusion that the momentum of the fastest out-going particle must coincide with that of the fastest incoming particle. Similary, one can match the momentum of the slowest out-going particle with that of the slowest incoming one. Energy-momentum conservation then dictates the absence of particle production. To summarize, the only possible outcome of a two particle scattering process is two particles with the same momenta as the incoming ones.

For an $n$-particle scattering process, we can once again act with $\mathcal{Q}$ on a certain incoming particle, moving it away from the scattering region of the other $n-1$ particles. Once all the interactions among these particles have occurred, one can consider the scattering of the out-going particles with that particular particle. By induction, we see that there is no particle production, and the final set of momenta must be identical to the initial set of momenta. Furthermore, the $n$-to-$n$ S-matrix is a product of $n(n-1)/2$ two-to-two S-matrices. $\square$

The factorization property gives rise to important characteristics of the two-to-two S-matrix. Indeed, we have found that all the possible ways of factorizing an $n$-to-$n$  S-matrix into two-to-two S-matrices must lead to the same result. In particular, let us consider a three-to-three scattering process. We can use the charge $\mathcal{Q}$ to separate either the first or the third particle. This leads to two different scenarios
\begin{figure}[ht]
\centering
\includegraphics[width=12cm]{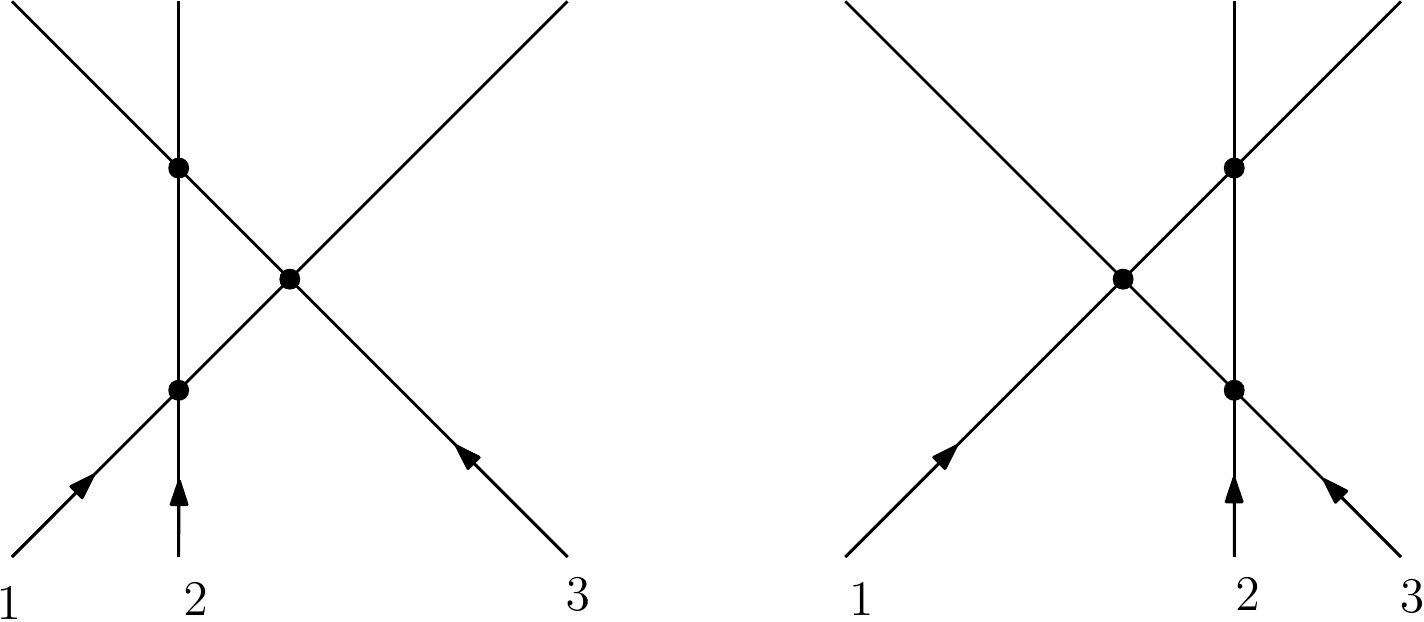}
\end{figure}

\noindent
The equivalence between the two processes leads to the Yang-Baxter equation
\begin{align}
\mathcal{S}_{23}\mathcal{S}_{13}\mathcal{S}_{12}=\mathcal{S}_{12}\mathcal{S}_{13}\mathcal{S}_{23}.\label{Yang-Baxter}
\end{align}
We note that the Yang-Baxter equation is a consequence of the existence of higher conserved charges, it is by no mean a sufficient condition for integrability.
\subsection{The two particle S-matrix}
\label{two-S}
The previous section has shown the importance of the two particle S-matrix. In this section we study its physical properties and analytic structure. 

\begin{figure}[ht]
\centering
\includegraphics[width=8.5cm]{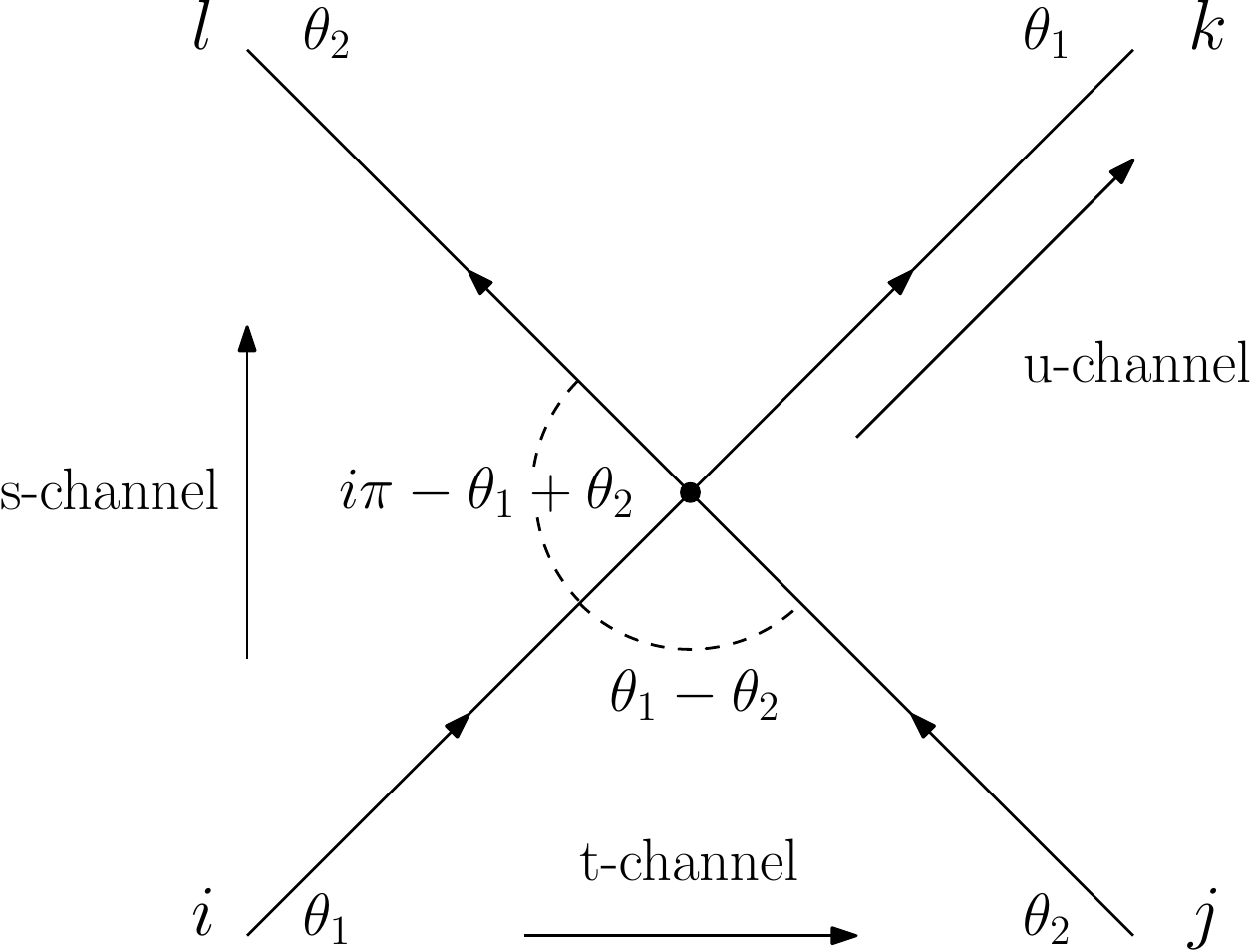}
\caption{The two particle scattering}
\label{2-2}
\end{figure}

Let us denote by $\theta_1$ and $\theta_2$ the rapidities of the incoming particle, with $\theta_1>\theta_2$ and by $\theta_3$ and $\theta_4$ the rapidities of out-going particles, with $\theta_3<\theta_4$. As established above, we only have non-trivial scattering for $\theta_4=\theta_1,\theta_3=\theta_2$. A two-particle elastic relativistic S-matrix is then given by
\begin{align}
|\theta_1,\theta_2\rangle_{i,j}^\text{in}=S_{ij}^{kl}(\theta_{12})|\theta_1,\theta_2\rangle_{k,l}^\text{out}\label{2-to-2}
\end{align}
and represented graphically in figure \ref{2-2}, the indices $i,j,k,l$ denote particle type and $\theta_{12}\equiv \theta_1-\theta_2$. When $S_{ij}^{kl}\propto \delta_i^k\delta_j^l$ we say that the S-matrix is \textit{diagonal}, otherwise we say that it is  \textit{non-diagonal}. One can also work in Maldelstam variables
\begin{align}
s&=(p_1+p_2)^2=m_i^2+m_j^2+2m_im_j\cosh(\theta_{12})\nonumber\\
t&=(p_1-p_3)^2=m_i^2+m_l^2-2m_im_l\cosh(\theta_{13})\\
u&=(p_1-p_4)^2=m_i^2+m_k^2-2m_im_k\cosh(\theta_{14})\nonumber
\end{align}
which are the square of the center-of-mass energies in the channels defined by the process $i + j \to k + l$ ($s$-channel), $i+\bar{l}\to k+\bar{j}$ ($t$-channel) and $i+\bar{k}\to l+\bar{j}$ ($u$-channel)
respectively. One finds that $u = 0$ and $t(\theta_{ 12} ) = s(i\pi − \theta_{ 12} )$, meaning the S-matrix depends only on one variable, say $S = S(s)$.

The $C,P,T$ invariance dictates the symmetry of S as a matrix
\begin{align}
S_{\bar{i}\bar{j}}^{\bar{k}\bar{l}}(s)=S_{ij}^{kl}(s),\quad S_{ij}^{kl}(s)=S_{ji}^{lk}(s),\quad S_{ij}^{kl}(s)=S_{lk}^{ji}(s).\label{CPT-invariance}
\end{align}
Let us now discuss analytic properties of S. One important property is crossing symmetry. It means the freedom to choose between the $s$- and $t$-channels without affecting the scattering amplitude
\begin{align}
S_{ij}^{kl}(s)=S_{\bar{l}i}^{\bar{j}k}(t).\label{crossing-symmetry}
\end{align}
As in a generic QFT \cite{Eden:1966dnq}, the S-matrix has a branch cut running from the two-particle threshold $s=(m_i+m_j)^2$ to infinity. Due to crossing symmetry \eqref{crossing-symmetry}, another cut runs from $(m_i-m_j)^2$ towards minus infinity. However, S has a distictive feature owed to the lack of particle production namely it does not have further cuts coming from multi-particle thresholds. With these two cuts, S is a single-valued, meromorphic function on the complex s-plane. The physical region situating just above the right cut defines the physical sheet of the Riemann surface for S.  By combining the unitarity condition in the physical region
\begin{align}
S_{ij}^{kl}(s)[S_{lk}^{nm}(s)]^*=\delta_i^n\delta_j^m\label{unitarity-condition}
\end{align} 
with the real analyticity property
\begin{align}
S_{ij}^{kl}(s^*)=[S_{ij}^{kl}(s)]^*\label{real-analyticity}
\end{align}
we can show that the two cuts are of square-root type \cite{Dorey:1996gd}. In principle, analytically continuing S through  one cut might end up on a different sheet than the one obtained through the other cut. Hence, the Riemann surface of the S-matrix consists in general of several sheets, possibly infinite.

The structure of this Riemann surface turns out to be more transparent if we map the s-plane to the rapidity plane
\begin{align*}
\theta_{12}&=\text{arccosh}\big(\frac{s^2-m_1^2-m_2^2}{2m_1m_2}\big)=\log\frac{s^2-m_1^2-m_2^2+\sqrt{[s-(m_1+m_2)^2][s-(m_1-m_2)^2]}}{2m_1m_2}.
\end{align*}
The right (left) cut is mapped into $\mathbb{R}+2i\pi \mathbb{Z}$ and $\mathbb{R}+i\pi+2i\pi \mathbb{Z}$ respectively. The physical sheet of the s-plane corresponds to the strip $0 \leq \Im(\theta_{ 12} ) \leq \pi$ in the theta-plane. Furthermore, the original cuts are opened up and $S(\theta)$ is analytic at the images $i\pi\mathbb{Z}$ of the branch points. In conclusion, $S(\theta)$ is a meromorphic function of $\theta$.  Unitarity \eqref{unitarity-condition} and crossing \eqref{crossing-symmetry} are written in rapidity variable as
\begin{align}
S_{ij}^{kl}(\theta)S_{lk}^{nm}(-\theta)=\delta_i^n\delta_j^m,\quad S_{ij}^{kl}(\theta)=S_{l\bar{i}}^{\bar{j}k}(i\pi-\theta).\label{unitarity-and-crossing}
\end{align}
The real analyticity in the s-plane \eqref{real-analyticity} implies that $S(\theta)$ is real when $\theta$ is purely imaginary. Multiplying the S-matrix by any function $F(\theta)$ which satisfies $F(\theta)F(-\theta) =1$
 and $F(i\pi -\theta) = F(\theta)$ will give an S matrix still obeying the Yang-Baxter equation, crossing and unitarity. Such function $F$ is called a CDD factor.
\begin{figure}[ht]
\centering
\includegraphics[width=16cm]{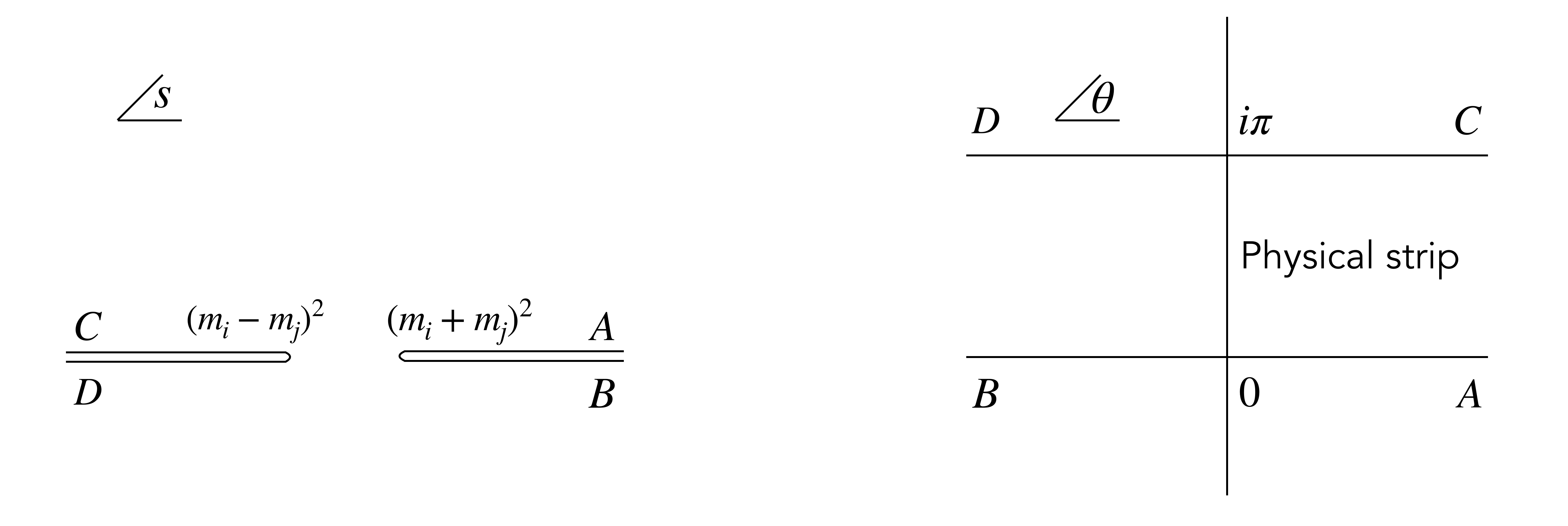}
\caption{Analytic structure of the two particle S-matrix on the $s$ and $
\theta$ plane.}
\end{figure}

\begin{wrapfigure}{r}{0.38\textwidth}
  \begin{center}
    \includegraphics[width=0.34\textwidth]{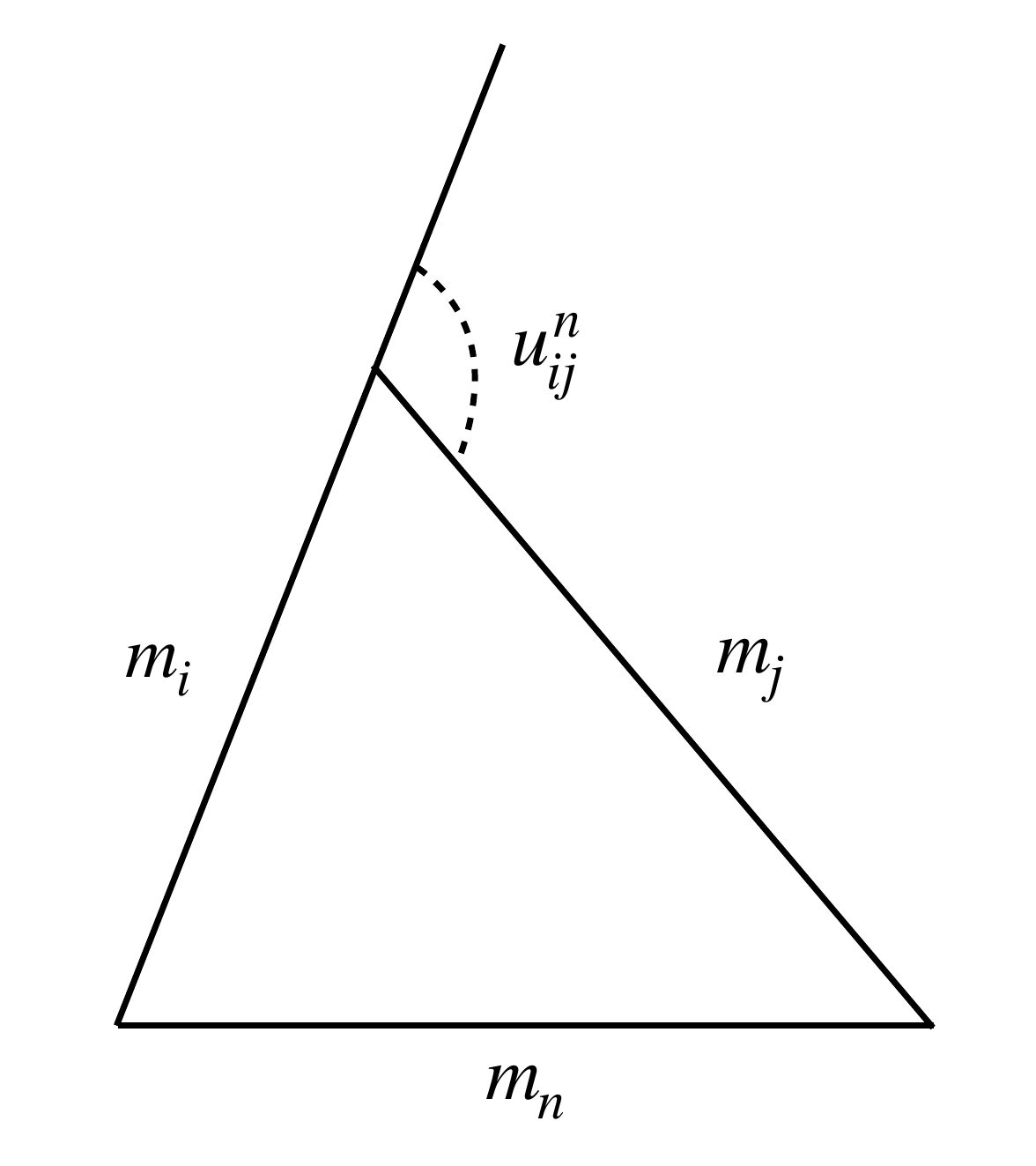}
  \end{center}
  \caption{The mass triangle}
  \label{mass-triangle}
\end{wrapfigure}

The two-particle S-matrix can also have simple poles related to bound states. In the s-plane, these poles are found between the two-particle thresholds i.e. $(m_i-m_j)^2$ and $(m_i+m_j)^2$. This interval is mapped into the interval $(0,i\pi)$ in the $\theta$-plane. Denote by $iu_{ij}^n$ a simple pole corresponding to a bound state $n$ formed by two particles $i$ and $j$. The mass of this bound state is given by 
\begin{align}
m_n^2=m_i^2+m_j^2+2m_im_j\cos u_{ij}^n.
\end{align}
This identity admits a simple geometric interpretation: the three masses are  three sides of a triangle and the real part of the pole is one of its outside angles, see figure \ref{mass-triangle}.

\noindent
This geometric point of view also hints at a democracy between bound states and elementary particles. For instance, one can regard $i$ and $j$ themselves as bound states formed respectively by $j,n$ and $i,n$ particles. If one can even find the scattering matrix between these particles, then the corresponding poles in the physical strip $u_{jn}^i$ and $u_{in}^j$ must satisfy
\begin{align}
u_{ij}^n+u_{jn}^i+u_{ni}^j=2\pi.
\end{align}
One can proceed and scatter this bound state with other particles and other bound states, looking for poles and new bound states. Consistency of the theory requires that this procedure closes upon itself, namely one should always end up with a finite set of particles. This is the idea of S-matrix bootstrap principle \cite{Zamolodchikov:1989zs}. The problem of finding S-matrix of bound states can be solved by the so-called fusion method \cite{KAROWSKI1979244,KAROWSKI198161,Ogievetsky:1987vv}. In the following section we will bootstrap the S-matrix of the $SU(N)$ chiral Gross-Neveu model.
\subsection{Finding the two-particle S-matrix: an example}
\label{finding-two-particle-S}
In many integrable models \cite{Zamolodchikov:1989fp,Ogievetsky:1987vv,BRADEN1990689,DELIUS1992365,MUSSARDO1992215,DELIUS1995445}, the S-matrix can often by fixed  by  a variety of
constraints: the Yang-Baxter equation \eqref{Yang-Baxter}, unitarity and crossing symmetry \eqref{unitarity-and-crossing}, global symmetries, the consistency of bound states and finally, by requiring that the bootstrap procedure closes. One can then check that the obtained S-matrix is indeed the correct one by computing the free energy  of the model (for instance by the Thermodynamic Bethe Ansatz method that will be described in the next sections) and ensuring that it
agrees with the correct results in various limits. We illustrate this idea for the $SU(N)$ chiral Gross-Neveu model.

To remind, each particle of the theory is in one-to-one correspondence with fundamental representations of the $SU(N)$ group . The particle corresponding to the Young tableau of one column and $a$ rows is a bound state of $a$ vector particles. According to the previous section, S-matrices between bound states can be referred from those between elementary particles. The S-matrix between vector particles is therefore all we need to find. 

Invariance under the $SU(N)$ group implies that the S-matrix can be expressed as linear combinations of projection operators. In the case of vector particles, only  the symmetric and antisymmetric representation appear in their tensor product. The S-matrix of two vector particles can thus be written as
\begin{align}
S_{\text{V}\text{V}}(\theta)=f^\text{S}_{\text{V}\text{V}}(\theta)\mathcal{P}_\text{S}+f^\text{A}_{\text{V}\text{V}}(\theta)\mathcal{P}_\text{A},\label{V-V-GN}
\end{align}
where the projection operators on the symmetric (S) and 
antisymmetric (A) tensors are given by
\begin{align*}
\mathcal{P}_\text{S}[a_i(\theta_a)b_j(\theta_b)]=\frac{1}{2}[a_i(\theta_a)b_j(\theta_b)+a_j(\theta_a)b_i(\theta_b)],\\
\mathcal{P}_\text{A}[a_i(\theta_a)b_j(\theta_b)]=\frac{1}{2}[a_i(\theta_a)b_j(\theta_b)-a_j(\theta_a)b_i(\theta_b)].
\end{align*}
The subscripts in $a_i$ and $b_j$ represent the $i$ and $j$ particles in the vector multiplets. The advantage of expressing the S-matrix in terms of projection operators is that the Yang-Baxter equation \eqref{Yang-Baxter} greatly simplifies. It can be shown that the two coefficients in \eqref{V-V-GN} must satisfy \cite{BERG1978125}
\begin{align}
\frac{f^\text{A}_{\text{V}\text{V}}(\theta)}{f^\text{S}_{\text{V}\text{V}}(\theta)}=\frac{\theta+2\pi i \Delta}{\theta-2\pi i \Delta},\label{A-S-GN}
\end{align}
where $\Delta$ is a parameter that cannot be fixed by the Yang-Baxter equation. Unitarity and crossing symmetry then requires that
\begin{align}
f^\text{S}_{\text{V}\text{V}}(\theta)=X(\theta)F^{\text{min}}_{\text{V}\text{V}},\quad \text{with}\quad F^{\text{min}}_{\text{V}\text{V}}=\frac{\Gamma(1-\frac{\theta}{2\pi i})\Gamma(\frac{\theta}{2\pi i}+\Delta)}{\Gamma(1+\frac{\theta}{2\pi i})\Gamma(-\frac{\theta}{2\pi i}+\Delta)}.
\end{align}
In this expression $X(\theta)$ is a CDD phase and $F^{\text{min}}_{\text{V}\text{V}}$ is called the minimal solution, because the corresponding S-matrix has no pole in the physical strip.

The CDD phase can be determined by looking at the bound states structure of model.  As there are  bound states in the 
antisymmetric representation of $SU(N)$, but not in the symmetric representation,
there must be a pole in $f^\text{A}_{\text{V}\text{V}}$ but not in $f^\text{S}_{\text{V}\text{V}}$. The relation \eqref{A-S-GN} tells us that this pole 
is situated at $\theta=2\pi i\Delta$, corresponding to a bound-state with mass
 $m_\text{A}/m_\text{V} = 2\cos\pi\Delta$. Such pole has been canceled by the  minimal solution and so the CDD factor must reintroduce it into the S-matrix
\begin{align*}
X(\theta)=\frac{\sinh(\theta/2+\pi i \Delta)}{\sinh(\theta/2-\pi i \Delta)}.
\end{align*}
Now that we have obtained the S-matrix of vector particles as a function of $\Delta$, we can apply the bootstrap procedure. By requiring that the bootstrap closes, we can fix this  parameter  to be $\Delta = 1/N$ \cite{BERG1978205,PhysRevD.20.897}. We conclude that the vector-vector S-matrix of the chiral $SU(N)$ Gross-Neveu model is given by
\begin{align}
\mathcal{S}_{\text{V}\text{V}}(\theta)=\frac{\sinh(\theta/2+\pi i/N)}{\sinh(\theta/2-\pi i /N)}\frac{\Gamma(1-\frac{\theta}{2\pi i})\Gamma(\frac{\theta}{2\pi i}+\frac{1}{N})}{\Gamma(1+\frac{\theta}{2\pi i})\Gamma(-\frac{\theta}{2\pi i}+\frac{1}{N})}\bigg(\mathcal{P}_\text{S}+\frac{\theta+2\pi i/N}{\theta-2\pi i/N}\mathcal{P}_\text{A}\bigg).\label{V-V-SU(N)-GN}
\end{align}
The S-matrices between other particles in the spectrum can be built from this S-matrix by fusion method.
%

\subsection{The Bethe equation of  SU(2) chiral Gross-Neveu model }
\label{SU(2)-Bethe}
With the two-particle S-matrix at hand, one can find the on-shell condition of an asymptotic state living in a finite, periodic but very large (compared to the inverse mass scale) volume $L$. We consider as example the  $SU(2)$ chiral Gross-Neveu model. There is only the vector multiplet in this theory, so the vector-vector S-matrix \eqref{V-V-SU(N)-GN} is all we need
\begin{align}
\mathcal{S}^{\text{SU(2)}}(\theta)=\frac{\theta-\pi i\mathcal{P}}{\theta-\pi i}S_0^{\text{SU(2)}},\quad S_0^{\text{SU(2)}}(\theta)=-\frac{\Gamma(1-\theta/2\pi i)}{\Gamma(1+\theta/2\pi i)}\frac{\Gamma(1/2+\theta/2\pi i)}{\Gamma(1/2-\theta/2\pi i)}.
\end{align}
The periodicity condition imposed on the wave function $|\Psi\rangle =|\theta_1,\theta_2,...,\theta_N\rangle $ when a particle of rapidity $\theta_j$ is brought around the circle and scatters with other particles reads
\begin{align*}
e^{-ip(\theta_j)L}\Psi=\prod_{k\neq j}\mathcal{S}^{\text{SU(2)}}(\theta_j,\theta_k)\Psi,\quad j=\overline{1,N}.
\end{align*}
In this equation, the product runs on the index $k$ and this rule is implicitly understood from now on.
Using the fact that the scattering matrix at coinciding rapidity is exactly minus the permutation operator, one can cast the above equation in the following form
\begin{align}
e^{-ip(\theta_j)L}\Psi=-T^\textnormal{SU(2)}(\theta_j)\Psi,\quad j=\overline{1,N}.
\label{transfer-matrix-GN-1}
\end{align}
where $T(u)=\tr_{\mathbb{C}_u^2}[\mathcal{S}(u,\theta_1)...\mathcal{S}(u,\theta_N)]$ is the transfer matrix. The advantage of writing Bethe equations in this form is that we can now regard the physical rapidities $\theta's$ as non-dynamical impurities on a spin chain. The argument $u$ of the transfer matrix plays the role of the rapidity of an auxiliary particle living in time direction. Up to a scalar factor, $T$ is also the transfer matrix of the XXX spin chain with Hamiltonian
\begin{align*}
H=-J\sum_{i=1}^{N}(\vec{\sigma}_i.\vec{\sigma}_{i+1}-1),
\end{align*}
where $\vec{\sigma}$ are Pauli matrices with periodicity  $\sigma_{N_f+1}=\sigma_1$. For $J > 0$ this is a model of a ferromagnet where spins
prefer to align, while for $J < 0$ we have an antiferromagnet where spins prefer to alternate.

 As a consequence of the Yang-Baxter equation, the transfer matrices at different values of the spectral parameters commute with each other $[T(u),T(v)]=0$. By developping the transfer matrix in powers of $u$, we can obtain a tower of commuting observables, one of which is the Hamiltonian. The transfer matrix can be diagonalized by the technique of algebraic Bethe ansatz \cite{Faddeev:1996iy}. Its eigenvalues are parametrized by a set of spin chain excitations $u_1,...,u_M$. Replacing these eigenvalues into \eqref{transfer-matrix-GN-1} one obtains the "physical" Bethe equation for the $SU(2)$ chiral Gross-Neveu model
\begin{gather}
1=e^{ip(\theta_j)L}\prod_{k\neq j}^NS_0^\text{SU(2)}(\theta_j-\theta_k)\prod_{m=1}^M
\frac{\theta_j-u_m+i\pi/2}{\theta_j-u_m-i\pi/2},\quad j=\overline{1,N}.\label{SU(2)-physical}
\end{gather}
The spin chain rapidities themselves must satisfy an  Bethe equation on the spin chain setting
\begin{align}
1=\prod_{j=1}^N\frac{u_k-\theta_j-i\pi/2}{u_k-\theta_j+i\pi/2}\prod_{l\neq k}^M\frac{u_k-u_l+i\pi}{u_k-u_l-i\pi},\quad k=\overline{1,M}.\label{SU(2)-auxiliary}
\end{align} 
We see that in the physical perspective, the rapdities $u's$ correspond to auxiliary particles with vanishing energy and momentum. By construction of algebraic Bethe ansatz, their number should  not exceed half the number of physical rapidities.
\subsection{Integrability in open systems}
\label{open-integrable-section}
The scattering theory of systems with a reflecting boundary was first investigated by Cherednik  in \cite{Cherednik:1985vs}. In addition to the bulk two-particle S-matrix the author proposed a reflection matrix that describes the 
reflection amplitudes at boundary. This matrix must satisfy unitarity and a boundary version of the Yang-Baxter  equation. The boundary analogy of crossing symmetry and bootstrap equation was later found in \cite{Ghoshal:1993tm} and \cite{Fring:1993mp}.  In this subsection we consider a simple situation where both the diagonal bulk scattering  matrix and the reflection matrix are diagonal. The boundary unitarity and boundary crossing-unitarity read in that case
\begin{gather}
R_a(\theta)R_a(-\theta)=1,\\
R_a(\theta)R_{\bar{a}}(\theta-i\pi)=S_{aa}(2\theta).\label{boundary-crossing-unitarity}
\end{gather}
In particular, $R$ is $2\pi i$ periodic. The boundary bootstrap equation that describes the reflection amplitude of a bound state $c$ of particles $a$ and $b$ is given by
\begin{align}
R_{c}(\theta)=R_a(\theta+iu_{ac}^b)R_b(\theta-\bar{u}_{bc}^a)S_{ab}(2\theta+i\bar{u}_{ac}^b-iu_{bc}^a),\label{boundary-bootstrap}
\end{align}
where $u$ denotes the fusion angle and $\bar{u}=i\pi-u$. Similar to the CDD ambiguity in the bulk S-matrix, the reflection matrix cannot be uniquely fixed from these constraints. For instance, one can multiply a  solution to these equations by a solution of the bulk bootstrap, unitarity and crossing equations to obtain another viable solution. Without additional requirements, the minimal solution i.e. the one with smallest possible number of poles and zeros often proves to be physically meaningful.

Let us illustrate this idea on the affine Toda theory of type $A_k$. This theory consists of $k-1$ particles $a=1,...,k-1$  ($\bar{a}=k-a$) with mass spectrum $
m_a=m\sin(\pi a/k)/\sin(\pi/k)$.
Particles $a$ and $b$ can form bound states at fusion angle $u_{ab}^c=(a+b)\pi/k$ and $2\pi-[(a+b)\pi/k]$. The purely elastic scattering consistent with this bound state structure was bootstrapped in \cite{BRADEN1990689,1990IJMPA...5.1025F}
\begin{gather}
S_{ab}(\theta)=F_{|a-b|}(\theta)F_{a+b}(\theta)\prod_{s=|a-b|/2+1}^{(a+b)/2-1}F^2_s(\theta),\quad F_\alpha(\theta)\equiv\sinh\big(\frac{\theta}{2}+\frac{i\pi \alpha}{2k}\big)/\sinh\big(\frac{\theta}{2}-\frac{i\pi \alpha}{2k}\big).
\end{gather}
Periodicity and boundary unitarity of the reflection factors require them to be products of the building blocks $F$ as well. One then relies on crossing-unitarity to find the smallest number of $F$ needed. Indeed, each pole or zero of $S_{aa}(2\theta)$ on the right hand
side of \eqref{boundary-crossing-unitarity} must appear in one of the two factors on the left hand side of the same equation. This condition sets a lower bound on the number of poles and zeros for the reflection factor $R_a(\theta)$. The minimal reflection factor \cite{Corrigan:1994ft}
\begin{align}
R_a(\theta)=\prod_{s=0}^{a-1}F_s(\theta)F_{k+1-s}^{-1}(\theta),\quad a=\overline{1,k-1}.
\end{align}
obtained this way turned out to also satisfy the boundary bootstrap equation \eqref{boundary-bootstrap}.
This solution will be relevant for our study of g-function of current-perturbed WZNW model in chapter \ref{non-diag-g-section}.
\newpage
\section{Thermodynamic Bethe Ansatz}
\label{TBA-old-section}
In this section we present a method to describe the thermodynamics of integrable models: the thermodynamic Bethe ansatz (TBA). It was discovered by Yang and Yang \cite{Yang:1968rm} to study the  Bose gas with delta function interaction at finite temperature. It was then extended to lattice integrable models such as Heisenberg spin chain \cite{PhysRevLett.26.1301,TAKAHASHI197281} and Hubbard model \cite{10.1143/PTP.47.69}. When applied to relativistic integrable quantum field theories, the TBA free energy density is equivalent to the ground state energy in finite volume \cite{Zamolodchikov:1987zf}. 

We start with a summary of the derivation of Zamolodchikov, applicable for any massive integrable quantum field theories with diagonal scattering matrix.
\subsection{The traditional derivation}
\label{tradition}
Assume that we have a relativistic, integrable QFT with a  single neutral particle of mass $m$. The question  we try to answer is whether we can compute the ground state energy $E_0(R)$ of the theory compactified on a circle of length $R$ from its S-matrix data?  The idea of Zamolodchikov  is to let the theory evolve under a very large imaginary time $L$. The Euclidean partition function in such toroidal geometry is dominated by the ground state energy
\begin{align}
Z(R,L)\approx e^{-LE_0(R)}.\label{L-channel-approx}
\end{align}
On the other hand, relativistic invariance allows the same partition function to be computed in the other channel, where time evolution is along the $R$-circle
\begin{align}
Z(R,L)=\tr e^{-RH(L)}=\sum_{n}e^{-RE_n(L)},\label{R-channel-trace}
\end{align}
where  $H(L)$ and $E_n(L)$ are respectively the Hamiltonian of the theory in volume $L$ and its energy levels. This means that if we succeed at extracting the free energy density at finite temperature $R$ from expression \eqref{R-channel-trace} then the ground state energy would simply be given by 
\begin{align}
E_0(R)=Rf(R). \label{free-energy-R}
\end{align} 
Why is it possible to evaluate the trace in \eqref{R-channel-trace}? The advantage of being  in very large volume $L$ is that we can construct a basis of asymptotic states in which individual particles are well separated. \\
Each asymptotic state is characterized by a set of rapidities $\theta_1,...,\theta_N$ which are subjected to  Bethe equations
\begin{align}
e^{im\sinh(\theta_j)L}\prod_{k\neq j}S(\theta_j-\theta_k)=1,\quad j=1,2,...,N.\label{mirror-BAE}
\end{align}
Once we find a solution of this system of equation, the energy of the corresponding state is given by $E_{|\vec{\theta}\rangle}(L)=m\cosh(\theta_1)+...+m\cosh(\theta_n)$. In this thesis, we will mostly consider \textit{fermionic} Bethe wave functions where the rapidities take pairwise distinguishing values. This happens when  $S(0)=-1$ ($+1$) and the particles are of Bose (Fermi) statistics.  Next, we have to decide which states appear in the sum \eqref{R-channel-trace} i.e. what is a complete basis of the Hamiltonian $H(L)$?  To answer this question, let us write the Bethe equations \eqref{mirror-BAE} in their logarithmic form
\begin{align}
m\sinh(\theta_j)+\frac{1}{L}\sum_{k\neq j}[-i\ln S(\theta_j-\theta_k)]=\frac{2\pi}{L} n_j,\quad j=1,2,...,N.\label{Bethenumbers}
\end{align}
The integers $n_j$ that appear on the right hand side are called Bethe numbers. For some specific models \cite{Yang:1968rm}, it can be rigorously shown that for each set of pairwise distinct integers $\lbrace n_1,...,n_N\rbrace$ the system of equations \eqref{Bethenumbers} admits a unique solution. Furthermore, the entirety of these states form a  complete basis of the Hamiltonian $H(L)$. Normally, we admit this as a hypothesis.

When $L$ tends to infinity, we can characterize the thermodynamic state by a smooth distribution $\rho_\text{p}(\theta)$ of  particle rapidities. In terms of  this distribution, the Bethe equation \eqref{Bethenumbers} reads
\begin{align}
m\sinh(\theta_j)+\int [-i\ln S(\theta_j-\theta)]\rho_\text{p}(\theta)d\theta=\frac{2\pi}{L}n_j.\label{density-cont}
\end{align}
This equation means in particular that each density in the space of rapidity induces a density on the lattice $\mathbb{Z}/L$ of Bethe quantum numbers. Inversely, the density of unoccupied Bethe quantum numbers  induces itself a density $\rho_\text{h}(\theta)$ in the space of rapidity. We refer to $\rho_\text{p}$ and $\rho_\text{h}$ respectively as particles and holes density.
\begin{figure}[ht]
\centering
\includegraphics[width=10cm]{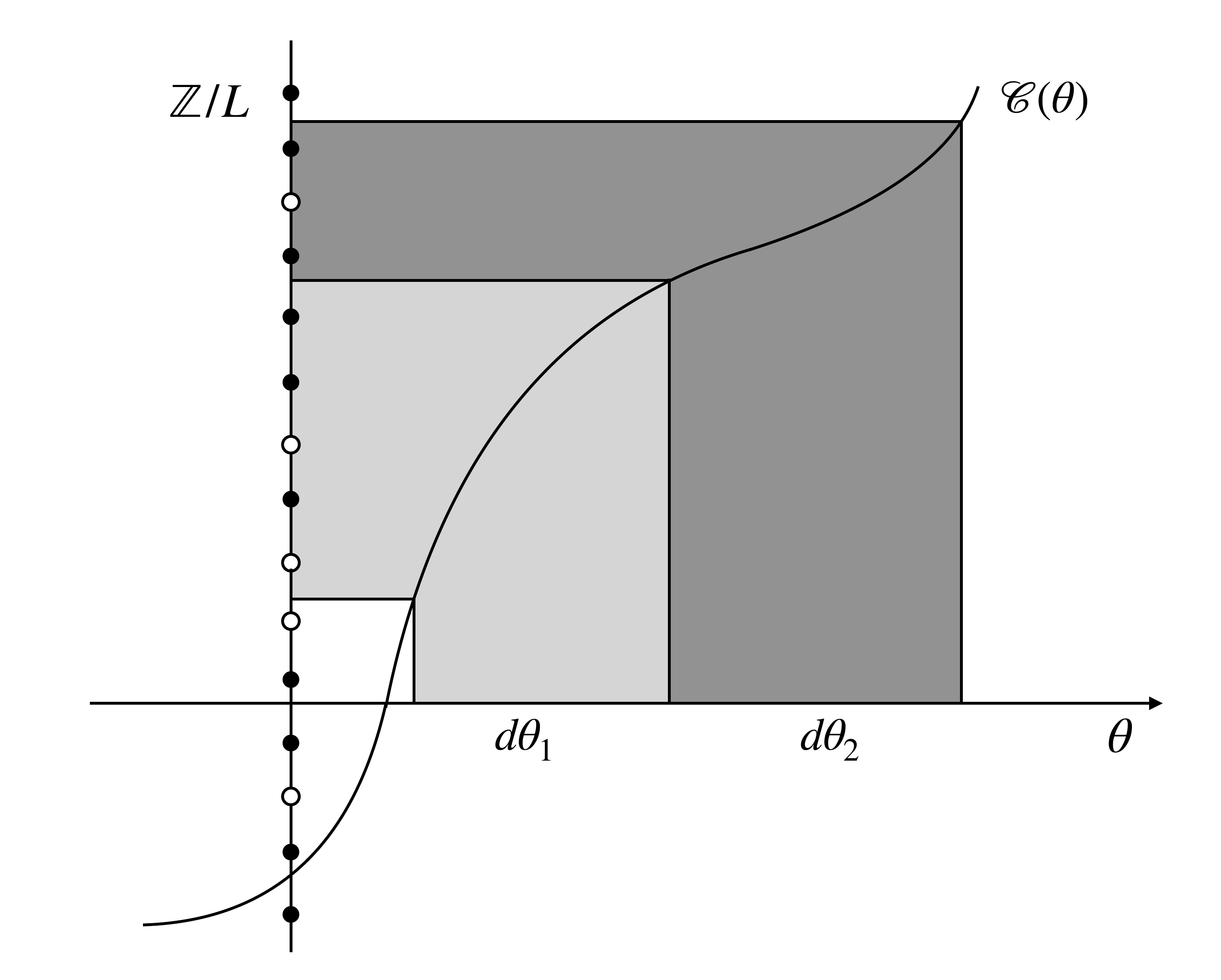}
\caption{Illustration of the counting function.}
\label{counting-function}
\end{figure}

\noindent
To express $\rho_\text{h}$ in terms of $\rho_\text{p}$ we define the counting function
\begin{align}
\mathscr{C}(\theta)=\frac{1}{2\pi}m\sinh(\theta)+\int \frac{d\eta}{2\pi}[-i\log S(\theta-\eta)]\rho_\text{p}(\eta).\label{counting-f}
\end{align}
It is straightforward from the definition that $\mathscr{C}(\theta_j)=n_j$. In the example shown in figure \ref{counting-function} the counting function hovers through the interval $d\theta_1$ and picks out 2 holes and 3 particles, marking 5 sites on the lattice in total. In the interval $d\theta_2$ it encounters only 1 hole and 2 particles. This is beacause the counting function is steeper in $d\theta_1$ than it is in $d\theta_2$. In other words, the derivative of the counting function gives  the total number of states and holes:
\begin{align}
\rho_\text{p}(\theta)+\rho_\text{h}(\theta)=\mathscr{C}'(\theta)=\frac{1}{2\pi}m\cosh(\theta)+\int K(\theta-\eta) \rho_\text{p}(\eta)d\eta,\label{constraint}
\end{align}
where $K(\theta)=-i\partial_\theta \log S(\theta)$.\\
We can then find the particle and hole densities of the thermodynamic state by minimizing the corresponding free energy $E-S/R$. The energy is simply given by
\begin{align}
E=L\int m\cosh(\theta)\rho_\text{p}(\theta)d\theta.
\end{align}
Regarding the entropy, in an infinitesimal interval $d\theta$, the $\rho_\text{p}(\theta)d\theta$ particles  and the $\rho_\text{h}(\theta)d\theta$ holes  can be re-shuffled without changing the energy. The number of ways to shuffling them is equal to the number of anagrams made from $\rho_\text{p}(\theta)d\theta$ letters \textit{s} and $\rho_\text{h}(\theta)d\theta$ letters $h$
\begin{align}
\Omega(d\theta)=\frac{[\rho_\text{p}(\theta)d\theta+\rho_\text{h}(\theta)d\theta]!}{[\rho_\text{p}(\theta) d\theta]![\rho_\text{h}(\theta) d\theta]!}.
\end{align}
We can take the interval $d\theta$ to be very large compared to $1/L$ (but still very small compared to 1) so that the number of states and holes in this interval are sufficiently large. Applyingthe Stirling approximation,  we find that the entropy of the thermodynamic state is given by
\begin{align}
S=L\int_{-\infty}^{+\infty} d\theta [ (\rho_\text{p}+\rho_\text{h})\ln(\rho_\text{p}+\rho_\text{h})-\rho_\text{p}\ln\rho_\text{p}-\rho_\text{h}\ln\rho_\text{h}].
\end{align}
By requiring that the functional derivative of $E-S/R$ with respect to $\rho_\text{p}$ vanishes under the constraint \eqref{constraint}, we find that
\begin{align}
Rm\cosh(\theta)+\ln\frac{\rho_\text{p}(\theta)}{\rho_\text{h}(\theta)}-\int \frac{d\eta}{2\pi}\ln\frac{\rho_\text{p}(\eta)+\rho_\text{h}(\eta)}{\rho_\text{h}(\eta)}K(\eta-\theta)=0.\label{equation-for-rho}
\end{align}
Clearly this equation involves only the relative ratio between the density of particles and holes. This is expected because at the beginning we did not have a chemical potential coupled to the number of particles. Let us define $\epsilon=-\ln(\rho_\text{p}/\rho_\text{h})$, then equation \eqref{equation-for-rho} becomes
\begin{align}
\epsilon(\theta)-Rm\cosh(\theta)+\int \frac{d\eta}{2\pi}\ln[1+e^{-\epsilon(\eta)}]K(\eta-\theta)=0.\label{pseudo-energy}
\end{align}
This is called TBA equation, and  $\epsilon$ is referred to as the pseudo-energy. Once we solve \eqref{pseudo-energy} for $\epsilon$, we can plug it into \eqref{constraint} to find $\rho_\text{p}$ and $\rho_\text{h}$ and deduce other physical quantities from them. It turns out that the free energy can be expressed as a function of the pseudo-energy.  To do this, let us multiply \eqref{pseudo-energy} with $\rho_\text{p}(\theta)$ and carry out the integration over $\theta$
\begin{align}
0=\int d\theta[Rm\cosh(\theta)-\epsilon(\theta)]\rho_\text{p}(\theta)-\int \frac{d\eta}{2\pi}\ln[1+e^{-\epsilon(\eta)}]\int d\theta K(\eta-\theta)\rho_\text{p}(\theta).
\end{align}
Using the density constraint \eqref{constraint}, the integration over $\theta$ on the second term can be obtained
\begin{align}
\int d\theta[Rm\cosh(\theta)-\epsilon(\theta)]\rho_\text{p}(\theta)=\int \frac{d\eta}{2\pi}\ln[1+e^{-\epsilon(\eta)}]\lbrace 2\pi[\rho_\text{p}(\eta)+\rho_\text{h}(\eta)]-m\cosh(\eta)\rbrace.\label{step1}
\end{align}
By using the definition of $\epsilon$ we can write the  free energy in the following form
\begin{align}
Rf(R)=\int\;d\theta[Rm\cosh(\theta)-\epsilon(\theta)]\rho_\text{p}(\theta)-\int\;d\theta[\rho_\text{p}(\theta)+\rho_\text{h}(\theta)]\ln[1+e^{-\epsilon(\theta)}].
\end{align}
By injecting equation \eqref{step1} into the above expression, we find that the only term that survives is 
\begin{align}
Rf(R)=E_0(R)=-\int\frac{d\theta}{2\pi}m\cosh(\theta)\ln[1+e^{-\epsilon(\theta)}].\label{ground-energy}
\end{align}
This is the ground-state energy of the system in volume $R$ that we seek. It can be shown (see for instance \cite{Leclair:1999ys}) that \eqref{ground-energy} is nothing but the free energy at inverse temperature $R$ of a non-interacting theory in which the energy of particles is given by $\epsilon/R$. \\
For \textit{bosonic} Bethe equations, particle rapidities can take coinciding values and similar analysis leads to the bosonic TBA equation and free energy
\begin{align}
\epsilon(\theta)&=Rm\cosh(\theta)+\int \frac{d\eta}{2\pi}\ln[1-e^{-\epsilon(\eta)}]K(\eta-\theta),\\
E_0(R)&=\int\frac{d\theta}{2\pi}m\cosh(\theta)\ln[1-e^{-\epsilon(\theta)}].
\end{align}
Generalization to theories with a non-degenerate mass spectrum $m_1,...,m_N$ is straightforward. Let us denote the S-matrix by $S_{ab}(\theta)$ for $a,b=1,2,...,N$ and $K_{ab}(\theta)=-i\partial_\theta \log S_{ab}(\theta)$. There is a pseudo-energy for each particle type and they are given by a system of $N$ TBA equations
\begin{align}
\epsilon_a(\theta)=Rm_a\cosh(\theta)\mp\sum_{b=1}^N\int\frac{d\eta}{2\pi}\ln[1\pm e^{-\epsilon_b(\eta)}]K_{ba}(\eta-\theta),\quad a=1,...,N.\label{TBA-system}
\end{align}
The ground-state energy of the theory is given by
\begin{align}
E_0(R)=\mp\sum_{a=1}^N m_a\int\frac{d\theta}{2\pi}\cosh(\theta)\ln[1\pm e^{-\epsilon_a(\theta)}].\label{ground-energy-system}
\end{align}
where the upper (lower) sign corresponds to the fermionic (bosonic) case. We stress that the unconventional order of the convolution in \eqref{pseudo-energy} and \eqref{TBA-system} comes directly from the minimization condition of the free energy. The analysis leading to this condition is completely general (it does not rely on any property of the S-matrix) and therefore this order must always be respected in the TBA formalism. At this stage, $K_{ab}(\theta)=K_{ba}(-\theta)$ as a consequence of unitarity  and this remark might seem redundant. As we shall see in section \ref{GN-TBA} however, for theories with non-diagonal scattering, the TBA formalism involves auxiliary particles which do not necessarily respect the unitary condition. In that case, it is of great importance to have the right order of convolution in the TBA equations.
\subsection{UV limit of the ground-state energy}
\label{uv-of-energy-section}
In this subsection we discuss the quantitative behavior of the solution of the TBA equation \eqref{pseudo-energy} as we vary the temperature $1/R$ from zero to infinity.   We cite equation \eqref{pseudo-energy} here for convenience 
\begin{align}
\epsilon(\theta)=Rm\cosh(\theta)-\int \frac{d\eta}{2\pi}K(\eta-\theta)\ln[1+e^{-\epsilon(\eta)}].\label{TBA-equation-original}
\end{align}
When $R$ tends to infinity, the driving term $Rm\cosh(\theta)$ dominates the right hand side of this expression. One can therefore perform a low-temperature limit expansion of the pseudo energy
\begin{align}
\epsilon(\theta)\approx Rm\cosh(\theta)-\int \frac{d\eta}{2\pi}K(\eta-\theta)\ln[1+e^{-Rm\cosh(\eta)}]+...
\end{align}
In the limit $R\to 0$, we know that the ground-state energy must become proportional to the effective central charge of the CFT describing the short-distance behavior of the theory \cite{PhysRevLett.56.742}
\begin{align}
\lim_{R\to 0} RE_0(R)=-\frac{\pi}{6}(c-12\Delta-12\bar{\Delta}),
\end{align}
where $\Delta,\bar{\Delta}$ are lowest dimensions of the CFT, these are zero for unitary theory but negative otherwise. Let us see if the finite-temperature effective central charge obtained from TBA
\begin{align}
c(R)=\frac{3}{\pi^2}\int_{-\infty}^{+\infty}d\theta \log[1+e^{-\epsilon(\theta)}]Rm\cosh(\theta)\label{effective-central-charge}
\end{align}
can be matched with the CFT central charge in the UV limit. By taking the derivative of \eqref{TBA-equation-original} with respect to $\theta$  one sees that the pseudo-energy becomes flat in the interval $|\theta|<< \log (2/Rm)$, as $R\to 0$. Denote this constant value of $\epsilon$ by $\epsilon(\infty)$, which satisfies an algebraic equation
\begin{align}
\epsilon^\text{UV}=-\ln(1+e^{-\epsilon^\text{UV}})\int\frac{d\theta}{2\pi}K(\theta).
\end{align}
It then turns out \cite{Andrews:1984af,Bazhanov:1987zu} that the integral \eqref{effective-central-charge} can be expressed in terms of the so-called Roger dilogarithm function 
\begin{align}
\lim_{R\to 0}c(R)=\textnormal{Li}_{\text{R}}([1+e^{\epsilon^\text{UV}}]^{-1})\quad \text{with}\quad \text{Li}_\text{R}(x)\equiv \frac{6}{\pi^2}[\text{Li}_2(x)+\frac{1}{2}\log(x)\log(1-x)].\label{CFT-limit-of-E0}
\end{align}
Two important properties of this function are
\begin{align}
\text{Li}_\text{R}(x)+\text{Li}_\text{R}(1-x)=1,\quad \text{Li}_\text{R}(x)+\text{Li}_\text{R}(y)=\text{Li}_\text{R}(xy)+\text{Li}_\text{R}\big(\frac{x(1-y)}{1-xy}\big)+\text{Li}_\text{R}\big(\frac{y(1-x)}{1-xy}\big).
\end{align}
Generalization of \eqref{CFT-limit-of-E0} to a theory with $N$ masses takes the form
\begin{align}
\lim_{R\to 0} c(R)=\sum_{a=1}^N \textnormal{Li}_{\text{R}}([1+e^{\epsilon_a^\text{UV}}]^{-1})\quad \text{where}\quad \epsilon_a^\text{UV}=\sum_{b=1}^N -\ln[1+e^{-\epsilon_b^\text{UV}}]\int\frac{d\theta}{2\pi}K_{ab}(\theta).
\end{align}
The conformal limit \eqref{CFT-limit-of-E0} of the ground-state energy provides a substantial check for the S-matrix derived in section \ref{finding-two-particle-S}. Indeed, the inclusion of extra CDD factors will modify the kernel $K$ appearing in TBA equation \eqref{TBA-equation-original} and affect its constant solution. As a result, the high temperature limit of the ground-state energy will not match the effective central charge of the corresponding CFT. This line of idea has been extensively exploited in \cite{Klassen:1989ui}, resulting in interesting relations between the central charge, dilogarithm function and constant solutions of TBA equations. See also \cite{Kirillov:1992tw,Kirillov:1993ih,Kuniba:1992ev} for more in-depth studies on the interplay between dilogarithm and CFT.\\
For illustration purpose  we consider the perturbation of the minimal model $\mathcal{M}_{2,2n+3}$ by its $\Phi_{1,3}$ operator of dimensions $[-(2n-1)/(2n+3),-(2n-1)/(2n+3)]$. This CFT has central charge $-2n(6n+5)/(2n+3)$ and effective central charge $2n/(2n+3)$. The perturbed theory is integrable  with $n$ masses interacting via an elastic S-matrix
\begin{align}
S_{ab}(\theta)=F_{|a-b|/(2n+1)}(\theta)\big[\prod_{k=1}^{\min(a,b)-1}F_{(|a-b|+2k)/(2n+1)}(\theta)\big]^2F_{(a+b)/(2n+1)}(\theta),\quad a,b=1,2,...,n,
\end{align}
where
\begin{align}
F_\alpha(\theta)\equiv \frac{\sinh \theta +i\sin \pi \alpha}{\sinh\theta-i\sin \pi \alpha}.
\end{align}
The constant values of the pseudo-energies are given by
\begin{align}
e^{\epsilon_a^\text{UV}}=\sin\big[a\pi/(2n+3)\big]\sin\big[(a+2)\pi/(2n+3)\big]/\sin^2\big[\pi/(2n+3)\big].
\end{align}
We recover from this solution the effective central charge of $\mathcal{M}_{2,2n+3}$ thanks to the identity
\begin{align}
\sum_{a=1}^n \textnormal{Li}_{\text{R}}\big[\frac{\sin^2(\pi/(2n+3))}{\sin^2((a+1)\pi/(2n+3)}\big]=\frac{2n}{2n+3}.
\end{align}

\subsection{Non-diagonal scattering and string hypothesis}
For theories with non-diagonal S-matrix, the TBA formalism is considerably more complicated. The analysis is usually model dependent and as an example we will derive the TBA equations for the $SU(2)$ chiral Gross-Neveu model. To remind, the spectrum of this theory contains only one particle in the vector multiplet of $SU(2)$ group. The Bethe equations for a wavefunction $|\theta_1,\theta_2,...,\theta_N\rangle$ in a periodic space of length $L$ have been derived in section \ref{SU(2)-Bethe}. We cite equations \eqref{SU(2)-physical} and \eqref{SU(2)-auxiliary} here for the ease of following 
\begin{align}
1&=e^{ip(\theta_j)L}\prod_{k\neq j}^NS_0^\text{SU(2)}(\theta_j-\theta_k)\prod_{m=1}^M
\frac{\theta_j-u_m+i\pi/2}{\theta_j-u_m-i\pi/2},\quad j=\overline{1,N},\label{physical-eq}\\
1&=\prod_{j=1}^N\frac{u_k-\theta_j-i\pi/2}{u_k-\theta_j+i\pi/2}\prod_{l\neq k}^M\frac{u_k-u_l+i\pi}{u_k-u_l-i\pi},\quad k=\overline{1,M}.\label{aux-eq}
\end{align}
The next step in the TBA formalism is to find a complete set of solutions of this system. For theories with diagonal S-matrix described in the previous section, we simply took all the (physical) rapidities  to be real. Here the situation is different: even if the physical rapidities are real,  the auxiliary ones can take complex values. For instance, there are complex solutions for auxiliary rapidities when $N=5$ and $M=2$. Let us assume the existence of these solutions when the number of physical rapidities grows arbitrarily large. Without loss of generality,  suppose that there exists  a state with $\Im(u_1 ) > 0$. Then the first product in \eqref{aux-eq} would vanish as $N$ tends to infinity.
The only way to compensate this zero is by having a pole in one of the terms in the second product of \eqref{aux-eq}. In other words, there must be another auxiliary rapidity, say $u_2$ with $u_2=u_1-i\pi$.\\
The appearance  of the rapidity $u_2$ has resolved the problem with the Bethe equation for $u_1$, but another issue arises. By multiplying the Bethe equation for $u_1$ and $u_2$ to eliminate their singular terms and by denoting $u_{12}=(u_1+u_2)/2$, we find that
\begin{align}
1=\prod_{j=1}^N\frac{u_{12}-\theta_j-i\pi}{u_{12}-\theta_j+i\pi}\prod_{l=3}^M\frac{u_{12}-u_l+3i\pi/2}{u_{12}-u_l-i\pi/2}\frac{u_{12}-u_l+i\pi/2}{u_{12}-u_l-3i\pi/2}.\label{string-1-2}
\end{align}
There are two scenarios: if $u_{12}$ is real then we can consider \eqref{string-1-2} as part of our system of Bethe equations, replacing the original equations of $u_1$ and $u_2$. Once we solve it for $u_{12}$ then $u_1$ and $u_2$ are given by $u_{1,2}=u_{12}\pm i\pi/2$.
On the other hand, if the imaginary part of $u_{12}$ is not zero then  the above arguments that have been applied for $u_1$ can now be applied for  $u_{12}$ as well . If $\Im(u_{12})>0$ then there must be a pole in the second product of \eqref{string-1-2}. This pole can not be at $u_l=u_{12}-i\pi/2$ however because in that case $u_l$ would coincide with $u_2$. Therefore there must be a rapidity, say $u_3$ which is equal to $u_{12}-3i\pi/2$. Inversely, if $\Im(u_{12})<0$ then $u_3=u_{12}+3i\pi/2$. We can now repeat the above procedure, multiplying the Bethe equations for $u_1,u_2,u_3$ together and write it in terms of the new rapidity $u_{123}=(u_1+u_2+u_3)/3$. Either $u_{123}$ is real or the procedure continues and we generate a bigger configuration.

To summarize, we have found that when the number of physical rapidities tends to infinity, if there are complex auxiliary rapidities then they must organize themselves into string patterns. A string is characterized by its real center $u$ and its length (number of its constituents)
\begin{align}
\lbrace u_Q\rbrace\equiv \lbrace u-(Q+1-2j)i\pi/2\;|\;j=1,2,...,Q\rbrace.
\end{align}
For the construction of string solutions in other model, see for instance  \cite{takahashi_1999} for the XXZ spin chain, or \cite{Saleur:2000bq} and \cite{Arutyunov:2012zt} for more sophisticated structures.

Now if we want to take the thermodynamic limit, we must send not only the number of physical particles but also the number of auxiliary magnons to infinity as well. In that regime, there is a plot hole in our analysis: the infinite product of magnon S-matrices with complex rapidities can mimic the role of a pole and our construction of magnon strings is no longer justified. Although solutions of this type (and other singular structures) can exist we might assume that they are rather uncommon. In other words, we can say that most solutions are of string type or more precisely, the dominant contribution to the free energy comes mostly from them. For example, in the XXX spin chain there exist solutions that do not approach the expected string forms in the thermodynamic limit \cite{Woynarovich:1981ca,Woynarovich_1982,Babelon:1982mc}, but the free energy is  correctly recovered by taking only string configurations into consideration \cite{doi:10.1080/00018738300101581}. The process of constructing string solutions and the assumption that they are the only  solutions of Bethe equations that are relevant in the thermodynamic limit is known in the literature as \textit{the string hypothesis}. For a detailed discussion and further references on the string hypothesis, see for instance \cite{doi:10.1142/9789812798268_others01}.\\
With the string hypothesis, a relevant system of Bethe equations for the $SU(2)$ chiral Gross-Neveu model involves: $N_0$ number of real physical rapidities $\theta_j$ with $j=\overline{1,N_0}$, $N_Q$ number of $Q$-string real centers $u_{Q,l};\;l=\overline{1,N_Q}$ for each $Q$ from $1$ to $+\infty$ 
\begin{align}
-1&=e^{ip(\theta_j)L}\prod_{k=1}^{N_0}S_{00}(\theta_j-\theta_k)\prod_{Q=1}^{+\infty}\prod_{l=1}^{N_Q}S_{0Q}(\theta_j-u_{Q,l}),\quad j=\overline{1,N_0},\label{SU(2)-physical-str}\\
(-1)^Q&=\prod_{j=1}^{N_0}S_{Q0}(u_{Q,l}-\theta_j)\prod_{P=1}^{+\infty}\prod_{m=1}^{N_P}S_{QP}(u_{Q,l}-u_{P,m}),\quad Q=\overline{1,+\infty}, l=\overline{1,N_Q}.\label{SU(2)-aux-str}
\end{align}
where the scattering phases involving strings are, by construction, the products of the scattering phases of their constituents
\begin{align}
S_{00}(\theta)&\equiv S_0^{\text{SU(2)}}(\theta)=-\frac{\Gamma(1-\theta/2\pi i)}{\Gamma(1+\theta/2\pi i)}\frac{\Gamma(1/2+\theta/2\pi i)}{\Gamma(1/2-\theta/2\pi i)},\nonumber\\
S_{0Q}(\theta-u_Q)&\equiv \prod_{u_j\in \lbrace u_Q\rbrace}\frac{\theta-u_j+i\pi/2}{\theta-u_j-i\pi/2},\quad S_{Q0}(u_Q-\theta)=\prod_{u_j\in \lbrace u_Q\rbrace}\frac{u_j-\theta-i\pi/2}{u_j-\theta+i\pi/2},\label{S-strings}\\
S_{PQ}(u_Q-u_P)&\equiv \prod_{j\in \lbrace u_Q\rbrace}\prod_{k\in\lbrace u_P\rbrace}\frac{u_j-u_k+i\pi}{u_j-u_k-i\pi}.\nonumber
\end{align}
We note that the total number of auxiliary particles i.e. $\sum_Q QN_Q$ should not exceed half the number of physical particles. \\
This construction is reminiscent of the S-matrix bootstrap procedure discribed in section \ref{two-S}. In the XXX spin chain language, the strings can be interpreted as bound states, having less energy than the total energy of individual real magnons. Furthermore, building the S-matrix between these bound states from the S-matrix between fundamental excitations can be regarded as spin chain equivalence of the fusion method. Despite the analogies, one should not take this idea too seriously. For instance, our choice of the "S-matrix" between physical rapidity and strings clearly violates unitarity: $S_{0Q}(\theta)S_{Q_0}(-\theta)\neq 1$.  As we shall explain in the next section, the reason for this choice of "S-matrix" is purely technical. With that being said, it is important to keep in mind that auxiliary particles are mathematical artifacts in the TBA formalism and they are not subjected to any physical constraint \textit{a priori}.

\subsection{TBA equations and  Y-system for SU(2) chiral Gross-Neveu model}
\label{GN-TBA}
\begin{figure}[ht]
\centering
\includegraphics[width=8cm]{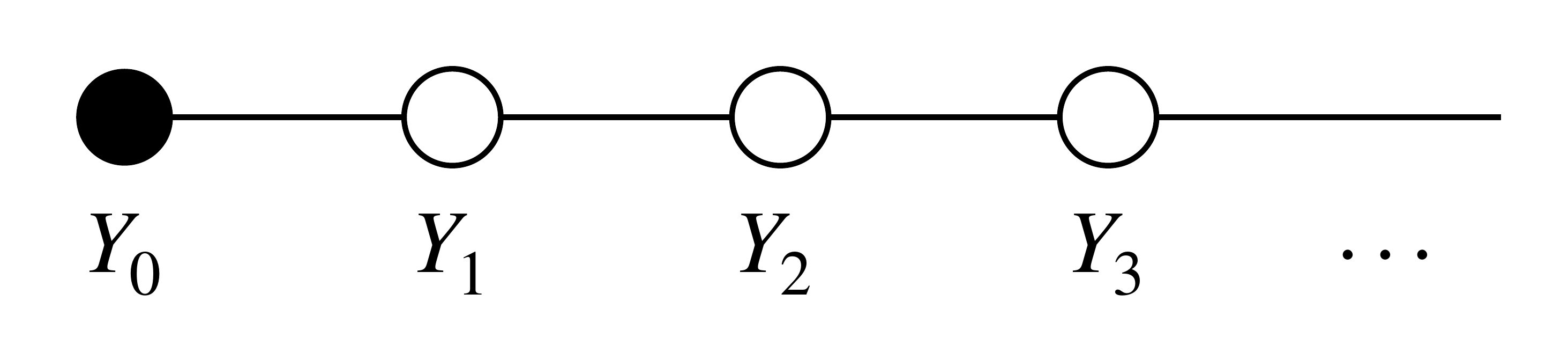}
\caption{The Y-system of $SU(2)$ chiral Gross-Neveu model, the black node stands for the physical Y-function while the $n^\text{th}$ white node represents that of the auxiliary $n$-string.}
\label{Y-of-SU(2)}
\end{figure}

With the Bethe equations \eqref{SU(2)-physical-str} and \eqref{SU(2)-aux-str} at our disposal, we can proceed to the next steps in the TBA formalism: taking logarithm of Bethe equations to find Bethe quantum numbers, passing to the continuum limit and  defining the counting function for each species of particle. Most of these steps are mechanical: one basically treats the theory as having an infinite number of particles interacting via elastic S-matrices given in \eqref{S-strings}. There is however a subtlety in defining the counting function for auxiliary strings that is worth mentioning. To guarantee that the density of states and holes are positive, the counting function must always be monotonically increasing (see equation \eqref{constraint}). For the physical particle, this poses no problem as the particle momentum serves as the leading term for the corresponding counting function \eqref{counting-f}. For auxiliary strings, the leading term is the spin chain momentum. The problem is that there are two equivalent ways of defining this quantity without modifying physically-relevant quantities. These two choices are of opposite values and only one of them  \footnote{one could nevertheless work with the other, the price to pay is that the TBA equations cannot be written in a uniform fashion} is an increasing function of rapidity: the one given  in \eqref{S-strings}. Take for instance 1-string then
\begin{align}
\partial_u p(u)>0\quad \text{with}\quad e^{ip(u)N}\equiv \prod_{j=1}^N\frac{u-\theta_j-i\pi/2}{u-\theta_j+i\pi/2}.
\end{align}
With this remark, we are now ready to write down the TBA equations for the theory. There is an infinite number of pseudo energies: $\epsilon_0$ for the physical rapidity and $\epsilon_Q$ for $Q$-string for $Q=\overline{1,\infty}$
\begin{align}
\epsilon_n(\theta)=Rm\cosh(\theta)\delta_{n,0}-\sum_{m=0}^\infty \log[1+e^{-\epsilon_m}]\star K_{mn} (\theta),\quad n=\overline{0,\infty},\label{TBA-equations-of-SU(2)}
\end{align}
where $K_{nm}(\theta)=-i\partial_\theta \log S_{nm}(\theta)$. The free energy only depends explicitly on the physical pseudo-energy
\begin{align}
Rf(R)=-\int\frac{d\theta}{2\pi}m\cosh(\theta)\log[1+e^{-\epsilon_0(\theta)}].\label{free-energy-density-su(2)}
\end{align}
The system  \eqref{TBA-equations-of-SU(2)} of infinitely many coupled equations can be cast into an equivalent form called Y-system, which has the advantage of being local
\begin{align}
\log Y_n+RE\delta_{n,0}=s\star [\log(1+Y_{n+1})+\log(1+Y_{n-1})],\quad n=\overline{0,\infty}.\label{simplified-SU(2)-TBA}
\end{align}
In this expression $Y_0=e^{-\epsilon_0}$, $Y_n=e^{\epsilon_n}$ for $n\geq 1$,  $Y_{-1}=0$ and $s$ is a simple kernel given in appendix \ref{GN-TBA-appendix}. The structure Y-systems can usually be represented  by graphs. In our case, it is the infinite Dynkin diagram of A-type, see figure \ref{Y-of-SU(2)}. 

\begin{figure}
\centering
\includegraphics[width=7cm]{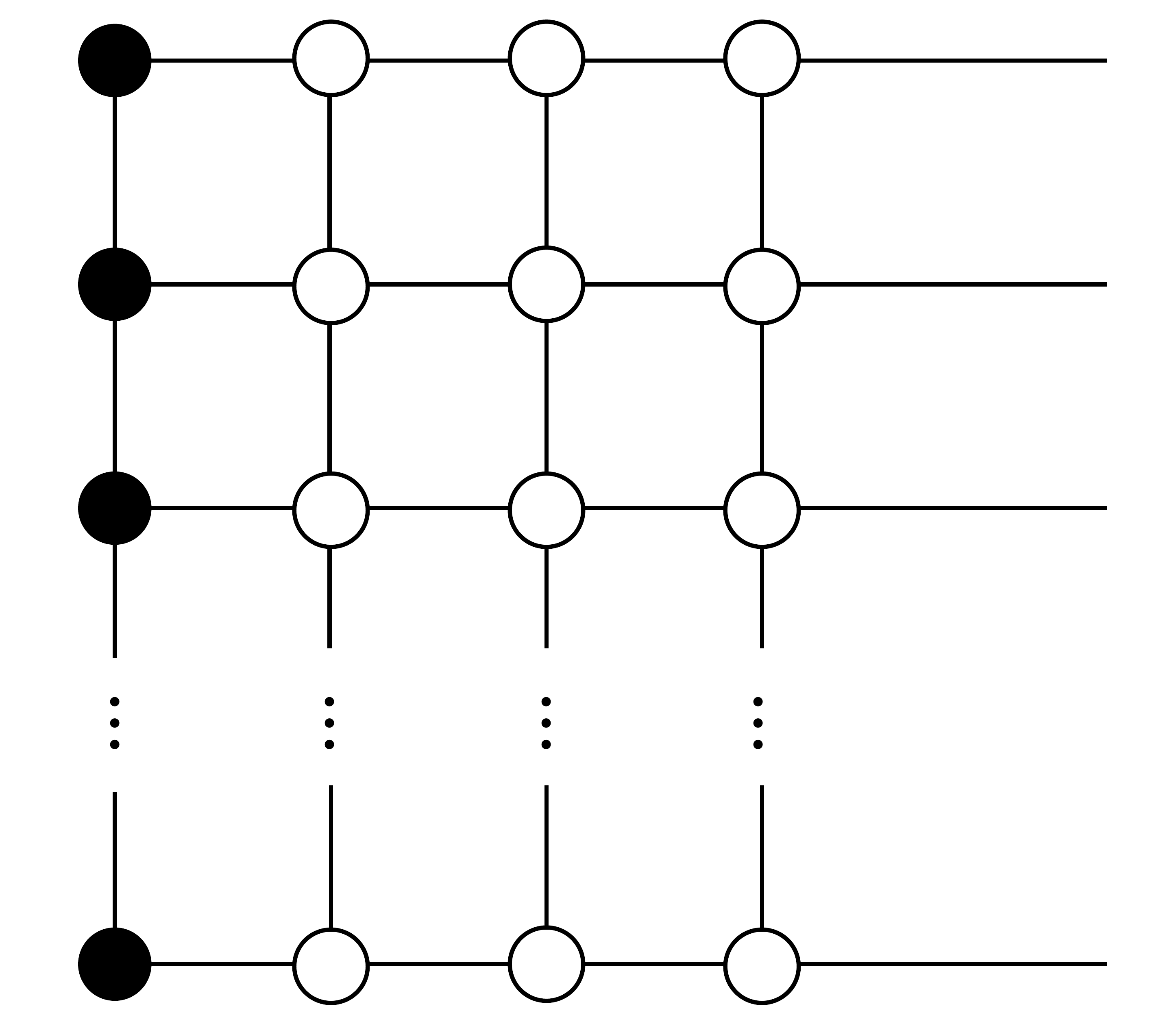}
\caption{Y-system of $SU(N)$ chiral Gross-Neveu model. There are $N-1$ rows and an infinite number of columns.}
\label{SU(N)-Y}
\end{figure}

In general the Y-functions live on a direct product of two Dynkin diagrams:  a finite one which represents the symmetry of the model and an infinite one which encodes the irreducible representations of this symmetry. For instance the Y-system of the $SU(N)$ chiral Gross-Neveu model is shown in figure \ref{SU(N)-Y}. To make the group theoretical meaning behind these graphs more precise, we recall that auxiliary strings of $SU(2)$ model are bound states of the XXX spin chain. A bound state of $Q$ constituents carry a spin of $Q/2$ and can be identified with an irreducible representations of $SU(2)$. The same structure holds for higher ranks: Y-functions correspond to inequivalent non-singlet irreducible representations of $SU(N)$, represented by Young diagrams of maximal height $N-1$. The totality of these diagrams form the grid in figure \ref{SU(N)-Y} if we draw a square around every node. For an in-depth review of Y-systems, see for instance \cite{Kuniba:2010ir}.

Let us come back to the $SU(2)$ chiral Gross-Neveu model and look at its Y-system in more details. Two special regimes are of interest. At zero temperature the solutions of TBA equations become constants
\begin{gather}
Y_n^{\text{IR}}=(n+1)^2-1,\quad n\geq 0.\label{IR-GN-not-twist}
\end{gather}
As we turn on the temperature, a plateau structure starts to develop for each Y-function inside the region  from $-\log[2/(mR)]$ to $\log[2/(mR)]$. Ouside of this region  Y-functions retain their IR values while the tops of the plateaux flatten out at height
\begin{gather}
Y_n^{\text{UV}}=(n+2)^2-1,\quad n\geq 0.\label{UV-GN-not-twist}
\end{gather}
There are two consistency checks for these stationary solutions. In the zero temperature limit one would expect the behavior of a non-interacting gas. In particular, as the physical particle belongs to the vector representation of $SU(2)$, the leading contribution to the free energy should be
\begin{align*}
\lim _{R\to 0}Rf(R)=-2\int\frac{d\theta}{2\pi}m\cosh(\theta)e^{-Rm\cosh(\theta)}.
\end{align*} 
By replacing the leading term of the physical Y-function into the expression \eqref{free-energy-density-su(2)} we see that this is indeed the case\footnote{For higher order matching between TBA and Luscher correction, see \cite{Ahn:2011xq}}. At the UV, the Casimir energy computed from TBA should match the central charge of the unperturbed CFT, as explained in subsection \ref{uv-of-energy-section}
\begin{gather}
\lim_{R\to 0}c(R)=\sum_{n\geq 0}\text{Li}_\text{R}(\frac{1}{1+Y_n^{\text{IR}}})-\sum_{n\geq 0}\text{Li}_\text{R}(\frac{1}{1+Y_n^{\text{UV}}})=1.
\end{gather}
The particle densities can also be easily computed in the UV limit \cite{Klassen:1990dx}. Let us denote by $D_0=N_0/L$ the density of physical particle and $D_a=N_a/L$ that of string of length a, then
\begin{align*}
\lim_{R\to 0} \pi RD_0(R)=\log(1+Y_0^{\text{UV}}),\quad \lim_{R\to 0} \pi RD_a(R)=\log(1+Y_a^{\text{IR}})-\log(1+Y_a^{\text{UV}}).
\end{align*}
We find that the density of physical particle is exactly twice the total density of auxiliary particles in this limit
\begin{align*}
\lim_{R\to 0} \pi RD_0(R)=\log 4,\quad \lim_{R\to 0}\sum_{a=1}^\infty \pi R a D_a(R)=\log 2.
\end{align*}

For later discussions of g-function, we will add to the TBA equations \eqref{TBA-equations-of-SU(2)} a chemical potential coupled to the $SU(2)$ symmetry. Denote by $\mu$ the chemical potential of the physical particle. In the spin chain language, $\mu$ can be thought of as the strength of an external magnetic field. The auxiliary particle corresponds to spin flipping and is assigned a chemical potential of $-2\mu$. A string of $n$ auxiliary particles have chemical potential $-2n\mu$. The TBA equations of chiral $SU(2)$ Gross-Neveu now read
\begin{gather}
\begin{aligned}
\log Y_0(\theta)&=-Rm\cosh(\theta)+\mu+\sum_{n=0}^{\infty} K_{n,0}\star\log(1+Y_n)(\theta),\\
\log  Y_n(\theta)&=-2n\mu+\sum_{m=0}^\infty K_{m,n}\star\log(1+Y_m)(\theta),\quad n\geq 1\;.
\end{aligned}
\end{gather}
This inclusion of chemical potential does not affect the structure of Y system. It does affect however the asymptotic values of Y-functions. Write $2\mu=-\log \kappa$ with $\kappa$ usually known as the twist parameter. The IR and UV values of Y-functions are given by (to be compared with \eqref{IR-GN-not-twist} and \eqref{UV-GN-not-twist})
\begin{gather}
1+\mathcal{Y}_n^{\text{IR}}(\k)=[n+1]_{\k}^2,\quad 1+\mathcal{Y}_n^{\text{UV}}(\k)=[n+2]_{\k}^2\; ,\label{IR-UV-GN-twist}
\end{gather}
where the $\k$-quantum numbers are defined as
\begin{align*}
[n]_\k\equiv (1+\k+...+\k^{n-1})/\k^{(n-1)/2}.
\end{align*}
We can repeat the above analysis for this twisted theory. At zero temperature, the double degeneracy of up/down spin is lifted 
\begin{align*}
Y_0^{\text{IR}}(u)=e^{-Rm\cosh(u)}\sqrt{1+\mathcal{Y}_1^{\text{IR}}(\k)}=[2]_\k e^{-Rm\cosh(u)}.
\end{align*}
In the UV limit the particle densities are now given by
\begin{align*}
\lim_{R\to 0} \pi RD_0(R,\mu)=2\log (1+\k)-\log \k,\quad \lim_{R\to 0}\sum_{a=1}^\infty \pi R a D_a(R,\mu)=\log (1+\k).
\end{align*}
The scaled free energy density
\begin{align*}
c(R,\mu)\equiv -\frac{6R^2f(R)}{\pi}=\frac{3}{\pi^2}\int mR\cosh(\theta) \log[1+Y_0(\theta)]d\theta-\frac{6}{\pi}\sum_{a=0}^\infty\mu_a RD_a(R,\mu),
\end{align*}
where $\mu_0=\mu,\;\mu_n=-2n\mu$ for $n\geq 1$, can again be computed in the UV limit with help of Roger dilogarithm function
\begin{align}
\lim_{R\to 0}c(R,\mu)=1-\frac{6\mu^2}{\pi^2}.
\end{align}

The Y-system is derived from the TBA equations and as such contains \textit{a priori} less information. This is indeed the case as we have shown that the TBA equations with and without a chemical potential give rise to the same Y-system.  If one wishes to only use the Y-system to describe thermodynamic quantities then one must provide extra information on the Y-functions. As we saw above, this could be their IR values, or equivalently their large $\theta$ asymptotics. 
\subsection{Excited state energies from analytic continuation}
\label{excited-section}
In this subsection we discuss the extension of the TBA method to include excited state energies.  This seems impossible at first regard, given the unique footing of the ground state energy in the mirror trick proposed by Zamolodchikov. In \cite{Dorey:1996re} Dorey and Tateo proposed an alternate route to bypass this barrier: an analytic continuation in some parameter of the theory that connects different energy levels. This approach appeared in fact earlier in the case of the quantum anharmonic oscillator \cite{PhysRev.184.1231}. Let us illustrate the idea on toy model 
\begin{align}
H\psi =E\psi \quad \text{with}\quad H=\begin{pmatrix}
1&0 \\
0&-1
\end{pmatrix}+\lambda\begin{pmatrix}
0&1\\
1&0
\end{pmatrix}.\label{quantum-mechanical-problem}
\end{align}
The spectrum of this system is very simple: the ground state energy is $− \sqrt{1 + \lambda^ 2}$ and the only excited state energy is  $\sqrt{1+\lambda^2}$. If we allow  the \textit{coupling constant} $\lambda$ to take complex values then the ground state energy has branch points at $\lambda = \pm i$. As a consequence, if we start with a real-valued coupling constant and we go around one of these branch points and come back to our starting point then the energy flips its sign and we end up with the excited state energy.

The problem is not so simple in the TBA formalism, because we do not have an explicit formula for the ground state energy. It is instead expressed in terms of the pseudo energy, which in turn is determined by an integral equation \eqref{pseudo-energy}. Let us mimic such situation by rewriting the ground state energy of the toy model as an integral
\begin{align}
E(\lambda)=-\int_{-1}^1\frac{dz}{2\pi i}f(z)g(z)-1,\label{integral-rep-ground}
\end{align}
where
\begin{align}
f(z)=\frac{1}{z-i/\lambda} \quad \text{and}\quad  g(z)=2\lambda\sqrt{1-z^2}
\end{align}
respectively playing the role of $\ln(1+Y)$ and the energy in the expression of TBA ground state energy \eqref{ground-energy}. 
\begin{figure}
\centering
\begin{subfigure}{.5\textwidth}
  \centering
  \includegraphics[width=.9\linewidth]{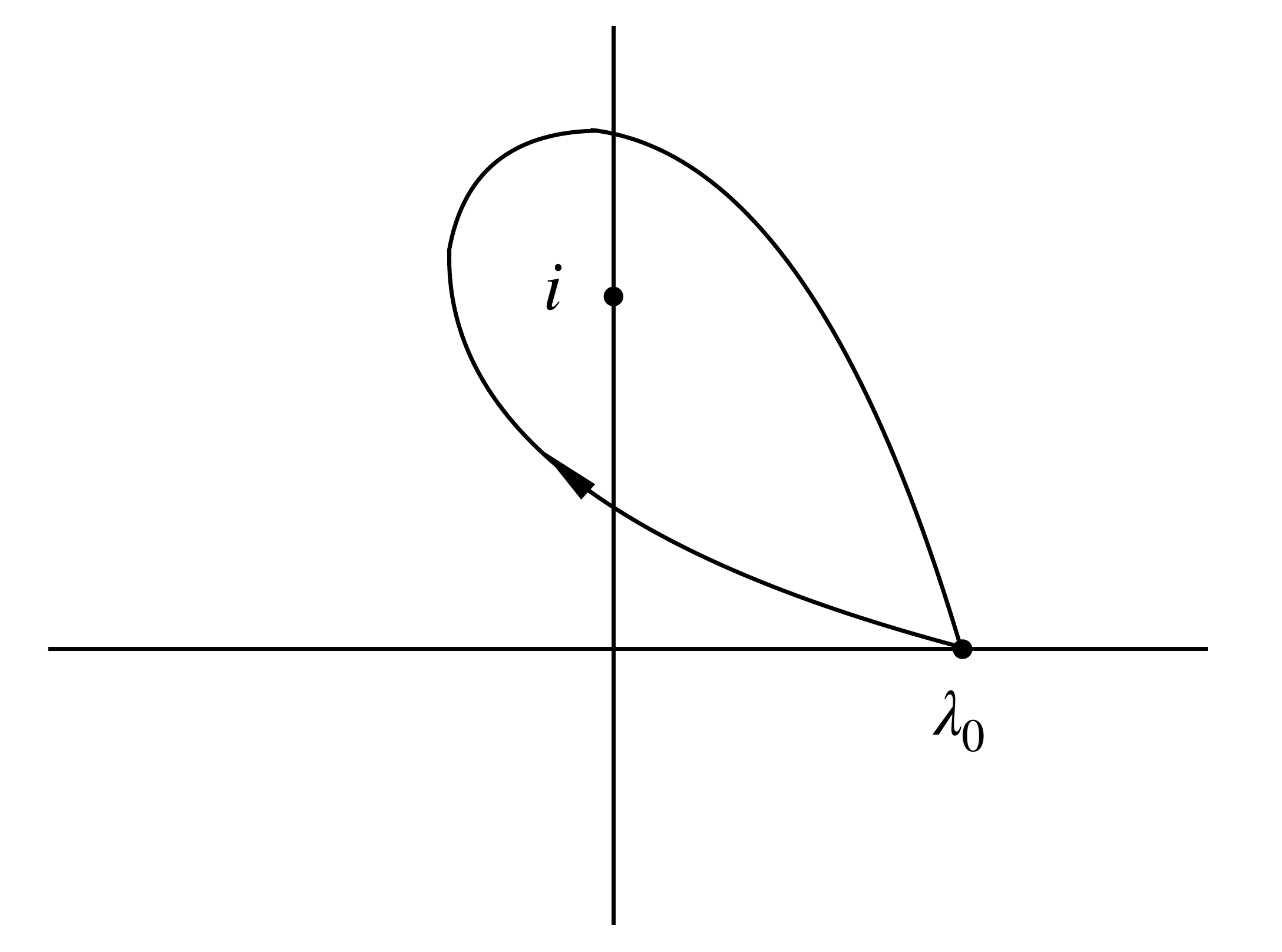}
  \caption{Movement of $\lambda$}
  \label{move-of-lambda}
\end{subfigure}%
\begin{subfigure}{.5\textwidth}
  \centering
  \includegraphics[width=.9\linewidth]{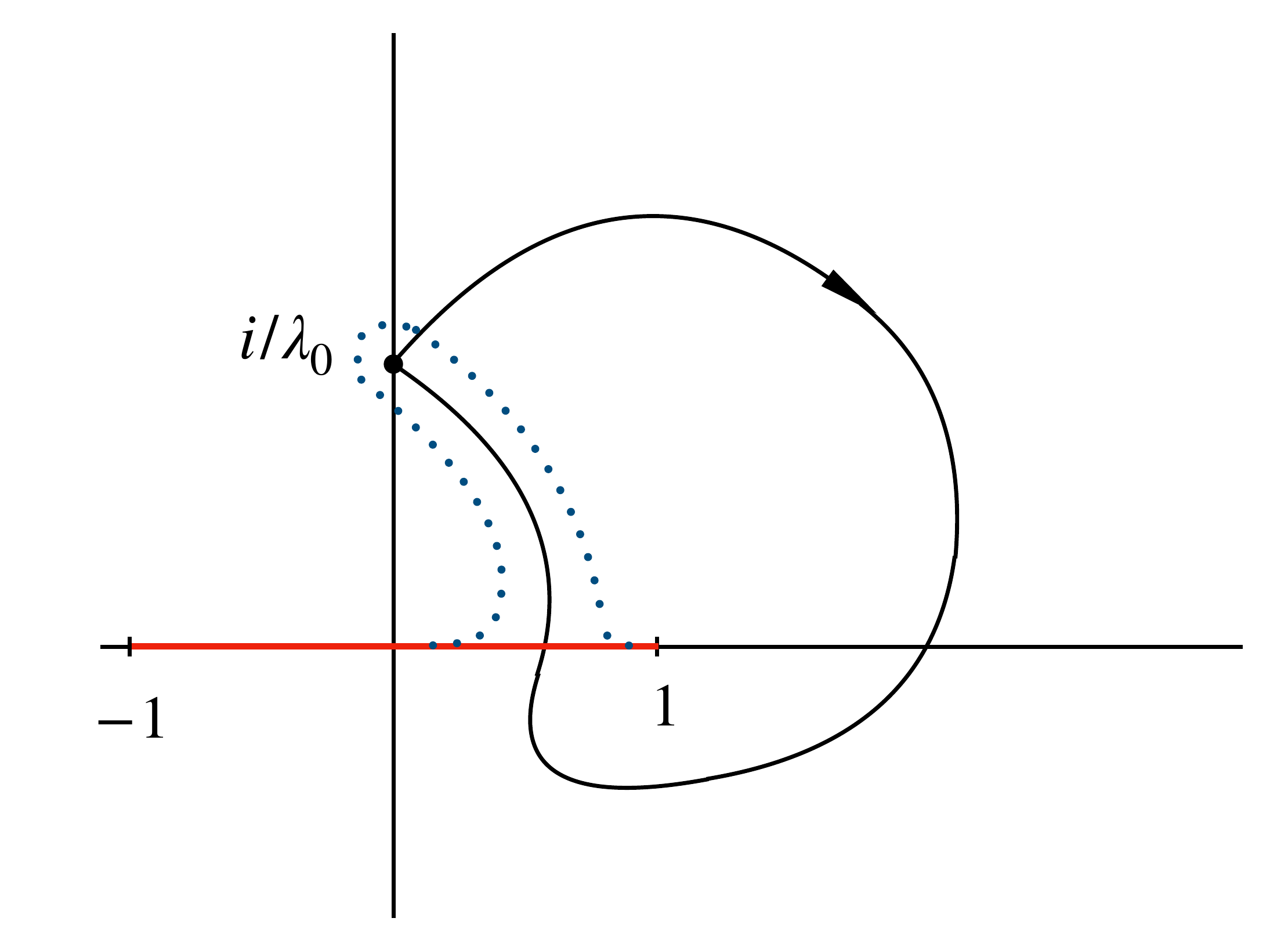}
  \caption{Movement of the pole $i/\lambda$}
  \label{move-of-pole}
\end{subfigure}
\caption{(a) When $\lambda$ is analytically continued around a branch point, the energy flips its sign. (b) The corresponding movement of the pole at $i/\lambda$. At some point, it crosses the integration contour and drags it along for continuity. Once the integration contour is taken back to the real line we gain a residual contribution.}
\label{fig:test}
\end{figure}

Let us analytically continue the expression \eqref{integral-rep-ground} to complex-valued $\lambda$. The integral is well-defined as long as the pole $i/\lambda$ is found outside of the integration domain. We start with a real-valued $\lambda_0$ of the coupling constant and move it around one of the branch point, for instance $\lambda=+i$. The only issue we encounter along the trajectory is when  we cross the imaginary axis at second time. There, the pole enters the integration contour, drags it and wraps it around the original pole $i/\lambda_0$. Upon coming back to our starting point, we pick up the residue at this pole and thus end up with the excited state energy
\begin{align}
E^c(\lambda_0)=E(\lambda_0) + g(i/\lambda_0).
\end{align}
In this procedure the explicit expressions of $f$ was not used. All we need to know is the relative position of its pole with respect to the integration contour of $E(\lambda)$. 

The same idea can be used to find excited state energies from the ground state one
\begin{align}
E_0(R)=\int\frac{d\theta}{2\pi}m\sinh(\theta)\frac{\partial_\theta Y(\theta)}{1+Y(\theta)},
\end{align}
where we have performed an integration by parts. We start by parametrizing the original TBA equation by some real parameter $\lambda$, for instance the inverse temperature $R$ or the mass scale of the theory.  Let us assume that in the process of analytically continuing this variable  we encounter some poles $\theta_j^*(\lambda)$ such that $Y(\theta_j^*)=-1$. If we move on the complex plane in such a way that some of these points cross the real line, then upon coming back, we will pick up their residues
\begin{align}
E^c(R)=i\sum m\sinh(\theta_j^*)-\int\frac{d\theta}{2} m\cosh(\theta)\log[1+Y^c(\theta)].\label{excited-energy}
\end{align}
The original TBA equation 
\begin{align}
\log Y(\theta)=-Rm\cosh(\theta)-\int\frac{d\eta}{2\pi i}\log S(\eta-\theta)\frac{\partial_\eta Y(\eta)}{1+Y(\eta)}.
\end{align}
is likewise affected by this analytic continuation 
\begin{align}
\log Y^c(\theta)=-Rm\cosh(\theta)-\sum_j\log S(\theta_j^*-\theta)+\int\frac{d\eta}{2\pi}K(\eta-\theta)\log[1+Y^c(\eta)].\label{excited-pseudo-energy}
\end{align}
The interpretation of excited energy \eqref{excited-energy} and the excited TBA equation \eqref{excited-pseudo-energy} will become clear if we define $\tilde{\theta}_j^*=\theta_j^*+i\pi/2$. Then $i\sinh(\theta_j^*)=\cosh(\tilde{\theta}_j^*)$ and $-\cosh(\theta_j^*)=i\sinh(\tilde{\theta}_j^*)$. The leading order in the large volume expansion of the equation $Y^c(\theta_j^*)=-1$ which determines the positions of the singular points is written in terms of these new variables as
\begin{align}
e^{im\sinh(\tilde{\theta}_j^*)R}\prod_k S(\tilde{\theta}_j^*-\tilde{\theta}_k^*)=-1,
\end{align}
which is nothing but the Bethe equation \eqref{mirror-BAE} at volume $R$. Furthermore, in this limit, the excited state energy \eqref{excited-energy} is simply 
\begin{align}
E^c(R)=\sum_j m\cosh(\theta_j^*).
\end{align}
That is, $\tilde{\theta}_j^*$ are precisely the rapidities of some state living in the mirror theory. The convoluted terms in the excited state energy \eqref{excited-energy} and excited TBA equation \eqref{excited-pseudo-energy} correspond to finite size corrections to the asymptotic expressions of this theory.

If we were to cross the real axis from the other direction then the sign in front of $im\sinh(\theta_j^*)$ and $\log S(\theta_j-\theta)$ in these equations would change. In that case, the mirror-physical conversion is defined as $\tilde{\theta}_j^*=\theta_j^*-i\pi/2.$

We note however that none of these two transformations is actually the \textit{honest} mirror transformation. A real mirror transformation interchanges space and time via a double Wick rotation $(x,t)\to (-it,ix)$ and thus $(p,E)\to (iE,-ip)$. It reads in terms of rapidity $\theta\to i\pi/2-\theta$. The above transformations are combinations of the honest mirror transformation with either the parity transformation or the time reversal transformation. By abuse of language and for simplicity we call them mirror transformation, however we will specify the rapidity conversion each time the terminology is used. The real mirror transformation will be used in a heuristic argument to obtain the average current in Generalized Hydrodynamics, see subsection \ref{equations-state-section}.
\newpage
\section{Application of form factors in finite volume}
The idea of the form factors program is construct fundamental building blocks for correlation functions in integrable quantum field theories. These building blocks can be determined from symmetry, their analytic structure and the knowledge of the two-particle S-matrix. The program was initially formulated for relativistic quantum field theories with excitations over the vacuum \cite{Smirnov:1992vz}. For the construction of form factors in $\mathcal{N}=4$ SYM see for instance \cite{Basso:2015zoa,Bajnok:2015hla}. More recently, there are propositions to study form factors built on top of a thermodynamic state \cite{Cubero:2018vyb, 2018JSMTE..03.3102D}. In this thesis, we focus on the applications of form factors in finite volume. 

We restrict our discussion to theories with a non-degenerate mass spectrum and we use the index $a$ to denote particle types.
\subsection{Motivation and axioms}
Facing the problem of computing a correlation function  $\langle \mathcal{O}_1(x_1)\mathcal{O}_2(x_2)...\mathcal{O}_n(x_n)\rangle$, one can take a reductionist stance, inserting as many resolutions of identity 
\begin{align}
1=\sum_n\sum_{a_1,...,a_n}\int\frac{d\theta_1...d\theta_n}{n!(2\pi)^n}|\theta_1,...,\theta_n\rangle_{a_1,...,a_n}{}_{a_1,...,a_n}\langle \theta_1,...,\theta_n|\label{resolution-identity}
\end{align}
as it takes to bring the problem down to two smaller tasks. The first task is to evaluate matrix elements of the type
\begin{align}
{}_{a'_1,...,a'_m}\langle \theta_1',...,\theta_m'|\mathcal{O}(0,0)|\theta_{1},...,\theta_{n}\rangle_{a_1,...,a_n}\equiv F^\mathcal{O}{}_{a'_1,...,a'_m;a_1,...,a_n}(\theta_1',...,\theta_m'\;|\;\theta_{1},...,\theta_{n}).\label{form-factor-generalized}
\end{align}
which are called the \textit{form factors} of the operator $\mathcal{O}$. The second task  is to perform the infinite sum involving these form factors. Some concrete examples following this line of idea can be found in \cite{Wu:1975mw,Babelon:1992sn,CastroAlvaredo:2000nk} for the Ising model and in \cite{Orland:2014mya,Cubero:2014sxa} for the Principal chiral model at large $N$. In this subsection we will only present the properties of form factors that are relevant to this thesis. For an in-depth review, see \cite{Smirnov:1992vz}.

\begin{figure}[ht]
\centering
\includegraphics[width=14cm]{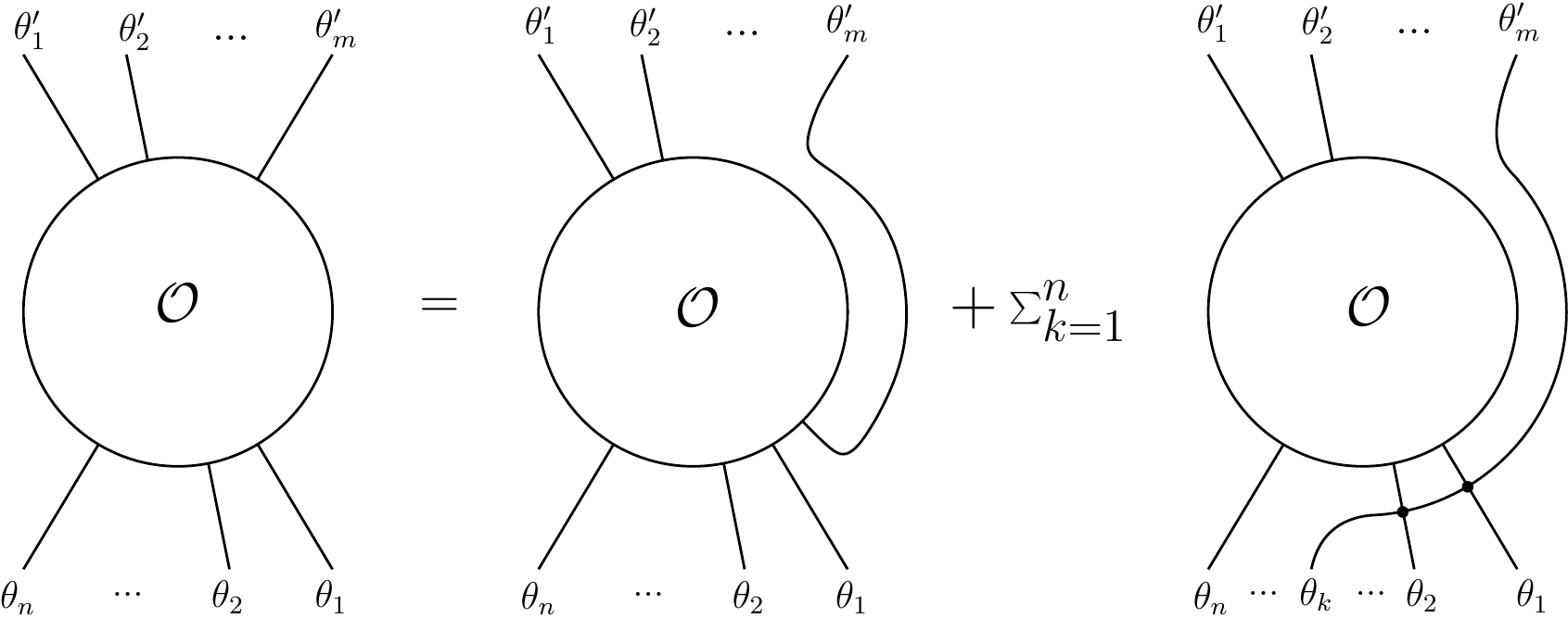}
\caption{The crossing relation \eqref{form-factor-crossing}.}
\label{crossing-FF-relation}
\end{figure}

As physical processes involve real rapidities, form factors are \textit{a priori} functions of real variables. However we can consider their analytic continuation on the complex plane, like we did with the two-particle S-matrix. The first simplification we can do with  \eqref{form-factor-generalized} is to move some rapidity from the bra to the ket. It can be shown (see for instance \cite{Babujian:1998uw}) that
\begin{align}
&F^\mathcal{O}{}_{a'_1,...,a'_m;a_1,...,a_n}(\theta_1',...,\theta_m'|\theta_{1}...\theta_{n})\nonumber\\
=&F^\mathcal{O}{}_{a'_1,...,a'_{m-1};a'_m,a_1,...,a_n}(\theta_1',...,\theta_{m-1}'|\theta'_m+i\pi,\theta_{1}...,\theta_{n})+\sum_{k=1}^n\delta_{a'_ma_k}\delta(\theta'_m-\theta_k)\nonumber\\
\times &\prod_{l=1}^{k-1}S_{a_la_k}(\theta_l-\theta_k)F^\mathcal{O}_{a'_1,...,a'_{m-1};a_1,...,a_{k-1},a_{k+1},...,a_n}(\theta_1',...,\theta_{m-1}'|\theta_{1}...,\theta_{k-1},\theta_{k+1},...,\theta_{n}),\label{form-factor-crossing}
\end{align}
This is called the \textit{crossing relation} and is illustrated in figure \ref{crossing-FF-relation}. 

\noindent
For $m=1$ there are two equivalent ways of writing the crossing relation
\begin{align}
&F^\mathcal{O}{}_{a'_1,;a_1,..,a_n}(\theta_1'|\theta_{1}..\theta_{n})\nonumber\\
=&F^\mathcal{O}{}_{a'_1,a_1,..,a_n}(\theta'_1+i\pi,\theta_{1},..,\theta_{n})+\sum_{k=1}^n\delta_{a'_1a_k}\delta(\theta'_1-\theta_k)\nonumber\\
\times &\prod_{l=1}^{k-1}S_{a_la_k}(\theta_l-\theta_k)F^\mathcal{O}_{a_1,...,a_{k-1},a_{k+1},...,a_n}(\theta_{1}...,\theta_{k-1},\theta_{k+1},...,\theta_{n}),\label{form-factor-crossing-2}\\
=&F^\mathcal{O}{}_{a_1,...,a_n,a'_1}(\theta_{1}...,\theta_{n},\theta'_1-i\pi)+\sum_{k=1}^n\delta_{a'_1a_k}\delta(\theta'_1-\theta_k)\nonumber\\
\times&\prod_{l=k+1}^{n}S_{a_la_k}(\theta_l-\theta_k)F^\mathcal{O}_{a_1,...,a_{k-1},a_{k+1},...,a_n}(\theta_{1}...,\theta_{k-1},\theta_{k+1},...,\theta_{n}).
\label{form-factor-crossing-3}
\end{align}
By  successively applying   \eqref{form-factor-crossing},  we can express any  form factor  in terms of elementary ones
\begin{align}
F^\mathcal{O}_{a_1,...,a_n}(\theta_1,...,\theta_n)\equiv\langle 0|\mathcal{O}(0,0)|\theta_1,...,\theta_n\rangle_{a_1,...,a_n}.\label{form-factor-elementary}
\end{align}
Elementary form factors that differ only in the order of their rapidities are related by the S-matrix
\begin{align}
F^\mathcal{O}_{a_1,..,a_j,a_{j+1},..,a_n}(\theta_1,..,\theta_j,\theta_{j+1},..,\theta_n)=S_{a_ja_{j+1}}(\theta_j-\theta_{j+1})F^\mathcal{O}_{a_1,..,a_{j+1},a_j,..,a_n}(\theta_1,..,\theta_{j+1},\theta_j,..,\theta_n).\label{ff-interchange}
\end{align}
By comparing the analytic part of \eqref{form-factor-crossing-2} and \eqref{form-factor-crossing-3}  we obtain the \textit{cyclic permutation} property
\begin{align}
F^\mathcal{O}_{a_1,...,a_n}(\theta_1+i\pi, ...,\theta_n)=F^\mathcal{O}_{a_2,...,a_n,a_1}(\theta_2,...,\theta_n,\theta_1-i\pi).\label{ff-periodic}
\end{align}
To evaluate the residue at the kinematical pole  let us write
\begin{align}
F_{a_1,a_2,...,a_n}(\theta_1+i\pi,\theta_2,...,\theta_n)\approx\frac{\Res_{\theta_{1}=\theta_2+i\pi}F^\mathcal{O}_{a_1,...,a_n}(\theta_1,\theta_2,...,\theta_n)}{\theta_1-\theta_2-i\epsilon},\label{residue-ff-1}\\
F_{a_2,...,a_n,a_1}(\theta_2,...,\theta_n,\theta_1-i\pi)\approx\frac{\Res_{\theta_{1}=\theta_2+i\pi}F^\mathcal{O}_{a_1,...,a_n}(\theta_1,\theta_2,...,\theta_n)}{\theta_1-\theta_2+i\epsilon}.\label{residue-ff-2}
\end{align}
Plugging \eqref{residue-ff-1} into \eqref{form-factor-crossing}, \eqref{residue-ff-2} into  \eqref{form-factor-crossing-2} and comparing their $\delta$-function parts we find 
\begin{align}
\Res_{\theta_{1}=\theta_2+i\pi}F^\mathcal{O}_{a_1,...,a_n}(\theta_1,\theta_2,...,\theta_n)=2iF^\mathcal{O}_{a_3,...,a_n}(\theta_3,...,\theta_n)\big[1-\delta_{a_1a_2}\prod_{j=3}^nS_{a_2a_j}(\theta_2-\theta_i)\big].\label{kinematical-pole}
\end{align}
This is usually served as a recursive relation between $n-$ and $(n-2)-$particle elementary form factors. Like the two-particle S-matrix, form  factors also have poles related to bound states. However these poles are not relevant to this thesis.
\begin{figure}[ht]
\centering
\includegraphics[width=10cm]{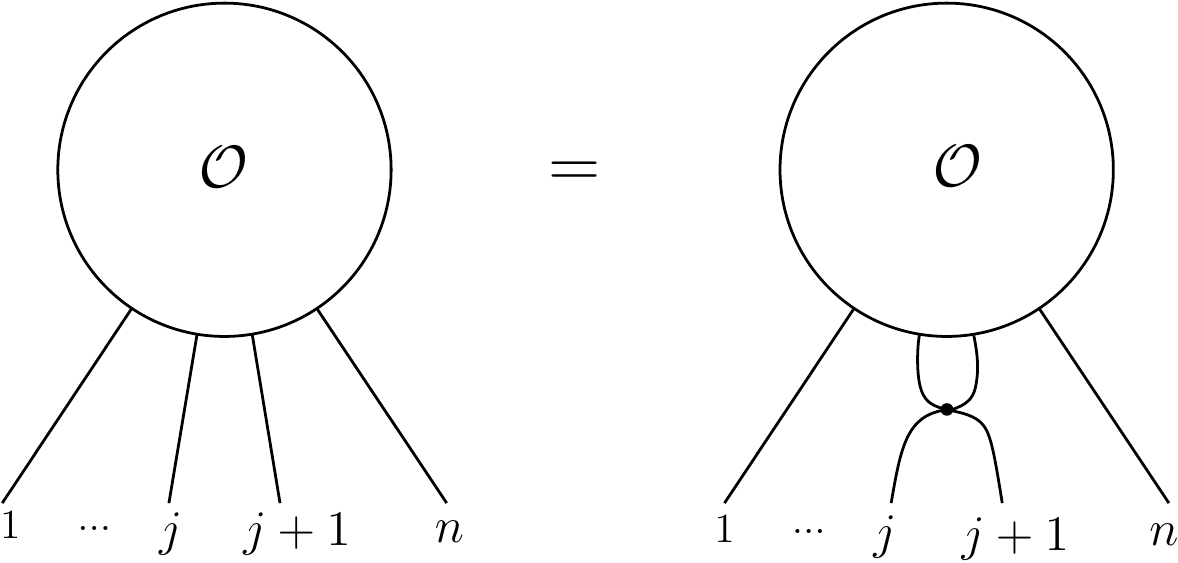}
\caption{The exchange relation \eqref{ff-interchange}}
\end{figure}
\begin{figure}[h]
\centering
\includegraphics[width=13.5cm]{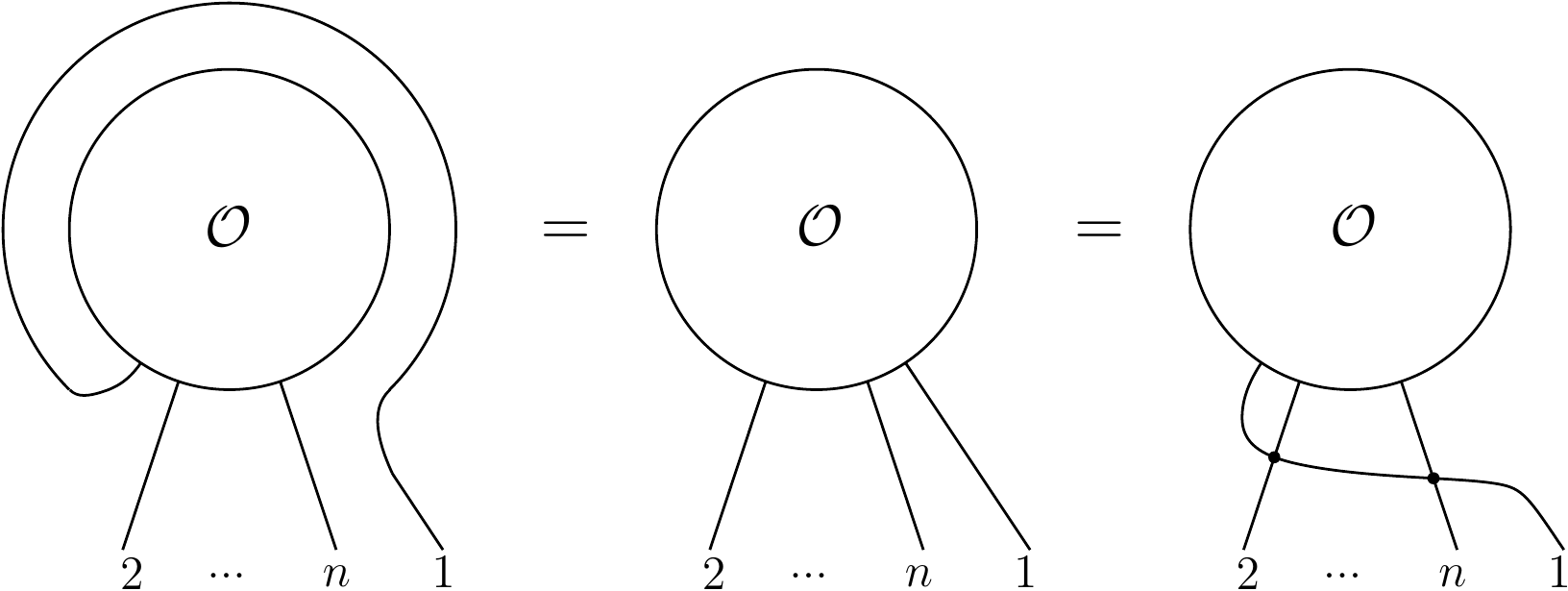}
\caption{The cyclic permutation property \eqref{ff-periodic}}
\end{figure}

\noindent

Finally, we note that if $\mathcal{O}$ carries a Lorenz charge $s$ then under a Lorenz boost $\Lambda$ the form factors of $\mathcal{O}$ transform as  $F^\mathcal{O}_{\vec{a}}(\theta_1+\Lambda,...,\theta_n+\Lambda)=e^{s\Lambda}F^\mathcal{O}_{\vec{a}}(\theta_1,...,\theta_n)$. In particular, if $\mathcal{O}$  is a scalar then $F^\mathcal{O}_{\vec{a}}(\theta_1,...,\theta_n)$ is a function of rapidity differences.

For notational simplicity, we consider in the following theories with only one particle type. Generalization to those with more than one particle type is straightforward.
\subsection{Connected and symmetric evaluation of diagonal form factors}
\label{conn-symm-section}
Once we have obtained the elementary form factors, we can in use the crossing relation \eqref{form-factor-crossing}, \eqref{form-factor-crossing-2} to write down any form factor. Subtlety arises however when some rapidities in the bra coincide with some in the ket of \eqref{form-factor-generalized}. Of relevance to the following sections are the diagonal form factors $F^\mathcal{O}(\theta_n,...,\theta_1\;|\;\theta_1,...,\theta_n)$. 

To avoid singularity in the crossing relation, one can regularize the corresponding elementary form factor by shifting each rapidity $\theta_j$ by a small amount $\epsilon_j$. It turns out that this limit depends on how the regulators are taken to zero. Generally one has the following structure \cite{Delfino:1996xp}
\begin{align}
F^\mathcal{O}(\theta_1+i\pi+\epsilon_1,...,\theta_n+i\pi+\epsilon_n,\theta_n,...,\theta_1)=\prod_{j=1}^n\frac{1}{\epsilon_j}\sum_{j_1=1}^n...\sum_{j_n=1}^na_{j_1...j_n}\epsilon_{j_1}...\epsilon_{j_n}+...
\end{align}
where $a_{j_1...j_n}$ is a totally symmetric tensor and the ellipsis denote terms that vanish when $\epsilon_i$'s tend to zero. The connected evaluation of the diagonal matrix element $\langle \theta_n,...,\theta_1|\theta_1,...,\theta_n\rangle$ is then defined as $F_\text{c}^\mathcal{O}(\theta_1,...,\theta_n)\equiv n!a_{12...n}$. We will refer to it simply as the \textit{connected form factor}. On the other hand, the \textit{symmetric form factor} is obtained when all the regulators are equal
\begin{align}
F_\text{s}^\mathcal{O}(\theta_1,...,\theta_n)=\lim_{\epsilon\to 0}F^\mathcal{O}(\theta_1+i\pi+\epsilon,...,\theta_n+i\pi+\epsilon,\theta_n,...,\theta_1)=\sum_{j_1=1}^n...\sum_{j_n=1}^na_{j_1...j_n}.
\end{align}
To see the relation between these two form factors let us consider as an example $n=2$ where
\begin{gather*}
F(\theta_1+i\pi+\epsilon_1,\theta_2+i\pi+\epsilon_2,\theta_2,\theta_1)=\frac{1}{\epsilon_1\epsilon_2}(a_{11} \epsilon_1^2+2a_{12}\epsilon_1\epsilon_2+a_{22}\epsilon_2^2),\\
F_\text{c}(\theta_1,\theta_2)=2a_{12},\quad F_\text{s}(\theta_1,\theta_2)=a_{11}+2a_{12}+a_{22}.
\end{gather*} 
The kinematical pole at fixed $\epsilon_2$ and $\epsilon_1=0$ is prescribed by relation \eqref{kinematical-pole}
\begin{align*}
\Res_{\epsilon_1=0}F(\theta_1+i\pi+\epsilon_1,\theta_2+i\pi+\epsilon_2,\theta_2,\theta_1)=i[1-S(\theta_1-\theta_2)S(\theta_1-\theta_2-i\pi-\epsilon_2)]F(\theta_2+i\pi+\epsilon_2,\theta_2)
\end{align*}
Developing the right hand side to first order in $\epsilon_2$ we obtain 
\begin{align*}
a_{22}=K(\theta_2-\theta_1)F_\text{c}(\theta_2)\quad \text{with}\quad K(\theta)=-i\partial_\theta\log S(\theta). 
\end{align*}
Similarly  $a_{11}=K(\theta_1-\theta_2)F_\text{c}(\theta_1)$ so that
\begin{align*}
F_\text{s}(\theta_1,\theta_2)=K(\theta_2-\theta_1)F_\text{c}(\theta_2)+K(\theta_1-\theta_2)F_\text{c}(\theta_1)+F_\text{c}(\theta_1,\theta_2)
\end{align*}
More generally, a symmetric form factor can be expressed in terms of connected form factors with smaller numbers of particles. The exact relation was found by induction in \cite{Pozsgay:2007gx}  
\begin{align}
F_\text{s}^\mathcal{O}(\theta_1,...,\theta_n)=\sum_{\alpha\subset \lbrace 1,2,..,n\rbrace}\mathcal{L}(\alpha|\alpha)F_\text{c}^\mathcal{O}(\lbrace \theta_j\rbrace_{j\in\alpha}).\label{symm-conn}
\end{align}
In this equation the sum runs over non-empty subsets of $\lbrace 1,2,..,n\rbrace$ and $\mathcal{L}(\alpha|\alpha)$ denotes the principal minor obtained by deleting the $\alpha$ rows and columns of the following matrix 
\begin{equation}
L(\theta_1,...,\theta_n)_{jk}=\delta_{jk}\sum_{l\neq j}K(\theta_j-\theta_l)-(1-\delta_{jk})K(\theta_j-\theta_k).\label{Laplacian-def}
\end{equation}
\subsection{Finite volume matrix elements}
The  form factor formalism is plagued with divergences and we see an example in the previous subsection. The existence of divergences is  due to the fact that  form factors are constructed in infinite volume. As a mean to regularize these divergences, one can first define and study form factors in finite volume, before sending the volume to infinity. This approach has proven to be not only a physically meaningful regularization but also a suitable method to obtain thermal observables as the TBA formalism itself is constructed in finite volume. 

In volume $L$, multi-particle states are labeled by a set of Bethe quantum numbers $I_1,...,I_n$. The particle rapidities are related to the quantum numbers through Bethe equations
\begin{align}
2\pi I_j=\Phi_j(\vec{\theta})\equiv mL\sinh(\theta_j)-i\sum_{k\neq j}\log S(\theta_j-\theta_k)\quad \text{for}\quad j=\overline{1,n}.
\end{align}
The Gaudin matrix corresponding to this state is defined as the Jacobian matrix of the change of variables from quantum numbers to rapidities 
\begin{align}
G(\vec{\theta})_{jk}\equiv \frac{\partial \Phi_j(\vec{\theta)}}{\partial \theta_k}=\delta_{jk}[mL\cosh(\theta_j)+\sum_{k\neq j} K(\theta_j-\theta_k)]-(1-\delta_{jk})K(\theta_j-\theta_k).\label{Gaudin-definition}
\end{align}
Notice that this is the sum of a diagonal matrix and the Laplacian matrix \eqref{Laplacian-def} that appeared in the previous section, relating  connected and symmetric form factors in infinite volume. In some spin chains, the determinant of this matrix is proportional to the norm of Bethe wave function \cite{Korepin:1982gg}. 

Our aim is to relate the finite matrix element $ \langle I_1',...,I_m'|\mathcal{O}|I_{1},...,I_{n}\rangle$  to its infinite counterpart  \eqref{form-factor-generalized}. If the two sets $\lbrace \theta'_1,...,\theta'_m\rbrace$ and  $\lbrace \theta_1,...,\theta_n\rbrace$ are disjoint then there is no singularity in the crossing relation and their relation is quite simple \cite{Pozsgay:2007kn}
\begin{align}
\langle \vec{I}'|\mathcal{O}(0)|\vec{I}\rangle_L=\frac{F^\mathcal{O}(\theta'_m+i\pi,...,\theta'_1+i\pi,\theta_1,..,\theta_n)}{\sqrt{\det G(\vec{\theta'})\det G(\vec{\theta})}}+O(e^{-mL}).\label{finite-non-diagonal}
\end{align}
This expression directly follows the comparison of two point functions in finite and infinite volume. When the two sets of rapidities are identical, Saleur \cite{Saleur:1999hq} conjectured that
\begin{align}
\langle \vec{I}'|\mathcal{O}|\vec{I}\rangle_{L}=\frac{1}{\det G(\vec{\theta})}\sum_{\alpha\in \lbrace 1,2,...,n\rbrace}F^\mathcal{O}_\text{c}(\lbrace \theta_j\rbrace_{j\in\alpha})\det G(\vec{\theta})_{\alpha|\alpha}+O(e^{-mL}),\label{diagonal-sum-conn}
\end{align}
where $\det G(\vec{\theta})_{\alpha|\alpha}$ is the principal minor obtained by selecting the $\alpha$-indexed rows and columns of the full Gaudin matrix \eqref{Gaudin-definition}. This formula is physically intuitive if we interpret $\det G(\vec{\theta})_{\alpha|\alpha}$ as  the norm of the partial state $\lbrace \theta_j\rbrace, j\in \alpha$ in the presence of other particles. 
\begin{figure}[ht]
\centering
\includegraphics[width=12cm]{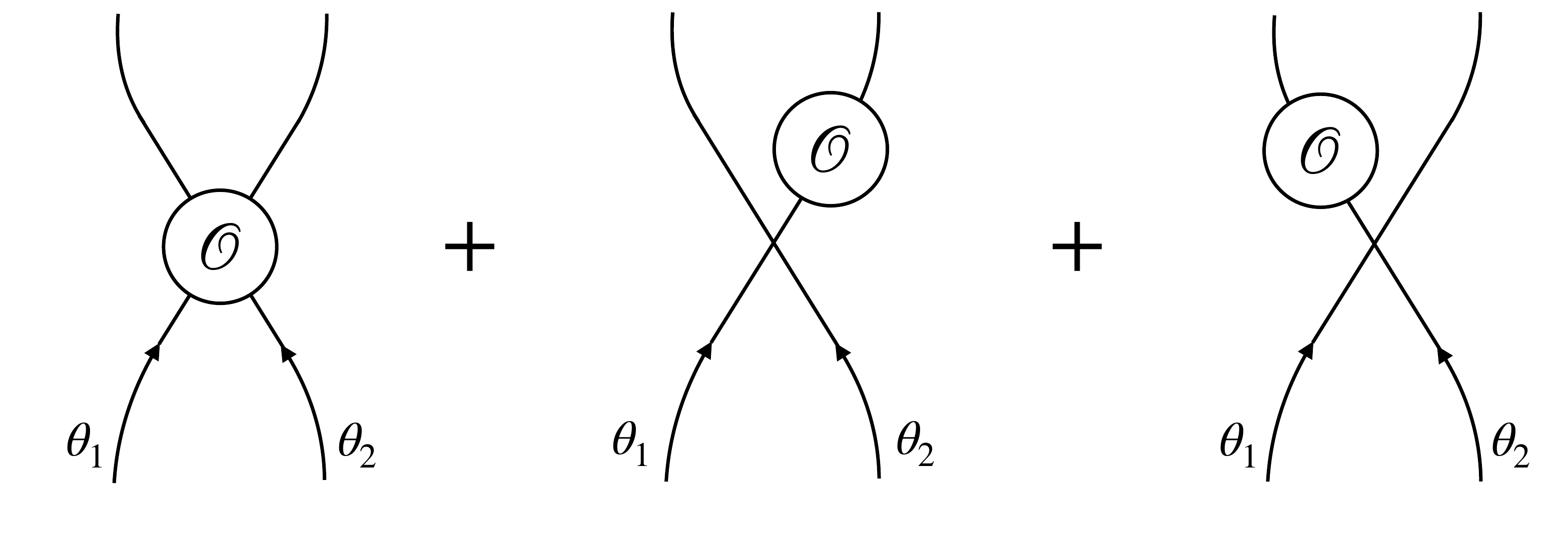}
\caption{Saleur's interpretation of the connected form factors. His proposal reads in the case of two particles $\langle \theta_2,\theta_1|\mathcal{O}|\theta_1,\theta_2\rangle=F_\text{c}^\mathcal{O}(\theta_1,\theta_2)+F_\text{c}^\mathcal{O}(\theta_1)\langle \theta_2|\theta_2\rangle_{\theta_1}+F_\text{c}^\mathcal{O}(\theta_2)\langle \theta_1|\theta_1\rangle_{\theta_2}$.}
\end{figure}

Using the relation \eqref{symm-conn} between connected and symmetric form factors, Pozsgay and Takacs showed \cite{Pozsgay:2007gx}  that \eqref{diagonal-sum-conn} is equivalent to
\begin{align}
\langle \vec{I}'|\mathcal{O}|\vec{I}\rangle_{L}=\frac{1}{\det G(\vec{\theta})}\sum_{\alpha\subset\lbrace 1,2,...,n\rbrace}F^\mathcal{O}_\text{s}(\lbrace \theta_j\rbrace_{j\in\alpha})\det G(\lbrace \tilde{\theta}_j\rbrace_{j\in{\bar{\alpha}}})+O(e^{-mL}),\label{diagonal-sum-symm}
\end{align}
where $G(\lbrace \tilde{\theta}_j\rbrace_{j\in{\bar{\alpha}}})$ is the Gaudin matrix of the subset of rapidities $\theta_j, j\in\bar{\alpha}$. The expression \eqref{diagonal-sum-conn} was rigorously proven by Bajnok and Wu \cite{Bajnok:2017bfg}. Their idea is to add a probe rapidity to the diagonal matrix element to make it non-diagonal and thus the formula \eqref{finite-non-diagonal} applies. When this rapidity is sent to infinity, one can use the asymptotic of the S-matrix and the clustering property of form factors to recover \eqref{diagonal-sum-conn}. 

\subsection{Leclair-Mussardo formula}
Despite its success in computing zero-temperature correlation functions, the form factor formalism faces serious challenges when it comes to finite-temperature observables. If one wishes to use the formalism to evaluate the following correlation function
\begin{align}
\langle \mathcal{O}_1(x_1)\mathcal{O}_2(x_2)...\mathcal{O}_n(x_n)\rangle_R=\frac{1}{Z}\tr [e^{-R\mathcal{H}}\mathcal{O}_1(x_1)\mathcal{O}_2(x_2)...\mathcal{O}_n(x_n)]\label{finite-temperature-correlation}
\end{align}
then after the insertions of identity \eqref{resolution-identity}, one has to carry out an infinite sum over form factors along with their respective thermal weight. Even without the form factors, summing over the thermal weight i.e. computing the partition function $Z$ itself is already a non-trivial task. While $Z$ can be computed using TBA, the traditional derivation relies on a thermodynamic state that minimizes the free energy rather than an honest summation. Due to this difficulty, only one point functions have so far been obtained by form factors. The expression goes under the name of Leclair-Mussardo  series \cite{Leclair:1999ys} 
\begin{equation}\label{lm}
\langle \mathcal{O}\rangle_R=\sum_{n=0}^\infty\frac{1}{n!}\int\frac{d\theta_1}{2\pi}...\frac{d\theta_n}{2\pi}\prod_{j=1}^nf(\theta_j) F_{\text{c}}^\mathcal{O}(\theta_1,...,\theta_n),
\end{equation}
where $f=1/(1+e^{\epsilon})$ is the TBA filling factor.  In practice, the formula is used with truncating after some terms, if excitation is small enough, this provides a fairly good approximation of the one point function. The validity of \eqref{lm} was first confirmed for non-interacting theories in \cite{Leclair:1996bf}. The authors of \cite{Leclair:1999ys} then argued that the effect of interaction can be mimicked by a mere replacement of bare energy by the TBA pseudo-energy and thus \eqref{lm} was conjectured to also hold for interacting theories. They also proposed a similar expression for two point functions, but given the daring nature of their argument, it is not surprising that their proposal was proven to be wrong \cite{Saleur:1999hq,CastroAlvaredo:2002ud,Doyon:2005jf,Doyon:2006pv,Chen_2014}.

Saleur gave a simple and yet convincing evidence \cite{Saleur:1999hq} to support \eqref{lm}. He considered the case where the operator is given by the density of some conserved charge $\Q=\int{\rm d}x\,Q(x,t)$. In this situation the corresponding one point function is well-known in the TBA formalism (see subsection \ref{equations-state-section} for more details)
\begin{equation}
\langle Q\rangle=\int\frac{{\rm d}p(\theta)}{2\pi}f(\theta)q^{\rm dr}(\theta),\label{TBA-one-point-charge}
\end{equation}
where $q(\theta)$ is the one-particle eigenvalue of $\Q$: $\Q|\theta\rangle=q(\theta)|\theta\rangle$ and the dressing operation is defined for any function $F$ as
\begin{equation}\label{dressing-def-unique}
F^{\rm dr}(\theta)=F(\theta)+\int\frac{{\rm d}\eta}{2\pi}K(\theta-\eta)f(\eta)F^{\rm dr}(\eta).
\end{equation}
On the other hand, as $\Q$ acts diagonally on the multi-particle basis, the right hand side of \eqref{diagonal-sum-conn} is proportional to the Gaudin determinant. By some simple matrix manipulations, it can be shown that the connected form factors of $Q$ are given by
\begin{equation}
F_{\rm c}^Q(\theta_1,...,\theta_n)=q(\theta_1)K(\theta_1-\theta_2)...K(\theta_{n-1}-\theta_n)p'(\theta_n) + {\rm perms},
\end{equation}
where perms. is understood as permutations with respect to the integer set $\{1,...,n\}$. Putting this into \eqref{lm}, we find complete agreement with the TBA result \eqref{TBA-one-point-charge}. For generic operators, one could also use \eqref{diagonal-sum-conn} to verify \eqref{lm} up to arbitrary order.

The first full-order proof of the Leclair-Mussardo series was given by Pozsgay \cite{Pozsgay:2010xd}. The idea is quite interesting: applying \eqref{diagonal-sum-conn} to the thermodynamic state that minimizes the TBA free energy. The ratio of the Gaudin determinants turns out to be simply proportional to the product of TBA filling factors in this limit and \eqref{lm} directly follows. This method works equally well if one starts with \eqref{diagonal-sum-symm} instead of \eqref{diagonal-sum-conn}. The end result is
\begin{align}
\langle \mathcal{O}\rangle_R=\sum_{n=0}^\infty\frac{1}{n!}\int\frac{d\theta_1}{2\pi}...\frac{d\theta_n}{2\pi}\prod_{j=1}^nf(\theta_j)w(\theta_j) F_{\text{s}}^\mathcal{O}(\theta_1,...,\theta_n),
\end{align}
where $w(\theta)=\exp[-\int \frac{d\eta}{2\pi}K(\theta-\eta)f(\eta)]$.

Concerning two point functions, although there have been multiple low-temperature expansions \cite{PhysRevB.68.104435,Essler:2007jp,PhysRevB.78.094411,Pozsgay:2010cr,Szecsenyi:2012jq}, no general structure has been extrapolated. Another strategy  is to use form factors with excitations over the thermodynamic state.  This approach has been applied for free theories in \cite{Doyon:2005jf,Doyon:2006pv,Chen_2014} and for interacting theories in a recent work \cite{Cubero:2018vyb}.  In yet another direction, the authors of \cite{Pozsgay:2018tns} proposed a Leclair-Mussardo formula for two point functions with space-like separation. They argued that in that case the product of the two operators can be considered as an composite object and it is the Leclair-Mussardo series for this bi-local operator that gives the two point function. The common point of \cite{Cubero:2018vyb} and \cite{Pozsgay:2018tns} is that they do not focus on the structure of the series itself but rather on the quantity that appears in the series. For \cite{Cubero:2018vyb} this is form factors built on top of a thermodynamic state, while it is form factors of the composite operator that \cite{Pozsgay:2018tns} tries to compute. Although these form factors were argued to obey a set of axioms, the task of bootstrapping them proves to be challenging.

\newpage
\section{Generalized hydrodynamics}
\label{GHD-intro-section}
In this section we introduce the most basic concepts of GHD: a hydrodynamics description of integrable models. Let us start the discussion by reminding where hydrodynamics fits in  the typical time evolution of a generic many-body system with short-range interaction. There are four time scales
\begin{itemize}
\item \textit{Microscopic regime}: at short time, individual particles of the gas propagate ballistically between collisions. This description is exact and the dynamics is reversible.

\item \textit{Boltzmann equation}: after sufficiently many collisions, the individual trajectories start to fill the single-particle phase space. One can then effectively describe the dynamics of the system using a probability distribution  of particle position and momentum.  The change from microscopic description to a density of states amounts to the irreversibility of this stage. 

\item \textit{Hydrodynamics}: at larger time, relaxation occurs and the system tends to maximize its entropy. Before this maximization of entropy takes place in the entire system, it can develop in sub-regions of mesoscopic size called  \textit{fluid cells}.  Inside each fluid cell resides a local thermodynamic states with maximum local entropy.  There are different scales within hydrodynamics corresponding to different orders in derivatives: the lowest order  is called  Euler scale, while second order gives rise to  Navier-Stokes terms that describe diffusion. 

\item \textit{Thermodynamics}: the spatial dependence of local states finally disappears and entropy maximization is realized in the entire system.
\end{itemize}
This picture serves as a guide on how to build a hydrodynamics theory of integrable system. First, we need to understand the characteristics of entropy-maximised states in the presence of an infinite number of conserved quantities. Maximal entropy states and their general properties will be discussed in subsection \ref{maximal-entropy-states}. In subsection \ref{equations-state-section} we will employ the TBA machinery to provide explicit expressions of quantities that characterize these states. Second, we need to know how the states of neighboring fluid cells differ from one another. This is described in subsection \ref{euler-hydrodynamics}, where we present Euler hydrodynamic equations\footnote{Diffusive effect is outside the scope of this thesis}. Finally we will solve the so-called partitioning protocol problem  as an application of the constructed formalism, 
\subsection{Maximal entropy states}
\label{maximal-entropy-states}
Let us denote by $\Q_i$ the set of  conserved charges present in the system and let us assume that they can be expressed as integrals of charge densities satisfying conservation laws
\begin{align}
\Q_i=\int dx Q_i(x,t),\quad \partial_tQ_i(x,t)+\partial_xJ_i(x,t)=0.\quad \p_t \Q_i=0. \label{conservation-law}
\end{align}
Looking at a subsystem of mesoscopic size, if the picture of hydrodynamics holds, we expect it to relax to some state while the rest of the system acts as an external bath. We would like to characterize this state by  a density matrix $\rho$ such that the expectation value of any  observable $\mathcal{O}$ is given by $\langle \mathcal{O}\rangle =\tr[\rho \mathcal{O}]$.
The maximization of the entropy $S(\rho) = −\tr[\rho \log \rho]$ must be subjected to the conservation laws. Let us denote by $\beta^ i$ and $\alpha$ the Lagrange parameters corresponding to the conserved charges $Q_i$ and the normalization of $\rho$,  the entropy maximization condition then reads
\begin{align}
\delta\tr[\rho(\log\rho+\sum \beta^i\Q_i+\alpha)]=0\Rightarrow \tr[\delta\rho(\log \rho+1+\alpha+\sum\beta^i\Q_i)]=0.\label{GGE-variation}
\end{align}
As a result, maximal entropy states are of  generalized Gibbs form $\rho\propto e^{-\sum \beta^i\Q_i}$. The generalized inverse temperatures $\beta^ i$  serve as a system of coordinates in the infinite-dimensional manifold of maximal entropy states. The average $\langle ...\rangle_ {\vec{\beta}}$ evaluated in these states satisfies the following property
\begin{align}
-\frac{\partial}{\partial \beta^i}\langle \mathcal{O}\rangle_{\vec{\beta}}=\int dx \langle \mathcal{O}Q_i(x,0)\rangle ^\text{c}_{\vec{\beta}},\label{GGE-tangent}
\end{align}
where the upper-script c denotes connected correlation functions. In the following, we will  use \eqref{GGE-tangent} as a definition of the inverse temperatures $\beta^i$ instead of the explicit Gibbs form. It is therefore constructive to define an inner product on the space of local observables
\begin{align}
(\mathcal{O}_1,\mathcal{O}_2)\equiv \int dx \langle \mathcal{O}_1(0,0)\mathcal{O}_2(x,0) \rangle^\text{c}_{\vec{\beta}}.\label{inner-product}
\end{align}
This inner product is positive semidefinite, since
\begin{align*}
\int dx \langle \mathcal{O}(0,0)\mathcal{O}(x,0) \rangle^\text{c}_{\vec{\beta}}=\lim_{L\to \infty}\frac{1}{L}\langle \Delta_\mathcal{O}^2\rangle _{\Beta}\geq 0,\quad \Delta_\mathcal{O}\equiv \int_0^L dx[\mathcal{O}(x,0)-\langle \mathcal{O}(0,0)\rangle_{\vec{\beta}}],
\end{align*}
with $L$ being the system size. We define the \textit{static covariance matrix}, denoted by $C_{i j}$ as the product between conserved densities. It is nothing but the Hessian matrix of the free energy  density  $F(\vec{\beta})=\lim_{L\to \infty}\log \tr[e^{-\sum \beta^i\Q_i}]/L$
\begin{align}
\frac{\partial F}{\partial \beta^i}=- \langle Q_i(0,0)\rangle_{\vec{\beta}} \Rightarrow C_{ij}=(Q_i,Q_j)=\frac{\partial^2 F}{\partial \beta^i\partial \beta^j}=(Q_j,Q_i)=C_{ji}.\label{C-matrix}
\end{align}
The positivity of $C$ implies that $F$ is a strictly convex function of $\vec{\beta}$. Let us denote by 
\begin{align}
\mathbf{Q}_i\equiv  \langle Q_i(0,0)\rangle_{\vec{\beta}},\quad \mathbf{J}_i\equiv  \langle J_i(0,0)\rangle_{\vec{\beta}}.
\end{align}
Due to the convexity of $F$, the map $\vec{\beta} \to \vec{\mathbf{Q}}$ from  inverse temperatures to averages of conserved charge densities is a bijection. This means, the set of averages of densities can also be used as a system of coordinates on the manifold of maximal entropy states. In view of the conservation laws \eqref{conservation-law}, this new coordinates system is particularly useful in describing the currents carried by the state. The dependence of the average currents on the average charge densities is referred  to as the \textit{equations of state} of the model
\begin{align}
\mathbf{J}_i=\mathbf{J}_i(\vec{\mathbf{Q}}).\label{equation-of-state}
\end{align}
We will find the explicit form of these equations in the next subsection using the ingredients of TBA. Before that, there is  a general property satisfied by the average currents that can be derived from nothing but the conservation laws. Considering the average currents as functions of the inverse temperatures, let us define the following matrix 
\begin{align}
B_{ij}\equiv \partial \mathbf{J}_i/\p \beta^j=(J_i,Q_j).\label{B-matrix}
\end{align}
Just like the static covariance matrix $C$, $B$ is also a symmetric \cite{Castro-Alvaredo:2016cdj,10.21468/SciPostPhys.6.4.049}.  To prove this property, we note that the conservation laws \eqref{conservation-law} implies the existence of a height field $\varphi_ i (x,t)$ such that
\begin{align}
d\varphi_i(x,t)=Q_i (x,t)dx − J_i (x,t)dt.
\end{align}
Then $\langle J_i (0,0) Q_j (x, t)\rangle^\text{c}_{\vec{\beta}}=-\langle [\partial_t\varphi_i](0,0)[\partial_x\varphi_j](x,t)\rangle^\text{c}_{\vec{\beta}}=\langle \varphi_i(0,0)[\partial_x\partial_t\varphi_j](x,t)\rangle^\text{c}_{\vec{\beta}}$ due to time-translation invariance $=-\langle (\partial_x\varphi_i)(0,0)(\partial_t\varphi_j)(x,t)\rangle^\text{c}_{\vec{\beta}}$ due to space-translation invariance \\$=\langle Q_i (0,0) J_j (x, t)\rangle^\text{c}_{\vec{\beta}}$. By integrating over $x$, we obtain the desired property.

The symmetry of $B$ has an important consequence on the equations of state \eqref{equation-of-state}. Introducing the flux Jacobian matrix
\begin{align}
A_i^j=\partial \mathbf{J}_i/\partial\mathbf{Q}_j.\label{flux-jacobian}
\end{align}
According to the chain rule and the symmetry of the matrix $C$
\begin{align}
\sum_k\frac{\partial\mathbf{J}_i}{\partial \mathbf{Q}_k}\frac{\partial\mathbf{Q}_k}{\partial \beta^j}=\sum_k\frac{\partial\mathbf{J}_j}{\partial \mathbf{Q}_k}\frac{\partial\mathbf{Q}_k}{\partial \beta^i} \quad \Leftrightarrow \quad AC=CA^\text{t}.\label{symmetry-of-A-matrix}
\end{align}
This means that $A$ is symmetric under the inner product induced by the inverse matrix of $C$ (note that $C$ is positive) $\langle \vec{v},\vec{w}\rangle\equiv \vec{v}C^{-1}\vec{w}$. 
Indeed $\langle \vec{v},A\vec{w}\rangle\equiv \vec{v}C^{-1}A\vec{w}=\vec{v}A^\text{t}C^{-1}\vec{w}=A\vec{v}C^{-1}\vec{w}\equiv\langle A\vec{v},\vec{w}\rangle$.  As a result, $A$ is diagonalizable and has real eigenvalues. We will discuss the physical interpretations of its eigenvectors and eigenvalues in the following subsections.
\subsection{Equations of state and hydrodynamic matrices}
\label{equations-state-section}
We continue our discussion of maximal-entropy states. Using TBA, we present in this subsection the explicit expression  for the average charge densities. We also sketch the heuristic derivation \cite{Castro-Alvaredo:2016cdj}  of the average densities using mirror transformation.  A more rigorous proof based on form factors will be delivered in section \ref{equations-of-state-new-derivation}. Once we have obtained the average charge densities and average currents, the hydrodynamic matrices $A,B,C$ directly follow.

The TBA for generalized Gibbs ensemble \cite{Mossel:2012vp} is almost identical to the traditional TBA. The only modification is that the source term in the TBA equation 
\begin{align}
\epsilon(\theta)=w(\theta)-\int\frac{d\eta}{2\pi}K(\theta-\eta)\log[1+e^{-\epsilon(\eta)}]\label{TBA-equation-of-GGE}
\end{align}
is now given by $w(\theta)=\sum \beta^iq_i(\theta)$ with $q_i$ being the one-particle eigenvalue of the conserved charge $\mathcal{Q}_i$: $\Q|\theta\rangle=q_i(\theta)|\theta\rangle$. The free energy density is again given by 
\begin{align}
F(\vec{\beta})=\frac{1}{L}\log\tr[ e^{\sum_j -\beta^i \Q_i}]= \int\frac{d(\theta)}{2\pi}p'(\theta)\log[1+e^{-\epsilon(\theta)}]\label{GGE-free-energy}
\end{align}
The average charge densities are obtained by differentiating $F$ with respect to $\beta^i$, as per \eqref{C-matrix}
\begin{align}
\mathbf{Q}_i=-\frac{\partial F}{\partial \beta^i}=\int\frac{dp}{2\pi}f(\theta)\partial_{\beta_j} \epsilon(\theta),\quad \text{with}\quad f=1/(1+e^\epsilon).
\end{align}
The derivative of the pseudo-energy is the dressed charge eigenvalue $q_j^\text{dr}$, where  the dressing operation was defined in \eqref{dressing-def-unique}.
The notion of TBA-dressed quantities will appear repeatedly in our discussion of GHD. In particular, the \textit{density of particles} and \textit{density of holes} can be written as
\begin{align}
\rho_\text{p}(\theta)=(p')^\text{dr}(\theta)f(\theta)/(2\pi),\quad \rho_\text{p}(\theta)+\rho_\text{h}(\theta)=(p')^\text{dr}/(2\pi).\label{densities-as-dressed}
\end{align}
There is a simple property satisfied by the dressing operation 
\begin{align}
\int d\theta \psi(\theta)f(\theta)\chi^\text{dr}(\theta)=\int d\theta \psi^\text{dr}(\theta)f(\theta)\chi(\theta)\label{symmetry-of-dressing}
\end{align}
Using this symmetry relation we can write the average charge densities in two equivalent forms
\begin{align}
\mathbf{Q}_i=\int\frac{d\theta}{2\pi}p'(\theta)f(\theta)q_i^\text{dr}(\theta)=\int d\theta \rho_\text{p}(\theta)q_i(\theta).\label{charge-average-TBA}
\end{align}
The TBA technique does not provide however (at least not directly) the average currents. In the following we present a trick based on mirror transformation to obtain them.

In relativistic quantum field theories, the mirror transformation $\gamma$ interchanges space and time through a double Wick rotation $(x, t) \to (-it, ix)$. It reads in terms of rapidity $\theta \to i\pi/2-\theta$ and as a result, momentum and energy $p(\theta)=m\sinh\theta$, $E(\theta)=m\cosh\theta$ becomes $(p,E)\to (iE,-ip)$. Under mirror transformation, one can expect charge densities and currents to be likewise exchanged. As we already know the average charge densities,  it suffices to apply the mirror transformation to the state  to obtain the average currents. Let us make this statement more precise.

The expectation value of an observable $\mathcal{O}$ in the state characterized by a source term $w(\theta)$ transforms as $\langle \mathcal{O}^\gamma\rangle_w=\langle \mathcal{O}\rangle_{w^\gamma}$ where $w^\gamma(\theta)\equiv w(i\pi/2-\theta)$. Denoting by $Q[q]$ and $J[q]$ the average charge density and average current as functions of the one particle eigenvalue $q(\theta)$. In the mirror theory, the average current becomes the average charge density
\begin{align}
(J[q])^\gamma=iQ[q^\gamma].
\end{align}
Using the fact the mirror transformation squares to identity, we can write
\begin{align}
\langle J[q]\rangle_w=\langle \lbrace(J[q])^\gamma\rbrace^\gamma\rangle_w=\langle iQ[q^\gamma]\rangle_{w^\gamma}.
\end{align}
It then follows from \eqref{charge-average-TBA} that
\begin{align}
\mathbf{J}_i=\int\frac{d\theta}{2\pi}E'(\theta)f(\theta)q_i^\text{dr}(\theta).\label{current-average-TBA}
\end{align}
We can also rewrite this expression in the following form 
\begin{align}
\mathbf{J}_i=\int d\theta v^\text{eff}(\theta)\rho_\text{p}(\theta)q_i(\theta)\quad \text{where}\quad v^\text{eff}(\theta)\equiv \frac{(E')^\text{dr}(\theta)}{(p')^\text{dr}(\theta)}.\label{v-eff}
\end{align}
Although we have obtained this result by mean of an heuristic argument, the form \eqref{v-eff} is rather convincing. At a fixed rapidity $\theta$ the density of particles is by definition $\rho_\text{p}(\theta)$. Each particle of this rapidity carries a charge $q_i(\theta)$ and propagates at  bare velocity $E'(\theta)/p'(\theta)$. Undergoing collisions with other particles in the gas, it acquires an effective velocity, which is given by the dressed version of the bare velocity. 

Let us now derive the matrices $A,B,C$ previously introduced.  To remind, $B$ and $C$ are obtained by differentiating  the average currents and average charge densities, respectively, as per \eqref{B-matrix} and \eqref{C-matrix}. Their computations are similar and we will carry out the case of $C$
\begin{align}
C_{ij}&=\frac{\partial ^2 F}{\partial \beta^j\beta^k}=\int\frac{dp}{2\pi}\bigg\lbrace f(\theta)[1-f(\theta)]\partial_{\beta^k} \epsilon(\theta)\partial_{\beta^j} \epsilon(\theta)+ f(\theta)\partial_{\beta^k}\partial_{\beta^j} \epsilon(\theta)\bigg\rbrace.
\end{align}
One can eliminate the second derivative of the pseudo energy 
\begin{align}
&\partial_{\beta_k}\partial_{\beta_j}\epsilon(\theta)=\int\frac{d\eta}{2\pi}K(\theta,\eta)\big\lbrace f(\eta)[1-f(\eta)]\partial_{\beta_k} \epsilon(\eta)\partial_{\beta_j} \epsilon(\eta)+f(\eta)\partial_{\beta_k}\partial_{\beta_j} \epsilon(\eta)\big\rbrace,\label{second-derivative}
\end{align}
by integrating this expression with the particle density measure. Using the fact that $\rho_\text{p}=f(p')^\text{dr}/(2\pi)$ one can then write the charge covariance as compactly as \cite{Mossel:2012vp,Doyon_2017}
\begin{align}
C_{ij}=\int d\theta\rho_\text{p}(\theta)[1-f(\theta)]q_i^\text{dr}(\theta)q_j^\text{dr}(\theta).\label{C-matrix-TBA}
\end{align}
The same manipulations give
\begin{align}
B_{ij}=\int d\theta v^\text{eff}(\theta)\rho_\text{p}(\theta)[1-f(\theta)]q_i^\text{dr}(\theta)q_j^\text{dr}(\theta).\label{B-matrix-TBA}
\end{align}
Once we have obtained $B$ and $C$, the flux Jacobian matrix $A$ directly follows in view of the relation \eqref{symmetry-of-A-matrix}
\begin{align}
\int d\theta v^\text{eff}(\theta)\rho(\theta)[1-f(\theta)]q_j^\text{dr}(\theta)q_j^\text{dr}(\theta)=\int d\theta\rho(\theta)[1-f(\theta)]\sum_k A_i^kq_k^\text{dr}(\theta)q_j^\text{dr}(\theta).
\end{align}
By completeness on the index $j$, we find that $A$ has a continuous spectrum formed by the allowed effective velocities of the particles
\begin{align}
\sum_k A_i^k q_k^\text{dr}(\theta)=v^\text{eff}(\theta)q_i^\text{dr}(\theta).\label{eigenvalues-of-A}
\end{align}
\subsection{Local entropy maximisation and Euler hydrodynamics} 
\label{euler-hydrodynamics}
Consider a generic situation where the system starts with an inhomogenous and non-stationary initial state. This means the average values of observables $\langle \mathcal{O}(x,t)\rangle$ depend explicitly on space  and time.  The main  assumption of  GHD is that this average can be evaluated in a maximal entropy state with $(x,t)$-dependent inverse temperatures
\begin{align}
\langle \mathcal{O}(x,t)\rangle\approx \langle \mathcal{O}(0,0)\rangle _{\vec{\beta}(x,t)}.\label{local-entropy-max}
\end{align}
Due to homogeneity  and stationarity of maximal entropy states, the position of the operator on the right hand side is inconsequential and we chose $(0,0)$ as a point of reference. We stress  that the same local state $\vec{\beta}(x,t)$ describes any local  observable at $(x,t)$.

\begin{figure}[ht]
\centering
\includegraphics[width=16cm]{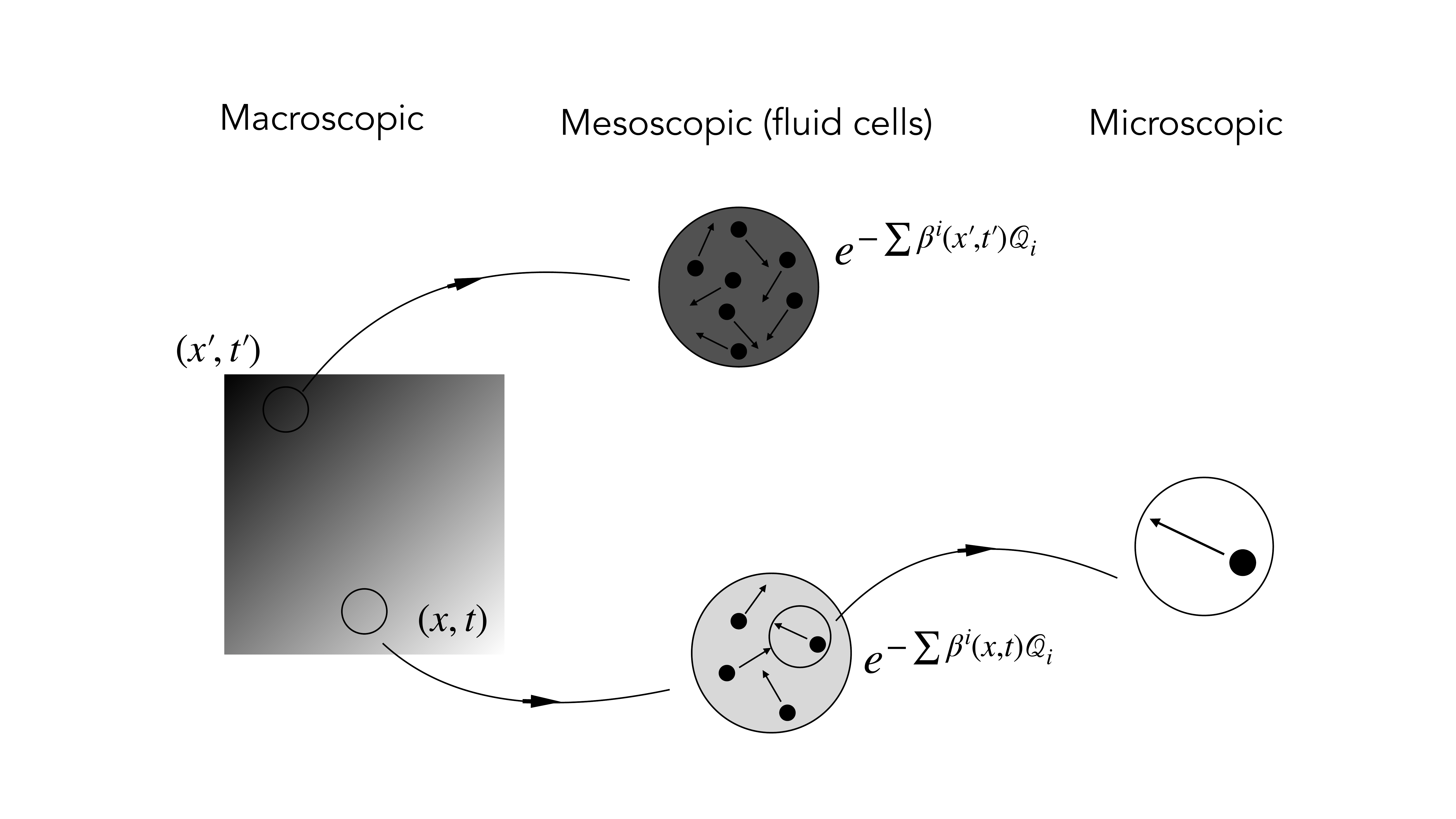}
\caption{The local entropy maximization hypothesis.}
\label{separation-scales}
\end{figure}

This assumption of local entropy maximization lying at the heart of GHD is a strong standpoint to take and is very hard to be rigorously proven for generic systems. Nevertheless one can expect that after long enough time, the scales in which significant variations in the local averages occur are infinitely large with respect to the microscopic scales and yet still infinitely small compared to laboratory scales. We will refer to these intermediate scales where local entropy maximization takes place as fluid cells. The separation between different scales  is illustrated  in figure \ref{separation-scales}.

Next, we want to know how the profiles of neighboring fluid cells differ from one another.  Consider a contour $[X_1, X_2]\times[T_1, T_2]$ inside which the assumption of local entropy maximization is valid. The integration of the continuity equation \eqref{conservation-law} over this contour gives
\begin{align}
\int_{X_1}^{X_2} dx[Q_i(x,T_2)-Q_i(x,T_1)]+\int_{T_1}^{T_2}dt[J_i(X_2,t)-J_i(X_1,t)]=0.
\end{align}
Denoting $\mathbf{Q}_i(x,t)=\langle Q_i(0,0)\rangle_{\vec{\beta}(x,t)}$, $\mathbf{J}_i(x,t)=\langle J_i(0,0)\rangle_{\vec{\beta}(x,t)}$ and applying \eqref{local-entropy-max}
\begin{align}
\int_{X_1}^{X_2}dx[\mathbf{Q}_i(x,T_2)-\mathbf{Q}_i(x,T_1)]+\int_{T_1}^{T_2}dt[\mathbf{J}_i(X_2,t)-\mathbf{J}_i(X_1,t)]=0.\label{maximal-entropy-integration}
\end{align}
By reversing the logic, we can express this relation in its differential form
\begin{align}
\partial_t\mathbf{Q}_i(x,t)+\partial_x\mathbf{J}_i(x,t)=0.\label{continuity-average-local}
\end{align}
Remember that the set of average charge densities $\vec{\mathbf{Q}}(x,t)$ encodes all information about the local state at $(x,t)$, just like the inverse temperatures $\vec{\beta}(x,t)$ do. Moreover, the average currents are functions of  the average charge densities, which is the virtue of the equations of state \eqref{equation-of-state}. In particular, we can insert  the flux Jacobian \eqref{flux-jacobian} into  \eqref{continuity-average-local} and rewrite as a wave equation
\begin{align}
\partial_t\mathbf{Q}_i(x,t)+\sum_j A_i^j(x,t)\partial_x\mathbf{Q}_j(x,t)=0,\quad \text{where}\quad  A_i^j (x,t) \equiv A_i^j (\vec{\mathbf{Q}}(x,t)).\label{Euler-hydro-equation}
\end{align}
We refer to \eqref{continuity-average-local} and equivalently \eqref{Euler-hydro-equation} as  \textit{Euler hydrodynamic equations}. The general strategy of GHD is to solve these equations for the profile of the local state at every point $(x,t)$. One can then obtain the average of any observable $\mathcal{O}(x,t)$  by using $\langle \mathcal{O}(x,t)\rangle  = \langle \mathcal{O}(0, 0)\rangle _{\vec{\beta}(x,t)}$.

In practice, one can exploit the fact that  the flux Jacobian matrix $A$  is diagonalizable, as per \eqref{eigenvalues-of-A} to facilitate the solving of Euler hydrodynamic equations.  Denoting by $R$ the matrix that diagonalises $A$: $RAR^{-1}=v^{\text{eff}}$. If we manage to write $R$ as the Jacobian of some function $\vec{n}(x,t)$ of $\vec{\mathbf{Q}}(x,t)$ 
\begin{align}
\partial n_i/\partial\mathbf{Q}_j=R_i^j.\label{normal-mode-jacobian}
\end{align}
Then  it follows from \eqref{Euler-hydro-equation} that
\begin{align}
\partial_tn_i(x,t)+v^\text{eff}\partial_xn_i(x,t)=0.\label{Euler-normal-mode}
\end{align}
This equation has a simple interpretation: the functions $\vec{n}$ are normal modes being convectively transported at velocity $v^\text{ eff}_i$. As $R$ is invertible, they can be considered as new coordinates of the maximal entropy state in addition to the average charge densities and the inverse temperatures. Moreover, it turns out that  equation \eqref{normal-mode-jacobian} can be explicitly solved and the normal modes are simply given by the product of the one particle eigenvalue  and the TBA filling factor
\begin{align}
n_i(\theta,x,t)=q_i(\theta)f(\theta,x,t).\label{normal-mode-TBA}
\end{align}
The derivation of this identity is technical and is given in appendix \ref{GHD-appendix}.
\subsection{The partitioning protocol}
\label{partition-section}
After a rather abstract formalism we present in this subsection a concrete application of GHD.   We consider a situation where we bring into contact two systems initially described by two homogenous states. We then let the whole system evolve unitarily and we seek to understand its dynamics. For instance, we can ask what is the current of the steady state arising at the contact point?  

Parametrizing the states by their average charge densities, we seek to solve the Euler hydrodynamic equation \eqref{Euler-hydro-equation} with the initial condition 
\begin{align}
\langle Q_i(x,0)\rangle=\begin{cases}
\mathbf{Q}_i^{(l)}\quad x<0\\
\mathbf{Q}_i^{(r)}\quad x>0
\end{cases}
\end{align}\\
\begin{figure}[ht]
\centering
\includegraphics[width=12cm]{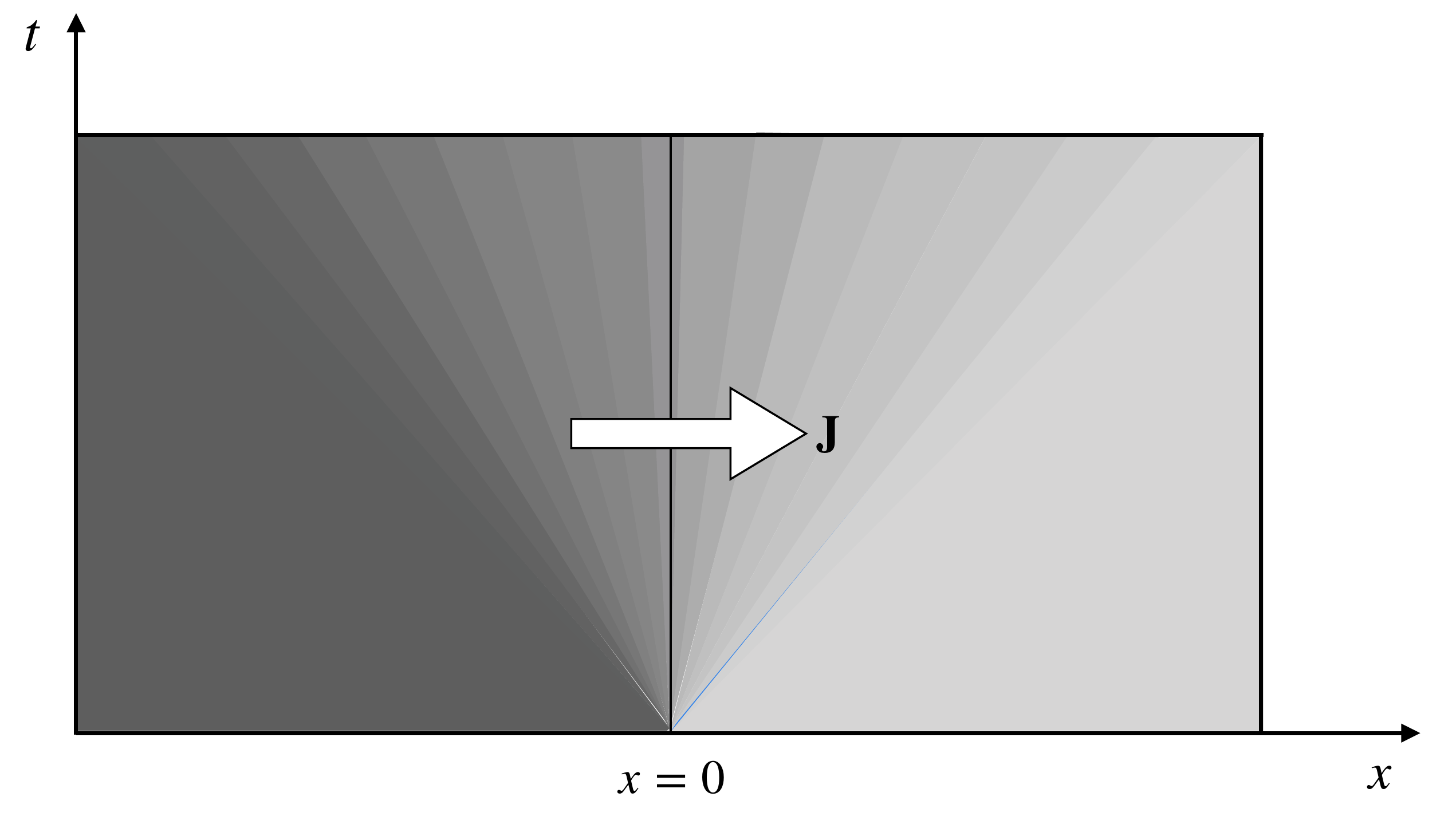}
\caption{The partitioning protocol}
\end{figure}
As both the equation and the initial state are scale invariant, we can assume that the solution itself depends only on the ratio $\xi = x/t$: $\mathbf{Q}_i(x,t)=\mathbf{Q}_i(\xi)$. The initial conditions translates into  the asymptotics of the solution
\begin{align}
\lim_{\xi\to -\infty}\mathbf{Q}_i(\xi)=\mathbf{Q}_i^{(l)},\quad \lim_{\xi\to +\infty}\mathbf{Q}_i(\xi)=\mathbf{Q}_i^{(r)}.\label{asymptotic-of-Q}
\end{align}
Moreover the Euler hydrodynamic equation \eqref{Euler-hydro-equation} is an ordinary differential equation in this parametrization
\begin{align}
\sum_j \big[A_i^j(\xi)-\xi\delta_i^j\big]\frac{d\mathbf{Q}_j}{d\xi}=0\label{Euler-hydrodynamic-ray}
\end{align}
According to the discussion at the end of the previous subsection, this eigenvalue problem can be cast into diagonal form by going to the frame of normal modes. Furthermore, equation \eqref{normal-mode-TBA} states that normal modes are proportional to the TBA filling factor, with the proportional factor being space-time independent. Equation \eqref{Euler-normal-mode} therefore becomes
\begin{align}
[v^\text{eff}(\theta,\xi)-\xi]\frac{\partial f(\theta,\xi)}{\partial \xi}=0.
\end{align}
That is, at fixed value of the rapidity, the TBA filling factor $f(\theta,\xi )$ is constant in the light ray parameter $\xi$ except at $\xi =\xi^*(\theta)$ satisfying
\begin{align}
\xi^*(\theta)=v^\text{eff}(\theta,\xi^*(\theta)).\label{f-shock}
\end{align}
The asymptotic conditions \eqref{asymptotic-of-Q} are hence fully sufficient to determine $f$
\begin{align}
f(\theta,\xi)=\begin{cases}
f^{(l)}(\theta)\quad \xi<\xi^*(\theta)\\
f^{(r)}(\theta)\quad \xi>\xi^*(\theta)
\end{cases}\label{f-solution}
\end{align}
This result survives an important test. Denoting by $T_L$ and $T_R$ the respective temperature of the initial left and right subsystem. It was found in \cite{Bernard_2012} that if we send both temperatures to infinity with their ratio kept fixed: $T_l=\kappa T_R$ then the current of the steady state at the junction between the two halves tends to 
\begin{align}
\lim_{T_L\to \infty}\mathbf{J}(\xi=0)=\frac{\pi c}{12}T_L^2(1-\kappa^2)\label{UV-current-average}
\end{align}
where $c$ is the central charge of the CFT describing the UV limit of the theory. Equations \eqref{f-solution} and \eqref{f-shock} can be numerically solved using standard TBA ingredients. Once $f$ is obtained with sufficient precision, we can compute the average current as per \eqref{current-average-TBA}. For the sinh-Gordon model, the numerical result was found to be in perfect agreement with the prediction \eqref{UV-current-average} \cite{Castro-Alvaredo:2016cdj}.
\chapter{Closed systems}
In this chapter we show that some thermal quantities introduced previously can be  obtained in a uniform fashion. The list includes the TBA equation, the excited state energies, the Leclair-Mussardo series for one point function and the cumulants of conserved charges in a GGE. Concerning the last quantity, only the first \eqref{charge-average-TBA} and second cumulant \eqref{C-matrix-TBA} have appeared in the literature. Our result for higher cumulants is new.

To our knowledge, the derivation of the TBA equation about to be presented here first appeared (in a slightly different form) in the work \cite{doi:10.1063/1.1396836,doi:10.1063/1.1501444,doi:10.1143/JPSJ.73.1171} of Kato and Wadati for the Lieb-Liniger model and the XXX Heisenberg ferromagnetic spin chain. Our derivation of TBA excited state energies and Leclair-Mussardo formula is however new.
\section{The TBA equation}
\label{TBA-section-new}
In this section we compute the partition function $Z(\vec{\beta},L)\equiv\tr[ e^{-\sum_j \beta^j \Q_j}]$ of a GGE and recover the TBA equation \eqref{TBA-equation-of-GGE}. Here, $L$ denotes the system volume and $\Q_j$ are conserved charges. In contrast to the traditional derivation presented in section \ref{TBA-old-section}, we approach the problem in a more straightforward manner. Namely we directly write the partition function as an infinite sum over a complete basis of the Hamiltonian. The wave functions in this basis are characterized by Bethe quantum numbers and they diagonalize all the conserved charges at once. In the thermodynamic limit the discrete sums over quantum numbers can be approximated by  continuous integrals over particle rapidity. This change of variable involves the determinant of the Gaudin matrix  that has been briefly mentioned in the previous chapter. These steps are standard in the low-temperature expansion and have been intensively used in the literature, both for free and interacting theories. Usually, one can only write down explicit expressions up to three or four particles. Our approach is different: instead of developing an analytic series we write the Gaudin determinant as a sum over diagrams prescribed with fixed Feynman rules. These diagrams are of tree type and their  combinatorial structure allows their generating function to be determined by a simple integral equation. This equation is nothing but the TBA equation. We present these steps one by one in the following subsections.  

Our method works for all theories in which the S-matrix is diagonal and is not necessarily a function of rapidities difference. We will also comment on theories with non-diagonal S-matrix at the end of this section.
\subsection{The sum over mode numbers}
The first step in computing the partition function $
Z(\vec{\beta},L)=\tr[ e^{\sum_j -\beta^j \Q_j}]$ is to choose a complete basis that diagonalizes all the conserved charges. We follow Bethe's hypothesis and label each multi-particle wave function by a set of quantum numbers $|n_1,n_2,...,n_N\rangle$. We illustrate our method for theories in which these numbers take integer values and are pairwise different. In order to know the corresponding eigenvalue of conserved charges, one has to convert these quantum numbers into particle rapidities $\theta_1,...,\theta_N$. This conversion is known as Bethe-Yang equations, and is written in term of the total scattering phase $\Phi$ defined as
\begin{gather}
\Phi_j(\theta_1,...,\theta_N)\equiv Lp(\theta_j)-i\sum_{k\neq j}^N\log S(\theta_j,\theta_k)=2\pi n_j, \quad j=\overline{1,N}.\label{Bethe-Yang}
\end{gather}
If we denote by $q_j(\theta)$ the one-particle eigenvalue of the conserved charge $\Q_j$ then its eigenvalue corresponding to the wave function $|n_1,n_2,...,n_N\rangle$ is implicitly given by
\begin{align}
{\Q}_{j}|n_1,n_2,...,n_N\rangle=\sum_{k=1}^Nq_j(\theta_k)|n_1,n_2,...,n_N\rangle.\label{action-of-conserved-charges}
\end{align}
By defining $w(\theta)\equiv \sum_j \beta^j q_j(\theta)$, we can write the partition function  as a formal sum over Bethe quantum numbers
\begin{align}
Z(\vec{\beta},L)=\sum_{N\geq 0}\;\sum_{\substack{n_1,n_2,...,n_N\in \mathbb{Z}\\n_1<n_2<...<n_N}}e^{-w(n_1,n_2,...,n_N)}.
\end{align}
In this expression, $w$ is an implicit function of mode numbers: one has to solve the Bethe-Yang equations \eqref{Bethe-Yang} for rapidities and replace
$ w(n_1,n_2,...,n_N)$ by $w(\theta_1)+w(\theta_2)...+w(\theta_N)$. In particular, $w$ is a completely symmetric function of mode numbers.

In order to transform this discrete sum to an integral over phase space, we first have to remove the constraint between mode numbers.  This can be done by inserting $1-\delta$ terms which kill configurations with coinciding mode numbers. To this end we obtain an unrestricted sum
\begin{align}
Z(\vec{\beta},L)=\sum_{N\geq 0}\;\frac{1}{N!}\sum_{n_1,n_2,...,n_N\in\mathbb{Z}}e^{-w(n_1,n_2,...,n_N)}\prod_{j<k}^N(1-\delta_{n_j,n_k}).
\end{align}
Once expanded, the Kronecker delta symbols glue mode numbers together into clusters of equal value. Let us see this effect in the two and three-particle sectors
\begin{align*}
\frac{1}{2!}\sum_{n_1,n_2}(1-\delta_{n_1,n_2})e^{-w(n_1,n_2)}=\frac{1}{2!}\sum_{n_1,n_2}e^{-w(n_1,n_2)}&-\frac{1}{2!}\sum_{n_1}e^{-w(n_1,n_1)}\\
\frac{1}{3!}\sum_{n_1,n_2,n_3}(1-\delta_{n_1,n_2})(1-\delta_{n_1,n_3})(1-\delta_{n_2,n_3})e^{-w(n_1,n_2,n_3)}&=\frac{1}{3!}\sum_{n_1,n_2,n_3}e^{-w(n_1,n_2,n_3)}\\
&-\frac{1}{2}\sum_{n_1,n_2}e^{-w(n_1,n_1,n_2)}+\frac{1}{3}\sum_{n_1}e^{-w(n_1,n_1,n_1)}.
\end{align*}
In general, we have an unrestricted sum over mode numbers with multiplicities $(n_1^{(r_1)},...,n_N^{(r_N)})$. Such tuple defines an (unphysical) Bethe state with $r_1+...+r_N$ particles. This state is a linear combination of plane waves with momenta $r_jp(\theta_j),\; j=1,...,N$ and thermal weight $
w(n_1^{r_1},...,n_N^{r_N})=r_1w(\theta_1)+...+r_Nw(\theta_N)$. 
The set of rapidities $\vec{\theta}$ is now given by Bethe-Yang equations with multiplicities.  There are two modifications we need to add to the total scattering phase $\Phi_j$ in the usual Bethe-Yang equations \eqref{Bethe-Yang}. First, each probe particle with rapidity $\theta_j$ winds  around the world and scatters $r_k$ times with $r_k$ particles of rapidity $\theta_k$ for each $k\neq j$. Second,  it also scatters with the $r_j − 1$ other particles with the same rapidity via the trivial S-matrix $S(\theta_j,\theta_j)=-1$
\begin{align}
\Phi_j(\vec{\theta},\vec{r})\equiv Lp(\theta_j)-i\sum_{k\neq j}^N r_k\log S(\theta_j,\theta_k)+\pi(r_j-1)=2\pi n_j,\quad j=\overline{1,N}.\label{Yang-Yang-multiplicities}
\end{align}
The relevance of these unphysical states and their Bethe equations has been already pointed out by Woynarovich \cite{Woynarovich:2010wt} and by Dorey \textit{et al} in \cite{Dorey:2004xk}. In short, one replaces a sum over physical states by a sum over all possible states with proper coefficients such that in the end, unphysical states are canceled out. Finding these coefficients is a purely combinatorial exercise 
\begin{align}
Z(\vec{\beta},L)=\sum_{N\geq 0}\frac{(-1)^N}{N!}\sum_{n_1,...,n_N\in\mathbb{Z}}\sum_{r_1,..,r_N\in\mathbb{N}}\prod_{j=1}^N\frac{(-1)^{r_j}}{r_j}e^{-w(n_1^{r_1},...,n_N^{r_N})}.
\end{align} 
One way to verify this formula is to consider the non-interacting case where the thermal weight $ w(n_1^{r_1},...,n_N^{r_N})$ simply decomposes into $r_1w(n_1)+...+r_Nw(n_N) $ and hence
\begin{align}
Z(\vec{\beta},L)&=\sum_{N\geq 0}\frac{1}{N!}\sum_{n_1,...,n_N\in\mathbb{Z}^N}\prod_{j=1}^N\sum_{r\in \mathbb{N}}\frac{(-1)^{r+1}}{r}e^{-rw(n_j)}=\sum_{N\geq 0}\frac{(-1)^N}{N!}\sum_{n_1,...,n_N\in\mathbb{Z}^N}\prod_{j=1}^N\log[1+e^{-w(n_j)}]\nonumber\\
&=\sum_{N\geq 0}\frac{1}{N!}\bigg[\sum_{n}\log[1+e^{-w(n)}]\bigg]^N=\exp\bigg[\sum_{n}\log[1+e^{-w(n)}]\bigg].\label{free-fermion-partition}
\end{align}
Without interaction, the mode number is nothing but $Lp/2\pi$. It suffices to perform a change of variable to bring the partition function into the known form
\begin{align*}
\log Z(\vec{\beta},L)=L\int \frac{dp}{2\pi}\log[1+e^{-w(p)}].
\end{align*}
In the case of free fermions, the multiplicities $r_j$ have obvious meaning. The vacuum energy is a sum of all fermionic loops including those winding $r$ times around the space circle. The weight of an $r$-winding loop consists of a Boltzmann factor $e^{-rRE}$, a sign $(−1)^r$ due to the Fermi statistics and a combinatorial factor $1/r$ counting for the $Z_r$ cyclic symmetry. It is natural to interpret the multiplicities $r_j$ as winding, or wrapping, numbers also in the case of non-trivial scattering.
\subsection{From mode numbers to rapidities}
The discrete sum over the allowed values of the total scattering phases $\Phi_j(\vec{\theta},\vec{r})$ for given wrapping numbers can be replaced, up to exponentially small in $L$ terms, by an integral
\begin{align}
\sum_{n_1,n_2,...,n_N}\approx\int\frac{d\Phi_1}{2\pi}\int\frac{d\Phi_2}{2\pi}...\int\frac{d\Phi_N}{2\pi}.
\end{align}
Since the thermal weight takes a simpler form as a function of rapidities, we are going to perform a change of variables from total scattering phases $\Phi_j$ to rapidities $\theta_j$
\begin{align}
Z(\vec{\beta},L)=\sum_{N\geq 0}\frac{(-1)^N}{N!}\sum_{r_1,..,r_N\in\mathbb{N}}\prod_{j=1}^N\frac{(-1)^{r_j}}{r_j}\int\frac{d\theta_je^{-r_jw(\theta_j)}}{2\pi}\det G[\theta_1^{(r_1)},...,\theta_N^{(r_N)}].
\end{align}
Aside from the Jacobian, the structure of this series is identical to that of free fermions theory  \eqref{free-fermion-partition}.  In other words, all non-trivial information about the interacting theory is encoded in this matrix
\begin{align}
G_{jk}[\theta_1^{(r_1)},...,\theta_N^{(r_N)}]\equiv \frac{\partial \Phi_j}{\partial\theta_k}=\big[Lp'(\theta_j)+\sum_{l\neq j} r_lK(\theta_j,\theta_l)\big]\delta_{jk}-r_kK(\theta_k,\theta_j)(1-\delta_{kj}).\label{Gaudin}
\end{align}

\subsection{The Gaudin determinant and its diagrammatic expansion}
\label{Gaudin-expansion-section}
Let us denote for brevity
\begin{align*}
p'_j\equiv p'(\theta_j),\quad K_{jk}=K(\theta_j,\theta_k).
\end{align*}
Inspecting the expansion of the Gaudin determinant for $N = 1, 2, 3$
\begin{align*}
\det G[\theta^{(r)}]&=Lp',\\
\det G[\theta_1^{(r_2)},\theta_2^{(r_2)}]&=L^2p'_1p'_2+Lp'_1r_1K_{21}+Lp'_2r_2K_{12},\\
\det G[\theta_1^{(r_1)},\theta_2^{(r_2)},\theta_3^{(r_3)}]&=L^3p'_1p'_2p'_3\\
&+L^2p'_2p_3r_2K_{12}+L^2p'_2p'_3r_3K_{13}+L^2p'_1p'_3r_1K_{21}\\
&+L^2p'_1p'_3r_3K_{23}+L^2p'_1p'_2r_1K_{31}+L^2p'_1p'_2r_2K_{32}\\
&+Lp'_1r_1r_3K_{31}K_{23}+Lp'_1r_1r_2K_{21}K_{32}+Lp'_1r_1^2K_{21}K_{31}\\
&+Lp'_2r_1r_2K_{31}K_{12}+Lp'_2r_2r_3K_{13}K_{32}+Lp'_2r_2^2K_{12}K_{32}\\
&+Lp'_3r_1r_3K_{31}K_{23}+Lp'_3r_2r_3K_{12}K_{23}+Lp'_3r_3^2K_{13}K_{23},
\end{align*}
we see that there are no cycles of the type $K_{12}K_{21}$ or $K_{12}K_{23}K_{31}$. 
In order to apply the matrix-tree theorem we consider the matrix 
\begin{align}
\tilde{G}_{jk}=r_kG_{jk}.\label{scaled-Gaudin-matrix}
\end{align}
This newly defined matrix is then the sum of a diagonal matrix  and a Laplacian matrix (one in which the elements in each row or column sum up to zero): $\tilde{G}_{jk}=\tilde{D}_j\delta_{jk}+\tilde{K}_{jk}$ where
\begin{gather}
D_j=Lr_jp'(\theta_j),\quad \tilde{K}_{jk}=\delta_{jk}\sum_{l\neq j}r_lr_kK(\theta_j,\theta_l)-(1-\delta_{kj})r_jr_kK(\theta_k,\theta_j).
\end{gather}
Before stating the result of the matrix-tree theorem, we need to introduce some terminologies in graph theory. A \textit{spanning forest} of  a directed graph $\Gamma$ is a  subgraph $\mathcal{F}$ of $\Gamma$ fulfilling the following three conditions
\begin{itemize}
\item $\mathcal{F}$ contains all vertices of $\Gamma$,
\item $\mathcal{F}$ does not contain cycles,
\item for any vertex of $\Gamma$ there is at most one oriented edge of $\mathcal{F}$ ending at this vertex.
\end{itemize}
The vertices with no incoming lines are called roots. Any forest $\mathcal{F}$ can be decomposed into connected components called directed trees. Each tree contains one and only one root. Let us prove this property, starting with the existence. Assume on the contrary that there is at least one coming edge entering every vertex of a tree $\mathcal{T}$. Starting with an arbitrary vertex $v_0$, we can find another vertex $v_1$ such that $\langle v_1\to v_0\rangle$ is an edge of $\mathcal{T}$. We can then repeat the procedure to create a chain of vertices $v_n\to v_{n-1}\to...\to v_1\to v_0$. Due to the finiteness of $\mathcal{T}$, this chain must close upon itself, creating a cycle and thus violating the second hypothesis. To prove the uniqueness of the root, assume again on the contrary that there are at least two roots $r$ and $\tilde{r}$ in the same tree $\mathcal{T}$. The connectedness of $\mathcal{T}$ guarantees the existence of a path $(r,v_0,v_1,..,v_n,\tilde{r})$ joining the two roots. By definition of roots, the edge joining $r$ and $v_0$ must be of direction $r\to v_0$. According to the third property, the edge connecting $v_0$ and $v_1$ must then come from $v_0$ to $v_1$. One keeps going and finds eventually that last edge of the path is of direction $v_n$ to $\tilde{r}$, contradicting the assumption that $\tilde{r}$ is a root.

The matrix-tree theorem \cite{Chaiken82acombinatorial} states that the determinant of $\tilde{G}$ can be written a sum over  directed forests that span the totally connected graph with vertices labeled by $j=1,...,N$
\begin{align}
\det \tilde{G}=\sum_{\mathcal{F}}\prod_{v_j \text{ roots}}D_j\prod_{\langle jk \rangle \;\text{branches}} r_jr_kK(\theta_k,\theta_j).\label{det-Gaudin-tree}
\end{align}
That is, the weight of a forest $\mathcal{F}$ is a product of factors $D_j$ associated with the roots and factors $K_{kj}$ associated with the oriented edges $\langle jk\rangle  =\langle v_j \to v_k\rangle$ of $\mathcal{F}$. The expansion in spanning forests for $N=1,2,3$ is depicted in Fig. 1
\begin{figure}[ht]
\centering
\includegraphics[width=14cm]{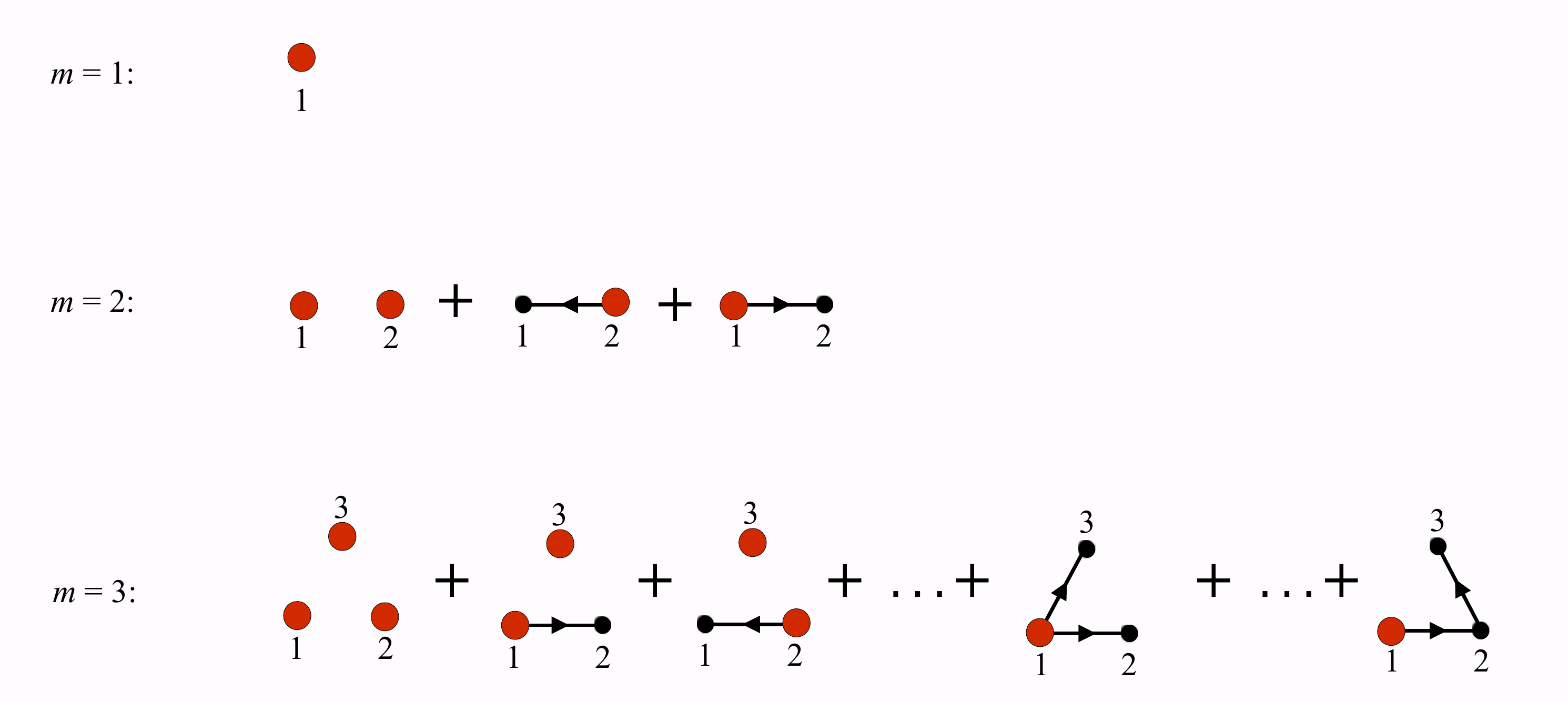}
\caption{The expansion of the determinant of the matrix $\bar{G}$ in
directed spanning forests for $N = 1, 2, 3$. Ellipses mean sum over the permutations of
the vertices of the preceding graph. Each vertex of a directed tree, except for the root,
has exactly one incoming edge and an arbitrary number of outgoing edges. The root can
have only outgoing edges. A factor $r_kr_jK_{kj}$ is associated with each edge $\langle jk\rangle$. A factor $D_j$ is associated with the roots of each connected tree, which is symbolised by a red dot.}
\end{figure}

With this expansion of the Gaudin determinant, can now write the partition function as
\begin{align}
Z(\vec{\beta})=\sum_{N\geq 0}\frac{(-1)^N}{N!}\sum_{r_1,..,r_N\in\mathbb{N}}\prod_{j=1}^N\int\frac{(-1)^{r_j}}{r_j^2}\frac{d\theta_j}{2\pi}e^{-r_jw(\theta_j)}\sum_{\mathcal{F}}\prod_{j \text{ roots}}Lr_jp'(\theta_j)\prod_{\langle jk \rangle } r_jr_kK(\theta_k,\theta_j).
\end{align}
The next step is to invert the order of the sum over graphs and the integral/sum over the coordinates $(\theta_j,r_j)$ assigned to the vertices. As a result we obtain a sum over the ensemble of tree graphs, with their symmetry factors, embedded in the space $\mathbb{R}\times \mathbb{N}$ where the coordinates $(\theta,r)$ of the vertices take values. The embedding is free, in the sense that the sum over the positions of the vertices is taken without restriction. One can think of these graphs as tree level Feynman diagrams obtained by applying the following Feynman rules
\vspace{0.2cm}
\begin{tcolorbox}[center,colback=white,width=12cm]
\begin{align}
\vcenter{\hbox{\includegraphics[width=1.1cm]{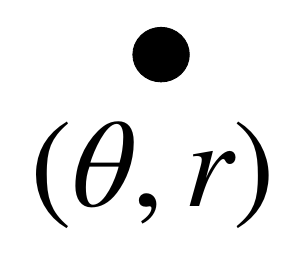}}}\quad\qquad&=\quad\frac{(-1)^{r-1}}{r^2}e^{-rw(\theta)}\nonumber\\
\vcenter{\hbox{\includegraphics[width=1.1cm]{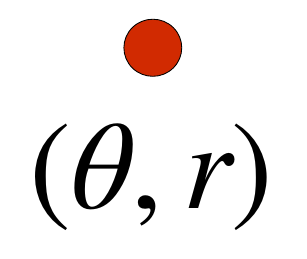}}}\quad\qquad&=\quad Lp'(\theta)\frac{(-1)^{r-1}}{r}e^{-rw(\theta)}\label{Feynmp}\\
\vcenter{\hbox{\includegraphics[width=4cm]{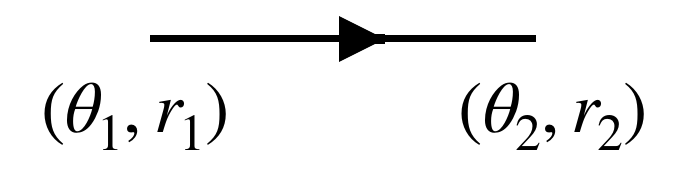}}}&=\quad r_1r_2K(\theta_2,\theta_1).\nonumber
\end{align}
\end{tcolorbox}
\subsection{Summing over the trees}
\begin{figure}[ht]
\centering
\includegraphics[width=10.5cm]{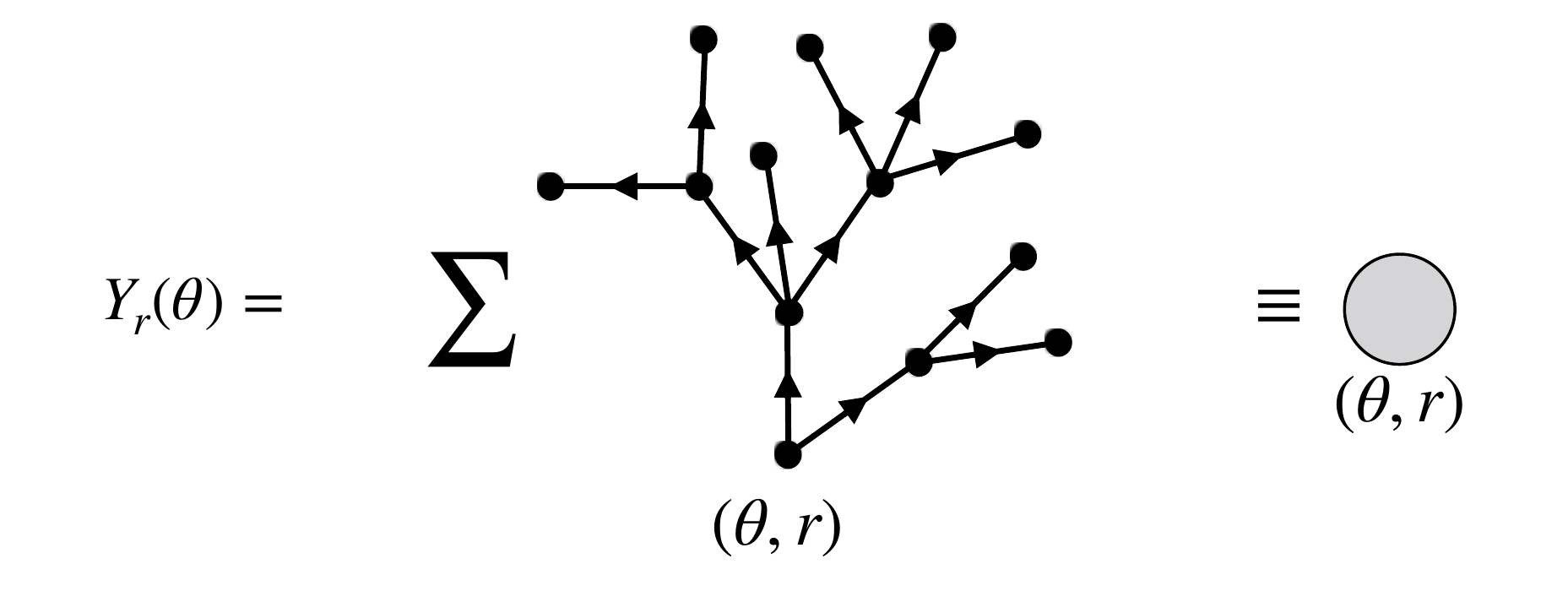}
\caption{The sum over all trees growing out of a fixed root. In defining $Y_r(\theta)$, we have extracted a factor $Lp'(\theta)r$ out of the weight \eqref{Feynmp} of the root. Thus, all vertices appearing in this diagram have the same weight.}
\label{Yr-definition}
\end{figure}

The sum over the embedded graphs is the exponential of the sum over connected ones. In this way we can write the free energy density as
\begin{align}
F(\vec{\beta})=\int \frac{d(\theta)}{2\pi}p'(\theta)\sum_{r\geq 1}rY_r(\theta),\label{free-energy-sum-trees}
\end{align}
where $Y_r(\theta)$ is the sum of all trees rooted at the point $(\theta,r)$, see figure \ref{Yr-definition}.

\begin{figure}[ht]
\centering
\includegraphics[width=14cm]{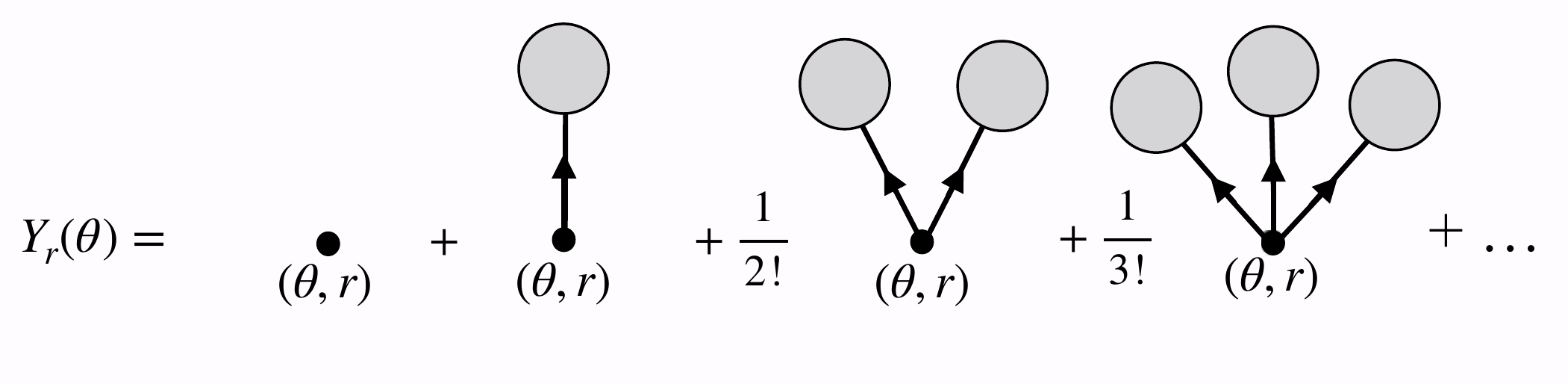}
\caption{Diagrammatic representation of the TBA equation.}
\label{diag-present-TBA}
\end{figure}

\noindent
As any partition sum of trees, it satisfies a simple non-linear equation (a Schwinger-Dyson equation in the QFT language), illustrated in figure \ref{diag-present-TBA}
\begin{align}
Y_r(\theta)=\frac{(-1)^r}{r^2}\big[e^{-w(\theta)}\exp\sum_{s} \int\frac{d\eta}{2\pi}sK(\eta,\theta)Y_s(\eta)\big]^r.\label{Schwinger-Dyson}
\end{align}

\noindent
By comparing this equation for arbitrary $r$ and for $r=1$ we find that $Y_r(\theta)=(-1)^rY_1^r(\theta)/r^2$.  The Schwinger-Dyson equation \eqref{Schwinger-Dyson} for $r=1$ is then the well-known TBA equation  if we identify $Y_1(\theta)=Y(\theta)=e^{-\epsilon(\theta)}$
\begin{gather}
Y(\theta)=e^{-w(\theta)}\exp\int\frac{d\eta}{2\pi}K(\eta,\theta)\log[1+Y(\eta)].
\end{gather}
Furthermore, the free energy density \eqref{free-energy-sum-trees} also takes its familiar form
\begin{align}
F(\vec{\beta})=\int\frac{d(\theta)}{2\pi}p'(\theta)\log[1+Y(\theta)].
\end{align}

Our machinery works equally well for \textit{bosonic} theories in which the mode numbers of a multi-particle state can take coinciding values. In this case, we simply need to remove all the minus signs in the combinatorial coefficients that appear in the expansion of the partition function
\begin{align}
Z(\vec{\beta},L)=\sum_{N\geq 0}\frac{1}{N!}\sum_{n_1,...,n_N\in\mathbb{Z}}\sum_{r_1,..,r_N\in\mathbb{N}}\prod_{j=1}^N\frac{1}{r_j}e^{-w(n_1^{r_1},...,n_N^{r_N})}.
\end{align} 
The Feynman rules for vertices are modified accordingly
\begin{align*}
\vcenter{\hbox{\includegraphics[width=1.1cm]{vertex.pdf}}}&=\;\frac{1}{r^2}e^{-rw(\theta)}\nonumber,\\
\vcenter{\hbox{\includegraphics[width=1.1cm]{root.pdf}}}&=\;Lp'(\theta)\frac{1}{r}e^{-rw(\theta)}.
\end{align*}
while the rule for propagators remain unchanged. It follows that $Y_r(\theta)=Y^r(\theta)/r^2$ and we recover the bosonic TBA equation
\begin{align*}
Y(\theta)=e^{-w(\theta)}\exp\big\lbrace-\int\frac{d\eta}{2\pi}K(\eta,\theta)\log[1-Y(\eta)]\big\rbrace,\quad F(\vec{\beta})=-\int\frac{d(\theta)}{2\pi}p'(\theta)\log[1-Y(\theta)].
\end{align*}
\subsection{A comment on theories with non-diagonal S-matrix}
We have provided a new derivation of the TBA equation for theories with diagonal S-matrix. Can we extend our method for theories in which the S-matrix is not diagonal, for instance the chiral $SU(2)$ Gross-Neveu model? Unfortunately we encounter a serious problem at the very first step of our formalism: the sum over mode numbers. For the chiral $SU(2)$ Gross-Neveu model, there are Bethe equations for the physical rapidity \eqref{SU(2)-physical} and there are also Bethe equations for auxiliary rapidity \eqref{SU(2)-auxiliary}. In deriving these equations, we have used the fact that the auxiliary rapidities correspond in fact to dynamical excitations on a spin chain where the physical rapidities are impurities. In particular the number of the auxiliary particles can never exceed half the number of physical particles. This constraint prohibits us from carrying out the infinite sum over mode numbers.

However, we saw above that the combinatorial structure of the trees is equivalent to the TBA equation. Therefore, we can reverse the logic and translate the TBA equations of the chiral $SU(2)$ Gross-Neveu model (which are obtained by the traditional method) into a sum over trees. These trees involve an infinite number of vertex types: they correspond to either the physical rapidity or the auxiliary strings. Going back further, we can say that if instead of summing over two species of mode number with constraint, we consider a free sum over an infinite number of mode number species then at the end, we would end up with the correct TBA equations. That is, if we deliberately commit two errors: first, removing the constraint between auxiliary and physical particles and second, including auxiliary strings in our summation then the net result turns out to be accurate. 

At this point, this is nothing more than a mere interpretation of the TBA equations with string solutions. In chapter \ref{non-diag-g-section} however, this line of thought will provide some useful information for the boundary entropy of a theory with non-diagonal scattering.
\newpage
\section{The charge statistics in GGE}
\label{charge-new-section}
In this section we use our diagrammatic formalism to obtain the cumulants of conserved charges in a GGE. As the results of this subsection will later be applied in the context of GHD, we restrict our consideration to S-matrices that depend on the difference of rapidities. In particular, the scattering kernel is symmetric: $K(\theta_1,\theta_2)=K(\theta_2,\theta_1)$ and the resulting graphs are unoriented.  The relevance of this symmetry will be explained at the end of subsection \ref{charge-covariance-new}.

The first two cumulants, namely the charge average and the charge covariance matrix has been obtained in subsection \eqref{equations-state-section} directly from the free energy. For the convenience of following we repeat here their expressions. The charge average is given by
\begin{align}
\frac{1}{L}\langle \Q_j\rangle=-\frac{\p F(\vec{\beta})}{\p \beta^i}=\int\frac{d\theta}{2\pi}p'(\theta)f(\theta)q_i^\text{dr}(\theta).\label{charge-average-repeat}
\end{align}
where  the dressing operation was defined in \eqref{dressing-def-unique}. 
The charge covariance matrix involves second derivatives of the pseudo energy which could be eliminated using the manipulation described in subsection \ref{equations-state-section}
\begin{align}
\frac{1}{L} \langle {\Q}_j {\Q}_k \rangle^\text{c}=\frac{\partial ^2 F(\vec{\beta})}{\partial \beta^j\beta^k}=\int \frac{d\theta}{2\pi}p'(\theta)f(\theta)[1-f(\theta)]q_j^\text{dr}(\theta)q_k^\text{dr}(\theta).\label{charge-covariance-TBA}
\end{align}
For the third cumulants, the same trick eliminates the third derivatives of the pseudo-energy but leaves the second ones
\begin{align}
\frac{1}{L} \langle {\Q}_j {\Q}_k {\Q}_l\rangle^\text{c}&=\int \frac{d\theta}{2\pi}p'(\theta)f(\theta)[1-f(\theta)]\times \bigg\lbrace [1-2f(\theta)]q_j^\text{dr}(\theta)q_k^\text{dr}(\theta)q_l^\text{dr}(\theta)\nonumber\\
&+ q_j^\text{dr}(\theta)\partial_{\beta_l}\partial_{\beta_k}\epsilon (\theta)+q_j^\text{dr}(\theta)\partial_{\beta_l}\partial_{\beta_k}\epsilon(\theta)+q_j^\text{dr}(\theta)\partial_{\beta_l}\partial_{\beta_k}\epsilon(\theta)\bigg\rbrace.\label{GGE-3}
\end{align}
This direct computation from the free energy is clearly impractical for higher cumulants. There is no general rule to write the obtained expression in terms of fundamental TBA quantities like the particle density, the filling factor or simple dressing operators. 
\subsection{Charge average}
As the conserved charges act diagonally on the basis of multi-particle states, we can calculate their averages  by following the exact same steps as before. The only point we need to pay attention to is their action on multi-wrapping states. In view of the interpretation of these states as linear combinations of plane waves, we propose a natural generalization of \eqref{action-of-conserved-charges}
\begin{align}
{\Q}_j|n_1^{r_1},...,n_N^{r_N}\rangle=\sum_{i=1}^N r_iq_j(\theta_i)|n_1^{r_1},...,n_N^{r_N}\rangle, \label{charge-multistate}
\end{align}
Consequently, the unnormalized charge average is given by 
\begin{align*}
&\tr[e^{-\sum \beta^i\Q_i}\Q_j]=\sum_{N\geq 0}\frac{(-1)^N}{N!}\sum_{n_1,...,n_N\in\mathbb{Z}}\sum_{r_1,..,r_N\in\mathbb{N}}\prod_{i=1}^N\frac{(-1)^{r_i}}{r_i}e^{-w(n_1^{r_1},...,n_N^{r_N})}\sum_{i=1}^N r_iq_j(\theta_i),\\
&=\sum_{N\geq 0}\frac{(-1)^N}{N!}\sum_{r_1,..,r_n\in\mathbb{N}}\prod_{j=1}^N\frac{(-1)^{r_j}}{r_j^2}\int\frac{d\theta_je^{-r_jw(\theta_j)}}{2\pi}\det \tilde{G}[\theta_1^{(r_1)},...,\theta_N^{(r_N)}]\sum_{i=1}^N r_iq_j(\theta_i).
\end{align*} 
Expanding the Gaudin determinant, we obtain a sum over forests with the same Feynman rules as \eqref{Feynmp} except for one modification: each forest now contains a vertex, the coordinate of which denoted by $(\theta,r)$, that is marked with a charge insertion and carries an extra weight of $rq_j(\theta)$ coming from \eqref{charge-multistate}. Contribution coming from un-inserted trees (vacuum diagrams) cancel with the partition function. As a result, the average of $\Q_j$ is a sum over trees with a vertex marked with charge insertion. 
\begin{figure}[ht]
\centering
\includegraphics[width=15cm]{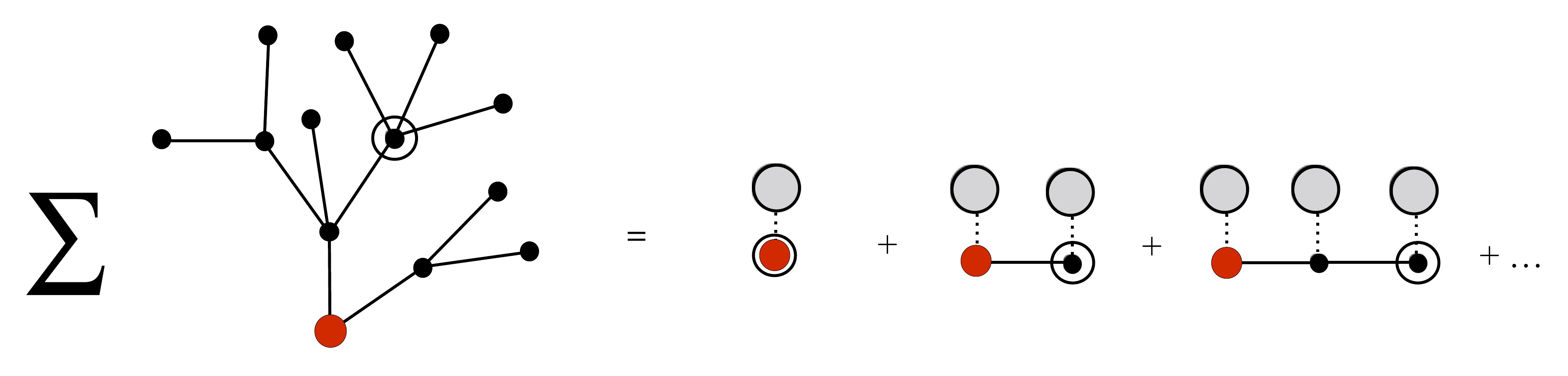}
\caption{Diagrammatic representation of the charge average. Circled vertex stands for the charge insertion.}
\end{figure}
\noindent
In order to perform the sum over these trees, we note that from the inserted vertex we can always trace a unique path- a \textit{spine} to the root of the tree\footnote{Existence comes from connectedness and uniqueness comes from the absence of loops in a tree.}. At each node $(\theta,r)$ inside this spine, we can sum up the trees growing out of it while absorbing the multiplicities $r^2$ coming from the two adjacent propagators. The nodes at the two ends of the spine receive a multiplicity from the charge (or momentum derivative) insertion and a residual multiplicity from one propagator. This results in the TBA filling factor on every node along the spine
\begin{align}
\sum_{r\geq 1} r^2Y_r(\theta)=\sum_{r\geq 1}(-1)^{r-1}Y^r(\theta)=\frac{Y(\theta)}{1+Y(\theta)}=f(\theta).\label{obtaining-the-filling}
\end{align}
Moreover, the sum over spines each of which carries a filling factor on its nodes is nothing but the explicit expansion of the dressing operation \eqref{dressing-def-unique}
\begin{align*}
q_j^\text{dr}(\theta)=q_j(\theta)+\int\frac{d\eta}{2\pi}K(\theta-\eta)f(\eta)q_j(\eta)+\int\frac{d\eta}{2\pi}\frac{d\zeta}{2\pi}K(\theta-\eta)f(\eta)K(\eta-\zeta)f(\zeta)q_j(\zeta)+...
\end{align*}
We thus recover the expression \eqref{charge-average-repeat}  of the charge average
\begin{align}
\frac{1}{L}\langle {\Q}_j\rangle=\int \frac{dp}{2\pi}f(\theta)q_j^\text{dr}(\theta).\label{LM-conserved}
\end{align}
\subsection{Charge covariance}
\label{charge-covariance-new}
The action of a product of two conserved charges $\Q_j$ and $\Q_k$ on a multi-wrapping states is factorized
\begin{align}
{\Q}_j{\Q}_k|\theta_1^{r_1},...,\theta_N^{r_N}\rangle =\Q_j\sum_{i=1}^N r_iq_k(\theta_i)|\theta_1^{r_1},...,\theta_N^{r_N}\rangle=\sum_{i=1}^N r_iq_j(\theta_i)\sum_{i=1}^N r_iq_k(\theta_i)|\theta_1^{r_1},...,\theta_N^{r_N}\rangle.\label{2charge-multistate}
\end{align}
Thus $\tr [e^{-\sum \beta _i \Q_i}{\Q}_j{\Q}_k]$ is given by  a sum over forests each one of which contains two vertices with charge insertions. Un-inserted diagrams again cancel with the partition function and the product average $\langle \Q_j\Q_k\rangle$ is the sum over inserted ones. They can be of two types: either a tree with two inserted vertices or two trees with one inserted vertex each. The sum over diagrams of second type is nothing but the product of charge averages $\langle \Q_j\rangle\langle \Q_j\rangle$.  Therefore the charge covariance $\langle \Q_j\Q_k\rangle^\textbf{c}$ is given by the sum over trees with two charge insertions.
\begin{figure}[ht]
\centering
\includegraphics[width=6cm]{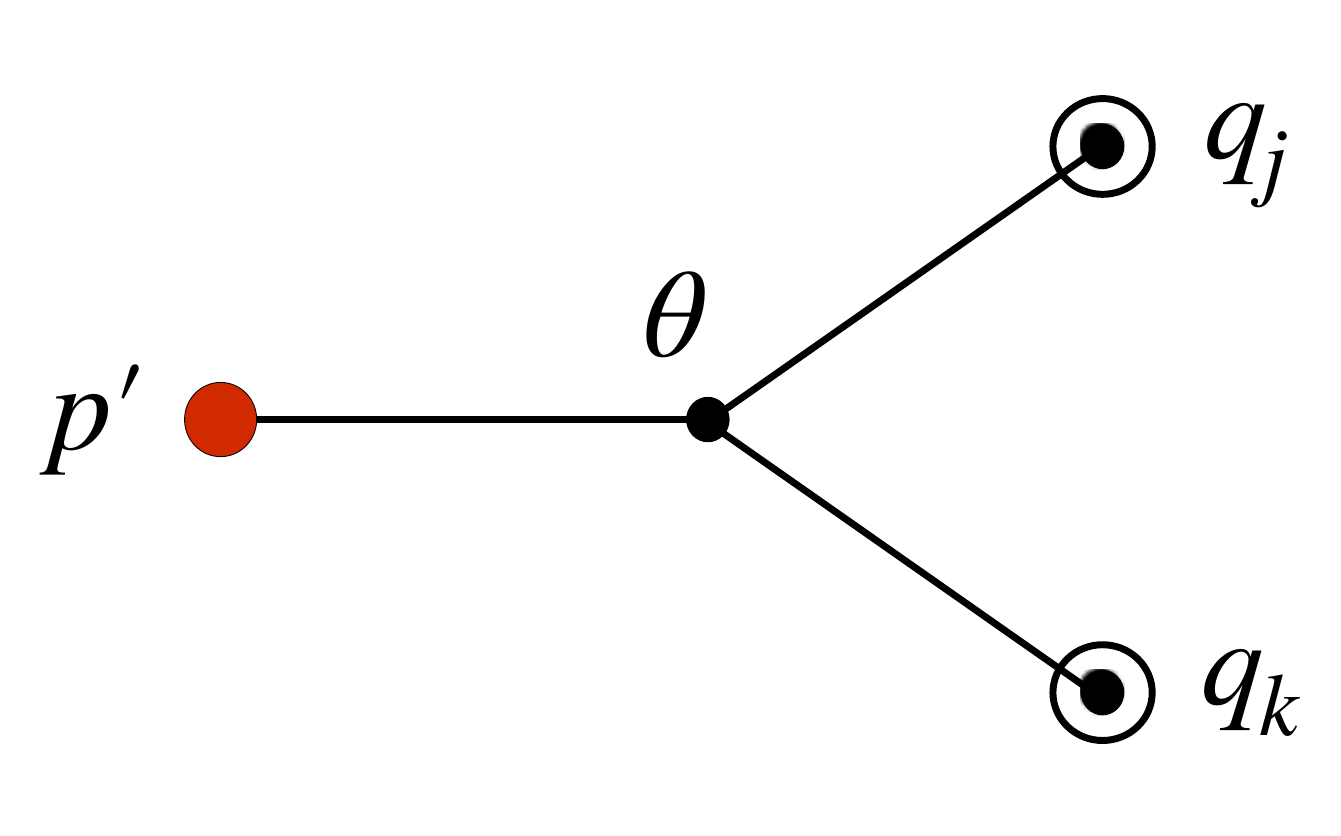}
\caption{Combinatorial structure of a tree with two leaves: there exists an internal vertex connected to the three external ones. We take this vertex as a reference point to sum over the trees.}\label{2-charges}
\end{figure}

\noindent
From each inserted vertex, one can find a unique path to the root of the tree. The two paths must join at some point $(\theta,r)$: a unique vertex linked to the root and likewise to the two leaves. Except for this vertex, all other vertices receive the filling factor $f$ as explained previously. At this vertex we can pull three multiplicities from the three adjacent propagators. This results in a special filling factor
\begin{align*}
\sum_{r\geq 1} r^3Y_r(\theta)=\frac{Y(\theta)}{[1+Y(\theta)]^2}=f(\theta)[1-f(\theta)].
\end{align*}
The charge covariance involves three dressed quantities corresponding to the three spines coming out of this intersection point. We recover the expression \eqref{charge-covariance-TBA}
\begin{align}
\frac{1}{L}\langle {\Q}_j{\Q}_k \rangle^\text{c}=\int\frac{d\theta}{2\pi}f(\theta)[1-f(\theta)](p')^\text{dr}(\theta)q_j^\text{dr}(\theta)q_k^\text{dr}(\theta).\label{two-point}
\end{align}
We can see here the reason why we chose to work with symmetric scattering kernels in this section. If that was not the case then we would need to define two dressing operations corresponding to the two directions on a spine. In this particular case, there is one spine coming toward $\theta$ and two spines coming outward. For higher cumulants, there are more propagators (spines) and it is tedious to specify their direction. We want to avoid this unnecessary complication by restricting our discussion to relativistically-invariant theories.
\subsection{Higher cumulants}
\begin{figure}
\centering
\includegraphics[width=13cm]{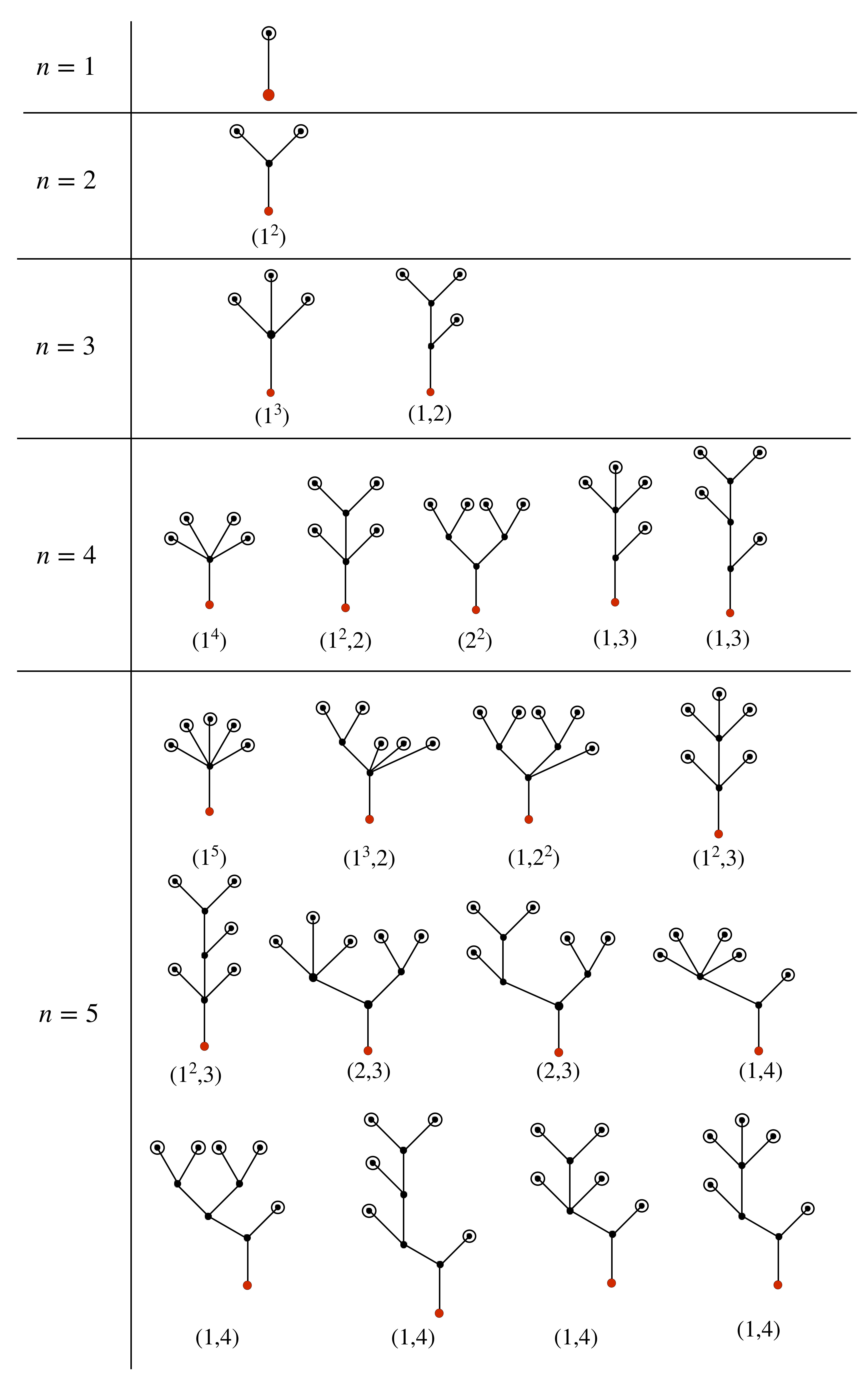}
\caption{Trees up to five leaves along with the partition used to generate them.}
\label{many}
\end{figure}

After understanding the explicit examples of the charge average and charge covariance, generalization to higher cumulants is straightforward. The $n^\text{th}$ cumulant is given by a sum over all tree-diagrams with $n+1$ external vertices : a root with $p'$ inserted and $n$ leaves carrying the $n$ conserved charges. Their internal vertices live in  phase space and will be integrated over. An external propagator connecting an internal vertex $\theta$ and an external vertex inserted with an operator $\psi$ has a weight $\psi^\text{dr}(\theta)$, here $\psi$ can either be the momentum derivative or the charge eigenvalues. An internal propagator connecting two internal vertices $\theta,\eta$ is assigned a weight $K^\text{dr}(\theta,\eta)$, where 
\begin{align}
K^\text{dr}(u,v)=K(u,v)+\int\frac{dw}{2\pi}K(u,w)f(w)K^\text{dr}(w,v).\label{dressed-propagator}
\end{align}  
The weight of an internal vertex $\theta$ of degrees $d$ is
\begin{align*}
\sum_{r\geq 1}(-1)^{r-1}r^{d-1}Y^r(\theta).
\end{align*}
We summarize these rules in the following
\begin{align}
\vcenter{\hbox{\includegraphics[width=3cm]{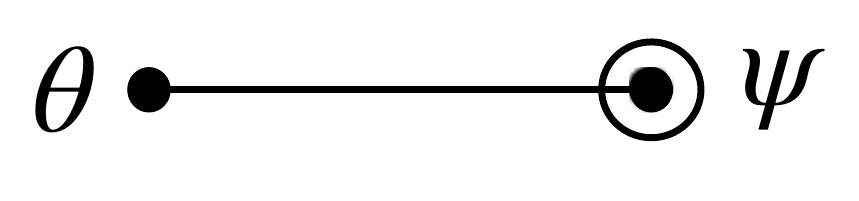}}}&=\psi^{\text{dr}}(\theta)\nonumber\\
\vcenter{\hbox{\includegraphics[width=3cm]{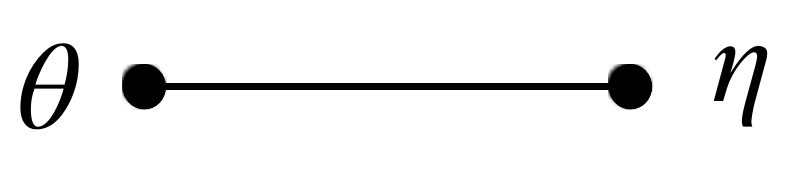}}}&= K^\text{dr}(\theta,\eta)
\label{reFeynman}\\
\vcenter{\hbox{\includegraphics[width=2.5cm]{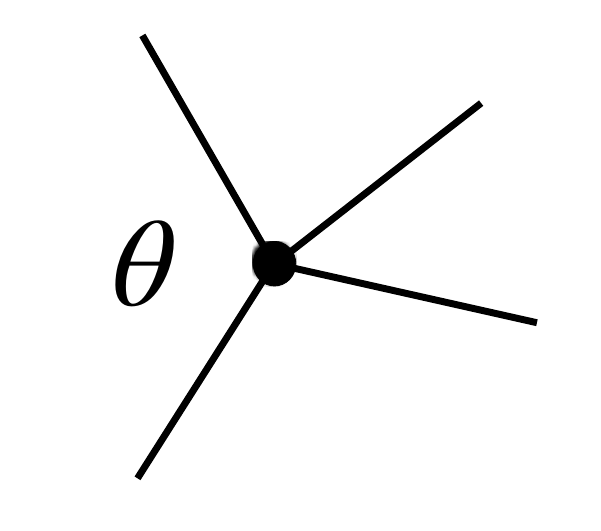}}}&=\sum_{r\geq 1}(-1)^{r-1}r^{d-1}Y^r(\theta)\nonumber
\end{align}
There is a simple recursive algorithm to generate all diagrams with $n$ leaves. For each partition of $n$ that is not the trivial one (i.e. $n=n$)
\begin{align}
n=\underbrace{a_1+...+a_1}_{\alpha_1}+\underbrace{a_2+...+a_2}_{\alpha_2}+...+\underbrace{a_j+...+a_j}_{\alpha_j},\quad a_1<a_2<...<a_j
\end{align} 
we choose $\alpha_1$ trees with $a_1$ leaves, ..., $\alpha_j$ trees with $a_j$ leaves. We then remove their roots and join them to a new common root. This algorithm translates into the following equation that determines the number $d_n$ of diagrams with $n$ leaves
\begin{align}
d_n=\sum_{\substack{p\in \mathcal{P}_n,\; |p|>1\\
p=(a_1^{\alpha_1},...,a_j^{\alpha_j})}}\prod_{i=1}^j\binom{d_{a_1}+\alpha_i-1}{\alpha_i}.
\end{align}
Some values of $d_n$ are given in table \ref{number-of-trees}. We also list all diagrams with up to $5$ leaves in figure \ref{many}.
\begin{table}[!htb]
\centering
  \begin{tabular}{ |c|c | c | c|c|c|c|c|c|c|c| }
    \hline
    $n$& $1$ & $2$ & $3$& $4$ & $5$ & 6 & 7 &8&9&10 \\ \hline
    $ d_n$& $1$ & $1$ & $2$& $5$ & $12$ & 33& 90& 261 & 766 & 2312\\ \hline
  \end{tabular}
\caption{Number of trees as function of their leaves.}
\label{number-of-trees}
\end{table}

\noindent
The third cumulant $\langle  {\Q}_j{\Q}_k{\Q}_l\rangle ^\text{c}$
can be read directly from the two diagrams with three leaves. The one on the left gives
\begin{align}
\int\frac{d\theta}{2\pi} f(\theta)[1-f(\theta)][1-2f(\theta)](p')^{\text{dr}}(\theta)q_j^{\text{dr}}(\theta)q_k^{\text{dr}}(\theta)q_l^{\text{dr}}(\theta).\label{second-diagram}
\end{align}
The one on the right involves three permutations of vertices
\begin{align}
\int\int\frac{d\theta}{2\pi}\frac{d\eta}{2\pi} (p')^{\text{dr}}(\theta)f(\theta)[1-f(\theta)] f(\eta)[1-f(\eta)]K^{\text{dr}}(\eta,\theta)\big[q_j^{\text{dr}}(\theta)q_k^{\text{dr}}(\eta)q_l^{\text{dr}}(\eta)\nonumber\\
+q_k^{\text{dr}}(\theta)q_j^{\text{dr}}(\eta)q_l^{\text{dr}}(\eta)+q_l^{\text{dr}}(\theta)q_j^{\text{dr}}(\eta)q_k^{\text{dr}}(\eta)\big].\label{three-first}
\end{align}
This result agrees with the expression \eqref{GGE-3} obtained from GGE free energy. Indeed, we can write the second derivative of pseudo-energy  in terms of the dressed propagator as follows
\begin{align}
\partial_{\beta^l}\partial_{\beta^k}\epsilon(\theta)=\int\frac{d\eta}{2\pi} f(\eta)[1-f(\eta)] K^{\text{dr}}(\theta,\eta)\partial_{\beta^k}\epsilon(\eta)\partial_{\beta^l}\epsilon(\eta).
\end{align}
Our formalism provides an intuitive picture of these cumulants: external vertices are the bare charges while internal vertices are virtual particles that carry anomalous corrections

\newpage
\section{The Leclair-Mussardo formula}
\label{lm-new-derivation}
The Pozsgay-Takacs formula provides an expansion of the diagonal matrix elements of a local operator in terms of the infinite-volume form factors with the same or lower number of particles
\begin{align}
\langle n_N,...,n_1|\mathcal{O}|n_1,...,n_N\rangle_L=\frac{1}{\det G(\theta_1,...,\theta_N)}\sum_{\alpha \cup \bar{\alpha}=\langle \theta_1,...,\theta_N\rangle}F_\text{c}^\mathcal{O}(\alpha)\det _{j,k\in\bar{\alpha}}G(\theta_1,...,\theta_N).
\end{align}
where the sum goes over all partitions of the rapidities $ {\theta_1, . . . , \theta_n}$ in to two complementary sets $\alpha$ and $\bar{\alpha}$ and $\det_{j,k\in\bar{\alpha}}G$ denotes the minor of the Gaudin matrix  obtained by deleting the lines and the columns that belong to the subset $\alpha$. With this result we will derive the Leclair-Mussardo formula from the tree expansion method.
For that we will need the diagonal matrix elements also for the multi-wrapping states $|n_1^{(r_1)},..., n_N^{(r_N)}\rangle$. We will make a very natural conjecture about this action namely \footnote{A relation between connected form factors with coinciding rapidities and those with generic rapidities was obtained in \cite{Pozsgay:2013jua}. It would be interesting to study the connection between that relation and our conjecture.}
\begin{align}
&\langle n_N^{(r_N)},...,n_1^{(r_1)}|\mathcal{O}|n_1^{(r_1)},...,n_N^{(r_N)}\rangle_L\nonumber\\
=&\frac{1}{\det G[\theta_1^{(r_1)},...,\theta_N^{(r_N)}]}\sum_{\alpha \cup \bar{\alpha}=\lbrace \theta_1,...,\theta_N\rbrace}\prod_{j\in\alpha}r_j F_\text{c}^\mathcal{O}(\alpha)\det_{j,k\in\bar{\alpha}} G[\theta_1^{(r_1)},...,\theta_N^{(r_N)}].\label{diagonal-multi-wrapping}
\end{align}
The logic behind this conjecture is that the action of the operator on a multi wrapping particle is the same as if it were single wrapping particle. The only difference is that the $r-$ wrapping particle appears $r$ times in the same time slice, the operator acts on each copy, which brings an overall factor of $r$. We should mention here that a discussion about the “multi-diagonal” matrix elements was presented in \cite{Bajnok:2017mdf}.

\begin{figure}[ht]
\centering
\includegraphics[width=13cm]{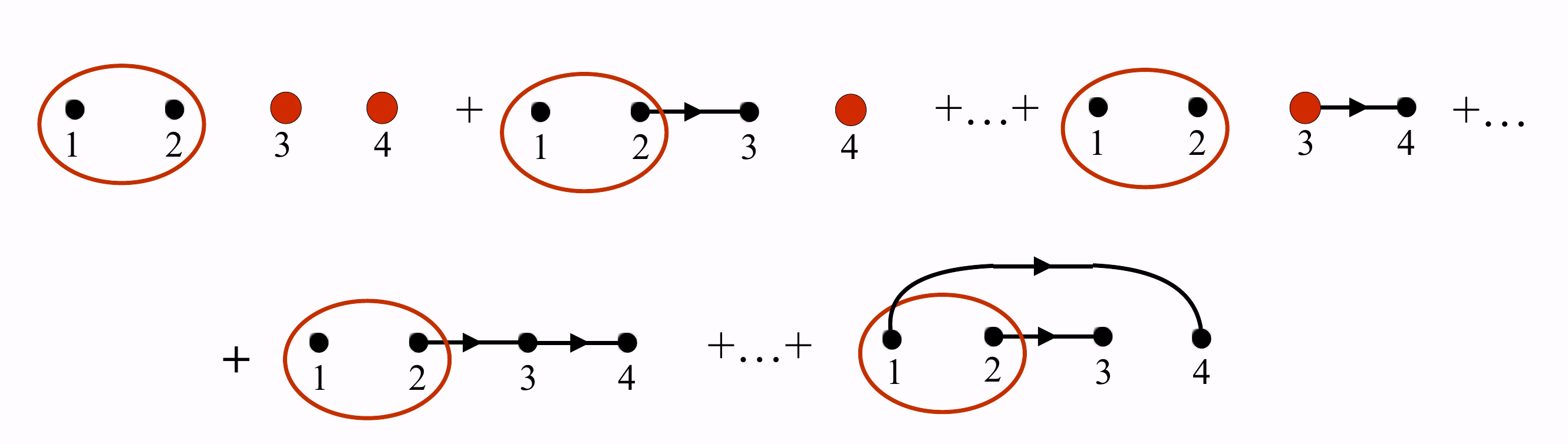}
\caption{The tree expansion for a principal minor of the Gaudin matrix $\det_{j,k\in \bar{\alpha}} \tilde{G}$ for $\alpha = \lbrace 1,2\rbrace$ and $\bar{\alpha} = \lbrace 3,4\rbrace$.}
\label{principal-example}
\end{figure}

Repeating the argument from the beginning of section 2.4, we can perform the sum over the complete set of states in the thermal expectation value of the operator $\mathcal{O}$
\begin{align}
\langle\mathcal{O}\rangle_{\vec{\beta}}=\frac{1}{Z(\vec{\beta},L)}\sum_{N\geq 0}\sum_{n_1<...<n_N}e^{-w(n_1,....,n_N)}\langle n_N,...,n_1|\mathcal{O}|n_1,...,n_N\rangle_L
\end{align}
Then, by inserting and expanding $1-\delta$ symbols we obtain a sum over mode numbers and rapidities where the multi-wrapping matrix element \eqref{diagonal-multi-wrapping} appears in each term of the sum. The Jacobian of the change of variables from mode numbers to rapidities compensates for its inverse on the right hand side of \eqref{diagonal-multi-wrapping}. As a result, the one point function of $\mathcal{O}$ takes the following form
\begin{align}
\langle\mathcal{O}\rangle_{\vec{\beta}}=&\frac{1}{Z(\vec{\beta},L)}\sum_{N\geq 0}\frac{(-1)^N}{N!}\sum_{r_1,...,r_N\in \mathbb{N}}\prod_{j=1}^N\frac{(-1)^{r_j}}{r_j^2}\int\frac{d\theta_je^{-r_j}w(\theta_j)}{2\pi}\nonumber\\
\times &\sum_{\alpha \cup \bar{\alpha}=\lbrace \theta_1,...,\theta_N\rbrace}\prod_{j\in\alpha}r_j^2 F_\text{c}^\mathcal{O}(\alpha)\det_{j,k\in\bar{\alpha}} \tilde{G}[\theta_1^{(r_1)},...,\theta_N^{(r_N)}].\label{one-point-function-integral-rep}
\end{align}
The next step is to apply the matrix-tree theorem for the diagonal minors of the (modified) Gaudin matrix in the last factor in the integrand in \eqref{one-point-function-integral-rep}. A minor obtained by removing all edges and all columns from the subset $\alpha \in \lbrace 1,..., N\rbrace$ of the matrix $\tilde{G}$ defined in eq.  has the following expansion
\begin{align}
\det_{j,k\in\bar{\alpha}} \tilde{G}=\sum_{\mathcal{F}\in \mathcal{F}_{\alpha,\bar{\alpha}}}\prod_{\substack{j\in\bar{\alpha}\\ j \text{ is a root}}} D_j\prod_{\langle jk\rangle \in \mathcal{F}}r_jr_kK(\theta_k,\theta_j)\label{all-minor-Gaudin}
\end{align}
The spanning forests $\mathcal{F} \in \mathcal{F}_{\alpha,\bar{\alpha}} $ are subjected to the three conditions in subsection \ref{Gaudin-expansion-section}, with the additional restriction that all vertices belonging to $\alpha$ are roots. The weight of these roots is one. An example is given in fig. \ref{principal-example}

The expansion \eqref{all-minor-Gaudin} follows directly from the expansion \eqref{det-Gaudin-tree} of the previous section which corresponds to the particular case $\alpha = \emptyset, \bar{\alpha} = \lbrace \theta_1, . . . , \theta_N\rbrace$. Indeed, the rhs of \eqref{all-minor-Gaudin} by retaining only the terms in the rhs of \eqref{det-Gaudin-tree} that contain the factor $\prod_{j\in\alpha} D_j$ and then dividing the sum by this factor.

Now we can proceed similarly to what we have done in the computation of the partition function, where rearranging of the order of summation allowed us to rewrite the sum as a series of tree Feynman diagrams. This time there will be two kinds of Feynman graphs: the “vacuum trees” and diagrams representing a vertex $F_\text{c}^\mathcal{O}$ with $n$ lines and a tree attached to each line. The weight of such tree is the same as the weight of the vacuum trees except for a factor of $r^2$ associated with the root. 
\begin{figure}[ht]
\centering
\includegraphics[width=13cm]{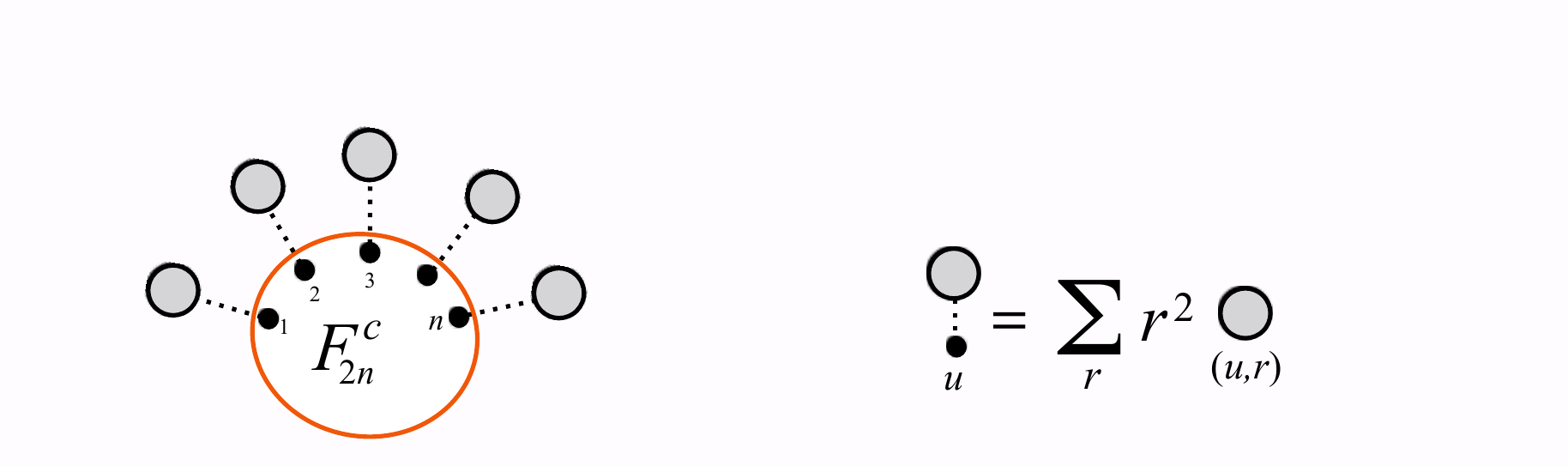}
\caption{The tree expansion for the thermal expectation value  of a local operator.}
\label{lm-diagram}
\end{figure}

\noindent
The sum over the vacuum trees cancels with the partition function and the sum over the surviving terms has the same structure as 
\begin{align}
\langle \mathcal{O}\rangle_{\vec{\beta}}=\sum_{n=0}^\infty\frac{1}{n!}\int\frac{d\theta_1}{2\pi}...\frac{d\theta_n}{2\pi}\prod_{j=1}^nf(\theta_j) F_{\text{c}}^\mathcal{O}(\theta_1,...,\theta_n)
\end{align}
which is depicted in fig. \ref{lm-diagram}. The factor $f (\theta)$ is obtained as the sum of all trees with a root at the point $\theta$, with extra weight $r^2$ associated with the root, as explained in \eqref{obtaining-the-filling}.
The difference of the sum over trees in the factor $f(u)$ compared with the sum over vacuum trees (2.34) is that there is an extra factor $r$ associated with the root reflecting the breaking of the $Z_r$ symmetry of the corresponding wrapping process.

\newpage
\section{Excited state energies}
\label{excited-new-section}
In subsection \ref{excited-section} we have interpreted the singular points in the TBA equation as rapidities of some state in the mirror theory. In this subsection we will recover the excited state energy \eqref{excited-energy} and excited state TBA equation \eqref{excited-pseudo-energy} by computing the partition in the presence of a mirror state $|\tilde{\boldsymbol{\theta}}^*\rangle=|\tilde{\theta}_1^*,...,\tilde{\theta}_M^*\rangle$ propagating along the spatial direction. We carry out the computation in the physical channel, summing over physical eigenstates $|\boldsymbol{\theta}\rangle=|\theta_1,...,\theta_N\rangle$. It is important to mention that the idea of using particles as defects and the defect TBA equations were derived in \cite{Bajnok:2007jg} while the idea  of deriving the exact Bethe ansatz equation from a partition function with defects already appeared in \cite{Basso:2017muf}.

\begin{figure}[ht]
\centering
\begin{subfigure}{.5\textwidth}
  \centering
  \includegraphics[width=.75\linewidth]{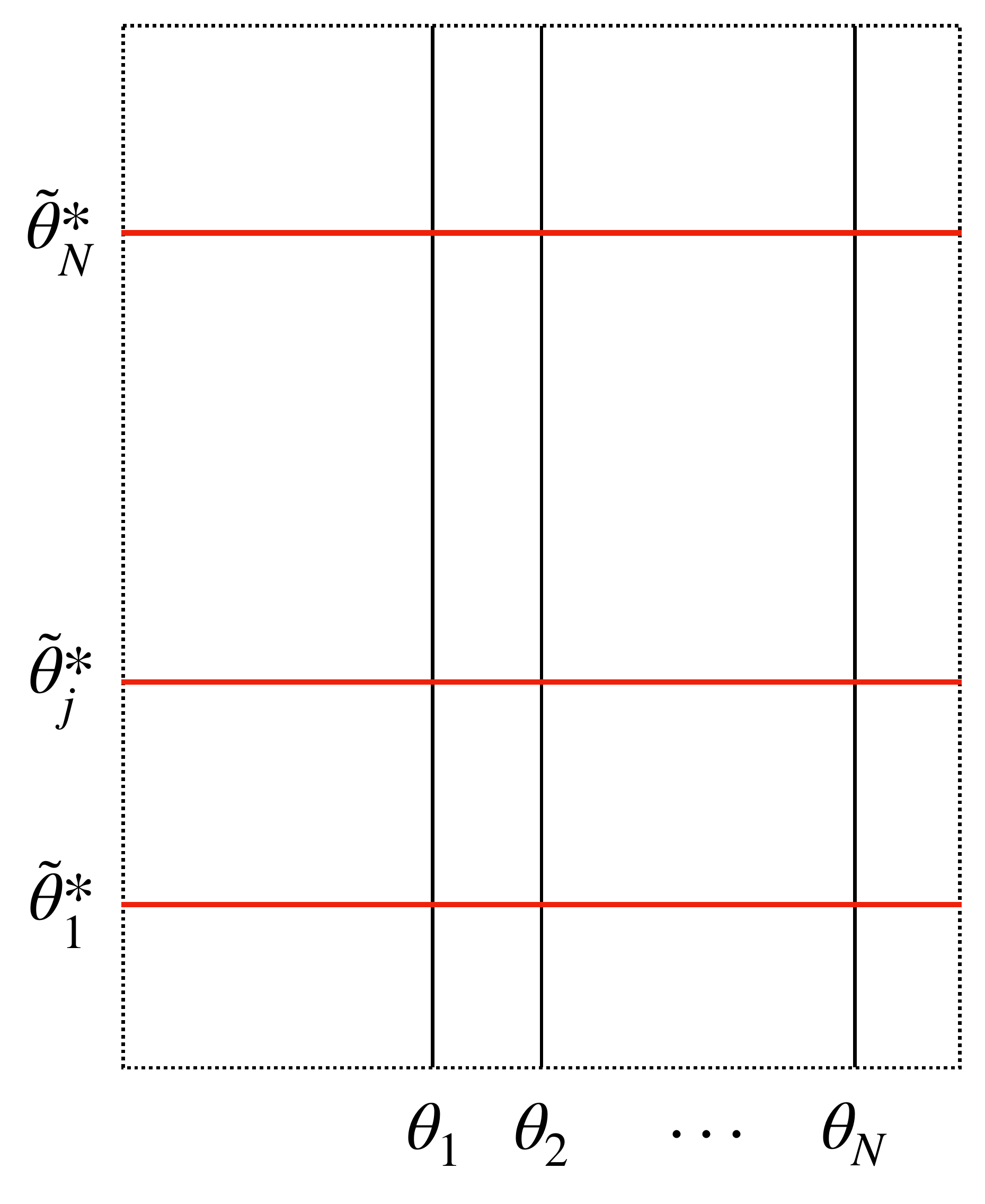}
  \caption{The partition function $Z(R,L,\boldsymbol{\tilde{\theta}})$}
  \label{Z}
\end{subfigure}%
\begin{subfigure}{.5\textwidth}
  \centering
  \includegraphics[width=.75\linewidth]{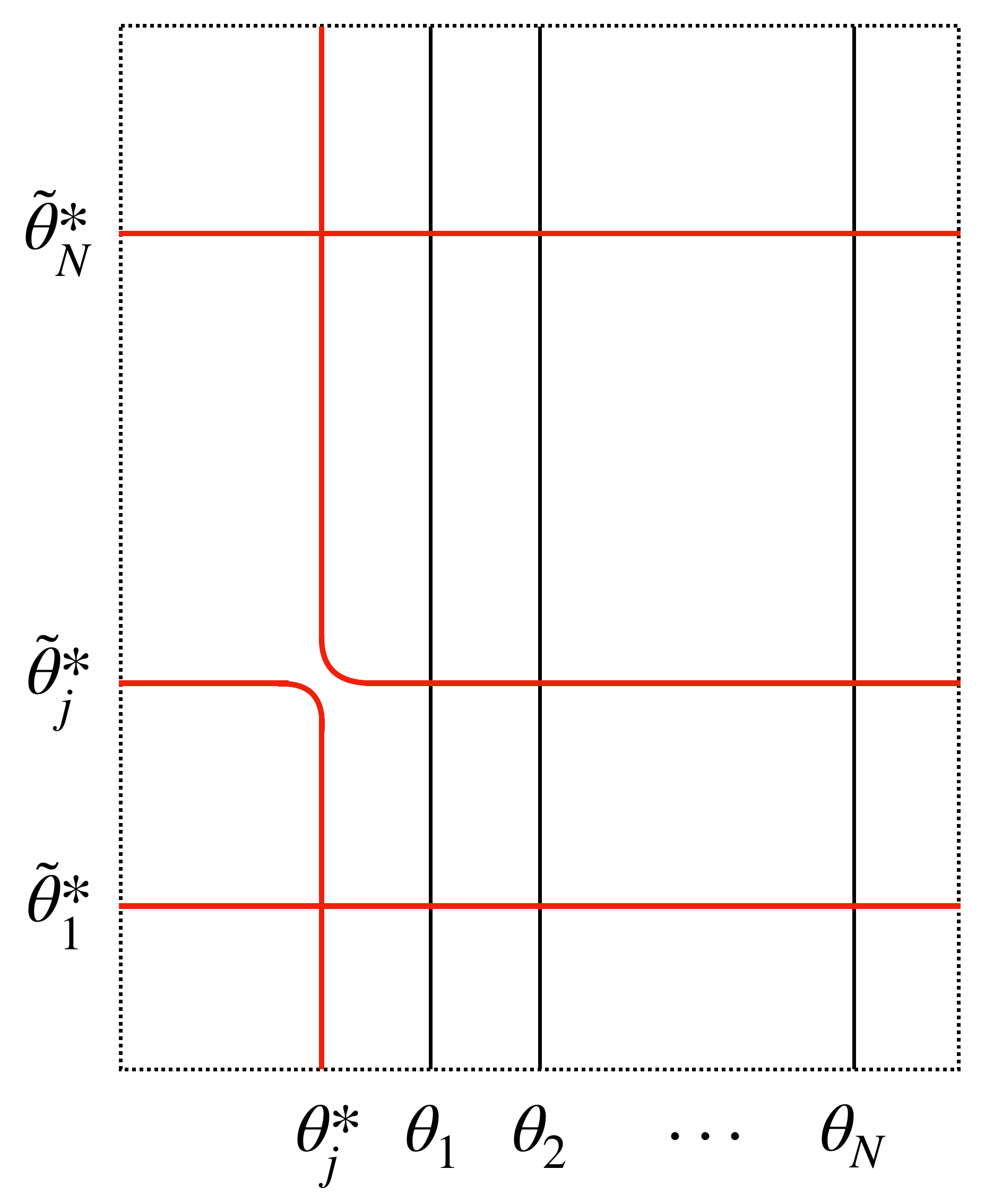}
  \caption{The partition function $Z_j(R,L,\boldsymbol{\tilde{\theta}})$}
  \label{Zj}
\end{subfigure}
\caption{Schematic representation of the partition function under the presence of a mirror state $\boldsymbol{\tilde{\theta}}^*$. A mirror particle can wraps around the time direction and acts like a physical particle. By requiring the equality between the two equivalent ways of computing the partition function, we recover the excited state TBA equation.}
\label{excited-TBA-diagrammatic}
\end{figure}

\noindent
For consistency with section \ref{excited-section}, we adopt the convention  $\theta^{\text{mirror}}=\theta^{\text{physics}}+i\pi/2$ in the following computation. The presence of a mirror state leads to  two modifications. First, the thermal weight of a physical particle now includes the interaction with mirror particles
\begin{align}
e^{-w(\theta,\tilde{\boldsymbol{\theta}}^*)}\equiv e^{-Rm\cosh(\theta)}\prod_{j=1}^M S(\theta,\tilde{\theta}_j^*-i\pi/2).
\end{align}
Second, the total scattering phases \eqref{Bethe-Yang} also  contain a similar contribution
\begin{align}
\Phi_j(\theta_1,...,\theta_N,\tilde{\boldsymbol{\theta}}^*)=Lp(\theta_j)-i\sum_{l=1}^M\log S(\theta_j,\tilde{\theta}_l^*-i\pi/2)-i\sum_{k\neq j}^N\log S(\theta_j,\theta_k).
\end{align}
This will modify the diagonal part of the corresponding Gaudin matrix. Consequently, the Feynman rules must be updated, while the rule for propagators remain unchanged

\begin{align*}
\vcenter{\hbox{\includegraphics[width=1.1cm]{vertex.pdf}}}&=\frac{(-1)^{r-1}}{r^2}e^{-rw(\theta,\tilde{\boldsymbol{\theta}}^*)}\nonumber,\\
\vcenter{\hbox{\includegraphics[width=1.1cm]{root.pdf}}}&=[Lp'(\theta)+\sum_{l=1}^M K(\theta,\tilde{\theta}_l^*-i\pi/2)]\frac{(-1)^{r-1}}{r}e^{-rw(\theta,\tilde{\boldsymbol{\theta}}^*)}.
\end{align*}
The combinatorial structure of the trees is also the same as before. The energy of the mirror state is obtained by adding to the free energy the contribution from the mirror particles that go around the spatial direction without scattering
\begin{align}
E(\boldsymbol{\tilde{\theta}}^*)&=\sum_{l=1}^M m\cosh(\tilde{\theta}_l^*)-\lim_{L\to \infty}\frac{1}{L}\int\frac{d\theta}{2\pi}[Lp'(\theta)+\sum_{l=1}^M K(\theta,\tilde{\theta}_l^*-i\pi/2)]\log[1+Y^c(\theta)] \nonumber\\
&=\sum_{l=1}^M m\cosh(\tilde{\theta}_l^*)-\int\frac{d\theta}{2\pi}Lp'(\theta)\log[1+Y^c(\theta)],\label{excited-energy-new}
\end{align}
where $Y^c$ solves for the excited state TBA equation
\begin{align}
Y^c(\theta)=e^{-w(\theta,\tilde{\boldsymbol{\theta}}^*)}\exp\int \frac{d\eta}{2\pi} K(\eta,\theta)\log [1+Y^c(\eta)].\label{excited-TBA}
\end{align}
The exact Bethe equations for the mirror state are obtained by the following requirement. Let $Z_j(R, L,\tilde{\boldsymbol{\theta}}^*)$ be the partition function with the $j$-th mirror particle winding once around the time circle before winding around the space circle. The configurations that contribute to $Z (R, L)$ and $Z_j (R, L)$ are depicted in Figs. \ref{Z} and \ref{Zj}. In order to compute the partition function$ Z_j (R, L)$ we notice that the configurations in Fig. \ref{Zj} can be simulated by pulling one of the mirror particles out of the thermal ensemble giving to its rapidity a physical value $\theta_j^*=\tilde{\theta}_j^*-i\pi/2$ . Indeed, since $S (\theta_j^* , \theta_j ^*) = −1$, the partition function in presence of such extra mirror particle is $-Z_j (R, L,\tilde{\boldsymbol{\theta}}^*)$. In this way $Z_j (R, L,\tilde{\boldsymbol{\theta}}^*)$ is given by the sum over all trees, with one extra tree having a root $\theta_j^*$ and $r = 1$. The generating function for such trees is $Y^c(\theta_j^*)$, while the contribution of the “vacuum” trees give the partition function: $Z_j = −Y^c(\theta_j^*)Z$. The periodicity in the space direction requires that $Z_j = Z$, which gives the exact Bethe-Yang equation
\begin{align}
Y^c(\theta_j^*)=-1,\quad j=1,2,...,M.\label{excited-Bethe}
\end{align}

\chapter{Open systems with diagonal scattering}
In the previous chapter we have applied our diagrammatic formalism to re-derive some well-known thermal quantities in periodic integrable systems (and also obtain a new result). In this chapter we will investigate the extension of our method to systems with integrable boundaries. We restrict our discussion to  $1+1$ dimensional field theories with a single massive excitation above the vacuum. The analysis of theories with non-diagonal bulk scattering matrix will be presented in the next chapter.
\section{Finite temperature g-function: definition} 
Consider the theory in  an open interval of   length
$L$, with two boundaries  $a$ and $b$.  
The theory is integrable with a two-to-two bulk scattering phase
$S(\theta,\eta)$ and reflection factors $R_a(\theta),R_b(\theta)$ at the boundaries. They satisfy a set of conditions (see subsection \ref{open-integrable-section}), among which  the unitarity condition
\begin{align}
S(\theta,\eta)S(\eta,\theta)=R_a(u)R_a(-u)=R_b(u)R_b(-u)=1.\label{unitarity-open-theories}
\end{align}
Following the spirit of the previous chapter, we consider a general scenario where the bulk S-matrix does not necessarily depend on the difference between rapidities. We however  assume a milder condition
\begin{align}
S(\theta,-\eta)S(-\theta,\eta)=1.
\label{assumption-on-S-matrix}
\end{align}
The relevance of this identity will become clear shortly. The partition  function of the theory at inverse temperature $R$ is given by the thermal trace
\begin{align}
Z_{ab}(R,L)=\tr\;e^{-H_{ab}(L)R},
\label{Z-open-R-channel}
\end{align}
where $H_{ab}(L)$ denotes the Hamiltonian defined on a segment 
of length $L$  with boundary conditions $a$ and $b$. If the two boundaries are removed then we recover the partition function of the periodic theory
\begin{align}
\label{Z-periodic-R-channel}
Z(R,L)=\tr\;e^{-H(L)R}.
\end{align}
which was computed in the previous chapters.
The  boundary entropy of the open system is  given by the difference between the two free energies
\begin{align}
\mathcal{F}_{ab}(R,L)\equiv\log Z_{ab}(R,L)-\log Z(R,L).
\end{align}
The g-function is defined as the contribution of a
\textit{single} boundary to the free energy.  To obtain it, we pull
the two boundaries far away from each other to avoid interference
\begin{align}
\log g_a(R)\equiv \frac{1}{2}\lim_{L\to\infty}\mathcal{F}_{aa}(R,L).\label{g-function-definition}
\end{align}

\begin{figure}[ht]
\centering
\includegraphics[width=12cm]{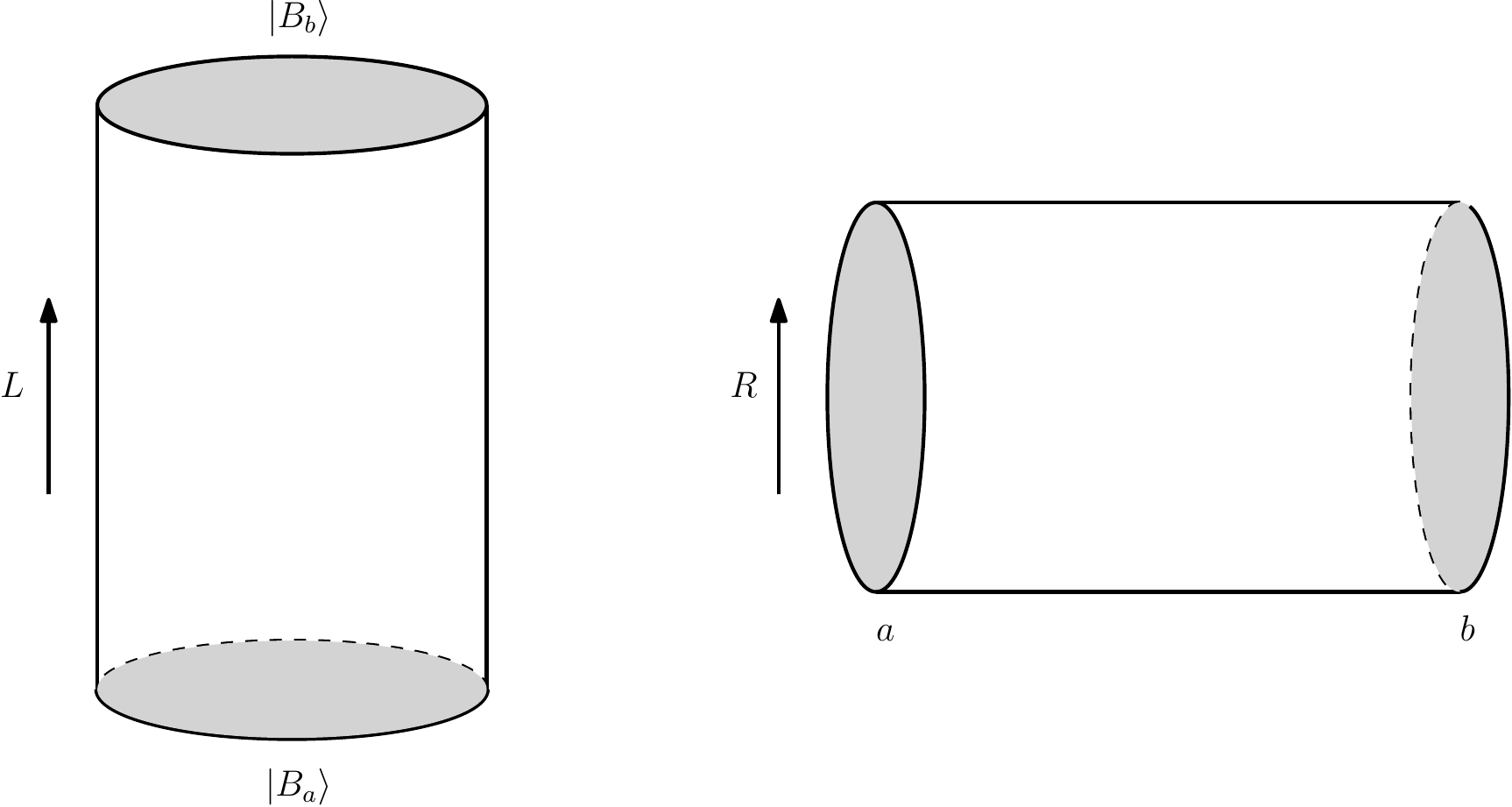}
  \caption{Two equivalent ways of evaluating the partition function on a cylinder.}
\end{figure}


 From the perspective of the  mirror theory, defined on a circle with circumference $R$, 
 the partition function with periodic boundary conditions \eqref{Z-periodic-R-channel}
 takes a similar form
\begin{align}
Z(R,L)=\tr\;e^{-\tilde{H}(R)L},\label{Z-periodic-L-channel}
\end{align}
 where the trace is in the Hilbert space  of the mirror theory.
 In contrast,  after a mirror transformation the thermal partition function with open boundary conditions   becomes  the overlap of    an initial state
$\langle B_a|$ and a final state  $|B_b\rangle$    defined on a circle 
of circumference $R$ after evolution 
  at mirror   time  $L$    \cite{Ghoshal:1993tm}. 
 Evaluated in the mirror theory, the partition function \eqref{Z-open-R-channel} reads
\begin{align}
Z_{ab}(R,L)=\langle B_a|e^{-\tilde{H}(R)L}|B_b\rangle.\label{Z-open-L-channel}
\end{align}
Although the partition function is the same, the physics is rather
different in the two channels.  In the mirror theory, the g-function provides information about overlapping of the boundary state and the ground state at finite volume.  To see this, we write \eqref{Z-open-L-channel} as a
sum over eigenstates $|\psi\rangle$ of the periodic Hamiltonian
$\tilde{H}(R)$
\begin{align}
\langle B_a|e^{-\tilde{H}(R)L}|B_b\rangle=\sum_{|\psi\rangle}\frac{\langle B_a|\psi\rangle}{\sqrt{\langle \psi|\psi\rangle}}e^{-L\tilde{E}(|\psi\rangle)}\frac{\langle \psi|B_b\rangle}{\sqrt{\langle \psi|\psi\rangle}}.
\end{align}
In the large $L$ limit, this sum is dominated by a single term corresponding to the ground state $|\psi_0\rangle$. The $g$-function is then given by the overlap between this state and the boundary state
\begin{align}
g_a(R)=\frac{\langle B_a|\psi_0\rangle}{\sqrt{\langle \psi_0|\psi_0\rangle}}.\label{overlap-ground}
\end{align}

\section{Known results}
The first attempt to compute the g-function \eqref{g-function-definition} was carried out  by  LeClair,  Mussardo,  Saleur and  Skorik \cite{LeClair:1995uf}, using the traditional TBA saddle point approximation.  They  obtained  an expression similar to the  bulk TBA free energy
\begin{align}
\log(g_ag_b)^\textnormal{saddle}(R)=\frac{1}{2}\int_{-\infty}^{+\infty}
\frac{d\theta}{2\pi}\, \Theta_{ab}(\theta)\log[1+e^{-\epsilon(\theta)}],\label{g-function-first-attempt}
\end{align}
where $\epsilon$ is the pseudo-energy at inverse temperature $R$ given by the familiar TBA equation
\begin{align}
\epsilon(\theta)=E(\theta)R-\int_{-\infty}^\infty
\frac{d\eta}{2\pi}K(\eta,\theta)\log(1+e^{-\epsilon(\eta)}).\label{familiar-TBA}
\end{align}
and the source term $\Theta$  involves the bulk scattering and the boundary reflection matrices
\begin{gather}
\Theta_{ab}(\theta)\equiv K_a(\theta)+K_b(\theta)-2K(\theta,-\theta)-2\pi\delta(u),\quad \text{where}\quad K_{a,b}(\theta)=-i\partial_\theta \log R_{a,b}(\theta)\label{Omega-term}.
\end{gather}
It was later shown by Woynarovich \cite{Woynarovich:2004gc} that another volume-independent contribution is  produced  by  the fluctuation around the TBA saddle point.  The result can be written as a Fredholm determinant
\begin{align}
\log(g_ag_b)^\textnormal{fluc}(R)=-\log\det(1-\hat{K}^+),\label{fluc}
\end{align}
where the kernel $\hat{K}^+$ has support on the positive real axis and its action is defined by
\begin{align}
\hat{K}^+ F(\theta)=\int_0^\infty\frac{d\eta}{2\pi}\big[K(\theta,\eta)+
K(\theta,-\eta)\big]\frac{1}{1+e^{\epsilon(\eta)}}F(\eta).\label{K-plus}
\end{align}
In particular, the fluctuation \eqref{fluc} around the saddle point is boundary independent. One can say its presence is of pure geometric origin. A major problem of Woynarovich's computation  is that it also predicts a similar term  for periodic systems, while it is known that  there is no such correction.

Dorey, Fioravanti,  Rim  and   Tateo \cite{Dorey:2004xk} took a different approach towards this problem.  They started with the definition of the partition function as a thermal sum over a complete set of states labelled by mode numbers.  In the infinite volume limit, this sum can be replaced by integrals over rapidities.  The integrands were explicitly worked
out for small number of particles.  Based on these first terms and the structure of TBA saddle point result \eqref{TBA}, the authors advanced a conjecture about the boundary-independent part of $g$-function.  Their proposal has the structure of a Leclair-Mussardo type series
\begin{align}
\log(g_ag_b) (R)&=\log(g_ag_b)^\textnormal{saddle}(R)\nonumber\\
&+\sum_{n\geq
1}\frac{1}{n}\prod_{j=1}^n\int_{-\infty}^{+\infty}\frac{d\theta_j}{2\pi}
\frac{1}{1+e^{\epsilon(\theta_j)}}K(\theta_1+\theta_2)K(\theta_2-\theta_3)...K(\theta_n-\theta_1)
\label{cluster-expansion}
\end{align}

Pozsgay \cite{Pozsgay:2010tv} (see also Woynarovich \cite{Woynarovich:2010wt}) argued that the same expression for $g$-function could be obtained from a refined version of TBA saddle point approximation. He noticed that the mismatch between \eqref{fluc} and the series in \eqref{cluster-expansion} is resolved if one uses a non-flat measure for the TBA functional integration.   
This  non-trivial measure comes from  the Jacobian  of the change of variables from mode number to rapidity,  and represents the continuum limit of the   Gaudin determinant.
   
The fluctuation around the saddle point involves only diagonal elements  of this Gaudin matrix, resulting in   the inverse power of  the Fredholm determinant $\det(1-\hat{K}^+)$.  On the other hand, the functional integration measure contains the off-diagonal elements as well, which constitute another Fredholm determinant $\det(1-\hat{K}^-)$. The kernel $K^-$ is defined in a similar way as $K^+$ 
\begin{align}
\hat{K}^- F(\theta)=\int_0^\infty\frac{d\eta}{2\pi}\big[K(\theta,\eta)-K(\theta,-\eta)\big]\frac{1}{1+e^{\epsilon(\eta)}}F(\eta).\label{K-minus}
\end{align} 
Pozsgay rewrote the result \eqref{cluster-expansion} in terms of these two Fredholm determinants
\begin{align}
\log(g_ag_b)(R)=\log(g_ag_b)^\textnormal{saddle}(R)+\log\det\frac{1-\hat{K}^-}{1-\hat{K}^+}.
\label{g-function-formula}
\end{align}
The two kernels $\hat{K}^\pm$ can be read off from the asymptotic Bethe equations.  For a periodic system, they happen to be the same and the effects from the fluctuation and the measure cancel each other.

It is important to distinguish the Jacobians in \cite{Dorey:2004xk} from the one in \cite{Pozsgay:2010tv}.  The former appear in each term of the cluster expansion while the latter is obtained from the thermodynamic state that minimizes the TBA functional action.  Put it simply, the Jacobian in \cite{Pozsgay:2010tv} is the thermal average of all the Jacobians in \cite{Dorey:2004xk}.

 In the next sections  we will derive  the expression \eqref{g-function-formula}  by evaluating the partition function in the $R$-channel, namely equation \eqref{Z-open-R-channel}, in the limit when $L$ is large.  In order to do that, we will insert a decomposition of the identity in  a complete basis of  eigenstates of the Hamiltonian $H_{ab}(L)$ and perform the thermal trace.   
\section{New derivation}
\subsection{The sum over mode number}
The $g$-function \eqref{g-function-definition} is extracted by taking the limit of  large  volume $L$. In this limit, we can diagonalize the Hamiltonian $H_{ab}(L)$ using
the technique of Bethe ansatz. Consider an $N$-particle eigenstate 
$|\vec{\theta}\rangle=|\theta_1,\theta_2,...,\theta_N\rangle$.  To obtain the Bethe Ansatz equations in presence of two boundaries, we follow a particle of rapidity $\theta_j$ as it propagates to a boundary and is reflected to the opposite direction.  It continues to the other boundary, being reflected for a second time and finally comes back to its initial position, finishing a trajectory of length $2L$.  During its propagation, it scatters with the rest of the particles twice,  once from the left and once from the right. This process translates into the quantization condition of the state $|\vec{\theta}\rangle $
\begin{align}
e^{2ip(\theta_j)L}R_a(\theta_j)R_b(\theta_j)\prod_{k\neq
j}^NS(\theta_j,\theta_k)S(\theta_j,-\theta_k)=1,\quad \forall
j=1,...,N.\label{propagation}
\end{align}
We can write these equations in logarithmic form by introducing a new set of variables: the total scattering phases $\Phi_1,\Phi_2,...,\Phi_n$ defined by
\begin{align}
\Phi_j(\vec{\theta})\equiv 2p(\theta_j)L-i\log [R_a(\theta_j)R_b(\theta_j)\prod_{k\neq j}^N
S(\theta_j,\theta_k)S(\theta_j,-\theta_k)],\quad \forall j=1,...,N.\label{Bethe-equations}
\end{align}
In terms of these variables, the quantization of the state $|\vec{\theta}\rangle$ reads
\begin{align}
\Phi_j(\vec{\theta})=2\pi n_j\quad \forall j=1,...,N\quad
\textnormal{with } n_j\in\mathbb{Z}.\label{quantization}
\end{align}
In the presence of boundary two particles having the same mode numbers
but of opposite signs are indistinguishable.  To avoid 
overcounting of states, we should put a positivity constraint on mode
numbers. A basis in the $N$-particle sector of the Hilbert space is then labeled by all 
sets of strictly increasing positive integers
 $0<n_1<\dots<n_N$.  The corresponding eigenvector of the Hamiltonian $H_{ab}(L)$ is characterised by a set of rapidities $0<\theta_1<\dots <\theta_N$, obtained
by solving the Bethe equations \eqref{Bethe-equations} and
\eqref{quantization}.

Inserting a complete set of eigenstates we write the partition function on a cylinder as
\begin{align}
Z_{ab}(R,L)=\sum_{N=0}^\infty\;\sum_{0<n_1<...<n_N}e^{-RE(n_1,...,n_N)}.
\label{sum-constraint}
\end{align}
In this equation, the energy $E$ is  an implicit  function of mode numbers $n_1,...,n_N$.   To find its explicit form, one needs to solve the Bethe equations for the corresponding rapidities $\theta_1,...,\theta_N$.  As a function of the rapidities, the energy is  equal to the sum of the energies of the individual particles $E(n_1,...,n_N)=E(\theta_1)+...+E(\theta_N)$.

In order to write the sum \eqref{sum-constraint} as an integral over rapidities, we first have to remove the constraint between the mode numbers.  We do this by inserting Kronecker symbols to get rid of unwanted configurations
\begin{align}
Z_{ab}(R,L)=\sum_{N=0}^\infty\frac{1}{N!}\;\sum_{0\leq n_1,...,n_N}\,\prod_{j<k}^N(1-\delta_{n_j,n_k})\prod_{j=1}^N(1-\delta_{n_j,0})e^{-RE(n_1,...,n_N)}.\label{sum-unconstraint}
\end{align}
The first Kronecker symbol introduces the condition that the mode numbers are all different,
and the second one eliminates the mode numbers equal to zero. Let us expand  in monomials  the first  factor containing Kronecker symbols,  which imposes the exclusion principle.   As shown in the previous chapter, the partition function \eqref{sum-unconstraint} can be written as a sum over all sequences    $(n_1^{r_1},...,n_N^{r_N})$ of non-negative, but otherwise unrestricted  mode numbers $n_i$ with multiplicities $r_i$. Each sequence $(n_1^{(r_1)},...,n_N^{(r_N)})$ in the sum corresponds to a state with $r_j$ particles of the same mode number $n_j$, for $j=1,2,...,N$. The total number of particles in such a sequence is $r_1+\dots + r_N$
\begin{align}
Z_{ab}(R,L)=\sum_{N=0}^\infty\frac{(-1)^N}{N!}\;\sum_{0\leq
n_1,...,n_N}\prod_{j=1}^N(1-\delta_{n_j,0})\sum_{1\leq
r_1,...,r_N}\frac{(-1)^{r_1+....+r_N}}{r_1....r_N}e^{-RE[n_1^{(r_1)},...,n_N^{(r_N)}]}.\label{sum-multiplicity}
\end{align}
The  rapidities $\theta_1, \dots, \theta_N$  of  a generalised Bethe states $(n_1^{r_1},...,n_N^{r_N})$ satisfy the  Bethe equations
\be
\Phi_j=2\pi n_j, \quad j=1,  \dots, N\, ,
\ee
where the total scattering phases $\Phi _j= \Phi _j[\theta_1^{(r_1)},\dots, \theta_N^{r_N)}]$ are defined  by  
\begin{align}
e^{ i \Phi_j}\equiv 
 e^{2i p(\theta_j)L } \times R_a(\theta_j)R_b(\theta_j)\times
[e^{i\pi}S(\theta_j,-\theta_j))^{r_j-1}\times\prod_{k\neq
j}^N(S(\theta_j,\theta_k)S(\theta_j,-\theta_k)]^{r_k}.
\label{Bethe-equations-multiplicity}
\end{align}
The energy of this state is given by $r_1E(\theta_1)+...+r_NE(\theta_N)$.
\subsection{From mode numbers to rapidities}
In the large $L$ limit, we can replace a discrete sum over mode numbers $n$ by a continuum integral over variables $\Phi$
\begin{align*}
\sum_{0\leq n_1,,,,n_N}=\int_0^\infty\frac{d\Phi_1}{2\pi}...\int_0^\infty
\frac{d\Phi_N}{2\pi}+\mathcal{O}(e^{-L}).
\end{align*}
We can then use equation \eqref{Bethe-equations-multiplicity} to pass from $(\Phi_1,...,\Phi_m)$ to rapidity variables $(\theta_1,...,\theta_N)$.  The only subtle point compared with the periodic case is the factor excluding the mode numbers $n_j=0$ from the sum
\eqref{sum-multiplicity}
\begin{align*}
\sum_{0\leq
n_1,...,n_m}\prod_{j=1}^m(1-\delta_{n_j,0})
=\int_0^\infty\frac{d\Phi_1}{2\pi}...\int_0^\infty\frac{d\Phi_m}{2\pi}
\prod_{j=1}^m(1-2\pi\delta(\Phi_j))+\mathcal{O}(e^{-L}).
\end{align*}
We would like to incorporate this factor into the Jacobian matrix $\partial_\theta\Phi$.  We can do this by first expanding the product as a sum over subsets $\alpha\subset\lbrace 1,2,...,N\rbrace$
\begin{align*}
\int_0^\infty\frac{d\Phi_1}{2\pi}...\int_0^\infty\frac{d\Phi_N}{2\pi}\sum_{\alpha}(-2\pi)^{|\alpha|}\delta(\Phi_\alpha)=&\sum_{\alpha}\prod_{j=1}^m\int_0^\infty\frac{d\theta_j}{2\pi}\bigg[\frac{\partial\Phi}{\partial \theta}\bigg]_{  \alpha,  \alpha}(-2\pi)^{|\alpha|}\delta(\theta_\alpha)\\
=&\prod_{j=1}^m\int_0^\infty\frac{d\theta_j}{2\pi}\det\bigg[\frac{\partial\Phi}{\partial
\theta}-2\pi\delta(\theta)\bigg].
\end{align*}
Here  $[\partial\Phi/\partial \theta]_{  \alpha,  \alpha}$ denotes the diagonal minor of the Jacobian matrix obtained by deleting its $\alpha$-rows and $\alpha$-columns. The sum over subsets is the the expansion of the   determinant of a  sum of two matrices. Hence  the unphysical state at $\theta=0$ can be eliminated by adding a term equal to $-2\pi\delta(\theta)$ to the diagonal elements of the Jacobian matrix when we change variables from $\Phi$ to $\theta$
\begin{align}
&G_{jk}[\theta_1^{(r_1)},...,\theta_N^{(r_N)}]\equiv\partial_{\theta_k}\Phi_j-2\pi\delta(\theta_j)\delta_{jk}\nonumber\\
=&\big[D_{ab}(\theta_j)+2r_jK(\theta_j,-\theta_j)+\sum_{l\neq
j}^Nr_l(K(\theta_j,\theta_l)+K(\theta_j,-\theta_l))\big]\delta_{jk}\nonumber\\
-&r_k[K(\theta_k,\theta_j)-K(\theta_k,-\theta_j)] \; (1- \delta_{jk}), \quad \forall j,k=1,2,...,N\label{Gaudin-multiplicity}
\end{align}
where
\begin{align}
D_{ab}(\theta)\equiv 2Lp'(\theta)+\Theta_{ab}(\theta).\label{sigma-term}
\end{align}
with $\Theta_{ab}$ given in \eqref{Omega-term}. In order to apply the matrix-tree theorem, we consider the following matrix
\begin{align}
\tilde{G}_{jk}\equiv
r_kG_{kj}=&\big[r_jD_{ab}(\theta_j)+2r_j^2\bar{K}_{jj}+\sum_{l\neq
j}^Nr_jr_l(K_{jl}+\bar{K}_{jl})\big]\delta_{jk}\nonumber\\
&-r_jr_k(K_{jk}-\bar{K}_{jk})\, (1- \delta_{jk}),\quad \forall j,k=1,2,...,N,
\label{Gaudin-hat-multiplicity}
\end{align}
where we have used the notation 
\begin{align}
K_{jk}=K(\theta_j,\theta_k) , \quad 
 \bar{K}_{jk}=K(\theta_j,-\theta_k)=K(-\theta_j,\theta_k) . 
\end{align}
In terms of this matrix, the partition function is written as
\begin{align}
Z_{ab}(R,L)=\sum_{m=0}^\infty\frac{(-1)^N}{N!}\sum_{1\leq
r_1,...,r_N}\prod_{j=1}^N\int_0^\infty\frac{d\theta_j}{2\pi}\frac{(-1)^{r_j}}{r_j^2}
e^{-r_jRE(\theta_j)}\det
\tilde{G}[\theta_1^{(r_1)},...,\theta_N^{(r_N)}].\label{partition-Gaudin-hat}
\end{align}
\subsection{Matrix-tree theorem}
The matrix-tree theorem for signed graphs
\cite{Chaiken82acombinatorial} allows us to write the determinant of
the matrix \eqref{Gaudin-hat-multiplicity} as a sum over graphs.  This
theorem as stated in \cite{Chaiken82acombinatorial} is quite technical
and we provide a brief formulation in the following together with two proofs, one combinatorial and one field-theoretical in the appendices. First, let us define
\begin{equation}
\begin{aligned}
K^\pm_{jk}&= K_{jk}\pm\bar{K}_{jk}.\label{K-def}
\end{aligned}
\end{equation}
Then the Gaudin-like  matrix \eqref{Gaudin-hat-multiplicity} takes the form ($j,k=1,2,...,N$)
\begin{align}
\tilde{G}_{jk}&=\big[r_jD_{ab}(\theta_j)+r_j^2(K^+_{jj}-K^-_{jj})+\sum_{l\neq
j}^mr_jr_l\, K^+_{jl}\big]\delta_{jk}-r_jr_k\, K^-_{jk} (1-\delta_{jk}).
\label{Laplacian-new}
\end{align}
\begin{wrapfigure}{r}{0.28\textwidth}
  \begin{center}
    \includegraphics[width=0.27\textwidth]{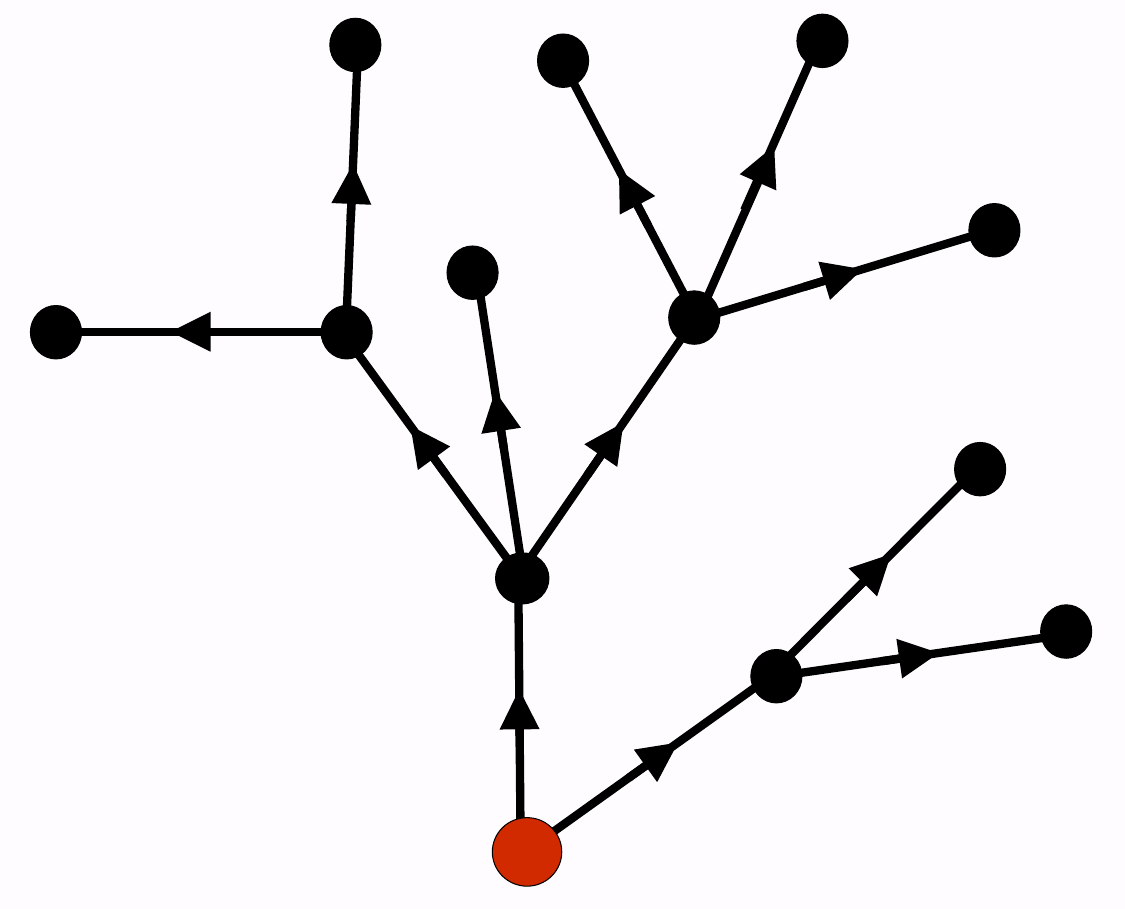}
  \end{center}
  \caption{A tree with $K^+$ edges}
  \label{tree-K+}
\end{wrapfigure}
The determinant of this matrix can be written as a sum over all
(not necessarily connected) graphs $\mathcal{F}$ having exactly  $m$ vertices
labeled by  $v_j$ with $ j=1,..., m$ and two types of edges, positive and negative,  which we denote by   $\ell_{jk}^\pm \equiv  \langle v_j\to v_k\rangle^\pm  $.  
The connected component of
each graph is either:
\begin{itemize}
\item A rooted directed tree with  
 only positive edges $\ell_{kl}^+= \langle v_k\to v_l\rangle^+$  oriented  so that the edge points to the vertex which is farther from  the root, as  shown in fig. \ref{tree-K+}. The weight of such a tree is a product of  a factor  $r_jD_{ab}(\theta_j)$ associated with the root  $v_j$ and  factors $r_lr_kK^+_{lk}$ associated with its  edges $\ell_{kl}^+$.
\item A positive   (fig.  \ref{loop-K+}) or a  negative  (fig.  \ref{loop-K-}) oriented  cycle 
with outgrowing trees. A  positive/negative  loop   is an oriented  cycle (including tadpoles which are cycles of length 1) entirely made of positive/negative  edges having the same orientation.  The outgrowing trees consist of positive  edges only.   The weight of a loop with outgrowing trees is the product of the weights of its edges, with 
the weight of an edge $ \ell_{kl}^\pm $  given by $r_lr_kK^\pm_{lk}$.  In addition, a negative  loop carries an extra minus sign.  This is why we will call  the   positive   loop bosonic  and the negative  loops fermionic.
\end{itemize}
\begin{figure}[ht]
\centering
\begin{subfigure}{0.30\textwidth}
\centering
\includegraphics[width=0.9\linewidth]{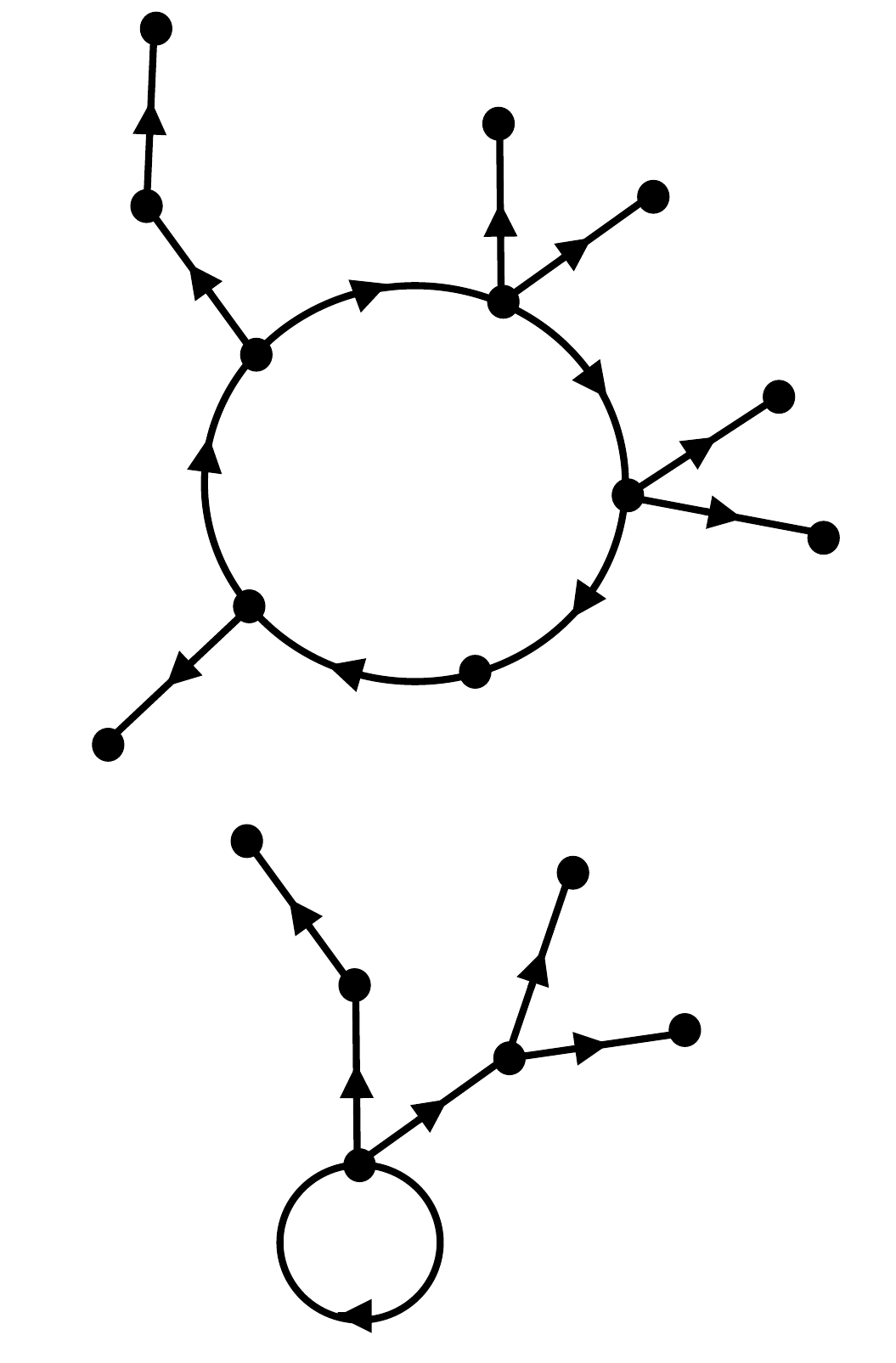}
\caption{$K^+$ loops}
\label{loop-K+}
\end{subfigure}
\hskip 5mm 
 \begin{subfigure}{0.3\textwidth}
 \centering
\includegraphics[width=0.9\linewidth]{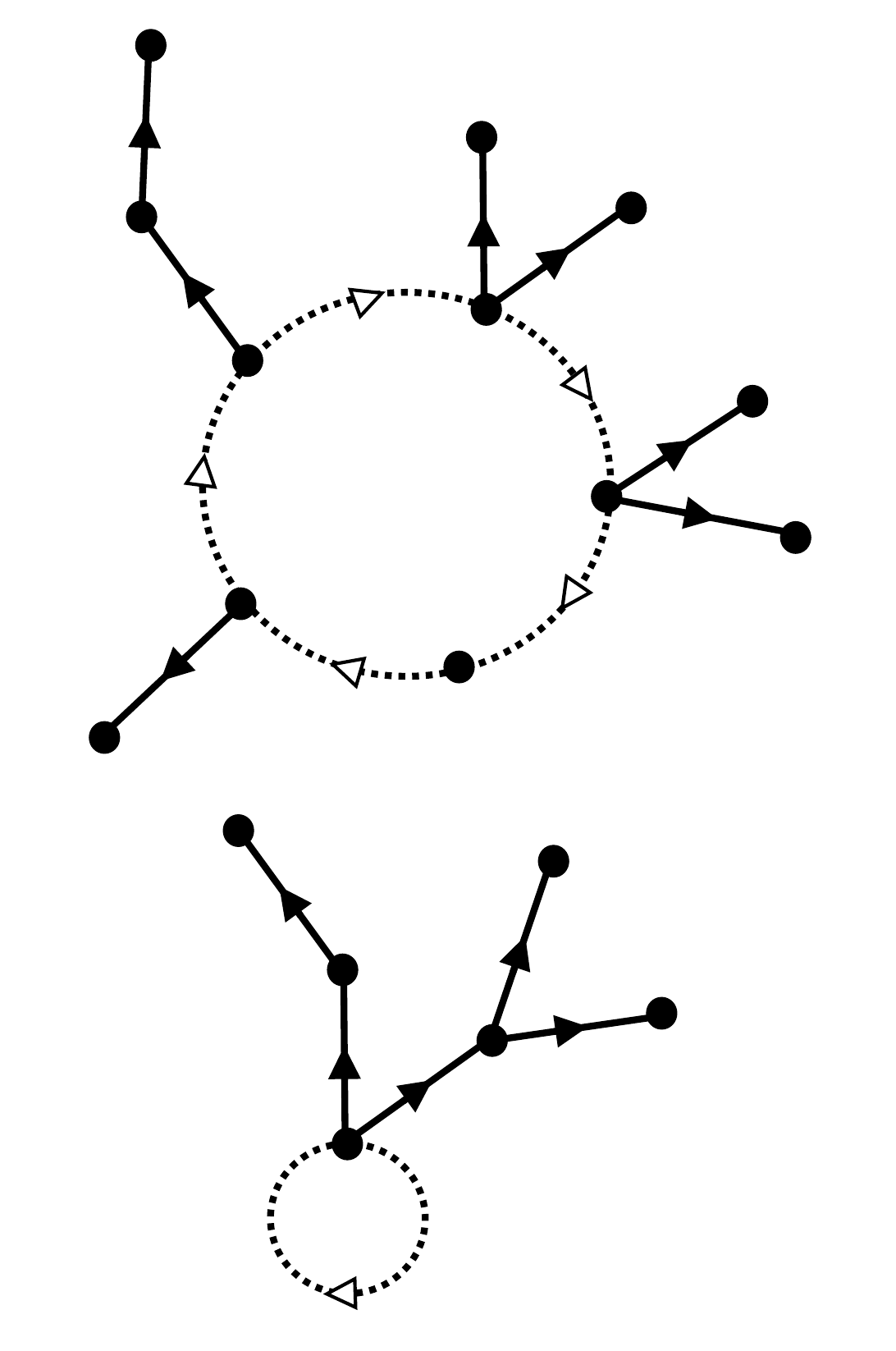}
\caption{$K^-$ loops}
\label{loop-K-}
\end{subfigure}
                 \caption{ \small   Examples of loops with out-growing trees. $K^-$ propagators are drawn as dashed lines. They appear only in a loop and such loop comes with a factor of $-1$. Tadpoles are loops of length $1$.}  
               \label{loops} 
\end{figure}
Summarising, we write the determinant of  the matrix \eqref{Laplacian-new} 
as
\begin{align}
\det \hat G_{jk}&=
\sum_\mathcal{F} W[\mathcal{F}],
\nonumber\\
W[\mathcal{F}]&= (-1)^{\#\text{negative loops}}
\prod_{v_j\in \text{roots}}r_jD_{ab}(\theta_j)\prod  _{e_{kl}^\pm \in\text{ edges}}   
r_lr_kK^\pm(\theta_l, \theta_k)
\label{graphexpansion}
\end{align}
with $K^\pm(\theta,\eta) = K(\theta,\eta) \pm K(\theta,-\eta)$. Equation \eqref{graphexpansion} allows us to express the Jacobian for  the integration measure as a sum over graphs whose  weights depend only on the coordinates $\{ \theta_j, r_j\}$ of
its  vertices. For a periodic system $K^+=K^-$ and the two families of loops cancel
each other, leaving only trees in the expansion of the Gaudin matrix.
\subsection{Graph expansion of the partition function}
Applying the matrix-tree theorem for each term in the series \eqref{partition-Gaudin-hat}, we obtain a graph expansion for the partition function
\begin{align}
Z_{ab}(R,L)=\sum_{N=0}^\infty\frac{(-1)^N}{N!}\sum_{1\leq r_1,...,r_N}\prod_{j=1}^N\int_0^\infty\frac{d\theta_j}{2\pi}\frac{(-1)^{r_j}}{r_j^2} e^{-r_jRE(\theta_j)}\sum_{\mathcal{F}}W[\mathcal{F}],
\end{align}
where the last sum runs over all graphs $\mathcal{F}$  with  $N$ vertices as
constructed above. The next step is to invert the order of the sum over graphs and the
integral/sum over the coordinates $\{\theta_j,r_j\}$ assigned to the
vertices.  As a result we obtain a sum over the ensemble of abstract
oriented tree/loop graphs, with their symmetry factors, embedded in
the space $\mathbb{R}^+\times\mathbb{N}$ where the coordinates $\theta,r$
of the vertices take values.  The embedding is free, in the sense that
the sum over the positions of the vertices is taken without
restriction.  As a result, the sum over the embedded  graphs is
the exponential of the sum over connected ones.  One can think of
these graphs as Feynman diagrams obtained by applying the Feynman
rules in Fig. \ref{Feynmp}.

The Feynman rules comprise  there kinds of vertices: \textit{root} vertices with only
outgoing bosons, \textit{bosonic} vertices with one incoming boson and an
arbitrary number of outgoing bosons, \textit{fermionic} vertices
with one incoming and one outgoing fermion, together with an
arbitrary number of outgoing bosons. The connected diagrams built from
these vertices are either trees (figure \ref{tree-K+}) or bosonic loops
 (fig.  \ref{loop-K+}) or fermionic loops  (fig.  \ref{loop-K-}).
 \begin{tcolorbox}[center,colback=white,width=13cm]
 \begin{align}
  \vcenter{\hbox{\includegraphics[width=1.75cm]{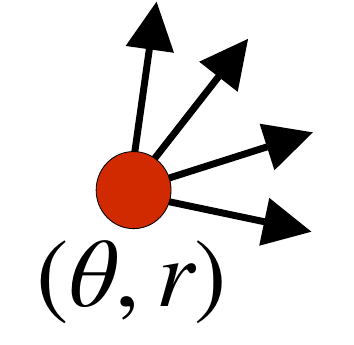}}}  \ \quad&=
   \  \  \  rD_{ab}(\theta)
    {(-1)^{r-1} \over r^2} \ e^{-r R E(\theta)}
\nonumber\\
 \vcenter{\hbox{\includegraphics[width=2cm]{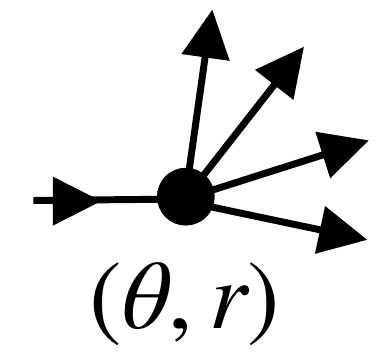}}}=\quad \vcenter{\hbox{\includegraphics[width=2cm]{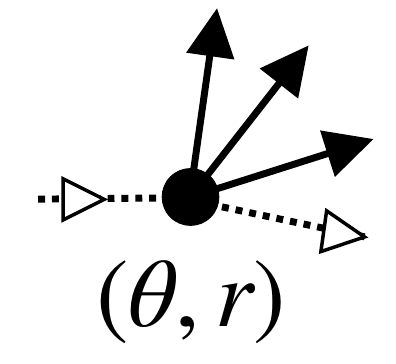}}} \ \quad  &= \  \  \   {(-1)^{r-1} \over r^2} \ e^{-r R E(\theta)}
\nonumber\\
\vcenter{\hbox{\includegraphics[width=4.5cm]{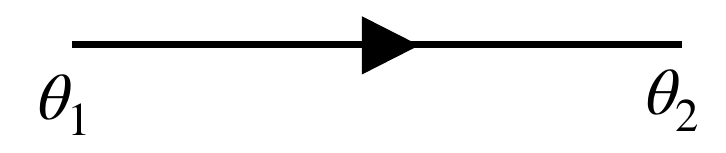}}}\ \quad   &= \ \ \ r_1 r_2 K^+(\theta_2, \theta_1)
\\
\vcenter{\hbox{\includegraphics[width=4.5cm]{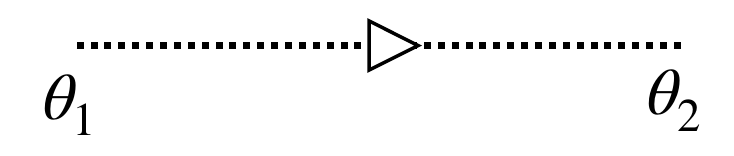}}}\ \quad   &= \ \ \ r_1 r_2 K^-(\theta_2, \theta_1)
\nonumber\\
 \vcenter{\hbox{\includegraphics[width=3cm]{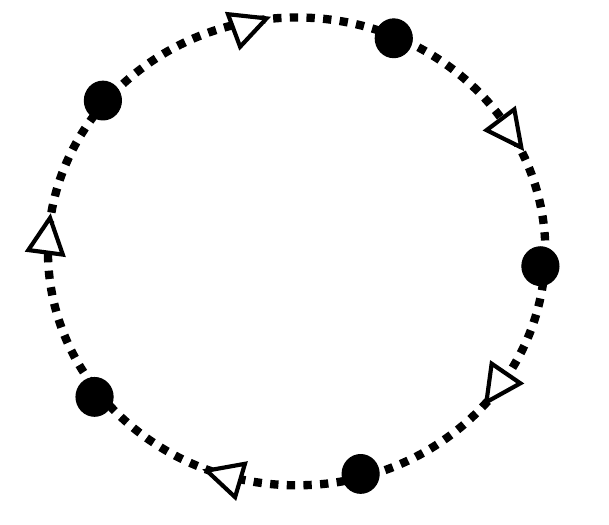}}} \quad &= \ \ \ -1\nonumber
\end{align}
\end{tcolorbox}

\noindent
The free energy is a sum over connected graphs
\begin{align}
\log Z_{ab}(R,L)=\int_0^\infty\frac{d\theta}{2\pi}D_{ab}(\theta)\sum_{r\geq
1}rY_r(\theta)+\sum_{n\geq 1}\mathcal{C}^\pm_n.\label{free-energy}
\end{align}
In this expression, $Y_r(\theta)$ denotes the  sum of  over all  trees   rooted at the point $(\theta,r)$ and $\mathcal{C}^\pm_n$ is the  sum  over the  Feynman  graphs  having a  bosonic/fermionic  loop  of length $n$.  We have defined $Y_r(\theta)$ in such a way that the  all vertices with $r$  outgoing lines, including the root, have the same weight.
\begin{align}
Y_r(\theta)&=\vcenter{\hbox{\includegraphics[width=7cm]{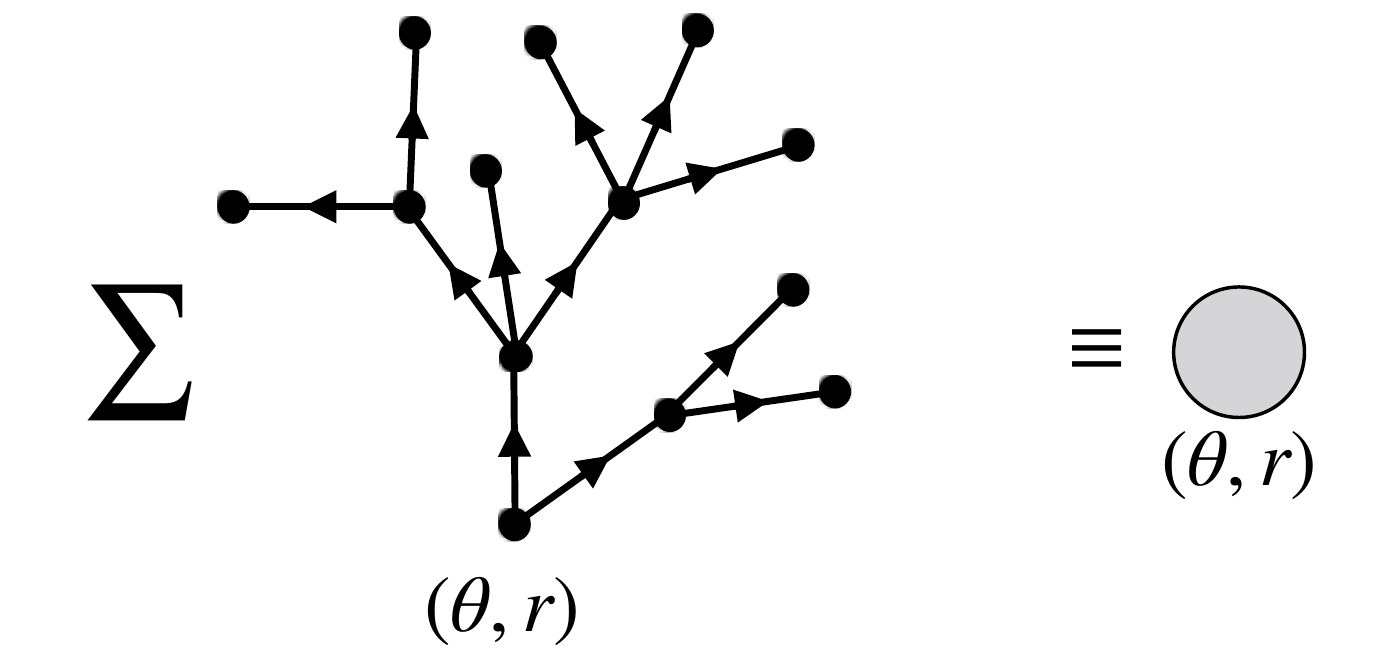}}}.\label{definition-of-trees}
\end{align}
\subsection{Summing over trees: saddle point TBA approximation}
We start by analyzing the part of free energy \eqref{free-energy} that comes from the tree-diagrams
\begin{align}
\log Z_{ab}(R,L)^\textnormal{trees}&=\int_0^\infty\frac{d\theta}{2\pi}D_{ab}(\theta)\sum_{r\geq 1}rY_r(\theta)
.\label{free-energy-trees}
\end{align}
Similar to the periodic case, the combinatorial structure of trees translates into an equation satisfied by $Y_r(\theta)$. The only difference with \eqref{Schwinger-Dyson} is that the integrals are evaluated over the positive axis
\begin{align}
Y_r(\theta)&=\frac{(-1)^{r-1}}{r^2}e^{-rRE(\theta)}\sum_{n=0}^\infty\frac{1}{n!}
\big[\sum_{s=1}^\infty\int_0^\infty sr\frac{d\eta}{2\pi}
K^+(\eta,\theta)Y_s(\eta)\big]^n\nonumber\\
&=\frac{(-1)^{r-1}}{r^2}\big[ e^{-RE(\theta)}\exp\sum_{s=0}^\infty\int_0^\infty
\frac{d\eta}{2\pi}  K^+(\eta,\theta)sY_s(\eta)\big] ^r,\label{TBA-1}
\end{align}
By comparing equation \eqref{TBA-1} for arbitrary $r$ and for $r=1$ we find the following simple relation between $Y_r$ and $Y_1$ 
\begin{align}
Y_r(\theta)=\frac{(-1)^{r-1}}{r^2}Y_1(\theta)^r.\label{relation}
\end{align}
Thanks to this simple identity, we can rewrite \eqref{TBA-1} for $r=1$ as an equation involving $Y_1$ only
\begin{gather}
Y_1(\theta)=e^{-RE(\theta)}\exp\int_0^\infty\frac{d\eta}{2\pi} 
K^+(\eta,\theta)\log[1+Y_1(\eta)].\label{TBA-2}
\end{gather}
This equation starts looking like the usual TBA equation. To bring it into the desired form, we note that the integral  in \eqref{TBA-2} can be extended to the real axis by using the parity of
the kernel $K^+(v,u)=K(v,u)+K(-v,u)$. This is the reason for our assumption \eqref{assumption-on-S-matrix} on the bulk S-matrix. Without it, one cannot relate \eqref{TBA-2} with the periodic TBA equation \eqref{familiar-TBA} and as a result, the subtraction \eqref{g-function-definition} cannot be carried out
\begin{gather}
Y(\theta)=e^{-RE(\theta)}\exp\int_{-\infty}^\infty\frac{d\eta}{2\pi}  K(\eta,\theta)\log[1+Y(\eta)],\label{TBA-periodic}
\end{gather}
where we have denoted $Y\equiv Y_1$ for short. In particular, we know that the free energy of a periodic system can be written in terms of $Y$ as
\begin{align}
\log Z(R,L)=L\int_{-\infty}^\infty\frac{d\theta}{2\pi}p'(\theta)\log[1+Y(\theta)].\label{free-energy-periodic}
\end{align}
Similarly, we can also extend the domain of integration in \eqref{free-energy-trees} to the real axis, using the parity of $D_{ab}(u,r)$.  By subtracting the periodic free energ \eqref{free-energy-periodic} from the tree part of the free energy \eqref{free-energy-trees}, we obtain the tree contribution to g-function
\begin{align}
g_{ab}(R)^\textnormal{trees}=\frac{1}{2}\int_{-\infty}^\infty\frac{d\theta}{2\pi}\Theta_{ab}(\theta)\log[1+Y(\theta)].\label{free-energy-trees-3}
\end{align}
This expression coincides with the saddle point contribution \eqref{g-function-first-attempt} obtained by the traditional TBA approach.
\subsection{Summing  over loops: Fredholm determinants}
We now turn to the sum over loops and show that they fill the missing
part of the $g$-function \eqref{g-function-formula} .  Let us define
\begin{align}
g_{ab}(R)^\textnormal{loops}=\sum_{n\geq
1}\mathcal{C}_n^\pm\label{free-energy-cycles}
\end{align}
For each $n\geq 1$, $\mathcal{C}_n^\pm$ is the partition sum of
$K^\pm$ loops of length $n$ with the trees growing out of these loops which can
be summed separately using the definition \eqref{definition-of-trees}
\begin{align*}
\mathcal{C}_n^\pm=\frac{\pm 1}{n}\sum_{1\leq
r_1,...,r_n}\int\limits_0^\infty\frac{d\theta_1}{2\pi}...\int\limits_0^\infty
\frac{d\theta_n}{2\pi}Y_{r_1}(\theta_1)....Y_{r_n}(\theta_n)r_2r_1K^\pm(\theta_2,\theta_1)....r_1r_nK^\pm(\theta_1,\theta_n).
\end{align*}
\begin{wrapfigure}{r}{0.38\textwidth}
  \begin{center}
    \includegraphics[width=0.36\textwidth]{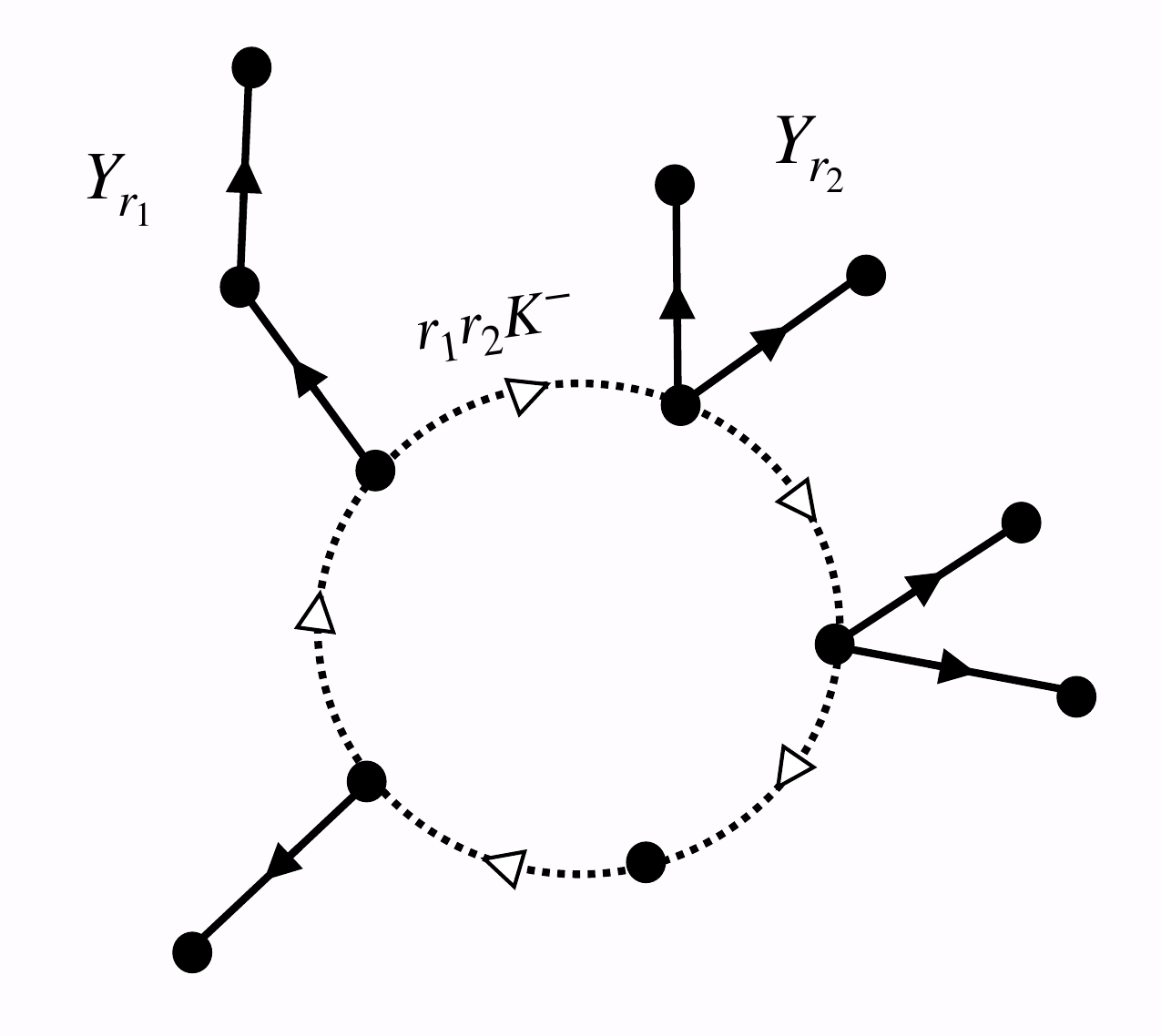}
  \end{center}
  \caption{A $K^-$ loop.}
\end{wrapfigure}
In this expression, the sign comes from fermion loop and $1/n$ is the usual loop symmetry factor.


Trees growing out of a vertex $(\theta_j,r_j)$ of the loop sum up to  $Y_{r_j}(\theta_j)$, by the definition \eqref{definition-of-trees}. The factor $r_{j-1}r_j$ coming from the propagator linking adjacent vertices $(\theta_{j-1},r_{j-1})$ and $(\theta_j,r_j)$ can be pulled into trees that grow out from them. Each tree therefore receives an extra factor of $r^2$ at their root. We can then use the relation \eqref{relation} to carry out the sum over $r$. To this end, we recover the TBA filling factor at each vertex of the loop
\begin{align}
\sum_{r\geq 1}r^2Y_r(\theta)=\frac{Y(\theta)}{1+Y(\theta)}=f(\theta).
\end{align} 
It follows that the sum over all positive (negative) loops of fixed length $n$ is simply given by
\begin{align}
\mathcal{C}_n^\pm =\frac{\pm 1}{n}\tr (\hat{K}^\pm)^n.
\end{align}
where the kernels $\hat{K}^\pm$ are defined in \eqref{K-plus} and \eqref{K-minus}.  We recover part of the g-function \eqref{g-function-formula} coming from fluctuations around the saddle point along with the non-trivial integration measure 
\begin{align}
g_{ab}(R)^{\textnormal{loops}}=\log\det\frac{1-\hat{K}^-}{1-\hat{K}^+}.\label{free-energy-cycles-2}
\end{align}
\section{Summary and outlook}
We propose a graph theory-based method to compute the $g$-function of a
theory with diagonal bulk scattering and diagonal reflection matrices.
The idea is to apply the matrix-tree theorem to write the Jacobians in
the cluster expansion of the partition function by a sum over graphs.
The $g$-function is then written as a sum over trees and loops.  The sum
over trees gives TBA saddle point result while the sum over loops
constitute the two Fredholm determinants.  

Similar determinant structure appears in the study of \textit{quantum quench}. In short, a quantum quench is a quantum mechanics problem where the initial state $|\Psi_0\rangle$ is not an eigenvector of the Hamiltonian $H$ that controls the time evolution. A common way to compute time independent observables in such system is through form factor expansion
\begin{align}
\langle \mathcal{O}(t)\rangle =\sum_{n,m}\langle \Psi_0|n\rangle \langle n|\mathcal{O}|m\rangle \langle m|\Psi_0\rangle e^{it(E_n-E_m)},
\end{align}
where $n$ and $m$ are two complete sets of normalized eigenstates of $H$. In order to evaluate this expression, one must know the overlap between the initial state and these eigenstates. For some families of initial states of the XXX and XXZ spin chain, it was found \cite{1742-5468-2014-6-P06011,1751-8121-47-34-345003,1751-8121-47-14-145003,1742-5468-2014-5-P05006,1742-5468-2018-5-053103} that this overlap can be written in closed form: a product of some local factor and the ratio of two determinants. These determinants are precisely the finite-particle analogies of the Fredholm determinants that appear in the g-function. Similar formulae also appears in the study of one point functions in AdS/dCFT \cite{deLeeuw:2015hxa,Buhl-Mortensen:2015gfd,deLeeuw:2016umh,deLeeuw:2018mkd,Gombor:2020kgu}.  We note that the known expressions for the overlap are valid for any value of the length of the corresponding spin chain. 

In \cite{Jiang:2019xdz} an expression for the excited state g-function was found by analytic continuation. In the large mirror volume limit ($R$ in our notation), the authors recovered the same structure of the integrable overlaps. It would be interesting to obtain this result from our formalism, following the same lines of idea of subsection \ref{excited-new-section}.
\chapter{Open systems with non-diagonal scattering}
\label{non-diag-g-section}
In the previous chapter we have obtained the expression \eqref{g-function-formula} for the finite temperature g-function of a theory with a single massive excitation. The generalization to theories with more than one particle interacting via diagonal S-matrix is straightforward. If we denote the particle types by $n$,  the scattering derivatives by $K_{nm}(\theta,\eta)\equiv -i\partial_\theta\log S_{nm}(\theta,\eta)$, the reflection derivative corresponding to a boundary condition of type $a$ by $K_n^a(\theta)\equiv -i\partial_\theta\log R_n^a(\theta)$, the Y-functions  at inverse temperature $R$ by $Y_n$ then
\begin{gather}
2\log g_a(R)=2\log g_a^\textnormal{trees}(R)+2\log g_a^\textnormal{loops}(R),\label{trees+loops}\\
2\log g_a^\textnormal{trees}(R)=\sum_{n}\int_{-\infty}^\infty\frac{d\theta}{2\pi}[K_{n}^a(\theta)-K_{nn}(\theta,-\theta)-\pi\delta(\theta)]\log[1+Y_n(\theta)],\label{trees-multi}\\
2\log g_a^\textnormal{loops}(R)=\log\det\frac{1-\hat{K}^-}{1-\hat{K}^+},\label{loops-multi}
\end{gather}
where the kernels $\hat{K}^\pm$ have support on $\mathbb{R}^+$ and their actions are given by
\begin{gather*}
\hat{K}_{nm}^\pm(F)(\theta)=\int_0^{+\infty}\frac{d\eta}{2\pi}K^\pm_{nm}(\theta,\eta)f_m(\eta)F(\eta),\\
\text{with }\quad K_{nm}^\pm(\theta,\eta)\equiv K_{nm}(\theta,\eta)\pm K_{nm}(\theta,-\eta),\quad f_m=\frac{Y_m}{1+Y_m}.
\end{gather*}
A general formula as \eqref{trees+loops} is not known for theories in which the bulk scattering is not diagonal. Let us first spell out why the methods of Pozsgay \cite{Pozsgay:2010tv} and our diagrammatic formalism are not justified in this case. The diagonalization by the Nested Bethe Ansatz technique involves particles of magnonic type, which are auxiliary particles with zero momentum and energy. The functional integration measure, as it is derived in \cite{Pozsgay:2010tv}, as well as the  summation over multiparticle states in our formalism or in \cite{Dorey:2004xk} treat the physical
and the auxiliary particles in exactly the same way. This is justified only for states with asymptotically large number of physical particles. For states with finite number of physical particles, which dominate in the IR limit, the solutions of the Bethe Ansatz equations do not obey the string hypothesis and moreover the number of the magnons and the number of the physical particles must respect certain constraint. On the other hand, finding and summing over the exact solutions for the auxiliary magnons is of course a hopeless task.

 In this chapter we demonstrate on a concrete example that assuming solutions in the form of Bethe strings and summing over unrestricted number of auxiliary particles nevertheless leads to a meaningful result for the boundary entropy, up to an infinite constant which can be subtracted. The subtraction is done by normalizing the g-function \eqref{trees+loops} by its zero temperature value.
 
The appearance of this infinite piece has an obvious explanation. When the theory is massive in the bulk, the boundary entropy must vanish at zero temperature. If the bulk scattering is diagonal, this condition  is automatically satisfied by the expression \eqref{trees+loops} as all Y-functions vanish in this limit. For non-diagonal bulk scattering however, magnonic particles decouple from the physical ones at zero temperature and the corresponding Y-functions retain non-zero values \footnote{For periodic boundary condition, this does not lead to a problem for the ground-state energy as the auxiliary particles have vanishing energy.}. In our diagrammatic formalism, this normalization amounts to subtracting the contribution from unphysical graphs made of these auxiliary particle. We denote in this chapter $g_{\text{IR}}\equiv g(\infty),\; g_{\text{UV}}\equiv g(0)$
\begin{align}
g(R)^\text{ren}= \frac{g(R)}{g_{\text{IR}}}.\label{proposition}
\end{align}
As a test for this proposal, we show that it is possible to match the ratio $g_{\text{UV}}/g_{\text{IR}}$ with a conformal g-function under certain assumptions.

The theories under study are the current-perturbed $SU(2)$ Wess-Zumino-Novikov-Witten (WZNW) model at positive integer levels $k$. The bulk scattering of these theories are not diagonal, as each particle carries quantum numbers that can be nontrivially exchanged during collisions. With the Nested Bethe Ansatz technique, one can trade the nondiagonal scattering for a diagonal one with extra magnonic particles: $SU(2)$ magnon and kink magnon. In the thermodynamic limit these particles can form bound states which are strings of evenly distributed rapidities on the complex plane. In particular $SU(2)$ magnons can form strings of arbitrary lengths, effectively leading to an infinite number of particles in the TBA formalism. In the derivation of the TBA free energy one ignores that the above is true only for asymptotically large number of physical particles. The price to pay, as we will show later, is that both expressions \eqref{trees-multi},\eqref{loops-multi} are logarithmic divergent in the IR and UV limit.

We regularize these divergencies by introducing a twist or equivalently a chemical potential to the TBA equations. The chemical potential makes the sum over the auxiliary magnons finite. The twist/chemical potential is added to the TBA equations for the sole purpose of regularizing the g-function and we do not discuss its effects on the UV limit of the theory. The only change induced by this modification in the formulae \eqref{trees-multi} and \eqref{loops-multi} is the asymptotic values of Y-functions. We are then able to express the IR and UV g-function as functions of the twist parameter and evaluate their ratio in the untwisted limit. For a specific choice of boundary condition it is given by  
\begin{align}
\bigg(\frac{g_{\text{UV}}}{g_{\text{IR}}}\bigg)^2=\sqrt{\frac{2}{k+2}}\times\frac{1}{\sin\frac{\pi}{k+2}},\label{final-result}
\end{align}
which coincides with a Cardy g-function \eqref{CFT-entropy}, namely $g_{k/2}$ for even $k$. Equation \eqref{final-result} is the main result of this chapter.

The chapter is organized as follows. In section 1 we present the basic features of $SU(2)$ WZNW CFT at level $k$ with emphasis on its Cardy g-functions. We also show how the boundary entropy of a massive perturbation of this CFT flows to its UV value when the temperature is sent to infinity. In section 2 we introduce the current perturbation of this CFT and its TBA equations. We show how various quantities can be extracted from solutions of the TBA equations in the UV or IR limit. We conjecture a specific set of diagonal reflection factors in section 3, fixing the values of $K_n^a(u)$ in equation \eqref{trees-multi}. With the data from section 2 and 3, we show in section 4 that the expressions \eqref{trees-multi} and \eqref{loops-multi} diverge in the UV and IR limit. We then show using  twisted TBA equations that these divergencies can be regularized, leading to \eqref{final-result} as the final result.

Although a general method for finding the g-function of a theory with non-diagonal bulk scattering is still missing, there are models with particular features that allow this quantity to be extracted via case-dependent techniques. In \cite{Dorey:2010ub}, the g-functions of perturbed unitary minimal models   were studied using the roaming trajectory of the staircase model \cite{Zamolodchikov:1991pc}. The latter is a theory with diagonal bulk scattering that depends on a free parameter. This parameter can be tuned with temperature to form a plateau RG flow with successive unitary minimal models on its steps. In another work, Pozsgay \cite{2018arXiv180409992P} computed the spin chain analogue of g-function for the XXZ spin chain using quantum transfer matrix and independent results on integrable overlaps. The TBA of this spin chain also involves an infinite number of magnon strings. Recently, the g-function of a model with supersymmetry was studied in view of its relation with a class of three point functions in planar $\mathcal{N}=4$ SYM \cite{Jiang:2019xdz}. In contrast to our setup, the Bethe equations considered in this work have the property that only the physical rapidities come in pair. Due to this reason, the resulting g-function is free of the above-mentioned divergence.  It could be observed from the results of  \cite{Dorey:2010ub, 2018arXiv180409992P, Jiang:2019xdz} that the Fredholm determinant structure \eqref{loops-multi} of the g-function is still relevant for theories with non-diagonal bulk scattering.
\section{The current perturbed $SU(2)_k$ WZNW theories}
The Wess-Zumino-Novikov-Witten (WZNW) model for a semisimple group G is defined by the action
\begin{align*}
S_{\text{WZNW}}=\frac{1}{4\lambda^2}\int_{S^2} d^2x\tr\partial_\mu g\partial^\mu g^{-1}+k\Gamma,
\end{align*}
where the Wess-Zumino term $\Gamma$ is 
\begin{align*}
\Gamma=\frac{1}{24\pi}\int_{B^2} d^3X\epsilon^{ijk}\tr \tilde{g}^{-1}\partial_i \tilde{g}\tilde{g}^{-1}\partial_j \tilde{g}\tilde{g}^{-1}\partial_k \tilde{g}.
\end{align*}
Here $g$ is a map from the two-sphere to $G$ and $\tilde{g}$ is its extension from the corresponding two-ball to the same group. Such an extension comes with an ambiguity of topological origin, leading to integer values of $k$.

At $\lambda^2=4\pi/k$ the global $G\times G$ symmetry is enhanced to a local $G(z)\times G(\bar{z})$ symmetry with two currents $J(z)=\partial_z g g^{-1}$ and $\bar{J}(\bar{z})=g^{-1}\partial_{\bar{z}}g$ separately conserved. These currents satisfy the current algebra $G_k$ while their bilinear satisfies the Virasoro algebra. The latter implies in particular conformal invariance and we refer to the theory at this coupling as the WZNW CFT of $G$ at level $k$.

In the following we consider the case $G=SU(2)$. The left(right) moving sector of this theory consists of $k+1$ irreducible representations $\mathcal{V}_\lambda$ of $SU(2)_k$ corresponding to its $k+1$ integrable  weights. The characters $\chi_\lambda=q^{-c/24}\tr_{\mathcal{V}_\lambda}q^{L_0}$ transform to one another under the modular transformation $\tau\to -1/\tau$. This transformation is encoded in the modular S-matrix of the theory
\begin{align*}
\chi_\lambda(q)=\sum_{\eta=0}^kS_{\lambda,\eta}\chi_\eta(\tilde{q}),\quad q\equiv e^{2i\pi\tau},\tilde{q}\equiv e^{-2\pi i/\tau}.
\end{align*}
It is explicitly given by
\begin{align}
S_{\lambda,\eta}=\sqrt{\frac{2}{k+2}}\sin\bigg[\frac{\pi(\lambda+1)(\eta+1)}{k+2}\bigg],\quad 0\leq \lambda,\eta\leq k,
\label{SU(2)-WZW-level-n-modular}
\end{align}
which is a real, symmetric matrix that satisfies $S^2=\mathbf{1}$. The central charge and conformal dimesnsions are obtained from the Sugawara construction of the energy-momentum tensor
\begin{align}
c=\frac{3k}{k+2},\quad h_\lambda=\frac{\lambda(\lambda+2)}{4(k+2)}.\label{WZW-n}
\end{align}
The fusion coefficients $\mathcal{N}_{\lambda,\eta}^\kappa$ denote how many times the field $\phi_\kappa$ appears in the operator product expansion of $\phi_\lambda$ and $\phi_\eta$. They satisfy the Verlinde formula
\begin{align}
\mathcal{N}_{\lambda,\eta}^\kappa=\sum_{\zeta}\frac{S_{\lambda,\zeta}S_{\eta,\zeta}S_{\zeta,\kappa}}{S_{0,\zeta}}.\label{Verlinde-formula}
\end{align}
Above is our quick summary of $SU(2)_k$ WZNW CFT data. Consider now this CFT on manifolds with boundaries. Two geometries are relevant for our discussion. First let us consider the upper half complex plane. The continuity condition through the real axis requires that the underlying symmetry involves only one copy of the algebra instead of two. A particular set of boundary states which are invariant under the Virasoro algebra as well as the affine $SU(2)$ Lie algebra is called Ishibashi states \cite{Ishibashi:1988kg}. These states are in one-to-one correspondence with bulk primaries and are denoted as $|\lambda\rrangle$ \footnote{The plus sign for the currents come from the fact that they are of spin 1.}
\begin{align}
(L_n-\tilde{L}_{-n})|\lambda\rrangle =(J^a_n+\tilde{J}^a_{-n})|\lambda\rrangle=0.
\end{align}
Other boundary states are obtained from linear combination of Ishibashi states
\begin{align}
|a\rangle=\sum_{\lambda}|\lambda\rrangle\llangle \lambda|a\rangle.\label{linear-combination-Ishibashi}
\end{align}
So, on the upper half complex plane, one is quite free to choose the boundary state. The situation is greatly different if the theory is restricted on an annulus. In this geometry let us consider two boundary states $|a\rangle$ and $|b\rangle$ of the form \eqref{linear-combination-Ishibashi} on its sides. Denote by $q=\exp(-\pi R/L)$ the modular parameter of this annulus. On one hand one can quantize this theory according to the Hamiltonian $H_{ab}$ with $a,b$ as boundary conditions
\begin{align}
Z_{ab}=\sum_{\lambda}n_{a,b}^\lambda\chi_\lambda(q)=\sum_{\lambda}n_{a,b}^\lambda\sum_\eta S_{\lambda,\eta}\chi_\eta(\tilde{q}),\quad \tilde{q}=\exp(-4\pi L/R),\label{partition-function-1}
\end{align}
where the non-negative integers $n_{a,b}^\lambda$ denote the number of copies of $\mathcal{V}_\lambda$ in the spectrum of $H_{ab}$. On the other hand one can consider the theory as evolving between two states $\langle a|$ and $|b\rangle$. The periodic Hamiltonian can be written in terms of Virasoro generators via a conformal mapping, leading to
\begin{align}
Z_{ab}=\langle a|\tilde{q}^{\frac{1}{2}(L_0+\bar{L}_0-c/12)}|b\rangle=\sum_{\eta}\langle a|\eta\rrangle\llangle \eta|b\rangle\chi_\eta(\tilde{q}).\label{partition-function-2}
\end{align}
\begin{figure}[ht]
\centering
\includegraphics[width=10cm]{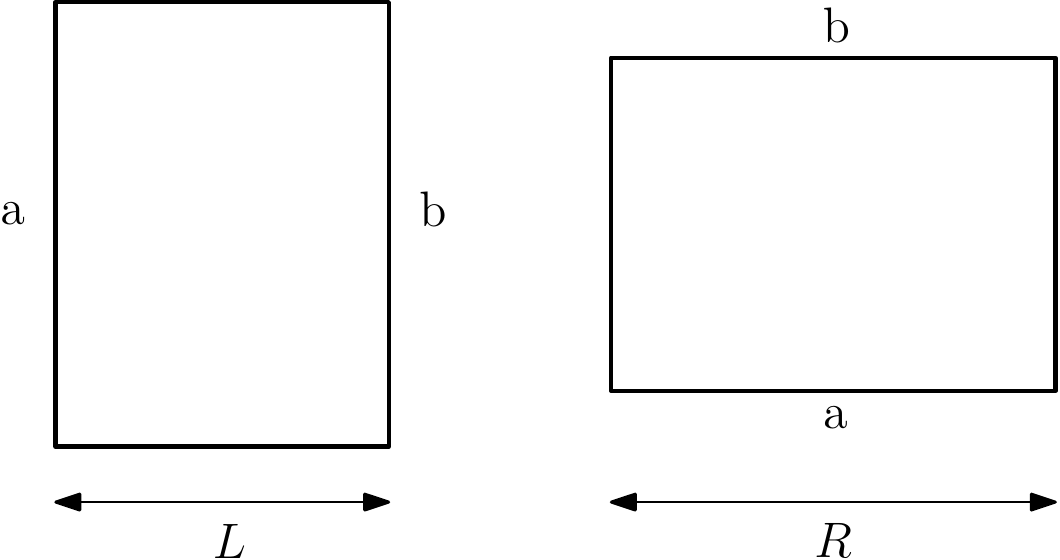}
\caption{Modular invariance of the annulus partition function.}
\end{figure}

For the theory under consideration, each representation $\lambda$ of the extended algebra appears only once in the spectrum. By identifying the two expressions \eqref{partition-function-1} and \eqref{partition-function-2} we obtain the following relation 
\begin{align}
\sum_\lambda n_{a,b}^\lambda S_{\lambda,\eta}=\langle a|\eta\rrangle \llangle \eta|b\rangle\Leftrightarrow n_{a,b}^\lambda=\sum_\eta S_{\eta,\lambda}\langle a|\eta\rrangle \llangle \eta|b\rangle,\label{Cardy-equation}
\end{align}
where we have used the fact that $S$ is real symmetric and $S^2=\mathbf{1}$. Relation \eqref{Cardy-equation} is referred to as Cardy equation \cite{Cardy:1989ir}, which sets the constraint on admissible boundary states on an annulus. A particular solution is given by
\begin{align*}
\langle a|\lambda\rrangle=\frac{S_{a,\lambda}}{\sqrt{S_{0,\lambda}}},\quad n=\mathcal{N},
\end{align*}
where the Cardy equation becomes the Verlinde formula \eqref{Verlinde-formula}. We refer to these boundary states as Cardy states and denote them by $|C_a\rangle,\; a=\overline{0,k}$.

Due to the difficulty in defining the Wess-Zumino term for a surface with boundary, our discussion of g-function should be taken at the operatorial level. We also stress that the  boundary states under consideration are invariant under the extended algebra. For boundary states which are only conformal invariant, see \cite{Gaberdiel:2001xm}. 
\subsection{Off-critical g-function}
Let us now assume that there is an 1+1 dimensional integrable massive quantum field theory which admits $SU(2)_k$ WZNW CFT as its UV fixed point. We consider such a theory on a cylinder of length $L$ and radius $R$ which plays the role of periodic Euclidean time or equivalently, inverse temperature. We further assume that we can define the boundary conditions in such a way that integrability is conserved.

One would expect from integrability that, at arbitrary temperature, it is possible to compute the bulk free energy and boundary entropy densities of the theory
\begin{align*}
Z_{ab}(R,L)=\exp[-LRf(R)]\times g_a(R) g_b(R).
\end{align*}
We also assume that in the conformal limit $R\to 0$ the two integrable boundary conditions can be identified with some CFT boundary sates $|a\rangle$ and $|b\rangle$ \footnote{This hypothesis is usually satisfied for integrable boundary conditions, see for instance \cite{Dorey:1999cj, Dorey:2005ak,Dorey:2010ub}}. Then in this limit the modular parameter $q$ tends to one and the contribution of vacuum state dominates other terms in the partition function \eqref{partition-function-1}
\begin{align*}
\lim _{R\to 0}Z_{ab}(R,L)=\chi_0(\tilde{q})\sum_{\lambda}n_{a,b}^\lambda S_{\lambda,0}.
\end{align*} 
Therefore the bulk free energy becomes proportional to the CFT central charge
\begin{align}
\lim_{R\to 0}R^2f(R)=-\frac{\pi c}{6} \label{CFT-energy}
\end{align} 
while the boundary contribution to the partition function is given by the sum
\begin{align*}
g_a(0)g_b(0)= \sum_{\lambda}n_{a,b}^\lambda S_{\lambda,0}.
\end{align*}
Apply now the Cardy equation \eqref{Cardy-equation}, one can identify the contribution of each boundary with the corresponding overlap with the Ishibashi state $|0\rrangle$. In particular, if the boundary state  happens to be a Cardy state, we expect the boundary entropy to flow to 
\begin{align}
g_a(0)=\frac{S_{a,0}}{\sqrt{S_{0,0}}}\label{CFT-entropy}
\end{align}
in the UV limit. 

It is the purpose of this chapter to give the exact expression for the boundary entropy $g_a$ at arbitrary temperature for a particular perturbation of $SU(2)_k$ WZNW CFT and to match its value with some Cardy g-function in the UV limit. First, we remind how to compute the bulk free energy $f(R)$ using the Thermodynamic Bethe ansatz technique. In particular we will verify the limit \eqref{CFT-energy}.
\section{The TBA equations and Y system}
The perturbation we are going to consider belongs to a larger family of perturbations of diagonal coset CFT's $G_k\times G_l/G_{k+l}$ where $G$ is simply-laced. It was first shown in \cite{AHN1990409} that it is possible to perturb this CFT while still preserving part of its symmetry. The perturbing operator is the branching between two scalar representations and the adjoint representation in the Goddard-Kent-Olive (GKO) construction. Moreover, for negative sign of the perturbing parameter, this perturbation leads to a massive field theory. In the same paper, a factorized scattering matrix for this field theory in the particular case of $G=SU(2)$ was proposed 
\begin{align}
\mathcal{S}=\mathcal{S}_{[k]}^\text{RSG}\otimes\mathcal{S}^\text{RSG}_{[l]},\label{S-matrix-k-l}
\end{align} 
where RSG stands for restricted sine-Gordon. Each factor in this tensor product is a S-matrix of the sine-Gordon theory at coupling $\beta^2/8\pi=(k+2)/(k+3)$ and $(l+2)/(l+3)$ respectively. The idea is to rely on the quantum group symmetry of the sine-Gordon S-matrix when the deformed parameter is a root of unity ($q=-\exp[-i\pi/(k+2)]$) to restrict the original multisoliton Hilbert space to direct sum of irreducible representations of this quantum group. As a result the infinite set of vacua of the sine-Gordon theory is truncated to $k+1$ $(l+1)$ vacua and a particle of the restricted theory is defined to be a kink interpolating adjacent vacua. This case was further studied in \cite{Zamolodchikov:1991vg} where a system of TBA equations was conjectured and shown to yield the correct central charge of the unperturbed CFT in the UV limit. It should be noted that the TBA system in \cite{Zamolodchikov:1991vg} was not based on the scattering reported in \cite{AHN1990409} but was instead taken as generalization of several known cases. A derivation of TBA equations for $G=SU(N)$ from the basis of scattering data was later carried out in \cite{HOLLOWOOD199443}. The extension to other simple Lie algebras was done in \cite{Babichenko:2003rf}. See also \cite{Bazhanov:1989yk} for a lattice realization.

We are interested in the limit $l\to \infty$ and $G=SU(2)$ where the coset CFT reduces to $SU(2)_k$ WZNW  and the branching operator becomes its Kac-Moody current. Moreover the second factor in the S-matrix \eqref{S-matrix-k-l} becomes the S-matrix of the chiral $SU(2)$ Gross-Neveu model.   Each particle of the theory  carries a $SU(2)$ quantum number and a kink quantum number. Each kink connects two adjacent vacua among $k+1$ vacua of the truncated sine-Gordon Hilbert space. One usually calls $(a,a+1)$ a kink and $(a+1,a)$ an anti-kink. When kinks are scattered, only the middle vacuum can be interchanged. For $k=1$ there are two vacua and the only scattering is between kink $(1,2)$ and anti kink $(2,1)$, which is trivial. The S-matrix for the kink scattering $K_{da}(\theta_1)+K_{ab}(\theta_2)\to K_{dc}(\theta_2)+K_{cb}(\theta_1)$ is the RSOS Boltzmann weight
\begin{align*}
\mathcal{S}_{[k]}^\text{kink}(\theta)\begin{pmatrix}
a&b\\
c&d
\end{pmatrix}(\theta)=&\frac{u(\theta)}{2\pi i}\bigg(\frac{\sinh(\pi a/p)\sinh(\pi c/p)}{\sinh \pi d/p)\sinh(\pi b/p)}\bigg)^{-\theta/2\pi i}\\
\times &\bigg[\sinh\bigg(\frac{\theta}{p}\bigg)\bigg(\frac{\sinh(\pi a/p)\sinh(\pi c/p)}{\sinh (\pi d/p)\sinh(\pi b/p)}\bigg)^{1/2}\delta_{db}+\sinh\bigg(\frac{i\pi-\theta}{p}\bigg)\delta_{ac}\bigg],
\end{align*}
where $a=\overline{1,k+1}$ is the vacuum index, $p=k+2$ and
\begin{align*}
u(\theta)&=\Gamma\bigg(\frac{1}{p}\bigg)\Gamma\bigg(1+\frac{i\theta}{p}\bigg)\Gamma\bigg(1-\frac{\pi+i\theta}{p}\bigg)\prod_{n=1}^\infty\frac{R_n(\theta)R_n(i\pi-\theta)}{R_n(0)R_n(i\pi)},\\
R_n(\theta)&=\frac{\Gamma(2n/p+i\theta/\pi p)\Gamma(1+2n/p+i\theta/\pi p)}{\Gamma((2n+1)/p+i\theta/\pi p)\Gamma(1+(2n-1)/p+i\theta/\pi p)}\;.
\end{align*}
The scattering matrix is the tensor product of the $SU(2)$ chiral Gross-Neveu S-matrix and  the kink S-matrix 
\begin{align}
\mathcal{S}^{\text{SU(2)}_k}(\theta)= \mathcal{S}_{[k]}^\text{kink}(\theta)\otimes\mathcal{S}^{\text{SU(2)}}(\theta).\label{level-k-s-matrix}
\end{align}
The TBA system for perturbed $SU(2)_k$ consists of two parts. The right wing consists of $SU(2)$ magnon bound states, exactly like the Gross-Neveu model. The left wing are formed of kink magnon bound states. There are a priori $k$ of them but the longest one does not contribute to the thermodynamic properties. This results in a reduced TBA system involving only $k-1$ kink magnons
\begin{align}
\log Y_n(u)+Rm\cosh(u)\delta_{n,k}=\sum_{m=0}^\infty K_{mn}\star\log(1+Y_m)(u),\quad n=\overline{1,\infty}.\label{level-k-TBA}
\end{align}
\begin{figure}[ht]
\centering
\includegraphics[width=14cm]{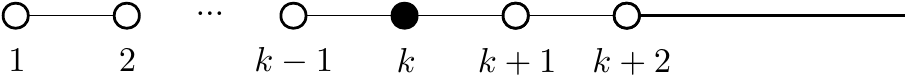}
\caption{Y system for current perturbed $SU(2)_k$ WZNW. The $k$'th node is the physical rapidity. The $j$'th node is kink magnon string of length $k-j$ for $1\leq j\leq k-1$. The $a$'th node is $SU(2)$ magnon string of length $a-k$ for $a\geq k+1$.}
\end{figure}
\noindent
The scattering kernels are given at the end of appendix \ref{higher-reduced}. For later discussion of  g-function we present here their convolutions with identity
\begin{gather*}
K_{ij}\star\mathbf{1}=\delta_{ij}-2\frac{\min(i,j)[k-\max(i,j)]}{k},\quad K_{ab}\star \mathbf{1}=\delta_{ab}-2\min(a-k,b-k),\\
K_{kj}\star\mathbf{1}=-K_{jk}\star\mathbf{1}=-\frac{j}{k},\quad K_{ka}\star\mathbf{1}=-K_{ak}\star\mathbf{1}=-1,\quad i,j\in\overline{1,k-1},a,b\in\overline{k+1,\infty},\\
K_{kk}\star\mathbf{1}=1-\frac{1}{2k}.
\end{gather*}
The TBA equations \eqref{level-k-TBA} can be transformed into an equivalent Y system.  The same kernel $s$ in \eqref{Fourier-s} connects the two wings to the physical node, despite different scattering structures on each wing. Again, we denote $\mathcal{Y}_n=Y_n^{2\delta_{n,k}-1}$
\begin{align}
\log \mathcal{Y}_n+RE\delta_{n,k}= s\star[\log(1+\mathcal{Y}_{n-1})+\log(1+\mathcal{Y}_{n+1})],\quad n=\overline{1,\infty}\label{level-k-Y-system}
\end{align}
The UV and IR solutions of this Y-system are given by
\begin{gather*}
1+\mathcal{Y}_n^{\text{UV}}=(n+1)^2,\quad n=\overline{1,\infty},\\
1+\mathcal{Y}_n^{\text{IR}}=\frac{\sin^2[(n+1)\pi/(k+2)]}{\sin^2[\pi/(k+2)]},\quad n=\overline{1,k-1},\quad 1+\mathcal{Y}_n^{\text{IR}}=(n-k+1)^2,\quad n=\overline{k,\infty},
\end{gather*}
From these values we can recover the central charge of the unperturbed CFT using Roger dilogarithm function
\begin{align}
\sum_{n=1}^\infty\textnormal{Li}_R\big(\frac{1}{1+\mathcal{Y}_n^{\text{IR}}}\big)-\textnormal{Li}_R\big(\frac{1}{1+\mathcal{Y}_n^{\text{UV}}}\big)=\frac{3k}{k+2}.
\end{align}
For a proof of this identity, see for instance \cite{Kirillov:1993ih}. We can add to the TBA equations \eqref{level-k-TBA} a chemical potential coupled to the $SU(2)$ symmetry only. The Y system \eqref{level-k-Y-system} is again protected. 
In the IR limit the left wing decouples from the right wing and is immune to the $SU(2)$ chemical potential. The IR values on the left wing is therefore unchanged, while on the right wing we have
\begin{align*}
1+\mathcal{Y}_n^{\text{IR}}(\k)=[n-k+1]_\k^2,\quad n=\overline{k,\infty}.
\end{align*}
In the UV limit all nodes are affected by the twist
\begin{align*}
1+\mathcal{Y}_n^{\text{UV}}(\k)=[n+1]_\k,\quad n=\overline{1,\infty}.
\end{align*}
As described in chapter 1, we can compute the particle densities in this limit
\begin{gather}
\lim_{R\to 0} \pi RD_0(R,\mu)=2\log \frac{1-\k^{k+1}}{1-\k}-k\log \k,\quad \lim_{R\to 0}\sum_{a=1}^\infty \pi R a D_a(R,\mu)=2\log \frac{1-\k^{k+1}}{1-\k}
 \label{particle-densities-level-k}
\end{gather}
as well as the scaled free energy density
\begin{align}
\lim_{R\to 0}c(R,\mu)=\frac{3k}{k+2}-\frac{6k\mu^2}{\pi^2}.
\end{align}
\section{The reflection factors}
Due to the factorization property of the S-matrix \eqref{level-k-s-matrix}, we can study the reflection factors for kink magnons and $SU(2)$ magnons independently. 

After the maximum string reduction procedure (see appendix \eqref{higher-reduced}), the effective scattering between $k-1$ kink magnon strings are very similar to the scattering of the A theories in the ADE family \cite{Klassen:1989ui}. The scattering phase between a kink magnon string of length $n$ and another one of length $m$, for $n,m=\overline{1,k-1}$ is given by
\begin{align}
S_{nm}(\theta)=\frac{\sinh\big(\frac{\theta}{k}+i\pi\frac{|n-m|}{2k}\big)}{\sinh\big(\frac{\theta}{k}-i\pi\frac{|n-m|}{2k}\big)}\frac{\sinh\big(\frac{\theta}{k}+i\pi\frac{n+m}{2k}\big)}{\sinh\big(\frac{\theta}{k}-i\pi\frac{n+m}{2k}\big)}\prod_{s=\frac{|n-m|}{2}+1}^{\frac{n+m}{2}-1}\bigg[\frac{\sinh\big( \frac{\theta}{k}+i\pi \frac{s}{k}\big)}{\sinh\big( \frac{\theta}{k}-i\pi \frac{s}{k}\big)}\bigg]^2\label{kink-magnon-scatterings}
\end{align}
where $\theta$ is the rapidity difference between the two string centers.
On the other hand, the $A_{k-1}$ S-matrix describes the purely elastic scattering of the coset CFT $SU(k)_1\times SU(k)_1/SU(k)_2$ ($\mathbb{Z}_{k}$ parafermions) perturbed by its $(1,1,\text{adj})$ operator \footnote{The central charge of this CFT is exactly the central charge of $SU(2)_k$ minus 1  : parafermion $\mathbb{Z}_k$ can be represented as $SU(2)_k/U(1)$.}. This massive perturbation consists of $k-1$ particles $n=1,...,k-1$ where $\bar{n}=k-n$ with mass spectrum
\begin{align*}
m_n=m\sin\big(\frac{\pi n}{k}\big)/\sin\big(\frac{\pi}{k}\big).
\end{align*}
The purely elastic scattering between these particles is
\begin{gather}
S_{nm}(\theta)=\frac{\sinh\big(\frac{\theta}{2}+i\pi\frac{|n-m|}{2k}\big)}{\sinh\big(\frac{\theta}{2}-i\pi\frac{|n-m|}{2k}\big)}\frac{\sinh\big(\frac{\theta}{2}+i\pi\frac{n+m}{2k}\big)}{\sinh\big(\frac{\theta}{2}-i\pi\frac{n+m}{2k}\big)}\prod_{s=\frac{|n-m|}{2}+1}^{\frac{n+m}{2}-1}\bigg[\frac{\sinh\big( \frac{\theta}{2}+i\pi \frac{s}{k}\big)}{\sinh\big( \frac{\theta}{2}-i\pi \frac{s}{k}\big)}\bigg]^2\label{A-series-scatterings}
\end{gather}
So indeed, despite the different underlying physics, the two scattering phases \eqref{kink-magnon-scatterings} and \eqref{A-series-scatterings} are identical up to a redefinition of rapidity variable
\begin{align*}
\theta^{\text{ kink magnon}}=\frac{k}{2}\times\theta^{\text{ A series}}.
\end{align*}
This suggests that we can use the minimal reflection factor already derived for A series \cite{Corrigan:1994ft},\cite{Dorey:2005ak} for the kink magnons. It is the solution of the boundary unitarity, crossing and bootstrap equations with a minimal number of poles and zeros
\begin{align*}
R_j(\theta)=\prod_{s=0}^{j-1}\frac{\sinh\big(\frac{\theta}{2}+i\pi\frac{s}{2}\big)}{\sinh\big(\frac{\theta}{2}-i\pi\frac{s}{2}\big)}\frac{\sinh\big(\frac{\theta}{2}-i\pi\frac{k-s+1}{2}\big)}{\sinh\big(\frac{\theta}{2}+i\pi\frac{k-s+1}{2}\big)},\quad j=\overline{1,k-1}.
\end{align*}
It satisfies in particular the following identity
\begin{align}
K_j\star \mathbf{1}-\frac{1}{2}K_{jj}\star\mathbf{1}-\frac{1}{2}=0,\quad j=\overline{1,k-1}.\label{A-series-boundary}
\end{align}
where $K_j=-i\partial\log R_j$, which greatly simplifies the form of the corresponding g-function in the next section. For parafermions, this set of relection factors is assigned with the fixed boundary condition or equivalently the vacuum representation of both $SU(k)_1$ and $SU(k)_2$. The g-function was found to be
\begin{align}
g_0^2=\frac{2}{\sqrt{k+2}\sqrt{k}}\sin\frac{\pi}{k+2}.\label{A-series-g-function}
\end{align}
On the other hand, we consider trivial reflection factors on the $SU(2)$ magnons. We do not aim at proving this point but we merely conjecture it based on the result of non linear $O(N)$ sigma model with boundary \cite{Aniceto:2017jor}, \cite{Gombor:2017qsy}, where similar magnon structure arises. What we are doing is to first diagonalize the bulk theory by nested Bethe Ansatz technique. We then treat the theory as one with diagonal scattering and find the reflection factors based on this bulk diagonal scattering. The standard way to do it would be to start with a set of reflection factors that satisfy the boundary Yang-Baxter equation. One then writes Bethe equations with these reflection factors and diagonalizes the corresponding two-row transfer matrix. 

To summarize, we conjecture the following set of TBA equations for current perturbed $SU(2)_k$ theories in the presence of boundaries
\begin{gather*}
\text{physical rapidity}\quad e^{iLm\sinh(\theta_{k,n})}R_{k}^2(\theta_{k,n})\prod_{j=1}^\infty\prod_{m}S_{kj}(\theta_{k,n}-\theta_{j,m})S_{kj}(\theta_{k,n}+\theta_{j,m})=-1\\
\text{Kink magnons}\quad R_{j}^2(\theta_{j,n})\prod_{l=1}^k\prod_{m}S_{jl}(\theta_{j,n}-\theta_{l,m})S_{jl}(\theta_{j,n}+\theta_{l,m})=-1,\quad j=\overline{1,k-1}\\
\text{SU(2) magnons}\quad \prod_{l=k}^\infty\prod_{m}S_{jl}(\theta_{j,n}-\theta_{l,m})S_{jl}(\theta_{j,n}+\theta_{l,m})=-1,\quad j=\overline{k+1,\infty}
\end{gather*}
We denote from now on the convolution of the boundary reflections with identity by $B_j=K_j\star \mathbf{1}$. They are given by \eqref{A-series-boundary} for kink magnons and are zero for $SU(2)$ magnons. For the physical rapidity, we leave it as a parameter.
\section{The divergence and the normalized g-function}
We now have all the necessary ingredients to study the UV and IR limit of the g-function of the current-perturbed $SU(2)_k$ theories. For convenience we repeat here the result \eqref{trees+loops}-\eqref{loops-multi} , with an equivalent form of the loop part that is more adapted to actual computation
\begin{align}
&2\log g(R)=2\log g^\textnormal{trees}(R)+2\log g^\textnormal{loops}(R),\label{g-function}\\
&2\log g^\textnormal{trees}(R)=\sum_{n}\int_{-\infty}^\infty\frac{d\theta}{2\pi}[K_{n}(\theta)-K_{nn}(\theta,-\theta)-\pi\delta(\theta)]\log[1+Y_n(\theta)],\label{tree-part}\\&2\log g^\textnormal{loops}(R)\nonumber\\
&=\sum_{n\geq 1}\frac{1}{n}\sum_{a_1,...,a_n\geq 0}\bigg[\prod_{j=1}^n\int\limits_{-\infty}^{+\infty}\frac{du_j}{2\pi}f_{a_j}(u_j)\bigg]K_{a_1a_2}(\theta_1+\theta_2)K_{a_2a_3}(\theta_2-\theta_3)...K_{a_na_1}(\theta_n-\theta_1).\label{equivalent-expression}
\end{align}
The goal of this section is to support to our proposition \eqref{proposition} by proving that it is possible to match the normalized UV g-function, namely $g_{\text{UV}}/g_{\text{IR}}$ with a conformal g-function \eqref{CFT-entropy} in some cases. While carrying out this normalization we encounter divergence in both IR and UV limit. We illustrate this phenomenon for the Gross-Neveu model and show how an appropriate regularization could lead to a finite ratio.
\subsection{Level 1- Gross Neveu model}
At zero temperature, the tree part \eqref{tree-part} of the g-function can be exactly evaluated. With the Y-functions given by constants in \eqref{IR-GN-not-twist}, the reflection kernels for $SU(2)$ magnons being zero and the scattering kernels $K_{nn}$ given in \eqref{aux-aux}, it turns out to be divergent in this limit
\begin{align}
2\log g^\textnormal{trees}_{\text{IR}}=\sum_{n=1}^\infty(n-1)\log\big[1+\frac{1}{n(n+2)}\big].\label{tree-GN-not-twist}
\end{align}
The tadpole (the $n=1$ term in the series \eqref{equivalent-expression}) suffers from a similar divergence 
\begin{align}
2\log g^\textnormal{tadpole}_{\text{IR}}&=\sum_{n=1}^\infty\frac{Y_n^{\text{IR}}}{1+Y_n^{\text{IR}}}\int_{-\infty}^{+\infty}\frac{du}{2\pi}K_{nn}(2u)=\sum_{n=1}^\infty\frac{1-2n}{2(n+1)^2}.\label{tadpole-GN}
\end{align}
This logarithmic divergence is present for higher order terms and for the infinite temperature limit alike. We believe it is a common feature among models with an infinite number of string magnons. 

As a regularization, we propose to use the twisted TBA solutions \eqref{IR-UV-GN-twist}. The tree part of the IR g-function can now be expressed in terms of the twist parameter $\k$
\begin{align}
2\log g^\textnormal{trees}_{\text{IR}}(\k)
=\sum_{n=1}^\infty(n-1)\log\big(1+\frac{1}{[n+1]_{\k}^2-1}\big)=-\log(1-\k^2).\label{g-GN-IR-tree}
\end{align}
To evaluate the loop part, we remark that for constant Y-functions the series \eqref{equivalent-expression} can be written as a determinant 
\begin{align}
2\log g^{\textnormal{loops}}_{\text{IR}}(\k)=-\frac{1}{2}\log\det[1-\hat{K}^{\text{IR}}(\k)],
\end{align}
where 
\begin{gather}
\hat{K}_{ab}^{\text{IR}}(\k)\equiv \sqrt{\frac{Y_a^{\text{IR}}(\k)}{1+Y_a^{\text{IR}}(\k)}\frac{Y_b^{\text{IR}}(\k)}{1+Y_b^{\text{IR}}(\k)}}\int_{-\infty}^{+\infty}\frac{d\theta}{2\pi}K_{ab}(\theta),\label{first-Fredholm}
\end{gather}
The factor $1/2$ comes from the change of variables $(\theta_1+\theta_2,\theta_2-\theta_3,...,\theta_n-\theta_1)\to (\tilde{\theta}_1,\tilde{\theta}_2,...,\tilde{\theta}_n)$. We show in appendix \ref{det-IR} that 
\begin{align}
\det[1-\hat{K}^\text{IR}(\k)]=(1-\k)^{-1}.\label{g-GN-IR-loop}
\end{align}
By combining the two contributions \eqref{g-GN-IR-tree} and \eqref{g-GN-IR-loop}, we obtain the IR g-function of Gross-Neveu model as a function of the twist parameter. In the untwisted limit $\k\to 1$ it behaves as
\begin{align}
\lim_{\k\to 1}2\log g_{\text{IR}}(\k)=-\log 2-\frac{1}{2}\log(1-\k).\label{g-GN-IR}
\end{align}

We can repeat the same analysis for the UV limit, using the corresponding twisted constant solution \eqref{IR-UV-GN-twist}. Compared to the IR limit we algo get contribution from the physical rapidity. The loop part can again be written as a determinant by replacing the IR by UV values in the matrix \eqref{first-Fredholm}. We show in appendix \ref{det-UV} that this determinant is again a very simple function of the twist parameter
\begin{align}
2\log g^\textnormal{trees}_{\text{UV}}(\k)&=(B_0-\frac{3}{4})\log\frac{(1+\k)^2}{\k}-\log(1-\k^3),\\
2\log g^\textnormal{loops}_{\text{UV}}(\k)&=\frac{1}{2}\log[2(1-\k)].
\end{align}
The UV value of g-function exhibits the same divergence as the IR one in the untwisted limit 
\begin{align}
\lim_{\k\to 1}2\log g_{\text{UV}}(\k)=(2B_0-1)\log 2-\log 3-\frac{1}{2}\log(1-\k)\label{g-GN-UV}
\end{align}
In particular their ratio is well defined
\begin{align}
\bigg(\frac{ g_{\text{UV}}}{g_{\text{IR}}}\bigg)^2=2^{2B_0}/3.\label{final-level-1}
\end{align}
The two Cardy g-functions \eqref{CFT-entropy} of $SU(2)_1$ CFT take the same value $g_1^2=g_2^2=1/\sqrt{2}$. For integrable boundary conditions, the reflection factor usually gives rational value for $B_0$ and our proposition \eqref{final-level-1} could not be matched with a Cardy g-function. We carry on our analysis to higher levels.
\subsection{Higher levels}
We first consider the IR limit, in which the left and right wing are decoupled. The former is not affected by the twist while the latter is identical to the IR of the Gross-Neveu model. Our choice of reflection factors with the property \eqref{A-series-boundary} eliminates the left wing from the tree part of the g-function. As a consequence we get the same result as the IR tree part of  Gross-Neveu model \eqref{g-GN-IR-tree}
\begin{align}
2\log g_{\text{IR}}^\textnormal{trees}(\k)=-\log(1-\k^2).\label{tree-IR-level-n}
\end{align}
The loop part is factorized into two determinants
\begin{align}
2\log g_{\text{IR}}^\textnormal{loops}=-\frac{1}{2}\log\det(1-\hat{K}_{1\to k-1})-\frac{1}{2}\log\det(1-\hat{K}_{k+1\to \infty}).
\end{align}
The finite  determinant involving the trigonometric Y-functions has been computed in \cite{Dorey:2005ak} while the infinite  determinant is exactly the same as that of IR Gross-Neveu 
\begin{align*}
\det(1-\hat{K}_{1\to k-1})=[\frac{4k}{k+2}\sin^2\frac{\pi}{k+2}]^{-1}
,\quad \det(1-\hat{K}_{k+1\to \infty})=(1-\kappa)^{-1}.
\end{align*}
By summing the two parts, we obtain the IR  g-function. Its behavior in the untwisted limit is
\begin{align}
\lim_{\k\to 1}2\log g_{\text{IR}}(\k)=-\log 2+\frac{1}{2}\log\frac{4k}{k+2}+\log\sin\frac{\pi}{k+2}-\frac{1}{2}\log(1-\k).\label{IR-level-k}
\end{align}

In the UV limit all Y-functions are twisted. Again only the right wing contributes to the tree part of g-function
\begin{align}
2\log g_{\text{UV}}^\textnormal{trees}(\k)=(B_k-1+\frac{1}{4k})\log [k+1]_\k^2-\log(1-\k^{k+2}).\label{tree-g-level-n}
\end{align}
The loop contribution is given by a determinant that connects the two wings. We compute this  determinant in appendix \ref{det-UV}.  Despite its complicated form, as the structure of scattering kernels on the left and right wing are different, the result is simple
\begin{align}
2\log g^\textnormal{loops}_{\text{UV}}(\k)=\frac{1}{2}\log 2k+\frac{1}{2}\log (1-\k).\label{loop-g-level-n}
\end{align}
The UV g-function is obtained by summing \eqref{tree-g-level-n} and \eqref{loop-g-level-n}
\begin{align}
\lim_{\k\to 1}2\log g_{\text{UV}}(\k)=(2B_k-2+\frac{1}{2k})\log (k+1)-\log(k+2)+\frac{1}{2}\log 2k-\frac{1}{2}\log (1-\k).\label{UV-level-k}
\end{align}

We see that the IR \eqref{IR-level-k} and UV \eqref{UV-level-k} values of g-function exhibit the same divergence in the untwisted limit. We can therefore extract their ratio
\begin{align}
\bigg(\frac{g_{\text{UV}}}{g_{\text{IR}}}\bigg)^2=(k+1)^{2B_k-2+\frac{1}{2k}}\times\sqrt{\frac{2}{k+2}}\times\frac{1}{\sin\frac{\pi}{k+2}}.\label{a-song-of-ice-and-fire}
\end{align}
To remind, the Cardy g-functions are given by
\begin{align}
g_\lambda^2=\sqrt{\frac{2}{k+2}}\times\frac{1}{\sin\frac{\pi}{k+2}}\times\sin^2\frac{(\lambda+1)\pi}{k+2},\;0\leq\lambda\leq k\label{g-cardy-level-n}
\end{align}
Therefore we can match our normalized UV g-function \eqref{a-song-of-ice-and-fire} with $g_{k/2}$ for even $k$ as long as the reflection factor of the physical rapidity satisfies $B_k=1-1/(4k)$. Let $k=2m$ then the corresponding bulk primary has conformal dimension
\begin{align*}
\Delta=\frac{m(m+2)}{8(m+1)}.
\end{align*}
\section{Conclusion}
In this chapter we propose the following procedure to study the g-function of a massive integrable theory with non-diagonal bulk scattering
\begin{itemize}
\item Diagonalize the theory using the Nested Bethe Ansatz technique.
\item Treat the newly obtained theory as diagonal with extra magnonic particles and apply the results \eqref{trees+loops}-\eqref{loops-multi} to compute its g-function.
\item Normalize the g-function by its zero temperature limit value.
\end{itemize}
We test our proposition for the current-perturbed $SU(2)$ WZNW CFTs. The TBA of these theories involves an infinite tower of magnon strings. As a consequence both the tree \eqref{trees-multi} and loop \eqref{loops-multi} part of the g-function diverge at zero and infinite temperature. This phenomenon is illustrated for the Gross-Neveu model in \eqref{tree-GN-not-twist},\eqref{tadpole-GN}. We conjecture that such divergence is present at arbitrary temperature. By considering the twisted TBA, we are able to compute these two limits of g-function as functions of the twist parameter $\K$. It is found that they exhibit the same divergence $-\frac{1}{2}\log(1-\K)$  in the untwisted  limit $\K\to 1$ \eqref{IR-level-k},\eqref{UV-level-k}. The normalized UV g-function is then well defined \eqref{a-song-of-ice-and-fire} and can be identified with a Cardy g-function of the unperturbed CFT under some assumption on the reflection factor of the physical rapidity and for even levels.

This normalization has a diagramatical interpretation. At zero temperature the boundary entropy is given by the sum of all graphs made exclusively of auxiliary magnons. The contribution of these graphs does not depend on the temperature and can be absorbed into the normalization of the partition function. No physical observable will involve such graphs.
\newpage
\chapter{Applications in Generalized Hydrodynamics}
In this chapter we present two applications of our diagrammatic formalism in the context of GHD. The first is a rigorous derivation based on form factors \cite{10.21468/SciPostPhys.6.2.023} of the average currents. As we saw in section \ref{GHD-intro-section} this quantity is a cornerstone in the GHD formalism: all transport-related observables are derived from it. The second is a conjecture \cite{Vu:2019pam} that allows the cumulants of  time-integrated currents  to be expressed as a sum over simple tree diagrams. They are almost the same diagrams that represent the cumulants of conserved charges in section \ref{charge-new-section}. Our conjecture thus highlights a remarkably simple duality between time-integrated currents and conserved charges, one that could potentially be extended to other transport properties as well.
\section{Equations of state from form factors}
\label{equations-of-state-new-derivation}
GHD is a framework to study the dynamics of integrable systems at the Euler scale\footnote{Diffusive effect is outside the scope of this thesis}. At such scale, generically, many-body systems are expected to be in a state where  local entropy maximization is realized. In such a state, physics is dominated by macroscopic processes protected by conserved charges, and the state potentially carry a current. In practice, this scale can be accessed by taking a scaling limit of infinitely many degrees of freedom (i.e. the ratio between a typical microscopic scale $l_{\rm mic}$, say the inter-particle length, and a typical macroscopic scale $l_{\rm mac}$ becomes zero: $\epsilon=l_{\rm mic}/l_{\rm mac}\to0$) while scaling the space-time simultaneously $(x,t)\to(\epsilon^{-1} x,\epsilon^{-1} t)$, which amounts to focusing on physics occurring at an emergent large scale called the  fluid cell. Note that depending on the exponent $\alpha$ of the scaling of $x$, $\epsilon^{-\alpha}x$, a different scaling limit can be obtained (e.g. diffusive scaling for $\alpha=1/2$ and super diffusive scaling for $1/2<\alpha<1$). 
The assumption of local entropy maximization provides us an efficient way to evaluate correlation functions at the Euler scale. In particular, the expectation value of a given local operator $\mathcal{O}$ is computed by $\langle \mathcal{O}(x,t)\rangle_{\rm Eul}=\mathrm{Tr}[\rho(x,t)\mathcal{O}]$ with $\rho(x,t)\propto \exp[-\sum_i\beta_i(x,t)\mathcal{Q}_i]$, where $\mathcal{Q}_i=\int{\rm d}x\,Q_i(x,0)$ are the conserved charges. This suggests that, at the Euler scale, in order to solve the macroscopic continuity equations $\p_t\langle Q_i(x,t)\rangle_{\rm Eul}+\p_x\langle J_i(x,t)\rangle_{\rm Eul}=0$, one only has to know the equilibrium form of the averages of densities $\langle Q_i\rangle_{\vec{\beta}}$ and currents $\langle J_i\rangle_{\vec{\beta}} $ as functions of Lagrange multipliers $\vec{\beta}$. The density average of a conserved charge can be be written as
\begin{equation}\label{avedensity}
\langle Q_i\rangle=\int\frac{{\rm d}p(\theta)}{2\pi}f(\theta)q_i^{\rm dr}(\theta),
\end{equation}
where $f$ is the TBA filling factor, $q_{i}$ is the one-particle eigenvalue of $\mathcal{Q}_i$ and the dressing operation was defined in \eqref{dressing-def-unique}.
Now, we need to know how $\langle J_i\rangle$ looks like in order to solve the macroscopic continuity equations. In \cite{Castro-Alvaredo:2016cdj,PhysRevLett.117.207201}, the exact expression of $\langle J_i\rangle$ was proposed that
\begin{equation}\label{avecurrent}
\langle J_i\rangle =\int\frac{{\rm d}E(\theta)}{2\pi}f(\theta)q_i^{\rm dr}(\theta)=\int{\rm d}\theta\,\rho_{\rm p}(\theta)v^{\rm eff}(\theta)q_{i}(\theta),
\end{equation}
where $v^{\rm eff}=(E')^\text{dr}/(p')^\text{dr}$ is the velocity of excitation over an equilibrium state.  The form of $v^{\rm eff}$ can be in fact considered as equations of state for GHD. Recall that equations of state are relations that relate the density averages $\langle Q_i\rangle$ and the current averages. Since it is precisely what $v^{\rm eff}$ is doing, making \eqref{avecurrent} different from \eqref{avedensity} by its very appearance in \eqref{avecurrent}, the functional form of the effective velocity determines the relation between the density and current averages.

The proof in \cite{Castro-Alvaredo:2016cdj}  exploits mirror transformation and has been sketched in subsection \ref{equations-state-section}. We note however that this proof is in fact incomplete as the assumed analyticity of  the TBA source term is not necessarily true for some GGEs. In \cite{PhysRevLett.117.207201}, the expression was extended to the XXZ spin-$\frac{1}{2}$ chain where strings are present without proof but with numerical verifications. Thus a full proof of the equations of state is still needed in GHD. So far, the validity of GHD, which is equivalent to that of the effective velocity, has been numerically confirmed for spin chains such as the XXZ spin-$\frac{1}{2}$ chain \cite{PhysRevB.96.020403,PhysRevLett.119.020602,PhysRevB.97.045407,PhysRevLett.119.220604,PhysRevB.96.115124,PhysRevB.97.081111} and the Fermi-Hubbard model \cite{PhysRevB.96.081118}, and it is believed that GHD correctly captures the long-wavelength dynamics of any Bethe solvable systems. Nonetheless, a down-to-earth proof of $v^{\rm eff}(\theta)$ is still highly-desired to complete the program of GHD, and it is the purpose of this section to report such a proof  for relativistic integrable field theories with diagonal S-matrix. With minor modifications, this proof has been shown to work equally well for integrable spin chains \cite{PhysRevX.10.011054}. Very recently, there is yet another proof that does not require explicit use of relativistic invariance \cite{Yoshimura:2020qxz}.

Our strategy relies on form factor expansions by means of the LeClair-Mussardo series \eqref{lm}. This series is universal in the sense that the expectation values of two operators differ only in their connected form factors. As a result, the task of establishing the equations of state boils down to a  direct comparison between the connected form factor of the charge densities and that of the currents. Such comparison can be done with help of the Pozsgay-Takacs relation \eqref{symm-conn} between the connected form factors and the symmetric ones. This relation involves principal minors of a Laplacian matrix and naturally admits a diagrammatic interpretation.
 \subsection{The connected and the symmetric form factors}
The connected and the symmetric evaluation of diagonal matrix elements play a pivotal role in our proof. For the convenience of following, we repeat here the Pozsgay-Takacs relation already introduced in subsection \ref{conn-symm-section}
\begin{equation}\label{symmconn}
F_{\rm s}^\mathcal{O}(\theta_1,\cdots,\theta_n)=\sum_{\alpha \subset \{1,\cdots,n\},
\alpha\neq\varnothing}\mathcal{L}(\alpha|\alpha)F_{\rm c}^\mathcal{O}(\{\theta_i\}_{i\in\alpha})
\end{equation}
where  $\mathcal{L}(\alpha|\alpha)$ is the principal minor obtained by deleting the $\alpha$ rows and columns of the Laplacian matrix 
\begin{equation}
L(\theta_1,\cdots,\theta_n)_{jk}=\delta_{jk}\sum_{l\neq j}K(\theta_j-\theta_l)-(1-\delta_{jk})K(\theta_j-\theta_k).\label{laplacian-L}
\end{equation}
Let us also repeat  the LeClair-Mussardo formula for the one point function of a local operator
\begin{equation}\label{lm-series}
\langle \mathcal{O}\rangle=\sum_{n=0}^\infty\biggl(\prod_{j=1}^n\int\frac{{\rm d}\theta_j}{2\pi}f(\theta_j)\biggr)F_{\rm c}^\mathcal{O}(\theta_1,\cdots,\theta_n).
\end{equation}
In the particular case of a conserved charge density $Q_j$, the corresponding connected form factor is given by 
\begin{equation}\label{connffq}
F_{\rm c}^{Q_i}(\theta_1,\cdots,\theta_n)=q_i(\theta_1)K(\theta_1-\theta_2)\cdots K(\theta_{n-1}-\theta_n)p'(\theta_n) + {\rm perm},
\end{equation}
where perm. is understood as permutations with respect to the integer set $\{1,\cdots,n\}$. 

We observe that the same structure holds for the current operator $J$ as well. Recalling \eqref{avecurrent}, we can also recast it into the similar form and expand as
\begin{align}\label{current2}
\langle J_i\rangle=\sum_{n=0}^\infty\biggl(\prod_{k=1}^n\int\frac{{\rm d}\theta_k}{2\pi}f(\theta_k)\biggr)q_i(\theta_1)K(\theta_1-\theta_2)\cdots K(\theta_{n-1}-\theta_n)E'(\theta_n),
\end{align}
This suggests that if the connected form factor of $J$ takes the following form, then \eqref{avecurrent} follows:
\begin{equation}\label{connffj}
F_{\rm c}^{J_i}(\theta_1,\cdots,\theta_n)=q_i(\theta_1)K(\theta_1-\theta_2)\cdots K(\theta_{n-1}-\theta_n)E'(\theta_n) + {\rm perm},
\end{equation}
which is the actual statement we are going to prove in order to establish \eqref{avecurrent}.

The relation \eqref{symmconn} between the symmetric and connected form factors can be understood graphically. The matrix $L$ whose minors appear in this relation is a Laplacian matrix. It is the discretized Laplacian operator  on a graph in which a weight $K(\theta_j-\theta_k)$ is assigned to the edge connecting $j$ and $k$. Although $L$ has a vanishing determinant, as the elements on each row sum up to zero, its principal minors can be expressed as a sum over forests. 

As our discussion in this chapter is restricted only to relativistically invariant theories, the scattering kernel $K$ is symmetric. Consequently,  the expansion of the principal minors of $L$ result in undirected graphs, in contrast to the previous chapters.  Let $\alpha$ be a subset of vertices $\{1,2,\cdots,n\}$. Then we have
\begin{equation}
\mathcal{L}(\alpha|\alpha)=\sum_{F\in\mathcal{
F}_\alpha}\prod_{e\in F}K_e,\label{all-minor-theorem}
\end{equation}
where the summation is performed over all forests of $n$ vertices each tree of which contains exactly one vertex from $\alpha$. The product runs over all edges of the forests. The result \eqref{all-minor-theorem} is exactly equivalent to \eqref{all-minor-Gaudin} and \eqref{det-Gaudin-tree} presented in previous chapters. The only difference is that we cannot refer to vertices of $\alpha$ as roots, due to the lack of direction on the edges. 

This is known as the all-minor version of the matrix-tree theorem. A particular case is given by considering principal minors of rank $n-1$ i.e. by taking $\alpha$ to be one-element subsets. The forests would then become trees.

\begin{figure}[!htb]
\centering
\includegraphics[width=12cm, clip]{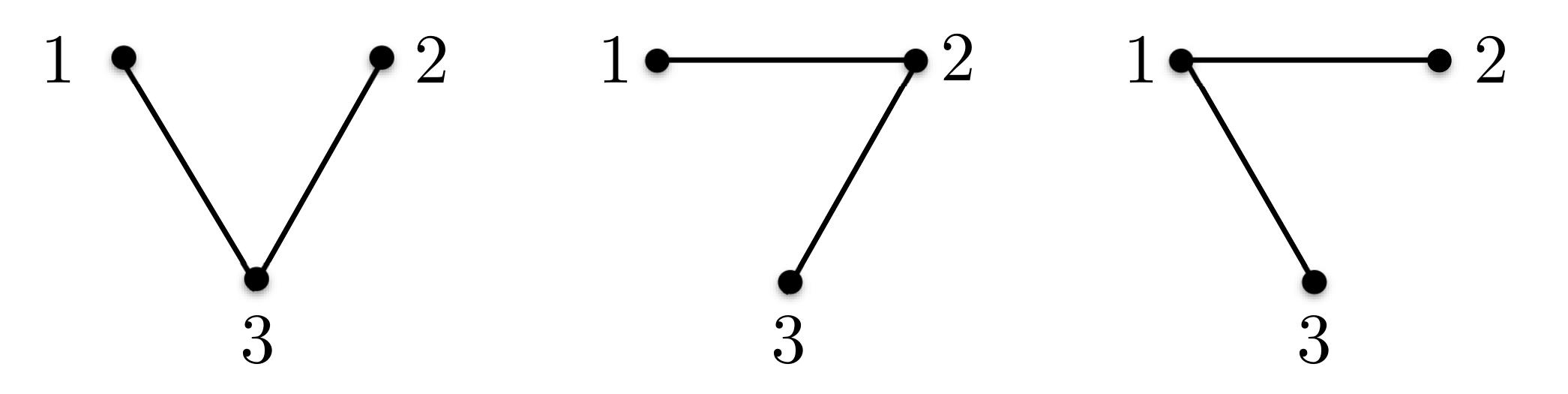}
\caption{Trees associated with a minor of rank 2.}
\label{fig1}
\end{figure}

\begin{figure}[!htb]
\centering
\includegraphics[width=12cm, clip]{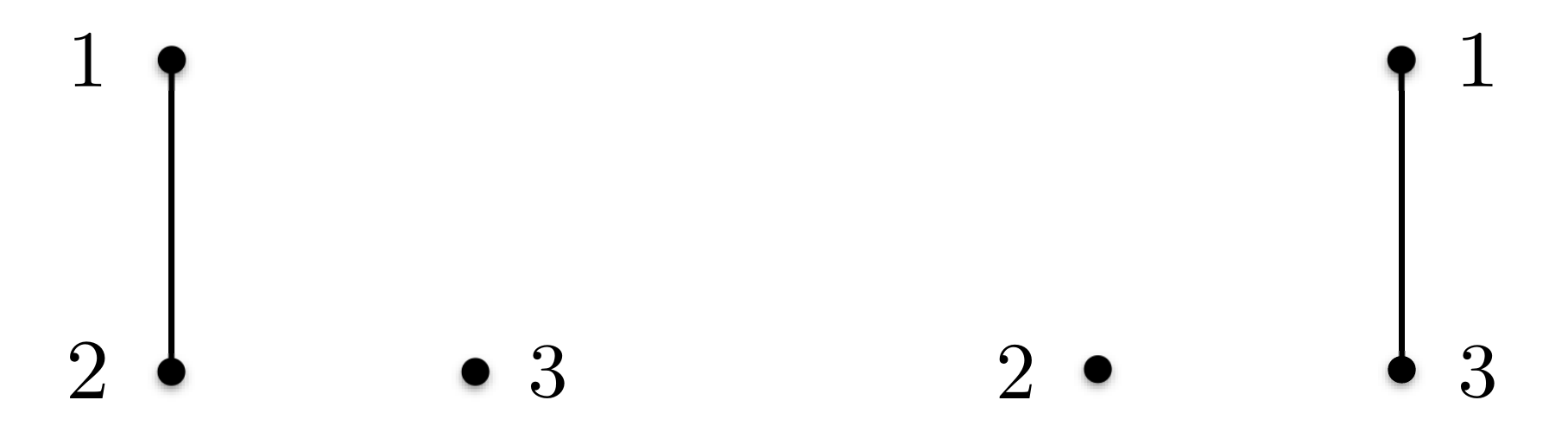}
\caption{Forests associated with a minor $\mathcal{L}(\{2,3\}|\{2,3\})$.}
\label{fig2}
\end{figure}

Let us illustrate the theorem in the case of three particles, where \eqref{laplacian-L} is given by
\begin{equation*}
L=\begin{pmatrix}
K_{12}+K_{13} & -K_{12} & -K_{13} \\
-K_{21} & K_{21}+K_{23} & -K_{23}\\
-K_{31} & -K_{32} & K_{31}+K_{32}
\end{pmatrix},\quad K_{ij}\equiv K(\theta_i-\theta_j).
\end{equation*}
All the principal minors of rank 2 are equal: $\mathcal{L}(1|1)=\mathcal{L}(2|2)=\mathcal{L}(3|3)=K_{21}K_{31}+K_{21}K_{32}+K_{23}K_{31}$. These terms are exactly the three trees spanning three vertices, see Fig.\ref{fig1}. Note that we are referring to \textbf{labelled} trees. In particular, the trees in Fig.\ref{fig1} are considered as being distinguished, despite their similar combinatorial structure. The principal minors of rank 1 are written as forests with two trees. For example, when $\alpha=\{2,3\}$ we have $\mathcal{L}(\alpha|\alpha)=K_{12}+K_{13}$, as in Fig.\ref{fig2}.

The matrix-tree theorem provides a simple interpretation of the relation \eqref{symmconn} between symmetric and connected form factors. For each  subset $\alpha$ of $\lbrace 1,2,...,n\rbrace$ we decorate the connected form factor $F_{\text{c}}^\mathcal{O}(\{\theta_i\}_{i\in\alpha})$ by trees growing out of the elements of $\alpha$. The decorations must guarantee that all $n$ vertices are covered. By summing over $\alpha$ and over all the possible decorations for each $\alpha$, we obtain the symmetric form factor $F_s^\mathcal{O}(\theta_1,..,\theta_n)$.
\subsection{Establishing the equations of state}
We are at the position to present a graph theoretic proof for the equations of state \eqref{connffj}. Our proof consists of three steps
\begin{itemize}
\item Obtain the symmetric form factor of the charge $F_{\rm s}^Q(\theta_1,\cdots,\theta_n)$ from the connected one \eqref{connffq} and the relation \eqref{symmconn}.
\item Compute the symmetric form factor of the current $F_{\rm s}^J(\theta_1,\cdots,\theta_n)$ from that of the charge, by using the continuity equation.
\item Find the connected form factor of the curent from the symmetric one, by going from the left hand side to the right hand side of equation \eqref{symmconn}.
\end{itemize}
The first and the last step are done with help of the matrix-tree theorem. In the first step, we represent the connected form factor of the charge 
\begin{equation}
F_{\rm c}^Q(\theta_1,\cdots,\theta_n)=q(\theta_1)K(\theta_1-\theta_2)\cdots K(\theta_{n-1}-\theta_n)p'(\theta_n) + {\rm perm},
\end{equation}
as $n!$ spines of length $n$ with the charge eigenvalue $q$ on one end and the momentum derivative $p'$ at the other end. Spines of length 1 with coinciding ends are allowed.

\begin{figure}[!htb]
\centering
\includegraphics[width=15cm, clip]{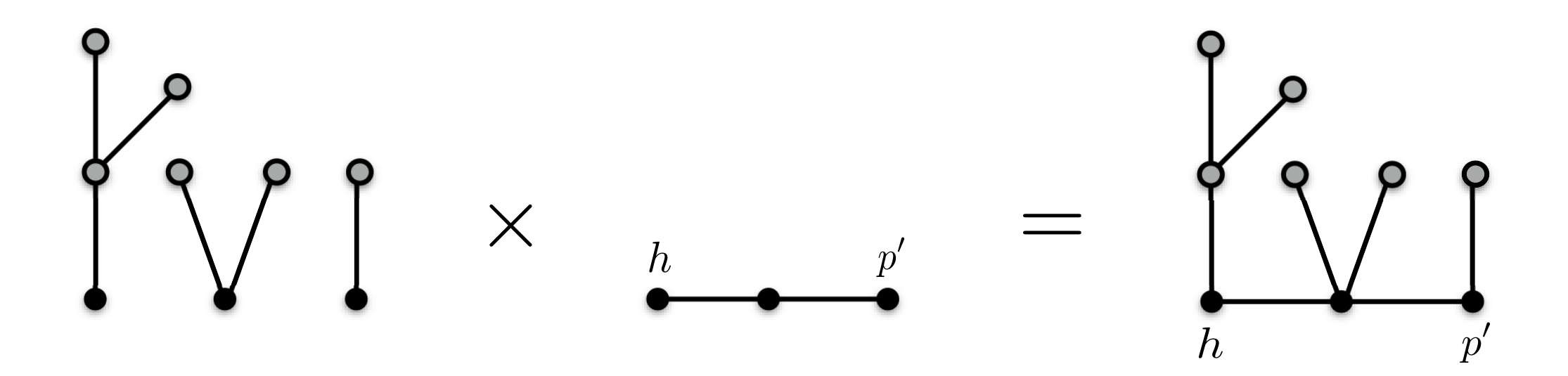}
\caption{Pictorial representation of one of terms in the RHS of \eqref{symmconn} when $\mathcal{O}$ is a conserved density. Each term (forest) of $\mathcal{L}(\alpha|\alpha)$ and each term in $F_{\rm c}(\{\theta_i\}_{i\in\alpha})$ form a spanning tree by merging together at vertices $\alpha$ represented by black dots.}
\label{fig3}
\end{figure}
The  corresponding symmetric form factor is obtained by decorating the spines with the trees, see Fig.\ref{fig3}. Because the trees have different labelings and the spines come from different permutations, each term in the symmetric form factor is a (labelled) tree with two marked points, no tree appears more than once. Vice versa, each tree with two marked points can be decomposed to a spine and a forest. Indeed, the connectedness guarantees the existence of a path between the two marked points. Moreover, the uniqueness of this path is ensured by the non-existence of cycles. We conclude that the symmetric form factor of the charge is given by the sum over all the trees of $n$ vertices, with the weights $q$ and $p'$ inserted at two arbitrary vertices. This sum factorizes into the sum over the weights and the sum over the unmarked trees
\begin{equation}\label{denfactor}
F_{\rm s}^Q(\theta_1,\cdots,\theta_n)=\sum_{j=1}^nq(\theta_j)\sum_{k=1}^np'(\theta_k)\sum_{T\in\mathcal{T}}\prod_{e\in \text{ edges of }\,T}K_e,
\end{equation}
Here, $\mathcal{T}$ denotes the set of all trees of $n$ vertices. The sum over these trees are exactly given by the principal minor of rank $n-1$ of the matrix \eqref{laplacian-L}. For instance, in the case of three particles
\begin{align*}
F_{\rm s}^Q(\theta_1,\theta_2,\theta_3)=[q(\theta_1)+q(\theta_2)+q(\theta_3)][p'(\theta_1)+p'(\theta_2)+p'(\theta_3)](K_{21}K_{31}+K_{21}K_{32}+K_{23}K_{31}).
\end{align*}

We now turn to the second step. In order to relate \eqref{denfactor} to the symmetric form factor of the current $F_{\rm s}^J(\theta_1,\cdots,\theta_n)$, where $J$ satisfies the continuity equation $\partial_t Q+\partial_x J=0$, we note that there is a relation between $F_{\rm s}^Q(\theta_1,\cdots,\theta_n)$ and $F_{\rm s}^J(\theta_1,\cdots,\theta_n)$ which is a simple consequence of the continuity equation
\begin{equation}\label{ffsymmconn}
F_{\rm s}^J(\theta_1,\cdots,\theta_n)=\frac{\sum_{k=1}^n E'(\theta_k)}{\sum_{k=1}^n p'(\theta_k)}F_{\rm s}^Q(\theta_1,\cdots,\theta_n),
\end{equation}
where we recall $E(\theta)=m\cosh\theta$ and $p(\theta)=m\sinh\theta$. To see this, we first observe
\begin{equation}
\langle\mathrm{vac}|J(x,t)|\overrightarrow{\theta},\overleftarrow{\theta'}\rangle =e^{-\ii m\sum_{k=1}^n\big[(\cosh\theta_k+\cosh\theta'_k)t-(\sinh\theta_k+\sinh\theta'_k)x\big]}\langle\mathrm{vac}|J(0,0)|\overrightarrow{\theta},\overleftarrow{\theta'}\rangle
\end{equation}
and thus,
\begin{equation}
\langle\mathrm{vac}|\partial_xJ(x,t)|\overrightarrow{\theta},\overleftarrow{\theta'}\rangle =\ii m\sum_{k=1}^n(\sinh\theta_k+\sinh\theta'_k)\langle\mathrm{vac}|J(x,t)|\overrightarrow{\theta},\overleftarrow{\theta'}\rangle.
\end{equation}
Using this, it then follows that
\begin{align}
F_{\rm s}^J(\theta_1,\cdots,\theta_n)&\equiv\lim_{\delta\to0}\langle\mathrm{vac}|J(x,t)|\overrightarrow{\theta},\overleftarrow{\theta'}\rangle
\n
&=\lim_{\delta\to 0}\frac{-i}{m\sum_k(\sinh\theta_k+\sinh\theta'_k)}\langle\mathrm{vac}|\partial_xJ(x,t)|\overrightarrow{\theta},\overleftarrow{\theta'}\rangle \n
&=\lim_{\delta\to 0}\frac{i}{m\sum_k(\sinh\theta_k+\sinh\theta'_k)}\langle\mathrm{vac}|\partial_tQ(x,t)|\overrightarrow{\theta},\overleftarrow{\theta'}\rangle \n
&=\lim_{\delta\to 0}\frac{\sum_k(\cosh\theta_k+\cosh\theta'_k)}{\sum_k(\sinh\theta_k+\sinh\theta'_k)}\langle\mathrm{vac}|Q(x,t)|\overrightarrow{\theta},\overleftarrow{\theta'}\rangle \n
&=\frac{\sum_kE'(\theta_k)}{\sum_kp'(\theta_k)}F_{\rm s}^Q(\theta_1,\cdots,\theta_n),
\end{align}
where we used the continuity equation when passing from the second line to the third line, and noted
\begin{equation}
\langle\mathrm{vac}|\partial_tQ(x,t)|\overrightarrow{\theta},\overleftarrow{\theta'}\rangle =-\ii m\sum_{k=1}^n(\cosh\theta_k+\cosh\theta'_k)\langle\mathrm{vac}|Q(x,t)|\overrightarrow{\theta},\overleftarrow{\theta'}\rangle,
\end{equation}
when moving from the third to the fourth line. Here, $\delta$ is defined as before in order to take the uniform limit $\theta'_j=\theta_j+\pi\ii+\delta$ of the symmetric evaluation.

Now, applying this relation to \eqref{denfactor}, it immediately follows that
\begin{equation}\label{currfactor}
F_{\rm s}^J(\theta_1,\cdots,\theta_n)=\sum_{k=1}^nq(\theta_k)\sum_{k=1}^nE'(\theta_k)\sum_{T\in\mathcal{T}}\prod_{e\in \text{ edges of } T}K_e,
\end{equation}
which is nothing but the summation over all the trees of $n$ vertices, this time with $q$ and $E'$ inserted at two arbitrary points. By reversing the logic of  the first step, we can write this as a sum over spines and decorating trees
\begin{equation}
F_{\rm s}^J(\theta_1,\cdots,\theta_n)=\sum_{\alpha \subset \{1,\cdots,n\},
\alpha\neq\varnothing}\mathcal{L}(\alpha|\alpha)F_{\rm c}^J(\{\theta_i\}_{i\in\alpha}),
\end{equation}
where the spines now have $q$ and $E'$ on two ends
\begin{equation}\label{currentconn}
F_{\rm c}^J(\theta_1,\cdots,\theta_n)=q(\theta_1)K(\theta_1-\theta_2)\cdots K(\theta_{n-1}-\theta_n)E'(\theta_n) + \mathrm{perms.}
\end{equation}
This is the desired formula for the current connected form factor. $\square$

Our proof makes use of the matrix-tree theorem to express all the determinants and minors in the relation \eqref{symmconn} between connected and symmetric form factors as sums over trees. We believe this is the natural language to understand this relation, as shown by the simplicity of the proof. One can of course argue that, because the matrix-tree theorem is two-fold, quantities which are expressed in terms of trees can be written as determinants of some matrices as well. As mentioned above, this is indeed true for the symmetric form factor of the charge or the current. For instance, \eqref{denfactor} can be equivalently written as
\begin{equation}\label{symm-algebra}
F_{\rm s}^Q(\theta_1,\cdots,\theta_n)=\mathcal{L}(1|1)
\sum_{j=1}^nq(\theta_j)\sum_{k=1}^np'(\theta_k)
\end{equation}
where $\mathcal{L}(1|1)$ is the principal minor, obtained by deleting the first row and column of the $n\times n$ matrix \eqref{laplacian-L}. One could try to derive \eqref{symm-algebra}, starting from \eqref{connffq} and \eqref{symmconn} with pure matrix manipulation instead of using the matrix-tree theorem.
\subsection{Conclusion}
In this section, we provided a graph theoretic proof of the equations of state used in GHD in the case of relativistic integrable quantum field theories without bound states. The proof applies to purely elastic scattering theories with one or multiple types of particles for which the corresponding LeClair-Mussardo formulae are known. Having the proofs for those cases, an obvious question would be if our approach can be applicable for theories where bound states and/or particles with internal degrees of freedom are present, such as the sine-Gordon model. This would be possible once we are able to extend the notion of connected form factor, or equivalently the LeClair-Mussardo formula for such theories. 

We exemplified the graph theoretic idea using relativistic integrable quantum field theories, but it also works for the nonrelativistic case, such as the Lieb-Liniger model, through taking appropriate non-relativistic limits \cite{PhysRevLett.103.210404}. Extension to integrable spin chains can be found in \cite{PhysRevX.10.011054}.
\section{Full counting statistics}
After having the average current of the non-equilibrium state, one can go further and ask whether it is possible to obtain the probabilities of  rare events with significant deflection from this mean value? In the large deviation theory \cite{TOUCHETTE20091}, these probabilities  are encoded in a rate function the Legendre transform of which is the generating function (also called the \textit{full counting statistics}) of the scaled cumulants 
\begin{align}
\lim_{t\to\infty}\frac{1}{t}\int_0^t dt_1...\int_0^t dt_n \langle J_1(0,t_1)...J_n(0,t_n)\rangle ^\textnormal{c}.\label{J-n}
\end{align}
A functional equation satisfied by the full counting statistics was obtained in \cite{10.21468/SciPostPhys.8.1.007} using linear fluctuating hydrodynamics. Although the individual cumulants can be in principal extracted from this equation, their expressions quickly become cumbersome as $n$ grows larger. We conjecture that these cumulants are simply given by the same diagrams that represent the cumulants of conserved charges, with only two modifications: the operator  at the root is the energy derivative $E'$ (instead of the momentum derivative) and each internal vertex $\theta$ of odd degree carries an extra sign of the effective velocity $\sgn[v^\text{eff}(\theta)]$.  We confirm our conjecture by a non-trivial matching with the result of \cite{Doyon:2019osx} up to the fourth cumulant.

This section is structured as follows: in the first part we remind  how the second cumulant was derived in \cite{SciPostPhys.3.6.039} using hydrodynamics approximation. The derivation in \cite{10.21468/SciPostPhys.8.1.007,Doyon:2019osx} of higher cumulants is beyond the scope of this thesis. In the second part we compare our diagrammatic proposal with the known analytic expressions of \cite{10.21468/SciPostPhys.8.1.007}. In the third part we present several directions in which  the conjecture could possibly be proven. Finally we discuss the impacts of this conjecture on other transport properties of GHD.
\subsection{Hydrodynamics approximation}
The second cumulant or the covariance matrix was first studied in \cite{SciPostPhys.3.6.039} and named the  \textit{Drude self-weight}
\begin{align}
D^\textnormal{s}_{ij} \equiv \lim_{t\to \infty}\int_0^t ds\langle J_i(0,s)J_j(0,0)\rangle^\textnormal{c}.
\end{align}
The name comes from its resemblance with the conventional Drude weight, a quantity that if is non-vanishing, signals a dissipationless transport in the system
\begin{align}
D_{ij}\equiv \lim_{t\to \infty}\int dx \langle J_i(x,t)J_j(0,0)\rangle^\textnormal{c}.
\end{align}
There are two ingredients in the derivation of the Drude self-weight in \cite{SciPostPhys.3.6.039}. The first is a  sum rule  that expresses the Drude self-weight  in terms of the charge-charge correlation function
\begin{align}
D^\textnormal{s}_{ij}=\int dx |x|\frac{1}{2}\big[\langle Q_i(x,t)Q_j(0,0)\rangle^\text{c}+\langle Q_j(x,t)Q_i(0,0)\rangle^\text{c}\big].\label{sum-rule}
\end{align}
For a proof of this identity and similar identities, see \cite{Mendl_2015}. The second ingredient is the large distance limit of this correlator. To obtain this limit, we first note that the charge-charge correlation function $S_{ij}(x,t)\equiv \langle Q_i(x,t)Q_j(0,0)\rangle^\text{c}$ satisfies, as a consequence of the Euler hydrodynamic equations \eqref{Euler-hydro-equation}
\begin{align}
\p_t S_{ij}(x,t)+\sum_k A_i^k(x,t)\p_x S_{kj}(x,t)=0,
\end{align}
with the initial condition $S_{ij}(x,0)\propto C$, where $A$ is the flux Jacobian matrix \eqref{flux-jacobian} and $C$ is the static covariance matrix \eqref{C-matrix}. Thus, in the hydrodynamic approximation, small $k$, large $t$, the Fourier transform of the charge-charge correlation function $ S_{ij}(k,t)\equiv \int dx e^{ikx} S_{ij}(x,t)$ can be approximated by
\begin{align}
S_{ij}(k,t)\approx\big( e^{iktA}C\big)_{ij}=\int \frac{d\theta}{2\pi} e^{ikt v^{\text{eff}}(\theta)}(p')^\text{dr}(\theta)f(\theta)[1-f(\theta)]q_i^\text{dr}(\theta)q_j^\text{dr}(\theta).
\end{align}
Replacing this into the sum rule \eqref{sum-rule} we obtain the Drude self-weight
\begin{align}
D^\textnormal{s}_{ij}=\int \frac{d\theta}{2\pi} (E')^\text{dr}(\theta)s(\theta)f(\theta)[1-f(\theta)]q_i^\text{dr}(\theta)q_j^\text{dr}(\theta),\label{Drude-s}
\end{align}
where we have denoted for short $s(\theta)=\sgn[v^\text{eff}(\theta)]$.

In \cite{10.21468/SciPostPhys.8.1.007,Doyon:2019osx} all the diagonal cumulants were studied at once by mean of their generating function
\begin{align}
F(\lambda)=\sum_{n=1}^\infty\frac{\lambda^n}{n!}c_n\quad \textnormal{with}\quad c_n=\lim_{t\to\infty}\frac{1}{t}\int_0^t dt_1...\int_0^t dt_n \langle J(0,t_1)...J(0,t_n)\rangle ^\textnormal{c}.\label{gen-function}
\end{align}
A functional equation satisfied by this function has been found by fluctuations from Euler-scale hydrodynamics. From this equation one can derive an explicit expression for each cumulant $c_n$ for any value of $n$. Nevertheless, such derivation requires special manipulation for each case. It seems  possible however that the individual cumulants can be derived from the same principle without considering the generating function, see the discussion at the end of this section. In the following we present the result of \cite{10.21468/SciPostPhys.8.1.007} for $c_{2,3,4}$ and show that they possess the same combinatorial structure as the  cumulants of the corresponding conserved charges.

The authors of \cite{Doyon:2019osx} also considered a generating function with different variables. Establishing a functional equation for such function would lead to non-diagonal cumulants. It would be interesting to see if this approach is in agreement with our conjecture.
\subsection{Comparison with diagrams}
\begin{figure}[ht]
\centering
\includegraphics[width=6.8cm]{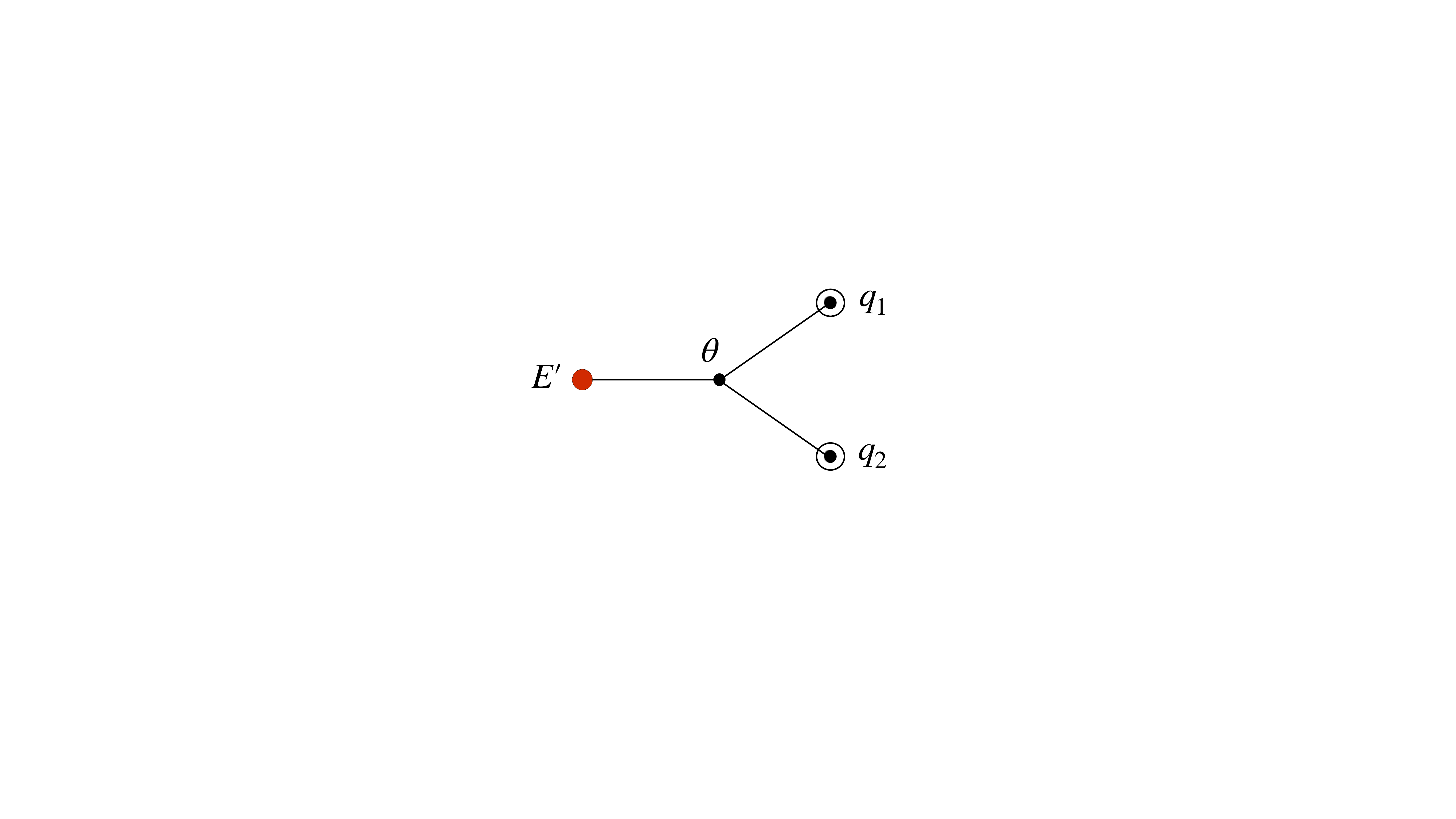}
\caption{The only tree with two leaves}
\label{second-cumulant-diagram}
\end{figure}

Let us first remind how the  cumulants of  conserved charges can be written as a sum over tree diagrams. In section \ref{charge-new-section} we shown that the $n^\text{th}$ cumulant  $\langle \mathcal{Q}_1...\mathcal{Q}_n\rangle^\text{c}$ is given by a sum over all tree-diagrams with $n+1$ external vertices : a root with momentum derivative inserted and $n$ leaves carrying the $n$ conserved charges. The internal vertices of these diagrams live in  phase space and will be integrated over. An external propagator connecting an internal vertex $\theta$ and an external vertex with an operator $\psi$ is assigned a weight $\psi^\text{dr}(\theta)$, here $\psi$ can either be the momentum derivative or the charges. An internal propagator connecting two internal vertices $\theta,\eta$ has a weight $K^\text{dr}(\theta,\eta)$, where 
\begin{align}
K^\text{dr}(u,v)=K(u,v)+\int\frac{dw}{2\pi}K(u,w)f(w)K^\text{dr}(w,v).
\end{align}  
An internal vertex $\theta$ of degrees $d$ has a weight
\begin{align}
\sum_{r\geq 1}(-1)^{r-1}r^{d-1}Y^r(\theta).
\end{align}
We summarize these rules in the following
\begin{align}
\vcenter{\hbox{\includegraphics[width=3cm]{interexter.pdf}}}&=\psi^{\text{dr}}(\theta)\nonumber\\
\vcenter{\hbox{\includegraphics[width=3cm]{interinter.pdf}}}&= K^\text{dr}(\theta,\eta)
\label{reFeynman-2}\\
\vcenter{\hbox{\includegraphics[width=2.5cm]{intervertex.pdf}}}&=\sum_{r\geq 1}(-1)^{r-1}r^{d-1}Y^r(\theta)\nonumber
\end{align}
\noindent
We now show that the result of \cite{10.21468/SciPostPhys.8.1.007} is correctly reproduced by our diagrams with the above mentioned modifications.\\
The Drude self-weight is given by \eqref{Drude-s} and can be represented as the diagram in figure \ref{second-cumulant-diagram} with energy derivative at its root and the sign of the effecive velocity at its internal vertex (of degree 3).

\begin{figure}[ht]
\centering
\includegraphics[width=13cm]{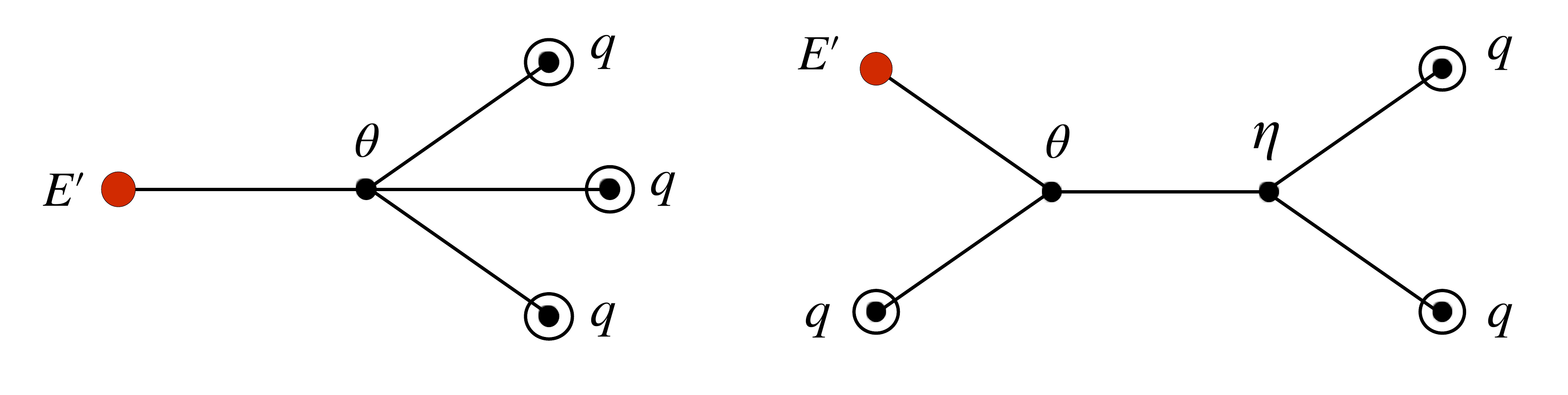}
\caption{Two trees with three leaves}
\label{three-leaves}
\end{figure}

\begin{figure}[ht]
\centering
\includegraphics[width=12cm]{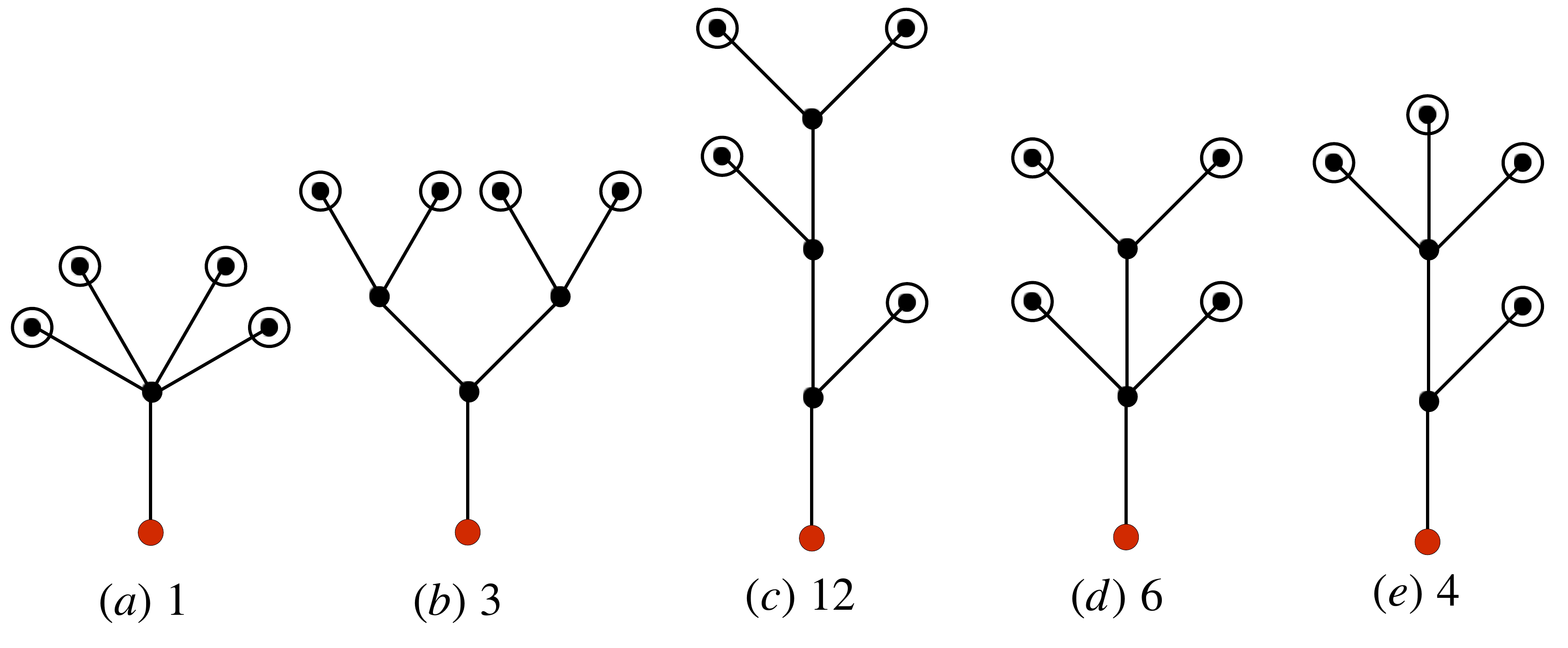}
\caption{Trees with four leaves along with their symmetry factors}
\label{4-tree}
\end{figure}

\noindent
The third cumulant was found to be
\begin{align}
c_3=\int\frac{d\theta}{2\pi} (E')^\text{dr}(\theta)f(\theta)&[1-f(\theta)]s(\theta)q^\text{dr}(\theta)\times \nonumber \\
&\times \big\lbrace [1-2f(\theta)][q^\text{dr}(\theta)]^2s(\theta)+3\big[(q^\text{dr})^2(1-f)s\big]^{*\text{dr}}(\theta)\big\rbrace,\label{third-cumulant}
\end{align}
where the star-dressing operator is defined as 
\begin{gather}
\psi^{*\text{dr}}(\theta)\equiv \psi^\text{dr}(\theta)- \psi(\theta) .\label{star-dr-def}
\end{gather}
The first term of \eqref{third-cumulant} is given by the left diagram in figure \ref{three-leaves}, again with energy derivative at the root. The internal vertex of this diagram is of degree 4 so there is no sign of the effective velocity. The second term is given by diagram on the right which comes with a symmetry factor of 3. The internal vertices are both of degrees 3 so each comes with a sign of the effective velocity. The matching is easily seen with the following writing of the star dressing operator in terms of the dressed propagator \eqref{dressed-propagator}
\begin{gather}
\psi^{*\text{dr}}(\theta)=\int\frac{d\eta}{2\pi}K^{\text{dr}}(\eta,\theta)f(\eta)\psi(\eta).\label{stardressing-equiv}
\end{gather}
This identity also reveals the physical picture behind our diagrams: the integration over internal vertices is nothing but the contribution from virtual particles that carry anomalous corrections  to the bare charges.
\noindent
The fourth cumulant is considerably more complicated and constitutes a highly non-trivial check for our conjecture. The original formula of  $c_4$ as it was derived in \cite{10.21468/SciPostPhys.8.1.007} is 
\begin{align}
&c_4=\int\frac{d\theta}{2\pi}(E')^\text{dr}(\theta)f(\theta)[1-f(\theta)]\times \bigg\lbrace \frac{Y(\theta)^2+6Y(\theta)+6}{[Y(\theta)+1]^2}s(\theta)[q^\text{dr}(\theta)]^4\nonumber\\
+&3s(\theta)\lbrace [(1-f)s(q^\text{dr})^2]^\text{dr}(\theta)\rbrace^2+12s(\theta)q^\text{dr}(\theta)\lbrace(1-f)sq^\text{dr}[(1-f)s(q^\text{dr})^2]^{\text{dr}}\rbrace^{\text{dr}}(\theta)\nonumber\\
+&6[f(\theta)-2)[q^\text{dr}(\theta)]^2[s(1-f)(q^\text{dr})^2]^\text{dr}(\theta)+4s(\theta)q^\text{dr}(\theta)[(1-f)(f-2)(q^\text{dr})^3]^\text{dr}(\theta)\bigg\rbrace.\label{c4}
\end{align}
For convenience, we show in figure \ref{4-tree} all diagrams with four leaves.

\noindent
Due to the identity \eqref{stardressing-equiv}, our trees are naturally expressed in terms of the star dressing operation. In order to compare them with \eqref{c4}, we repeatedly use the definition \eqref{star-dr-def} to make appear the dressing operation. We then show that the discrepancies cancel each other. The integration variable $\theta$ in the formula \eqref{c4} corresponds to the coordinate of the internal vertex closest to the root of each tree. These vertices are always of degree at least 3, therefore we can factorize a factor $f(\theta)[1-f(\theta)]$ from their weights. After this factorization, the contribution from the trees are (we omit the dependence on $\theta$)
\begin{itemize}
\item Tree (a)
\begin{align*}
\frac{Y^2-4Y+1}{(Y+1)^2}s(q^\text{dr})^4=\frac{Y^2+6Y+6}{(Y+1)^2}s(q^\text{dr})^4-5(1-f^2)s(q^\text{dr})^4
\end{align*}
\item Tree (b)
\begin{align*}
&3s\lbrace[(1-f)s(o^\text{dr})^2]^{*\text{dr}}\rbrace^2=3s \lbrace [(1-f)s(o^{\text{dr}})^2]^{\text{dr}}\rbrace^{2}\\
+&3s(1-f)^2(o^\text{dr})^4-6(1-f)(o^\text{dr})^2[(1-f)s(o^\text{dr})^2]^\text{dr}
\end{align*}

\item Tree (c)
\begin{align*}
&12sq^\text{dr}\lbrace(1-f)sq^\text{dr}[(1-f)s(q^\text{dr})^2]^{*\text{dr}}\rbrace^{*\text{dr}}=12sq^\text{dr}\lbrace(1-f)sq^\text{dr}[(1-f)s(q^\text{dr})^2]^{\text{dr}}\rbrace^{\text{dr}}\\
-&12(1-f)(q^\text{dr})^2[(1-f)s(q^\text{dr})^2]^\text{dr}-12sq^\text{dr}[(1-f)^2(q^\text{dr})^3]^\text{dr}+12(1-f)^2s(q^\text{dr})^4
\end{align*} 
\item Tree (d)
\begin{align*}
&6(1-2f)(q^\text{dr})^2[(1-f)s(q^\text{dr})^2]^{*\text{dr}}=6(f-2)(q^\text{dr})^2[(1-f)s(q^\text{dr})^2]^{\text{dr}}\\
-&6s(f-2)(1-f)(q^\text{dr})^4+18(1-f)(q^\text{dr})^2[(1-f)s(q^\text{dr})^2]^{\text{dr}}-18s(1-f)^2(q^\text{dr})^4
\end{align*}
\item Tree (e)
\begin{align*}
&4sq^\text{dr}[(1-f)(1-2f)(q^\text{dr})^3]^{*\text{dr}}=4sq^\text{dr}[(1-f)(f-2)(q^\text{dr})^3]^\text{dr}\\
-&4s(q^\text{dr})^4(1-f)(f-2)-12s(q^\text{dr})^4(1-f)^2+12sq^\text{dr}[(1-f)^2(q^\text{dr})^3]^\text{dr}
\end{align*}
\end{itemize}
The discrepancies indeed cancel each other.
\subsection{Comments on the conjecture}
There are two plausible ways to prove our conjecture.

First, one can try to derive the matrix elements of the product of total currents. One can then repeat the same steps of section 2 to perform their summation. The correct matrix elements must guarantee that the resulting diagrams have energy derivative at their root and sign of the effective velocity at their odd internal vertices. Concerning these two properties, the former is expected while the latter is more puzzling. Let us elaborate on this point.

In our proof of the current average, it was understood that the form factor of a current is very similar to that of the corresponding charge: both are given by trees, the only difference being the operator at the root. It is then natural that any average involving currents, if admits combinatorial structure of trees, would have the energy derivative at the roots.

As for the sign of the effective velocity, a naive guess would be to assign such sign for each bare propagator and for each external vertex. Most of them will cancel each other except for internal vertices of odd degrees. The flaw in this argument is that the weights of graph components should involve only bare quantities, like the ones in \eqref{Feynmp}. Only after the graphs are summed over do we have  renormalized (dressed) quantities, see \eqref{reFeynman}. The effective velocity is a dressed quantity and as such cannot be included in the  weight of bare propagators. In most cases however, the sign of the effective velocity coincides with that of the rapidity and the above modification could in principle be implemented.

Second, one can regard the combinatorial structure of the charge cumulants as a result of successive derivatives on the free energy \eqref{GGE-free-energy}. Simply speaking these derivatives generate branches and joints (internal vertices) of the trees. If one can prove the existence of a similar "free  energy" whose derivatives lead  to cumulants of the total transport, it is natural that the same combinatorial structure would arise. Such free energy should not be confused with the generating function \eqref{gen-function}: what we seek for is the derivative with respect to the GGE chemical potentials, not the auxiliary variable $\lambda$.

This approach seems possible in view of the following identity, proven in \cite{Doyon:2019osx}
\begin{align}
\int_0^t ds\langle J_i(0,s) \mathcal{O}(0,0)\rangle ^\text{c}=-\sum_{j}\sgn(A)_{ij}\frac{\partial}{\partial\beta_j}\langle \mathcal{O}(0,0)\rangle
\end{align}
for any local observable $\mathcal{O}$. Here $A$ is the flux Jacobian matrix, 
and the sign is defined as the sign matrix of its eigenvalues. If one can show that this identity is still valid when the local operator $\mathcal{O}$ is replaced by the product of the total currents then one would be able to obtain their cumulants from successive derivatives of the current average. 
\subsection{Summary and outlook}
Our systematic treatment not only reduces the computational complexity but also improves the  conceptual understanding of these cumulants. First, the simple combinatorial structure of the cumulants of total currents potentially translates into an analytic property of the full counting statistics. It is interesting  to find a new relation in addition to the one established in  \cite{10.21468/SciPostPhys.8.1.007}. Second, such structure provides hints about what the corresponding matrix elements would look like. For the current average, this line of idea has been exploited in recent work \cite{PhysRevX.10.011054}.  Explicit expressions of these matrix elements would have significant impact on the understanding of related quantities, for instance the Drude weight. Last but not least, the observed similarity between the two families of cumulants suggests that one could think of a  "free energy" that generates the time integrated currents in the same fashion that the usual TBA free energy generates  the conserved charges. 

In future work, we would like to see the extend of this combinatorial structure in dynamical correlation functions and related quantities. The study of large scale correlation functions in GHD has been addressed in \cite{Doyon:2017vfc}. For the charge-charge and charge-current correlation functions, the same combinatorial structure continues to hold, with the inclusion of a space-time propagator.  The situation is more subtle for the current-current correlator and the Drude weight. These quantities involve  the inverse of a dressed quantity  and it is currently not clear how such inversion could be represented in our formalism.
\chapter*{Conclusion}
In this thesis we propose a new method to compute thermodynamic
observables in integrable systems. The main idea is to use the matrix-tree theorem to express the Gaudin determinants as a sum over graphs. We have found two types of applications of this graph expansion. 

First, it can be used to directly evaluate the cluster expansion of thermodynamic quantities. In this context, the Gaudin determinant appears as the Jacobian of the change of variables from mode numbers to rapidities. This change of variables is the only approximation in our formalism and it is exact to all orders of powers in inverse volume. The new method is thus more powerful than the standard TBA, which is insensitive to all corrections of order 1 and lower. We have applied
it for a wide class of observable in theories with a diagonal S-matrix, confirming its versatility. There are however situations where it cannot be implemented. First, when a complete set of states is not known or is complicated. This happens for theories with a non-diagonal S-matrix and the standard TBA is more adapted to this situation as it only requires information about states that contribution to the thermodynamic limit. Nevertheless it is possible to interpret known TBA equations with strings in terms of diagrams. We have used this interpretation to study the boundary entropy of the corresponding theory and
although we have not obtained a complete answer we have made several important
observations. Second,  when the action of the observable on unphysical states can not be determined. These unphysical states are inserted into the cluster expansion to compensate the strict order between mode numbers. For the observables that we have considered, the action on these states is a natural generalization of the action on physical states. For more involved examples however, this task might not be straightforward or even impossible. Last but not least, there could be exotic Gaudin determinants for which a graph expansion is not known.

The second type of applications is to use the diagrammatic representation to replace the
algebraic manipulations of the Gaudin determinants. We have used this idea to derive the equations of state in GHD however we cannot make a general conclusion of when the diagrammatic representation is more useful than normal calculations.
\vspace{3mm}

There are several directions to explore with the new method
\begin{itemize}
\item Looking for situations where corrections of order $1/L$ or lower are needed. If the new method can be implemented in this case, its advantage over the standard TBA would be truly confirmed.
\item Computing finite size corrections in the hexagon form factor approach. There
might be simple setups where the form factors can be organized in a nice way and the exact
summation can be carried out. 
\item Interpreting known quantities in terms of diagrams and  finding the connection between quantities with similar diagrammatic structure.
\item Obtaining finite-particle matrix elements and form factors from  the thermodynamic
expression by applying the steps in reverse.
\item Explaining the origin of Gaudin determinants in spin chain scalar products \cite{Korepin:1982gg}. One could for instance investigate the  formation of trees in the algebra between the elements of the transfer matrix in the algebraic Bethe ansatz approach. 
\item Computing two point functions or one point function in open systems where the cluster expansion is sufficiently simple to be evaluated. 
\item Using the matrix elements of the current to derive other transport properties in GHD. These matrix elements can be obtained from the corresponding form factors and they also have a diagrammatic representation \cite{PhysRevX.10.011054}. 
\item Studying the graph expansion of other types of Gaudin determinant. It would be interesting to find an unconventional Gaudin determinant for which the graph expansion is not yet known in the mathematical literature.
\end{itemize}
\begin{appendices}
\chapter{TBA and Y system}
\section{Chiral SU(2) Gross-Neveu model}
\label{GN-TBA-appendix}
The Bethe equations for a state of $N$ physical rapidities $\theta_1,...,\theta_N$ and $M$ magnonic rapidities $u_1,...,u_M$ read
\begin{align*}
1&=e^{ip(\theta_j)L}\prod_{k\neq j}^NS_0(\theta_j-\theta_k)\prod_{m=1}^M\frac{\theta_j-u_m+i\pi/2}{\theta_j-u_m-i\pi/2}\\
1&=\prod_{j=1}^N\frac{u_k-\theta_j-i\pi/2}{u_k-\theta_j+i\pi/2}\prod_{l\neq k}^M\frac{u_k-u_l+i\pi}{u_k-u_l-i\pi}
\end{align*}
String solutions are formed of magnon rapidities equally spaced in distance of $i$. Let $u_{k,n}$ be the real center of a string of length $n$ then the ensemble of string rapidities are given by
\begin{align*}
u_{k,n}^a=u_{k,n}-i\pi\frac{n+1}{2}+i\pi a, \quad a=1,...,n
\end{align*}
The scattering phase between strings (and physical node) is the product between the scattering phases of their constituents
\begin{gather*}
S_{0n}(\theta,u_{k,n})=\frac{\theta-u_{k,n}+i\pi n/2}{\theta-u_{k,n}-i\pi n/2},\\
S_{nm}(u_{k,n},u_{l,m})=\frac{u_{k,n}-u_{l,m}+i\pi\frac{|n-m|}{2}}{u_{k,n}-u_{l,m}-i\pi\frac{|n-m|}{2}}\frac{u_{k,n}-u_{l,m}+i\pi\frac{n+m}{2}}{u_{k,n}-u_{l,m}-i\pi\frac{n+m}{2}}\prod_{s=\frac{|n-m|}{2}+1}^{\frac{n+m}{2}-1}\bigg[\frac{u_{k,n}-u_{l,m}+i\pi s}{u_{k,n}-u_{l,m}-i\pi s}\bigg]^2.
\end{gather*}
The corresponding scattering kernels are
\begin{gather}
K_{0n}(\theta-u)=-K_{n,0}(\theta-u)=-\frac{4\pi n}{4(\theta-u)^2+\pi^2n^2},\label{kernel-0-n}\\
K_{nm}(u)=K_{m,n}(u)=(1-\delta_{nm})K_{0,|n-m|}(u)+K_{0,n+m}(u)+2\sum_{s=\frac{|n-m|}{2}+1}^{\frac{n+m}{2}-1}K_{0,2s}(u).\label{kernel-n-m}
\end{gather}
Their Fourier transforms are simple
\begin{gather}
\hat{K}_{0n}(w)=-\hat{K}_{n0}(w)=-e^{-\pi n|w|/2}\label{phy-aux}\\
\hat{K}_{nm}(w)=\delta_{nm}+(e^{\pi|w|}+1)\frac{e^{-(n+m)\pi|w|/2}-e^{-|n-m|\pi|w|/2}}{e^{\pi|w|}-1}\label{aux-aux}
\end{gather}
Here we normalize the Fourier transformation as
\begin{align*}
\hat{f}(w)=\frac{1}{2\pi}\int_{-\infty}^{+\infty}f(t)e^{iwt}dt.
\end{align*}
For the physical-physical scattering
\begin{gather}
K_{00}(\theta)=\frac{1}{\pi}\sum_{l=0}^\infty-\frac{l+1}{(l+1)^2+\theta^2/4\pi^2}+\frac{l+1/2}{(l+1/2)+\theta^2/4\pi^2}\Rightarrow\hat{K}_{00}(w)=\frac{e^{-\pi|w|/2}}{2\cosh(\pi w/2)}\label{phy-phy}
\end{gather}
The above kernels control the TBA equations
\begin{align}
Y_n(u)=e^{-\delta_{0n}RE(\theta)}\exp\big[\sum_{m\geq 0}K_{m,n}\star\log(1+Y_m)(u)\big].\label{TBA}
\end{align}
By defining $\mathcal{Y}_n=Y_n^{-1}$ for $n\geq 1$ and $\mathcal{Y}_0=Y_0$ we can transform this to the Y-system  
\begin{align}
\log \mathcal{Y}_n+\delta_{n0}RE=\sum_{m=0}^\infty I_{mn}s\star\log(1+\mathcal{Y}_m).\label{Y}
\end{align}
where the kernel $s$ has the following Fourier transform
\begin{align}
\hat{s}(w)=\frac{1}{2\cosh(\pi w/2)}.\label{Fourier-s}
\end{align}
To prove that \eqref{TBA} leads to \eqref{Y} we first act by $-s$ to the TBA equation of $Y_1$
\begin{align}
\log Y_1&=K_{01}\star\log(1+Y_0)+K_{11}\star\log(1+Y_1)+\sum_{n\geq 2}K_{n1}\star\log(1+Y_n).\label{Y1}
\end{align}
With help of the following identities
\begin{gather*}
-s\star K_{01}=K_{00},\quad -s\star K_{11}=K_{1,0}-s,\quad s\star K_{n1}=K_{0n},\quad n\geq 2.
\end{gather*}
We can write \eqref{Y1} as
\begin{align*}
\log Y_0+RE=s\star\log (1+\frac{1}{Y_1})
\end{align*}
which is the first equation of Y system. Next, we act $s$ to the TBA equation of $Y_2$
\begin{align*}
\log Y_2&=K_{02}\star\log(1+Y_0)+K_{12}\star\log(1+Y_1)+K_{22}\star\log(1+Y_2)+\sum_{n\geq 3}K_{n2}\star\log(1+Y_n).
\end{align*}
this time we need the folowing identities
\begin{gather*}
s\star K_{02}=K_{01}+s,\quad s\star K_{12}=K_{11},\quad s\star K_{22}=K_{21}+s,\quad s\star K_{n2}=K_{n1},\quad n\geq 3.
\end{gather*}
From which we have
\begin{align*}
\log \mathcal{Y}_1 =s\star\log(1+Y_0)+s\star\log(1+\mathcal{Y}_2).
\end{align*}
For $n\geq 2$ we can show from the average property $s\star(K_{0,n-1}+K_{0,n+1})=K_{0n}$ that
\begin{align*}
s\star\log Y_{n+1}+s\star \log Y_{n-1}=\log Y_n+s\star\log(1+Y_{n+1})+s\star\log(1+Y_{n-1}).
\end{align*}
\section{Higher levels}
\label{higher-TBA}
\subsection{The scattering and the kernels}
The scalar factors and their exponential form \cite{HOLLOWOOD199443}
\begin{gather*}
S_0^{SU(2)}(\theta)=-\frac{\Gamma(1-\theta/2\pi i)}{\Gamma(1+\theta/2\pi i)}\frac{\Gamma(1/2+\theta/2\pi i)}{\Gamma(1/2-\theta/2\pi i)},\\
S_0^{[k]}(\theta)=\exp\bigg[\int_{-\infty}^{+\infty}\frac{dx}{x}e^{2i\theta x/\pi}\frac{\sinh[(k+1)x]\sinh x}{\sinh [(k+2)x]\sinh (2x)}\bigg].
\end{gather*}
The Bethe equations involving $N$ physical rapidities $\theta$, $M$ $SU(2)$ magnon rapidities $u$ and $P$ kink magnon rapidities $v$
\begin{gather*}
e^{-ip(\theta_j)L}=-\epsilon_j\prod_{i=1}^N S_0^{SU(2)}(\theta_j,\theta_i)S_0^{[k]}(\theta_j,\theta_i)\prod_{k=1}^M\frac{\theta_j-u_k+i\pi/2}{\theta_j-u_k-i\pi/2}\prod_{q=1}^P\frac{\sinh \dfrac{\theta_j-v_q+i\pi/2}{k+2}}{\sinh\dfrac{\theta_j- v_q-i\pi/2}{k+2}},\\
\prod_{j=1}^N\frac{u_k-\theta_j+i\pi/2}{u_k-\theta_j-i\pi/2}=\Omega_k\prod_{l=1}^M\frac{u_k-u_l+i\pi}{u_k-u_l-i\pi},\\
\prod_{j=1}^N\frac{\sinh\dfrac{v_q-\theta_j+i\pi/2}{k+2}}{\sinh\dfrac{v_q-\theta_j-i\pi/2}{k+2}}=\Omega_q\prod_{p=1}^P\frac{\sinh\dfrac{v_q-v_p+i\pi}{k+2}}{\sinh\dfrac{v_q-v_p-i\pi}{k+2}}.
\end{gather*}
with some constants $\epsilon_j,\Omega_k,\Omega_q$.
String solutions
\begin{gather*}
\text{ u strings of length } n=\overline{1,\infty}: \quad u_{k,n}^a=u_{k,n}-i\pi\frac{n+1}{2}+i\pi a, \quad a=1,...,n\\
\text{ v strings of length } m=\overline{1,k}: \quad v_{q,m}^b=v_{q,m}-i\pi\frac{m+1}{2}+i\pi b, \quad b=1,...,m
\end{gather*}
The scatterings between $SU(2)$ strings with themselves and between them and the physical rapidity are the same as before.
For kink magnon strings
\begin{gather*}
S_{0n}^{[k]}(\theta,v_{q,n})=\frac{\lbrace\theta-v_{q,m}+i\pi m/2\rbrace_k}{\lbrace\theta-v_{q,m}-i\pi m/2\rbrace_k}\\
S_{nm}^{[k]}(v_{q,n},v_{p,m})=\frac{\lbrace v_{q,n}-v_{p,m}+i\pi\frac{|n-m|}{2}\rbrace_k}{\lbrace v_{q,n}-v_{p,m}-i\pi\frac{|n-m|}{2}\rbrace_k}\frac{\lbrace v_{q,n}-v_{p,m}+i\pi\frac{n+m}{2}\rbrace_k}{\lbrace v_{q,n}-v_{p,m}-i\pi\frac{n+m}{2}\rbrace_k}\prod_{s=\frac{|n-m|}{2}+1}^{\frac{n+m}{2}-1}\bigg[\frac{\lbrace v_{q,n}-v_{p,m}+i\pi s\rbrace_{k}}{\lbrace v_{q,n}-v_{p,m}-i\pi s\rbrace_{k}}\bigg]^2
\end{gather*}
where we have noted for convenience
\begin{align*}
\lbrace u\rbrace_k  =\sinh\frac{u}{k+2}.
\end{align*}
The Fourier transforms of the kink magnon strings scattering kernel
\begin{gather*}
\hat{K}_{0n}^{[k]}(w)=-\frac{\sinh[(k+2-n)\frac{\pi w}{2}]}{\sinh(k+2)\frac{\pi w}{2}},\\
\hat{K}_{nm}^{[k]}(w)=\delta_{nm}-2\frac{\sinh\big[ \min(n,m)\frac{\pi w}{2}\big]\sinh\big[ (k+2-\max(n,m))\frac{\pi w}{2}\big]\cosh\frac{\pi w}{2}}{\sinh\big[(k+2)\frac{\pi w}{2}\big]\sinh\frac{\pi w}{2}}.
\end{gather*}
\subsection{Maximal string reduction and reduced TBA}
\label{higher-reduced}
At this point we have the raw TBA equations
\begin{gather*}
\log Y_{\tilde{n}}=\sum_{m=0}^k K_{mn}^{[k]}\star\log(1+Y_{\tilde{m}}),\quad n=\overline{1,k},\\
\log Y_0+RE=\sum_{n=0}^kK_{n0}^{[k]}\log(1+Y_{\tilde{n}})+\sum_{n=1}^\infty K_{n0}^{\text{SU(2)}}\log(1+Y_n),\\
\log(1+Y_n)=\sum_{m=0}^\infty K_{mn}\star\log(1+Y_m),\quad n=\overline{1,\infty}.
\end{gather*}
where we have used the tilde indices to denote kink rapidities, also $\tilde{0}=0$.
\begin{figure}[h]
\centering
\includegraphics[width=14cm]{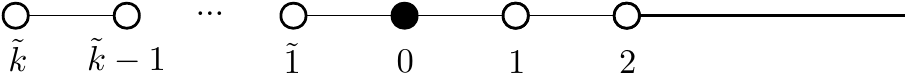}
\end{figure}

Maximal string reduction \cite{HOLLOWOOD199443}: $u$ string of length $k$ doesn't contribute in the thermodynamic limit. The $\tilde{k}$ string is frozen in the sense that $Y_{\tilde{k}}=\infty$. We look at the TBA equation for this string
\begin{align*}
\log Y_{\tilde{k}}=\sum_{m=0}^kK_{mk}^{[k]}\star\log(1+Y_{\tilde{m}}).
\end{align*}
Upon replacing $\log Y_{\tilde{k}}$ by $\log(1+Y_{\tilde{k}})$, we can effectively remove the $\tilde{k}$ node from our TBA system
\begin{align*}
\log(1+Y_{\tilde{k}})=\sum_{m=0}^{k-1}(1-K_{kk}^{[k]})^{-1}\star K_{mk}^{[k]}\star\log(1+Y_{\tilde{m}}).
\end{align*}
The reduced system (only the kink magnon related part) read
\begin{align*}
\log Y_{\tilde{n}}&=\sum_{m=0}^{k-1}\bigg[K_{kn}^{[k]}\star(1-K_{kk}^{[k]})^{-1}\star K_{mk}^{[k]}+K_{mn}^{[k]}\bigg]\star\log(1+Y_{\tilde{m}}),\quad n=\overline{1,k-1}\\
\log Y_0+RE &=\sum_{n=0}^{k-1}\bigg[K_{k0}^{[k]}\star(1-K_{kk}^{[k]})^{-1}\star K_{nk}^{[k]}+K_{n,0}^{[k]}\bigg]\star\log(1+Y_{\tilde{n}})+.....
\end{align*}
The following identity drastically simplifies this system
\begin{align}
K_{kn}^{[k]}\star(1-K_{kk}^{[k]})^{-1}\star K_{mk}^{[k]}+K_{mn}^{[k]}=K_{mn}^{[k-2]},\quad m,n=\overline{0,k-1}.\label{reduction-identity}
\end{align}
To summarize, the reduced TBA system for integrable perturbed $SU(2)_k$ is 
\begin{align*}
\log Y_n+\delta_{n,k}RE=\sum_{m,n} K_{mn}\star\log(1+Y_m),\quad n=\overline{1,\infty}
\end{align*}
where
\begin{gather*}
\hat{K}_{kn}(w)=-\hat{K}_{nk}(w)=-\frac{\sinh n\frac{\pi w}{2}}{\sinh k\frac{\pi w}{2}},\quad n=\overline{1,k-1}\\
\hat{K}_{nm}(w)=\delta_{nm}-2\frac{\sinh\big[ \min(n,m)\frac{\pi w}{2}\big]\sinh\big[ (k-\max(n,m))\frac{\pi w}{2}\big]\cosh\frac{\pi w}{2}}{\sinh\big[k\frac{\pi w}{2}\big]\sinh\frac{\pi w}{2}},\quad n,m=\overline{1,k-1}\\
\hat{K}_{kk}(w)=\frac{\sinh\frac{\pi w}{2}}{\sinh \pi w}\bigg(1+\frac{\sinh[(k-1)\frac{\pi w}{2}]}{\sinh[k\frac{\pi w}{2}]}\bigg)\\
\hat{K}_{kn}=-\hat{K}_{n,k}=-e^{ -(n-k)\pi|w|/2},\quad n=\overline{k+1,\infty}\\
\hat{K}_{nm}(w)=\delta_{nm}+(e^{\pi|w|}+1)\frac{e^{-(n+m-2k)\pi|w|/2}-e^{|n-m|\pi|w|/2}}{e^{\pi|w|}-1},\quad n,m=\overline{k+1,\infty}
\end{gather*}
\begin{figure}[h]
\centering
\includegraphics[width=14cm]{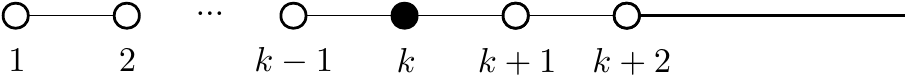}
\caption{Maximum string removed and indices rearranged}
\end{figure}
\subsection{Y system}
To transform this into the Y system, we notice that the universal kernel $s$ still satisfies the average property for the newly introduced hyperbolic kernels
\begin{align*}
\hat{s}(\hat{K}_{k,n-1}+\hat{K}_{k,n+1})=\hat{K}_{kn},\quad n=\overline{1,k-1}.
\end{align*}
As a result, deep in the left or right wing, there would be no problem. We just need to check for the three nodes $k-1,k,k+1$. We act with $-s$ on the TBA equations of $k\pm 1$
\begin{align*}
\log Y_{k-1}&=\sum_{n=1}^{k-2}K_{n,k-1}\star\log (1+Y_n)+K_{k-1,k-1}\star\log(1+Y_{k-1})+K_{k,k-1}\star\log (1+Y_k),\\
\log Y_{k+1}&=\sum_{n=k+2}^\infty K_{n,k+1}\star\log(1+Y_n)+K_{k+1,k+1}\star\log(1+Y_{k+1})+K_{k,k+1}\star\log(1+Y_k).
\end{align*}
We need the following identities
\begin{gather*}
-s\star K_{n,k-1}=K_{nk},\quad n\in\overline{1,k-2}, \quad -s\star K_{n,k+1}=K_{nk},\quad n\in \overline{k+2,\infty},\\
-s\star K_{k-1,k-1}=K_{k-1,k}-s,\quad -s\star K_{k+1,k+1}=K_{k+1,k}-s,\quad -s\star K_{k,k-1}-s\star K_{k,k+1}=K_{kk}.
\end{gather*}
Then it follows that
\begin{gather*}
-s\star\log Y_{k-1}-s\star\log Y_{k+1}=\log Y_k-s\star\log(1+Y_{k-1})-s\star\log(1+Y_{k+1}),\\
\Leftrightarrow \log Y_k=s\star\log (1+\mathcal{Y}_{k-1})+s\star(1+\mathcal{Y}_{k+1}).
\end{gather*}
The right wing is coupled to the physical node in the same way as Gross-Neveu. For the left wing, we act with s to the TBA equation of $Y_{k-2}$
\begin{align*}
\log Y_{k-2}=& K_{k,k-2}\star\log(1+Y_k)+K_{k-1,k-2}\star\log(1+Y_{k-1})+K_{k-2,k-2}\star\log(1+Y_{k-2})\\
+&\sum_{n=1}^{k-3}K_{n,k-2}\star\log(1+Y_n).
\end{align*}
With help of the following identities
\begin{gather*}
s\star K_{n,k-2}=K_{n,k-1},\quad n=\overline{1,k-3},\quad s\star K_{k,k-2}=s+K_{k,k-1},\\
s\star K_{k-1,k-2}=K_{k-1,k-1},\quad s\star K_{k-2,k-2}=s+K_{k-2,k-1}.
\end{gather*}
We obtain
\begin{gather*}
s\star\log Y_{k-2}=s\star\log(1+Y_k)+s\star\log(1+Y_{k-2})+\log Y_{k-1}\\
\Leftrightarrow \log \mathcal{Y}_{k-1}=s\star\log (1+Y_k)+s\star\log(1+\mathcal{Y}_{k-2}).
\end{gather*}
\chapter{Determinants}
We compute the determinants that appear in the main text. We have found these results by Mathematica. For simplicity we introduce the following notation $K_{ab}\equiv \int_{-\infty}^{+\infty}d\theta K_{ab}(\theta)/(2\pi)$.
\subsection{The IR determinant}
\label{det-IR}
We compute $\det(1-\hat{K})$ where 
\begin{align*}
\hat{K}_{ab}= K_{ab}\sqrt{\frac{Y_a^{\text{IR}}(\K)}{1+Y_a^{\text{IR}}(\K)}\frac{Y_b^{\text{IR}}(\K)}{1+Y_b^{\text{IR}}(\K)}}=[\delta_{ab}-2\min(a,b)]\frac{(1-\kappa)^2\sqrt{\kappa^a\kappa^b}}{(1-\kappa^{a+1})(1-\kappa^{b+1})},\quad  a,b\geq 1.
\end{align*}
This matrix can be implemented directly in Mathematica and we get five digit precision for twist parameters smaller than 1/2 using the first 30 magnon strings.\\
\begin{table}[!htb]
\centering
\captionsetup{justification=centering}
  \begin{tabular}{  |c | c|c|c|c| }
    \hline
    $\K=0.5$ & $\K=0.6$& $\K=0.7$ & $\K=0.8$ & $\K=0.9$ \\ \hline
   0.5 & 0.400001 & 0.30008 & 0.202126 & 0.121998 \\\hline
  \end{tabular}
  \caption{Approximation of $[\det(1-\hat{K})]^{-1}$ for some values of the twist parameter}
\end{table}

As the twist parameter tends to 1, more strings are needed to keep the precision. We can read from this numerical data that
\begin{align}
\det(1-\hat{K})=(1-\K)^{-1}.\label{IR-det}
\end{align}
This gives the loop part of IR g-function \eqref{g-GN-IR-loop}. There is a more elegant way to obtain this result. We remark that the matrix $\hat{K}$ can be written in a slightly different way without changing the determinant of $1-\hat{K}$
\begin{align*}
\hat{K}_{ab}=[\delta_{ab}-2\min(a,b)]\frac{Y_b^{\text{IR}}(\K)}{1+Y_b^{\text{IR}}(\K)},\quad  a,b\geq 1.
\end{align*}
By factorizing the second factor we can show that \eqref{IR-det} is equivalent to
\begin{align}
\frac{\det[2Y^\text{IR}(\K)+\text{Cartan}^\text{A}_{\infty}]}{\det(\text{Cartan}^\text{A}_\infty)}=\frac{1+\K}{1-\K}.
\end{align}
From the usual method of computing the determinant of Cartan matrix of A type, we can reformulate the problem as follows. Let $G_a$ be a sequence of numbers defined by the iterative relation
    \begin{align*}
    G_{a+1}+G_{a-1}=\big[2+2Y_a^{\text{IR}}(\K)\big]G_a,\quad G_0=0,G_1=1,
    \end{align*}
then
    \begin{align}
    \lim_{a\to\infty}\frac{G_a}{a+1}=\frac{1+\K}{1-\K}.\label{Janik}
    \end{align}
We owe this derivation to Romuald Janik. Unfortunately we can only verify numerically the asymptotic \eqref{Janik}.
\subsection{The UV determinant}
\label{det-UV}
We compute $\det(1-\hat{K})$ where 
\begin{align*}
\hat{K}_{ab}=K_{ab}\sqrt{\frac{Y_a^\text{UV}(\K) Y_b^\text{UV}(\K)}{[1+Y_a^\text{UV}(\K)][1+Y_b^\text{UV}(\K)]}},
\end{align*}
with
\begin{gather*}
K_{ab}=\delta_{ab}-2\frac{\min(a,b)[k-\max(a,b)]}{k},\; a,b\in \overline{1,k-1},\quad K_{ab}=\delta_{ab}-2\min(a-k,b-k),\; a,b\geq k+1\\
K_{ka}=-K_{ak}=-\frac{a}{k},\quad 1\leq a<k,\quad K_{kk}=1-\frac{1}{2k},\quad K_{ka}=-K_{ak}=-1,\quad a\geq k+1
\end{gather*}
and 
\begin{gather*}
\frac{Y_a^\text{UV}(\K)}{1+Y_a^\text{UV}(\K)}=\K^{a}\frac{(1-\K)^2}{(1-\K^{a+1})^2}\quad a\in\overline{1,k-1}\cup\overline{k+1,\infty},\quad \frac{Y_k^\text{UV}(\K)}{1+Y_k^\text{UV}(\K)}=\frac{(1-\K^{k})(1-\K^{k+2})}{(1-\K^{k+1})^2}.
\end{gather*}
Again we choose a cut-off on $SU(2)$ magnon string length of 30. This gives five digit precision for values of the twist parameter smaller than $1/2$.\\
\begin{table}[!htb]
\centering
\captionsetup{justification=centering}
  \begin{tabular}{ |c | c | c|c|c|c| }
    \hline
     & $\K=0.5$ & $\K=0.6$& $\K=0.7$ & $\K=0.8$ & $\K=0.9$ \\ \hline
    $k=2$ & 2 & 1.6 & 1.20032 & 0.80851 & 0.487993 \\\hline
    $k=3$ & 3 & 2.40001  & 1.80048 & 1.21276& 0.731989 \\\hline
    $k=4$ & 4 & 3.20001 & 2.40064 & 1.61701& 0.975986 \\\hline
    $k=5$ & 5 & 4.00001 & 3.0008 & 2.02126& 1.21998 \\\hline
    $k=6$ & 6 & 4.80001 & 3.60096 & 2.42552&  1.46398\\\hline
    $k=7$ & 7 & 5.60002 & 4.20112 & 2.82977& 1.70797 \\ \hline
    $k=8$ & 8 & 6.40002 & 4.80128 & 3.23402& 1.95197 \\\hline
    $k=9$ & 9 & 7.20002 & 5.40144 & 3.63828& 2.19597 \\\hline
  \end{tabular}
  \caption{Approximation of $[\det(1-\hat{K})]^{-1}$ for some  twist parameters and levels}
\end{table}

We predict from this data that $\det(1-\hat{K})=[2k(1-\K)]^{-1}$. This leads to the loop part of UV g-function \eqref{loop-g-level-n}.
\chapter{Normal modes of hydrodynamics}
\label{GHD-appendix}
In this appendix we derive the normal modes of Euler hydrodynamic equations. For the ease of following we repeat here the expressions we found for the hydrodynamic matrices 
\begin{align}
\sum_k A_i^j q_j^\text{dr}(\theta)&=v^\text{eff}(\theta)q_i^\text{dr}(\theta),\label{A-appendix}\\
B_{ij}&=\int d\theta v^\text{eff}(\theta)\rho_\text{p}(\theta)[1-f(\theta)]q_i^\text{dr}(\theta)q_j^\text{dr}(\theta),\label{B-appendix}\\
C_{ij}&=\int d\theta\rho_\text{p}(\theta)[1-f(\theta)]q_i^\text{dr}(\theta)q_j^\text{dr}(\theta).\label{C-appendix}
\end{align}
The computation of the normal modes is simpler if we work with integral operators representing these matrices. These operators act on the space of functions of rapidity and are defined as follows
\begin{align}
\sum_j A_i^j q_j(\theta)=(\mathcal{A}^tq_i)(\theta),\quad B_{ij}=q_i.\mathcal{B}q_j,\quad C_{ij}=q_i.\mathcal{C}q_j.\label{operators-A-B-C}
\end{align}
The dressing operation \eqref{dressing-def-unique} can also be expressed via an integral operator
\begin{align}
\psi^\text{dr}&=(1-\mt f )^{-1}\psi \quad \text{where}\quad  (\mt\psi)(\theta)\equiv \int\frac{d\eta}{2\pi } K(\theta-\eta)\psi(\eta).
\end{align}
The transpose in the definition of $\A$ allows the relation $B = AC$ among hydrodynamic matrices to transcend into the same relation among the corresponding integral operators
\begin{align*}
 q_i.\B q_j\equiv B_{ij}=A_i^kC_{kj}\equiv {\A}^t q_i.\C q_j=q_i.\A \C q_j.
\end{align*} 
Applying the dressing operation on the definition of $\A$ we obtain $(\A^tq_i)^\text{dr}(\theta)=v^\text{eff}(\theta)q_i^\text{dr}(\theta)$, which means 
\begin{align}
\A=(1-f\mt)^{-1}v^\text{eff}(1-f\mt)\label{integral-rep-A}
\end{align}
Other operators can be directly read off the expressions  \eqref{B-appendix} and \eqref{C-appendix}
\begin{align}
\B=(1-f\mt)^{-1}v^\text{eff}\rho_\text{p} (1-f)(1-\mt f)^{-1},\quad \C=(1-f\mt)^{-1}\rho_\text{p} (1-f)(1-\mt f)^{-1}.
\end{align}
To remind, the normal modes are defined in such a way that their Jacobian matrix with respect to average charge densities is given by the matrix that diagonalizes $A$
\begin{align*}
\p n_i/\p \Q_j= R_i^j\quad RAR^{-1}=v^\text{eff}.
\end{align*}
It follows from the integral representation \eqref{integral-rep-A} that the the integral operator $\mathcal{R}$ corresponding to the matrix $R$ is given by $ \mathcal{R}=1-f\mt$. The equation  that defines the normal modes, can be written as $-\p n_i /\p \beta^j=R_i^kC_{kj}$. In terms of integral operators, the right-hand side is 
\begin{align*}
R_i^kC_{kj}=\int d\theta q_i(\theta) RCq_j(\theta)=\int d\theta q_i(\theta)\rho_\text{p}(\theta)f(\theta)q_i^\text{dr}(\theta)=q_i.\rho_\text{p}(1-f)q_j^\text{dr}.
\end{align*}
We can take $n_i=q_i.n$ for some normal-mode function $n(\theta)$ that satisfies
\begin{align}
\p n/\p \beta^j=-\rho_\text{p}(1-f)q_j^\text{dr}.
\end{align}
We conclude from this equation that $n$ is given by the TBA filling factor $f$. $\square$

\end{appendices}

\end{document}